\documentclass[prb,showpacs,showkeys,preprintnumbers,superscriptaddress,a4paper,twocolumn]{revtex4-1}
\usepackage{graphicx}
\usepackage{subcaption}
\usepackage{float}
\usepackage{esvect}
\usepackage{bbold}
\usepackage[T1]{fontenc}
\usepackage[utf8]{inputenc}
\usepackage{graphicx}
\usepackage{amsmath,amssymb}
\usepackage{amstext}	
\usepackage{oldgerm}	
\usepackage[colorlinks,citecolor=blue,filecolor=blue,linkcolor=blue,urlcolor=blue]{hyperref}

\def\beq{\begin{equation}}
	\def\eeq{\end{equation}}
\def\bea{\begin{eqnarray}}
	\def\eea{\end{eqnarray}}
\def\ba{\begin{array}}
	\def\ea{\end{array}}

\begin{document}
\title {Study of Correlated Disorders and interaction in the Hofstadter Butterfly}
\author{Pooja Saini}
\affiliation{Institute of Physics, Bhubaneswar 751005, Odisha, India}

\author{Saptarshi Mandal}
\email{saptarshi@iopb.res.in}
\affiliation{Institute of Physics, Bhubaneswar 751005, Odisha, India}
\affiliation{Homi Bhabha National Institute, Training School Complex, Anushaktinagar, Mumbai 400094, India}

\author{Sanjay Gupta}
\email{guptajay2@gmail.com}
\affiliation{Department of Physics, Central University of Jharkhand, Cheri-Manatu, Ranchi-835222, India}

	\begin{abstract}
	
		
We investigate the impact of several quasiperiodic disorders and their continuous interpolation with the Aubry–André (AA) potential on the Hofstadter butterfly  using  mean field approximation at zero temperature for a two-dimensional square lattice. Weak disorder mildly smears the fractal spectrum, while strong quasiperiodic potentials destroy the butterfly and generate multiple energy gaps. The AA potential produces the strongest spectral restructuring, creating prominent gaps near half-filling. Interpolating AA with other quasiperiodic potentials reveals competing gap-opening mechanisms, ranging from AA-dominated gaps at small interpolation parameters to a robust half-filling gap generated by the competing disorders at large parameters. Entanglement entropy follows the area law at low and high magnetic fields but shows pronounced deviations at intermediate fields, with opposite trends for strong AA versus other quasiperiodic potentials. Localization analysis using IPR and NPR confirms enhanced localization with increasing disorder; the AA potential yields the largest IPR, with notable field dependence. Interpolation produces smooth crossovers between distinct localization regimes.
	\end{abstract}
	
	\maketitle
	\section{Introduction} The Hofstadter butterfly is a hallmark of condensed matter physics, describing the fractal energy spectrum of Bloch electrons in a two-dimensional lattice under a perpendicular magnetic field. Despite its apparent simplicity, several aspects remain unresolved due to limited experimental and exact theoretical results. The system can be viewed in two limits: in the weak-lattice regime, it reduces to free-electron Landau levels, while in the strong-lattice regime, a tight-binding model with Peierls substitution captures the physics, giving rise to the Hofstadter butterfly \cite{Hofs}. The spectrum depends critically on the magnetic flux ratio $\phi/\phi_0=p/q$, splitting each band into $q$ magnetic subbands for rational flux and forming a Cantor-set-like structure for irrational flux. Experimental studies have mainly focused on the spectral and transport properties, while direct access to the wavefunction structure remains challenging. Experimental evidence for the fractal nature of the Hofstadter spectrum has been observed across a variety of physical systems, including ultracold atom in optical lattices \cite{Jak, Aid, Miya}, graphene-based moiré superlattices \cite{Pono, Dean, Hunt, Yu, Yu1}, and GaAs/AlGaAs heterostructures with superlattices \cite{DP, Sch, Alb, Gei}, photons with the superconducting qubits \cite{Rou}, and a one-dimensional (1D) acoustic array \cite{Kuhl, Rich}. 
	
    The study of electronic behavior in disordered and quasi-periodic materials has been a central topic in condensed matter physics for many years. In perfectly periodic crystals, where atoms are arranged in a regular and repeating manner, Bloch’s theorem states that an electron’s wave function can be written as a plane wave modulated by a periodic function matching the lattice periodicity \cite{Bloch}. This indicates that the electron’s probability density is uniformly distributed across the lattice. In contrast, when disorder or quasi-periodicity is introduced, this periodic description breaks down, and electron wave functions can become spatially localized within the material. The influence of disorder on quantum Hall systems has been extensively investigated $\cite{Chal, Cain, Gal, Kram}$. In the continuum framework, this problem can be approached from two complementary perspectives: one may examine how the conductance evolves with increasing disorder strength at a fixed magnetic field, or, conversely, how it changes as the magnetic field is reduced while keeping the disorder constant. The evolution of Landau levels in a two-dimensional electron gas, as well as in lattice systems under a weakening magnetic field, has long been a subject of considerable discussion \cite{Huck, Ortu}. In real materials, however, the ideal Hofstadter spectrum is rarely observed because disorder \cite{Zhou, Kosh} and electron–electron (e–e) interactions \cite{Cza} are inevitably present. Understanding how disorder influences electronic conduction is a central topic in condensed matter physics, as it can both reveal and obscure novel quantum phenomena \cite{Chal, Cain, Gal}. Disorder may manifest in various forms, with Anderson disorder \cite{Ander} being one of the most prominent examples, where random on-site energies are assigned to each lattice site. In disordered systems, the presence of an external magnetic field generally weakens localization effects. For example, in weakly localized two-dimensional (2D) systems, applying a magnetic field leads to negative magnetoresistance, as the magnetic phase shift disrupts the constructive interference of electronic wave functions \cite{Ber}. Similarly, in the 2D Anderson model, a weak magnetic field tends to diminish localization for electronic states located near the centers of Landau bands \cite{And}. A different type of material, which is not  complete randomly disordered but also does not have perfect lattice periodicity, is called a quasi-periodic crystal or correlated disorder. Moreover, correlated disorders such as Rudin–Shapiro (RS) \cite{Phi}, Aubry Andre (AA) \cite{Aubry, Soko} and Fibonacci \cite{Kohm} typically blur the fine fractal features of the Hofstadter butterfly by introducing states within the energy gaps, and at sufficiently strong disorder, the characteristic pattern can disappear completely \cite{Zhou}.\\
	In general, electron–electron correlation restricts the motion of electrons within a lattice, often resulting in the opening of an energy gap, such systems are referred to as correlated insulators. When an external magnetic field is introduced, the behavior of these correlated electron systems becomes more intricate, making it challenging to predict their properties. Electron correlations can drive a transition from a metallic to an insulating phase, and this transition can also be influenced by an applied magnetic field \cite{Kaga, Mats, Mats1}. A well-known manifestation of this effect is the colossal magnetoresistance observed in manganites, where the resistivity changes by several orders of magnitude under an external magnetic field \cite{Ram, Dag, Sal}. Electron–electron correlations play a important role in shaping the Hofstadter spectrum \cite{Gud, Doh}. These interactions can smear the fine structure or open energy gaps in the spectrum, while typically preserving its overall self-similar, Cantor-set-like nature. Studies based on the single-band Hubbard model have shown that electron correlations mainly modify the band gaps and bandwidths without destroying the fundamental fractal characteristics. However, strong interactions may induce charge ordering and break translational symmetry, leading to the emergence of novel correlated phases such as incompressible nematic and ferrielectric states, which can significantly alter the fractal pattern. The effects of interactions on the Hofstadter butterfly have been extensively studied using various theoretical approaches, including mean-field approximation for both square lattices \cite{Gud, Doh} and honeycomb lattices hosting Dirac fermions \cite{Apa, Chak}, dynamical mean field theory (DMFT) $\cite{Tran}$ as well as exact diagonalization methods applied to the square lattice in a magnetic field $\cite{Cza}$.\\ 
    When both disorder and interaction are present in the Hofstadter system, the distorted butterfly pattern reappears, indicating the interplay between these two effects $\cite{Mandal1}$. In this article, we investigate the stability of the Hofstadter butterfly in the simultaneous presence of on-site Hubbard interaction and four types of corelated binary disorders: RS, Thue–Morse (TM) \cite{Thue}, Aubry Andre (AA) \cite{Aubry, Soko} and Fibonacci \cite{Kohm}.  The Fibonacci and TM sequences are canonical examples of quasiperiodic order, while the RS sequence more closely emulates random disorder in its spectral and transport properties. The AA model, in particular, offers a tunable localization–delocalization transition in one dimension, and its two-dimensional extensions in magnetic fields have revealed a variety of unconventional phenomena. We impose each disorder sequence on the on-site potential of a square lattice and incorporate electron–electron interactions through a single-orbital Hubbard term with on-site Coulomb repulsion $U$. Our primary goal is to determine how distinct disorder types, disorder strength and interpolation between the disorders reshape the Hofstadter spectrum, and under what conditions the butterfly structure can be restored or stabilized. Recently, entanglement has emerged as a powerful tool for understanding many-body systems, as it can effectively capture several fundamental properties such as phase transitions $\cite{Ami}$, topological order $\cite{Li}$, and edge states $\cite{Mandal}$. Earlier investigations of entanglement in the Hofstadter problem were primarily focused on bilayer configurations $\cite{Sch, Hua}$ or cylindrical geometries $\cite{Mora}$, where entanglement in momentum space was analyzed by tracing out one physical layer or one spatial direction. In such cases, analytical expressions for the momentum-space entanglement entropy could be derived. In contrast, our work focuses on real-space entanglement in a physical lattice, where a specific spatial region is integrated out to study how entanglement evolves with the applied magnetic field. In particular, we examine how the area law of entanglement entropy is modified. We further analyze the entanglement spectrum for the largest subsystem—corresponding to half of the total system size—and uncover characteristic patterns reminiscent of the Hofstadter butterfly. The combined effects of electron–electron interaction and disorders, interpolation between disorders on both entanglement entropy and entanglement spectrum are explored in detail. To study the localization behavior of the Hofstadter butterfly, we use the inverse participation ratio (IPR) \cite{Bell, Thou, weg}. The IPR ranges from zero to one, where higher values indicate stronger localization of the wave functions.  \\ 
	\section{Model and Method}
	\label{model}
	We introduce and discuss the model Hamiltonian and the framework for incorporating electronic interactions and various scheme of disorders. We consider a single-band Hubbard model on a square lattice within the tight-binding framework.
	
	To achieve a uniform magnetic field along the $z$-direction, we employ the Landau gauge $(0,Bx,0)$, where the vector potential components are $A_x=A_z=0$ and $A_y=Bx$. In this configuration, for the square lattice, the magnetic phase appears exclusively in the hopping terms along the $y$-direction. The Hamiltonian for the square lattice can be expressed as follows:
	\begin{widetext}
	\begin{eqnarray} 
		\label{main-h}
		\mathcal{H}&=&-\sum_{i, \alpha, \sigma}t_{i,\alpha}c^\dagger_{i\sigma}c^{}_{i+\delta_\alpha,\sigma}+h.c) -\sum_{i\sigma}(\epsilon_{i}-\mu)c^\dagger_{i\sigma}c^{}_{i\sigma} 
		+\sum_{i}U n_{i\uparrow} n_{i\downarrow} .
	\end{eqnarray}
    \end{widetext}
In our formulation, the index $i$ labels the sites of a two-dimensional square lattice. 
The operators $c^{\dagger}_{i\sigma}$ and $c_{i\sigma}$ create and annihilate, respectively, 
an electron with spin $\sigma$ at site $i$. We restrict the hopping amplitude $t$ to 
nearest--neighbor bonds. A convenient gauge choice is adopted such that the hopping along 
the $x$ direction remains unchanged, $t_{i,x}=t$, while the hopping along the $y$ direction 
acquires a Peierls phase from the external magnetic field, $t_{i,y}=t\, e^{-i\phi_i}$, with $\phi_i = \int \vec{A}\cdot d\vec{l} = \int A_y\,dy = B x_i$. Here $x_i$ is the $x$-coordinate of the site $i$. This choice ensures that the flux through each square plaquette is periodic under $B \rightarrow B + 2\pi$, giving the Hofstadter spectrum its 
$2\pi$ periodicity. Throughout, the magnetic field is measured in units of the flux quantum. Here, $\delta_{\alpha}$ denotes the nearest--neighbor displacement in the 
$\alpha = x, y$ direction. The last term in the Hamiltonian represents the on-site 
Hubbard interaction.The chemical potential $\mu$ is determined by fixing the total particle number to $N/2$. Operationally, this is achieved by averaging the energies of the $N/2$-th and $(N/2+1)$-th single-particle levels. Throughout this work, we set the hopping amplitude to $t=1$ for all bonds of the lattice. No assumption regarding magnetic ordering is imposed, and hence we retain the full spin structure in all subsequent expressions.\\
   The onsite Hubbard term in Eq. \ref{main-h} is decoupled using the UHF approximation, wherein correlations between spin-up and spin-down density fluctuations are ignored. This allows the interaction to be expressed in a simplified mean-field form.
	\begin{eqnarray}
		&&U(n_{i\uparrow}- \langle n_{\uparrow} \rangle)(n_{i\downarrow}- \langle n_{i\downarrow} \rangle)=0, or \nonumber \\ 
		&&Un_{i\uparrow}n_{i\downarrow} = U(n_{i\uparrow} \langle n_{\downarrow} \rangle + \langle n_{i\uparrow} \rangle n_{i\downarrow} - \langle n_{i\uparrow} \rangle \langle n_{i\downarrow}\rangle)
	\end{eqnarray}
	We will drop the last term because it does not contribute to the equation
	of motion. Once this approximation is
	implemented for the Hubbard interaction term, the on-site potential is modified as:
	\begin{eqnarray}
		&&\epsilon_{i\uparrow}^{'}=\epsilon_{i\uparrow}+U \langle n_{i\downarrow} \rangle,~\epsilon_{i\downarrow}^{'}=\epsilon_{i\downarrow}+U \langle n_{i\uparrow} \rangle 
	\end{eqnarray}
	And the corresponding decoupled Hamiltonian can be written as sum of 
	up and down components as follows.
	\begin{widetext}	
	\begin{eqnarray} 
		\label{MFH}
		H_{mf}&&=\sum_{i}\epsilon_{i\uparrow}^{'}n_{i\uparrow}-
		\sum_{i, \alpha}t_{i,\alpha}c^\dagger_{i\uparrow}c^{}_{i+\delta_\alpha,\uparrow}
		+\sum_{i}\epsilon_{i\downarrow}^{'}n_{i\downarrow}-\sum_{i, \alpha}t_{i,\alpha}c^\dagger_{i\downarrow}c^{}_{i+\delta_\alpha,\downarrow}\\
		H_{mf}&&=H_{i\uparrow}+H_{i\downarrow} 
	\end{eqnarray}
    \end{widetext}
	After separating the Hamiltonian into the up- and down-spin parts, we start the 
	self-consistent procedure by choosing an initial set of densities 
	$n_{i\uparrow}$ and $n_{i\downarrow}$. We then diagonalize the corresponding 
	mean-field Hamiltonian for each spin sector and obtain updated values of the 
	densities. This process is repeated until the new and old densities agree within 
	a chosen accuracy. Once convergence is reached, all physical quantities are calculated. In our numerical study we solve Eq.~\ref{MFH} self-consistently using the above phase-modified hopping. For each disorder configuration $\epsilon_i$, we begin with an initial distribution of $\langle n_{i\sigma} \rangle$, diagonalize the effective Hamiltonian, and generate updated densities $\langle n'_{i\sigma} \rangle$. The iteration continues until the two sets match within the required tolerance. The site potentials $\epsilon_i$ are assigned according to the four disorder types, mapped onto the lattice from their corresponding one-dimensional sequences.
\section{Quasi-periodic and Deterministic Disorder Sequences}
	\indent
	 We briefly introduce the four quasiperiodic corelated disorder which we used in the present work. First the value of the site potentials $\epsilon_{i}$'s are assigned by the binary sequence, which is wrapped around one-dimensional representation of the lattice. 
	 	\begin{figure}[!htb]
	 	\centering
	 	\includegraphics[width=0.6\linewidth]{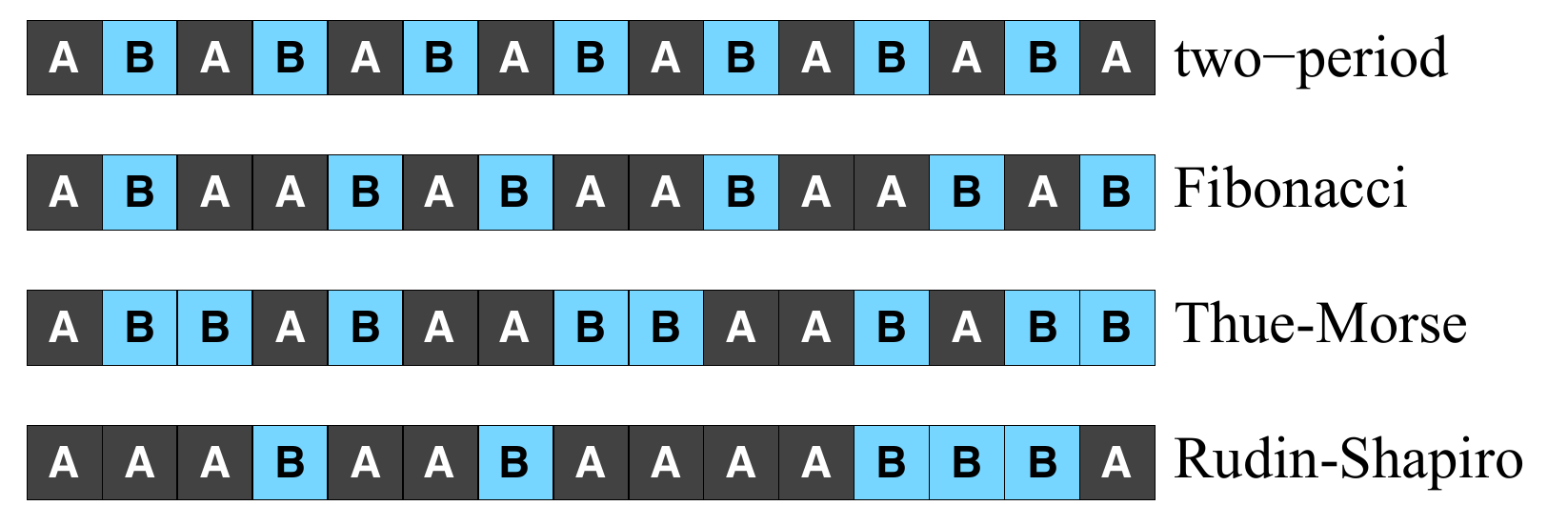}
	 	\caption{Schematic of the arrangement of the two values for the
	 		periodic and the deterministic aperiodic sequences.}
	 	\label{fig:gap_vs_U}
	 \end{figure}
 
\subsubsection{Rudin--Shapiro (RS) Disorder}
The Rudin--Shapiro (RS) sequence is generated using four symbols $A, B, C,$ and $D$, with the following substitution rules:
\[
A \rightarrow AB, \quad B \rightarrow AC, \quad C \rightarrow DB, \quad D \rightarrow DC.
\]
Here, $A, B, C,$ and $D$ represent the sequence for four successive sites along a one-dimensional chain, while $AB, AC, DB,$ and $DC$ denote the sequence for the next four sites.  
The Rudin--Shapiro sequence thus evolves as
\[\textit{ABACABDBABACDCACABACABDBDCDBABDB}\ldots\]
To obtain a binary Rudin--Shapiro sequence, we set $A=0$, $B=1$, $C=0$, and $D=1$. This binary version is referred to as $d=1$ throughout the manuscript.

\subsubsection{Fibonacci Disorder}
The Fibonacci disorder is deterministic in nature, neither periodic nor completely random. It follows the Fibonacci sequence defined by the substitution rules:
\[
A \rightarrow AB, \quad B \rightarrow A,
\]
where $A$ and $B$ represent two distinct lattice sites. Recursively, it can be expressed as:
\[
S_{N+1} = S_N S_{N-1} \quad (S_1 = B, \, S_2 = A).
\]
A typical Fibonacci chain for a given generation appears as:
\[
ABAABABAABAABABAABABA \ldots
\]
Using this Fibonacci sequence, we extend the concept of quasi-periodicity from one dimension to two dimensions by mapping the one-dimensional Fibonacci sequence onto a two-dimensional square lattice. Notably, the total number of lattice sites is not restricted to being a Fibonacci number.

\subsubsection{Thue-Morse (TM) Disorder}
The Thue--Morse (TM) sequence is another deterministic aperiodic sequence that exhibits self-similarity. It is defined by the substitution rules:
\[
A \rightarrow AB, \quad B \rightarrow BA,
\]
or recursively as:
\[
S_{N+1} = S_N \, \tilde{S}_N \quad (S_1 = A),
\]
where $\tilde{S}_N$ represents the string $S_N$ with the two symbols $A$ and $B$ interchanged.  
The binary representation of the TM sequence is given by:
\[
ABBABAABBAABABB...., \ldots
\]

\subsubsection{Aubry-André (AA) Disorder}
Next, the AA disorder is a well-known model used to study localization transitions in quasiperiodic systems. In this model, the onsite energy $\epsilon_i$ is given by
\[
\epsilon_i = \lambda \cos(2\pi \beta i + \phi),
\]
where $\lambda$ denotes the amplitude of the onsite potential modulation, $\phi$ is a relative phase, and $\beta$ is an irrational number (commonly chosen as the golden ratio) to introduce quasiperiodicity.

	\section{Details of subsystem for entanglement calculation}
	\label{subsys}
	After outlining the model and the quasi-periodicity, we now describe how the 
	entanglement quantities are computed in this work. All calculations are carried out 
	on a $42 \times 42$ square lattice with open boundary conditions. To examine how 
	entanglement varies with subsystem size, we select square regions of dimension 
	$n \times n$ inside the full lattice, with $n$ ranging from 3 to 21. Equivalently, 
	each subsystem encloses $m \times m$ plaquettes, where $m$ takes values from 2 to 20. 
	Figure~1(a) shows an example of a $4 \times 4$ region. The entanglement spectrum 
	is evaluated only for the largest subsystem corresponding to $m = 21$.
	
	For non-interacting fermions, the entanglement entropy can be extracted from the 
	correlation matrix restricted to the chosen subsystem. In the figures, the entanglement 
	entropy is displayed for different subsystem sizes labeled by $m$. The entanglement 
	spectrum refers to the eigenvalues of the reduced density matrix, which can be obtained 
	directly from the eigenvalues $\lambda_k$ of the subsystem correlation matrix. These 
	eigenvalues allow us to compute the von~Neumann entanglement entropy using the standard free-fermion formula.
	\begin{eqnarray}
		\mathcal{S}_{en}&&= \sum_{k} -\left(\lambda_k log \lambda_k+ (1-\lambda_k) log (1-\lambda_k)   \right)
	\end{eqnarray}
	\section{Effect of disorders and interaction on spectrum}
	
Figure \ref{fig:gap_vs_U} illustrates the dependence of the energy gap at half filling on the onsite Coulomb interaction $U$ for different correlated disorder sequences. For small $U$, the gap remains nearly closed, indicating that disorder-induced localization dominates. As $U$ increases beyond a critical value $U_c$, the electronic correlations begin to prevail, leading to the formation of a Mott–Hubbard gap. The critical interaction strengths are found to be  $U_c \approx 1.4 $ for the Aubry–André, $U_c \approx 1.6$ for the Fibonacci, and $U_c \approx 1.8$ for both the Rudin–Shapiro and Thue–Morse disorders. The comparatively smaller $U_c$ for the Aubry–André sequence indicates its higher sensitivity to correlation effects, while the larger $U_c$ values for the Rudin–Shapiro and Thue–Morse cases suggest stronger localization due to more random disorder configurations.

	\begin{figure}[!htb]
		\centering
		\includegraphics[width=0.6\linewidth]{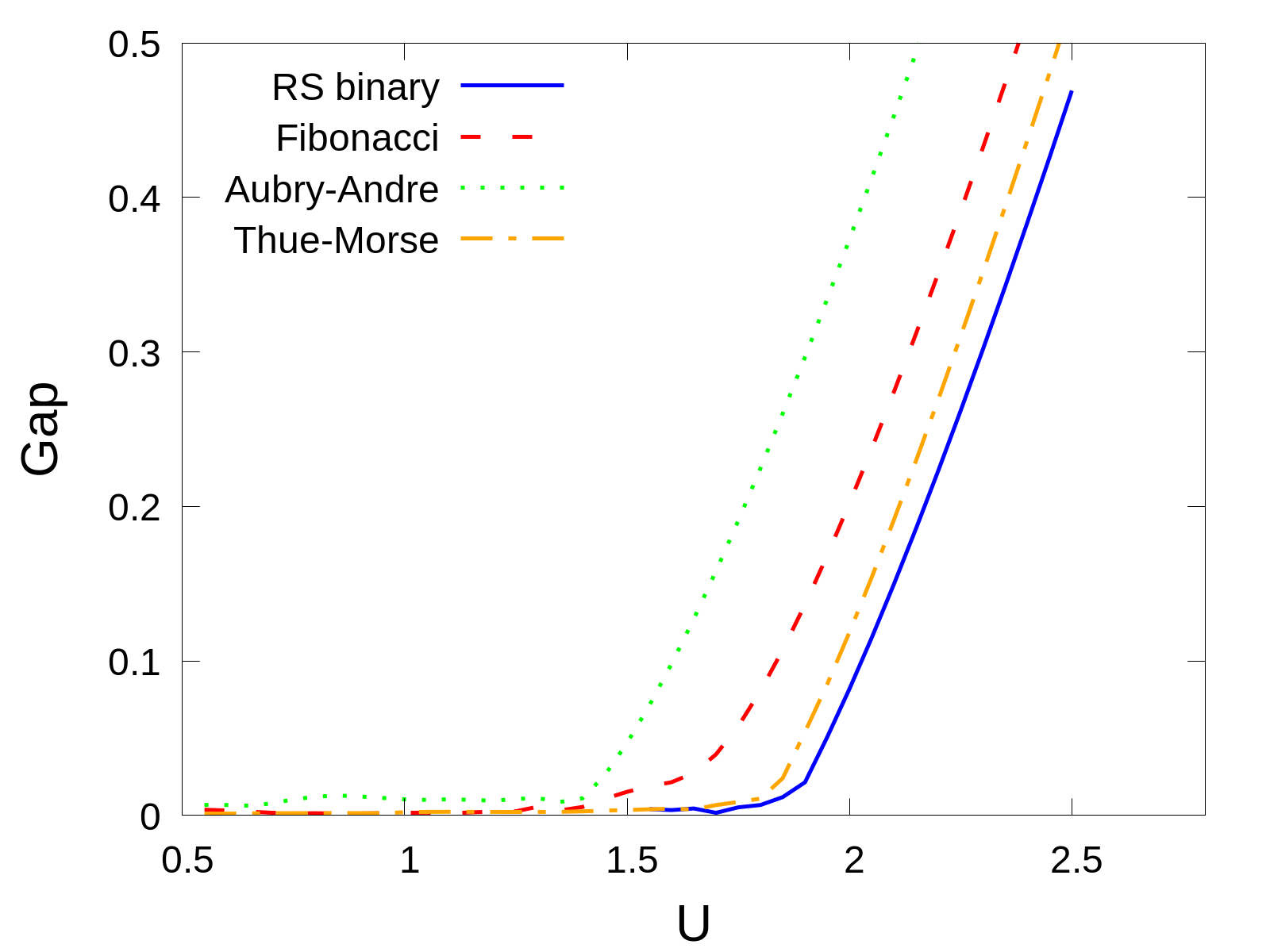}
		\caption{The gap at half filling as a function of coulomb interaction parameter $U$ , is shown for four different corelated disorders.}
		\label{fig:gap_vs_U}
	\end{figure}

\begin{figure*}[!htb]
		\centering
		\begin{subfigure}{.24\textwidth}
			\centering
		\includegraphics[width=\linewidth]{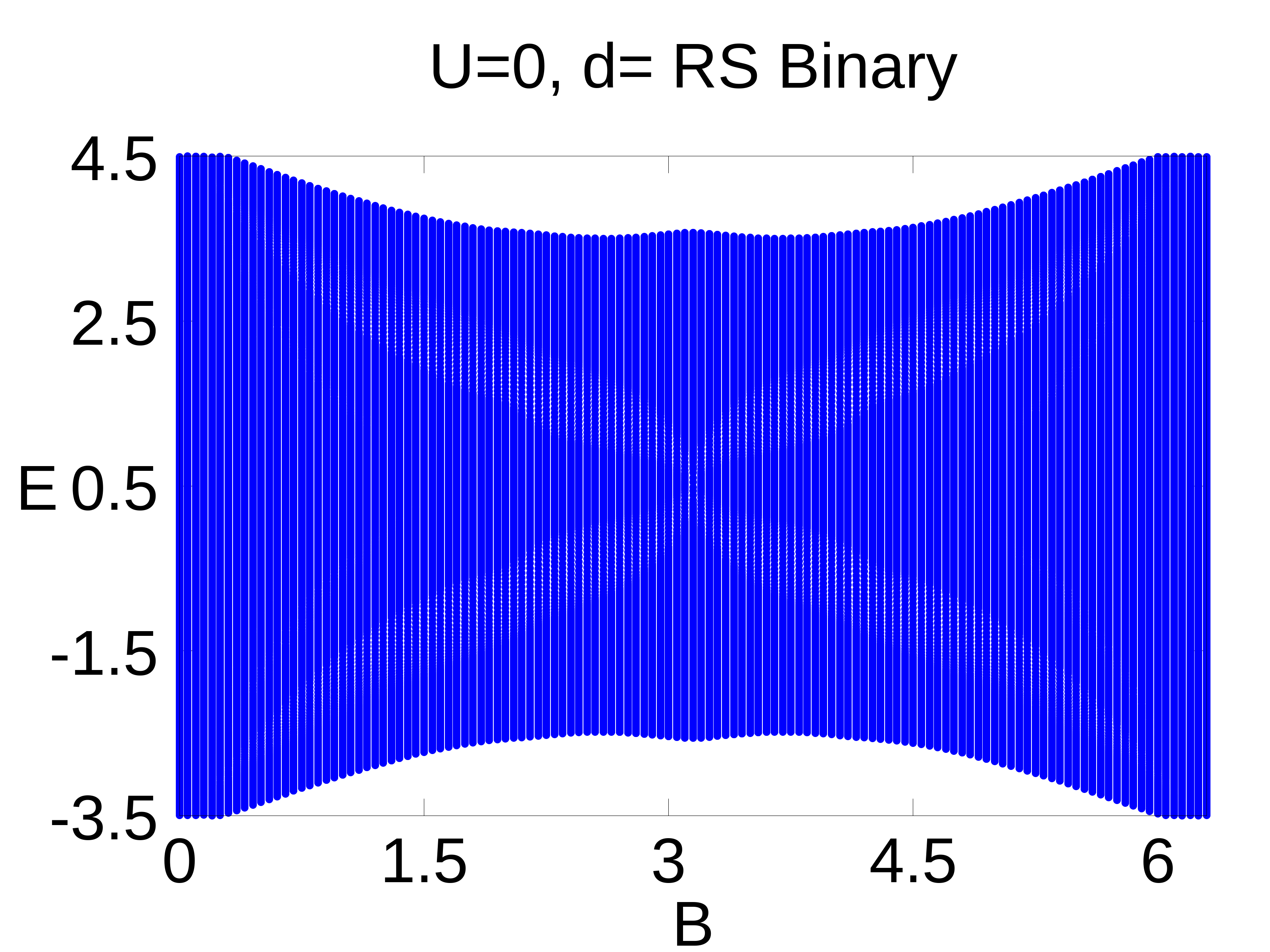}
			\subcaption{}
			\label{sq1}
		\end{subfigure}%
		\begin{subfigure}{.24\textwidth}
			\centering 
			\includegraphics[width=\linewidth]{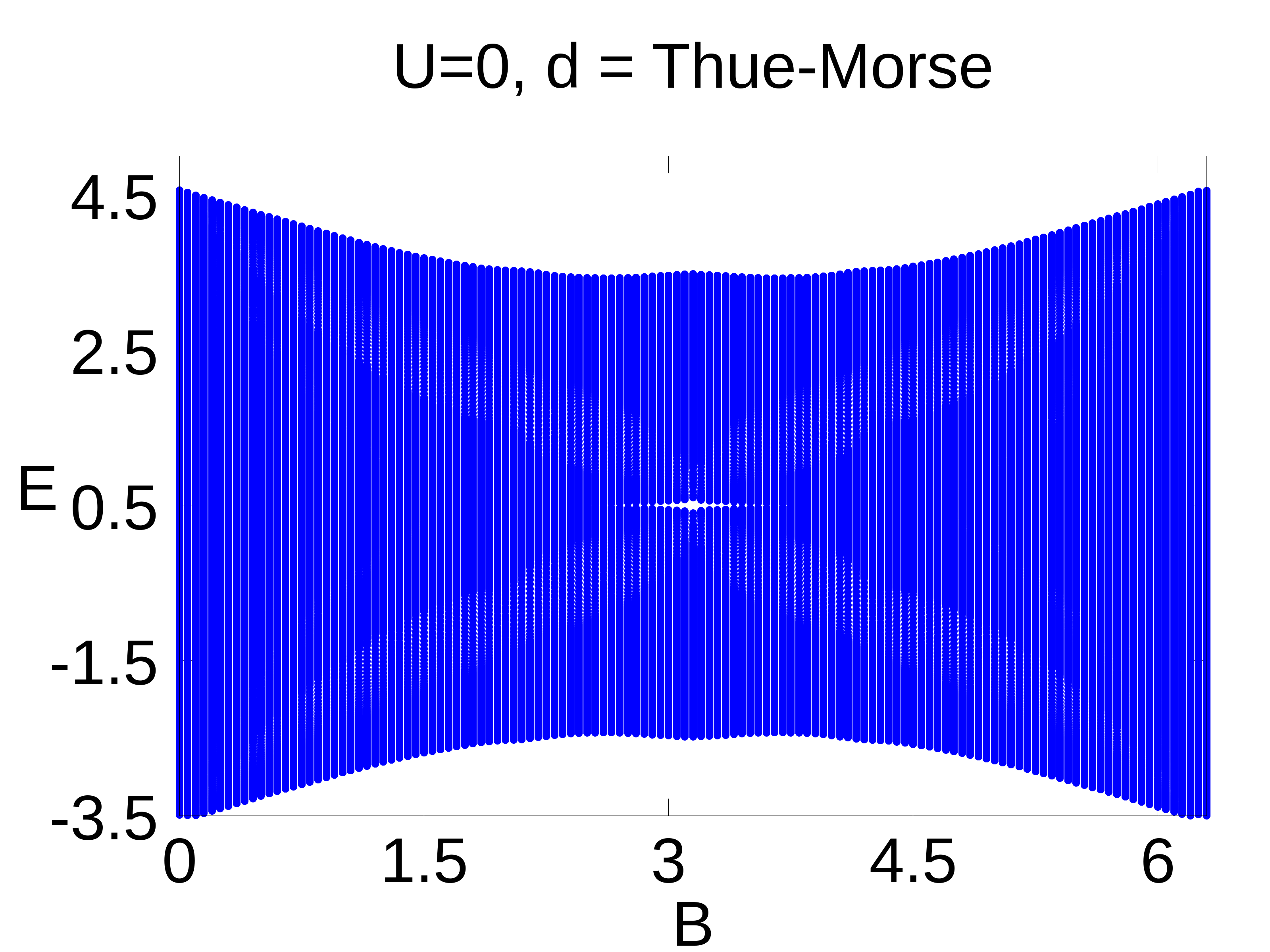}
			\subcaption{}
			\label{sq2}
		\end{subfigure}%
		\begin{subfigure}{.24\textwidth}
			\centering
			\includegraphics[width=\linewidth]{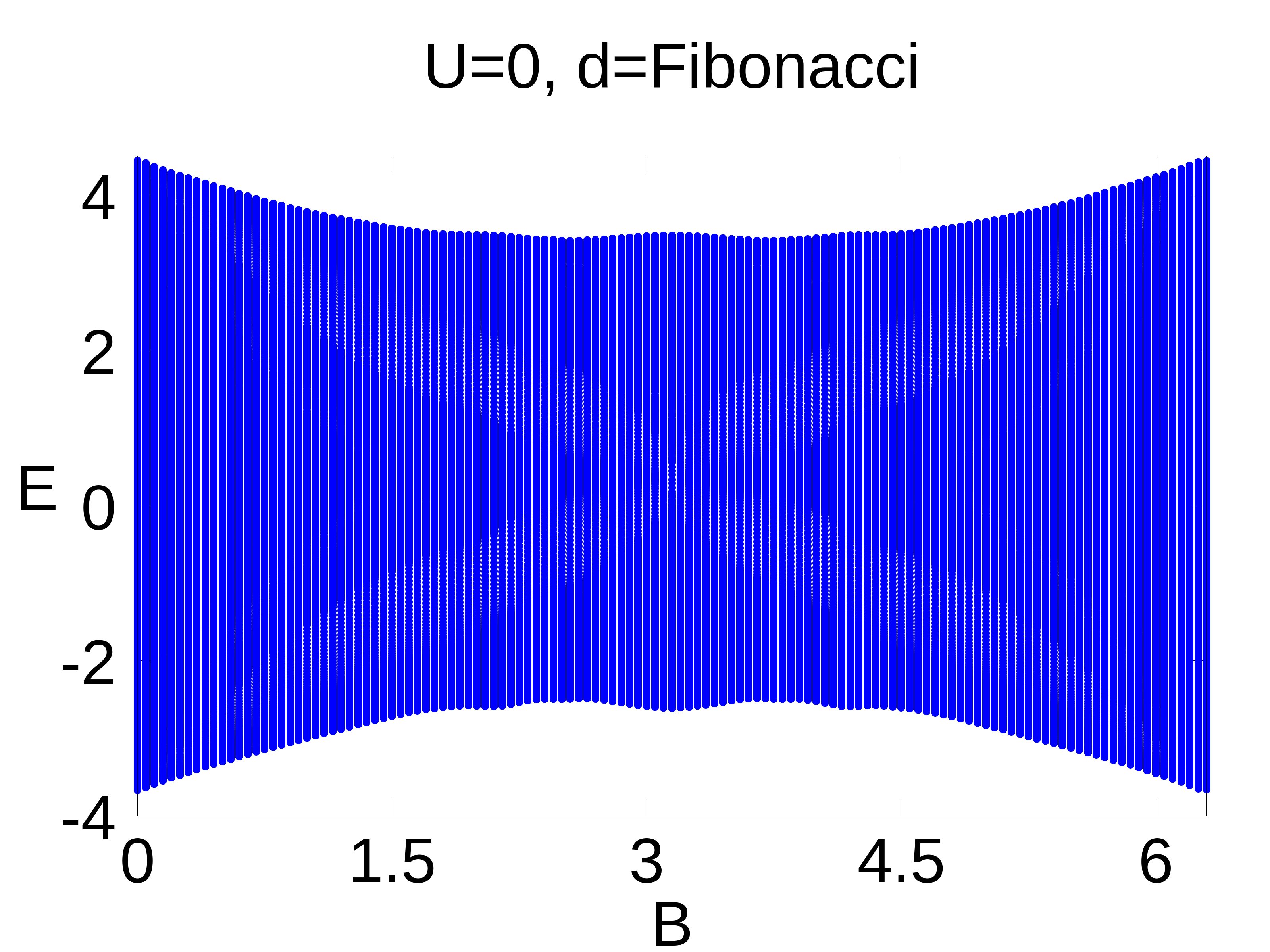}
			\subcaption{}
			\label{sq3}
		\end{subfigure}%
		\begin{subfigure}{.24\textwidth}
			\centering
			\includegraphics[width=\linewidth]{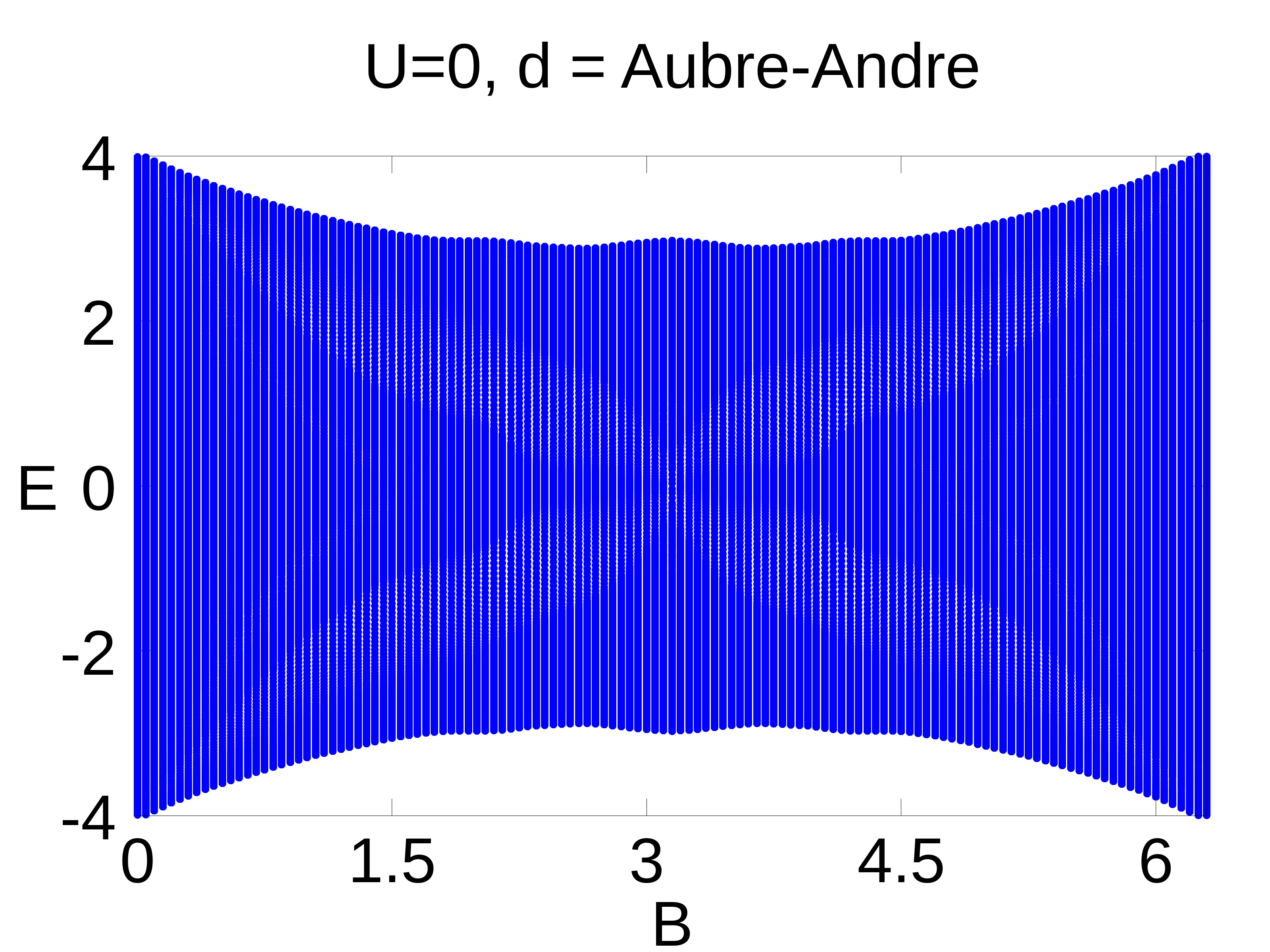}
			\subcaption{}
			\label{sq4}
		\end{subfigure}%

		\begin{subfigure}{.24\textwidth}
			\centering
			\includegraphics[width=\linewidth]{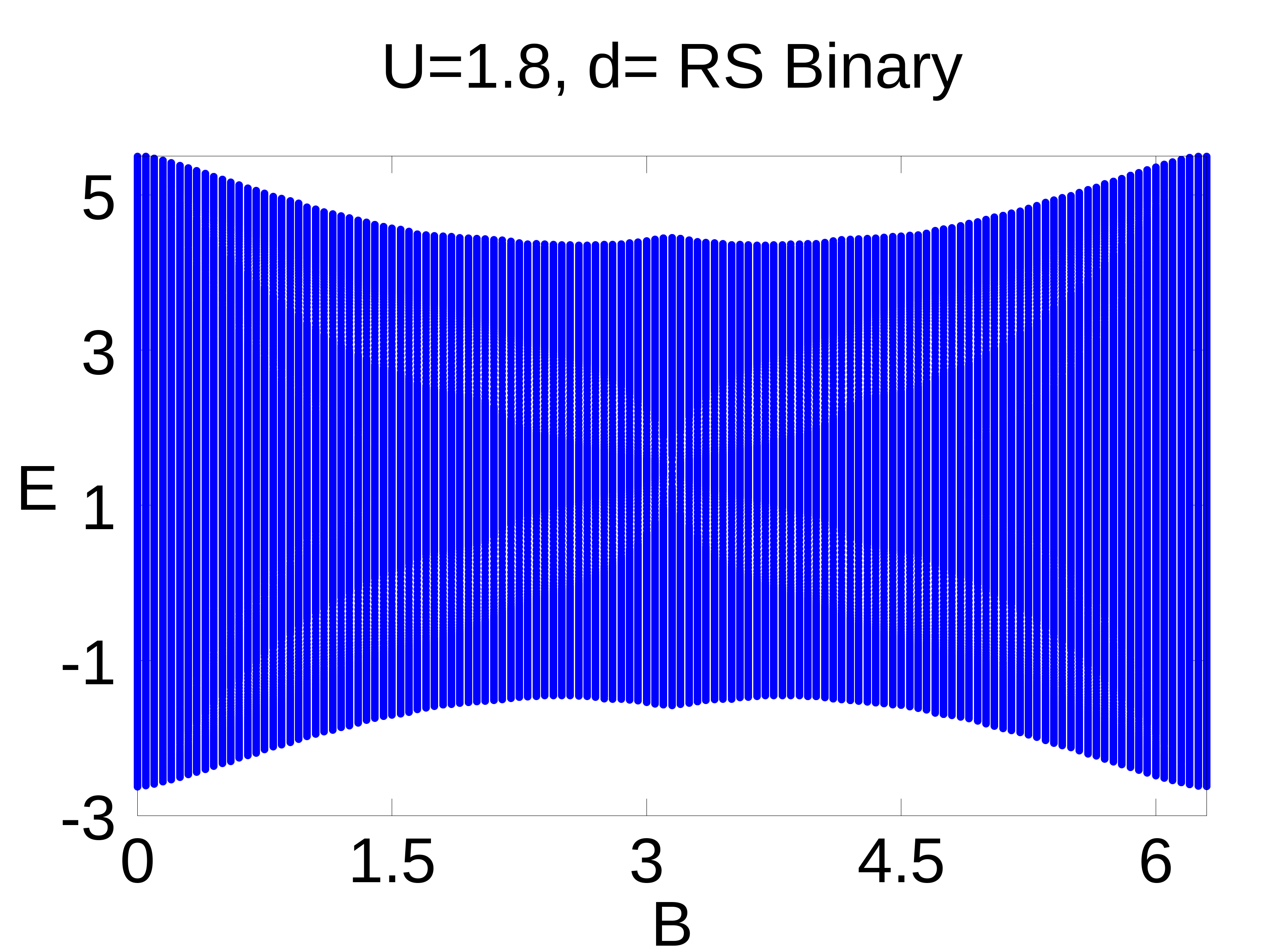}
			\subcaption{}
			\label{sq5}
		\end{subfigure}%
		\begin{subfigure}{.24\textwidth}
			\centering 
			\includegraphics[width=\linewidth]{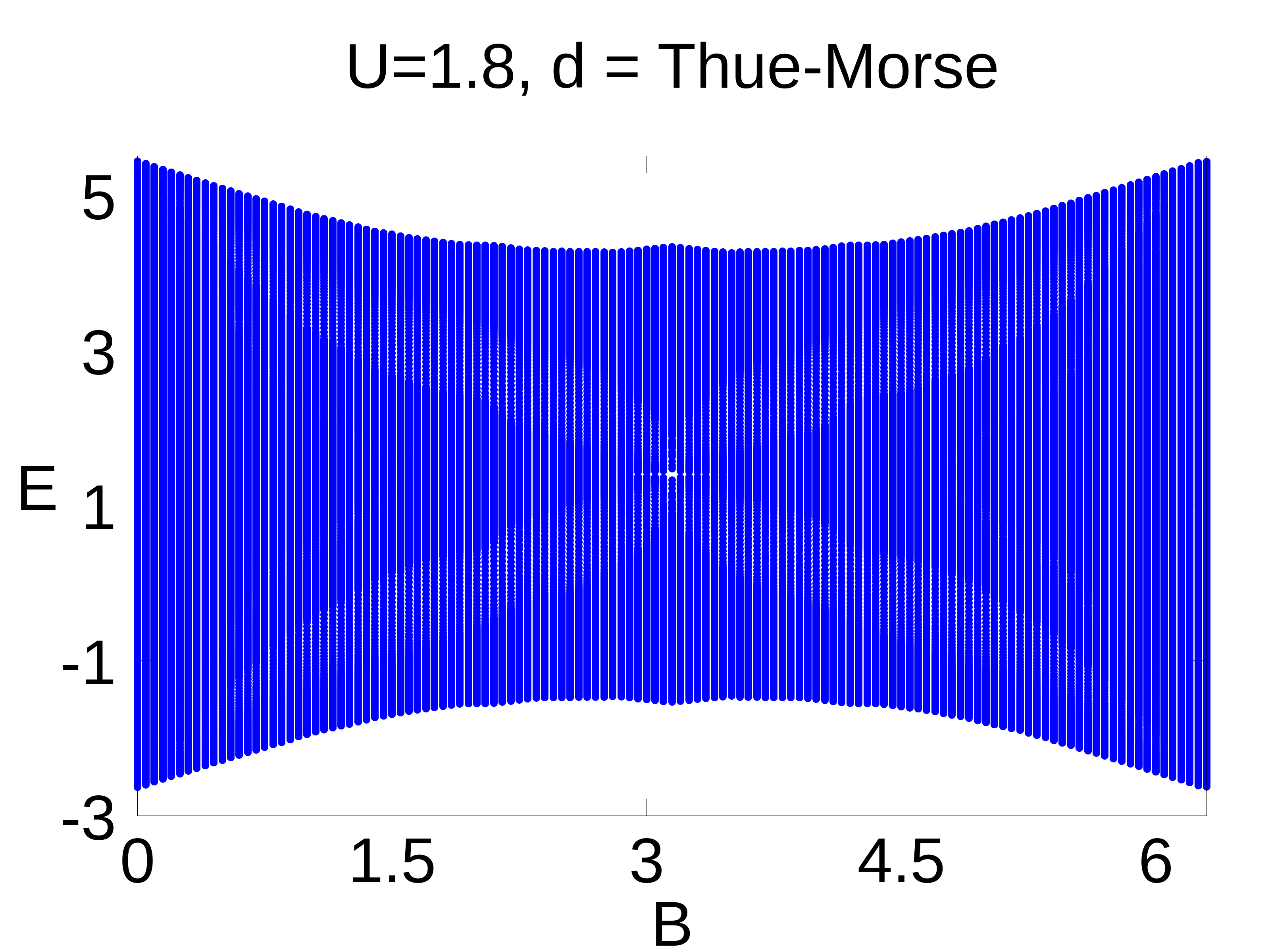}
			\subcaption{}
			\label{sq6}
		\end{subfigure}%
		\begin{subfigure}{.24\textwidth}
			\centering
			\includegraphics[width=\linewidth]{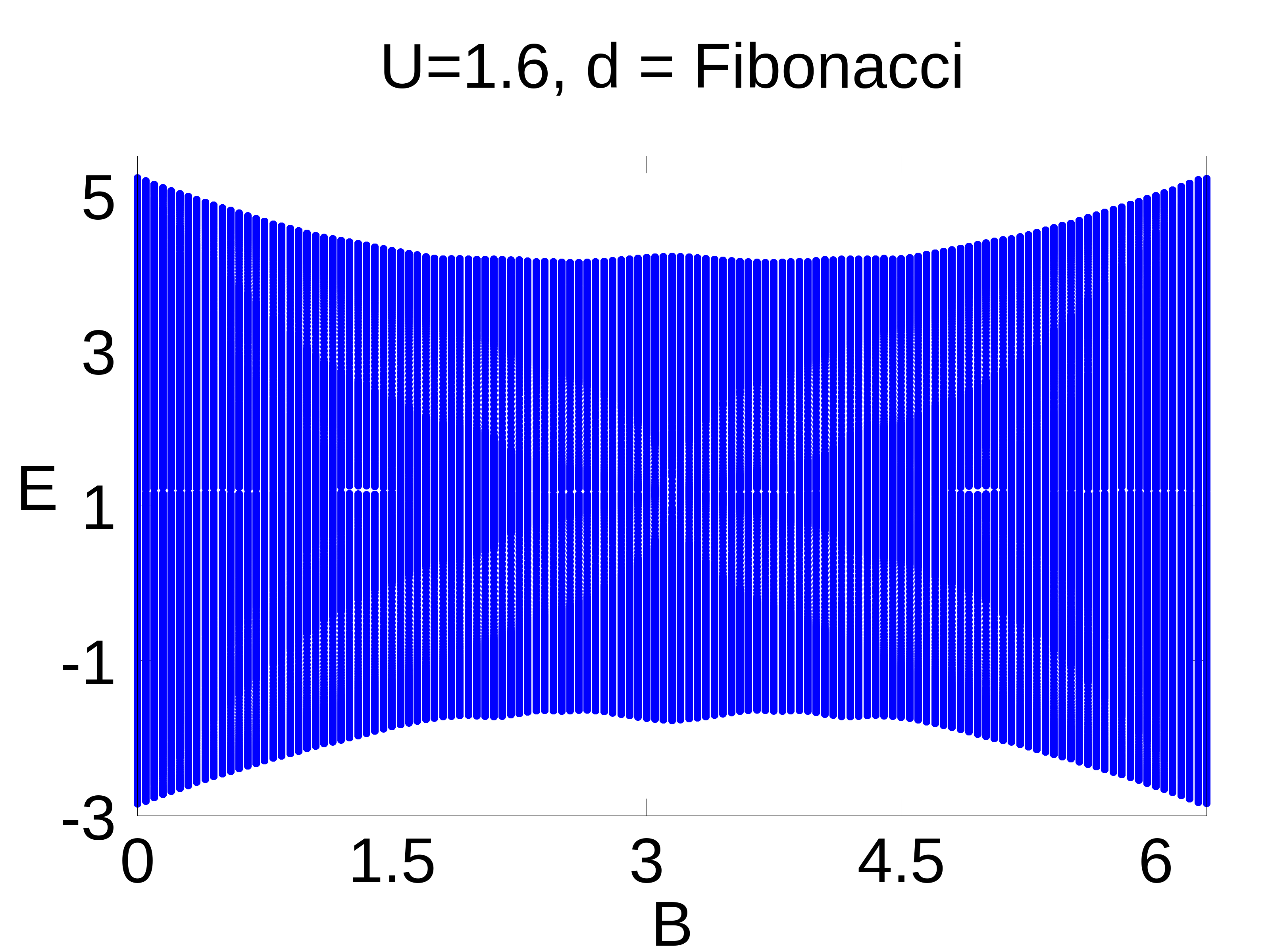}
			\subcaption{}
			\label{sq7}
		\end{subfigure}%
		\begin{subfigure}{.24\textwidth}
			\centering
			\includegraphics[width=\linewidth]{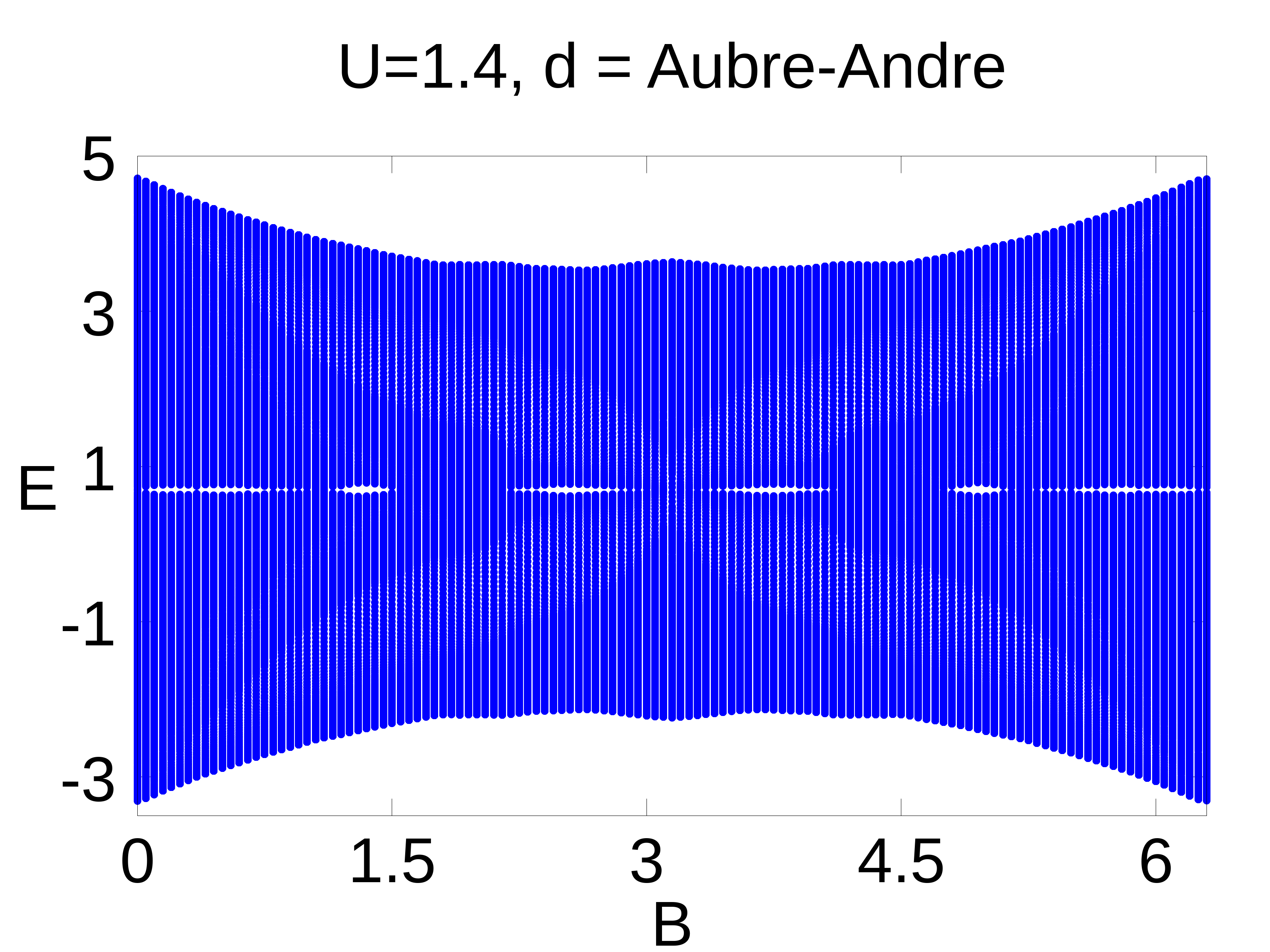}
			\subcaption{}
			\label{sq8}
		\end{subfigure}%
		
		\caption{The energy spectrum (E) is plotted as a function of the magnetic field (B) in a square lattice, illustrating the effects of various types of disorders and different combinations of disorder and interaction. Each plot is labeled to represent a specific configuration of disorder and interaction. As seen from the spectrum, all types of disorder tend to smear out the characteristic butterfly structure. The on-site interaction alone opens a clear gap in the spectrum \cite{Mandal}, whereas the simultaneous presence of disorder and interaction closes this gap and, to some extent, restores features of the original butterfly pattern. }
		\label{EBA}
	\end{figure*}

In Figure \ref{EBA}, we have shown the effect of various types of disorders and interaction for a square lattice. In the presence of disorders, the spectrum loses its original butterfly structure \cite{Mandal}. The quasi periodic disorders introduce additional states inside the gap, leading to a stronger breakdown of the fractal butterfly structure and producing noticeably thinner wings. Despite this distortion, the lowest-energy branch continues to exhibit reflection symmetry about $B = \pi$. When Thue-Morse disorder is present then spectrum gets a small central gap. In the case of Fibonacci disorder, the broad fractal structure of the Hofstadter butterfly remains largely intact, but the finer details become distorted. The quasi-periodicity introduced by these disorders alter the precise gap structure, and the states can transition from extended to quasi-localized, reflecting the intermediate nature of the disorder.

\begin{figure*}[!htb]
		\centering
		\begin{subfigure}{.24\textwidth}
			\centering
			\includegraphics[width=\linewidth]{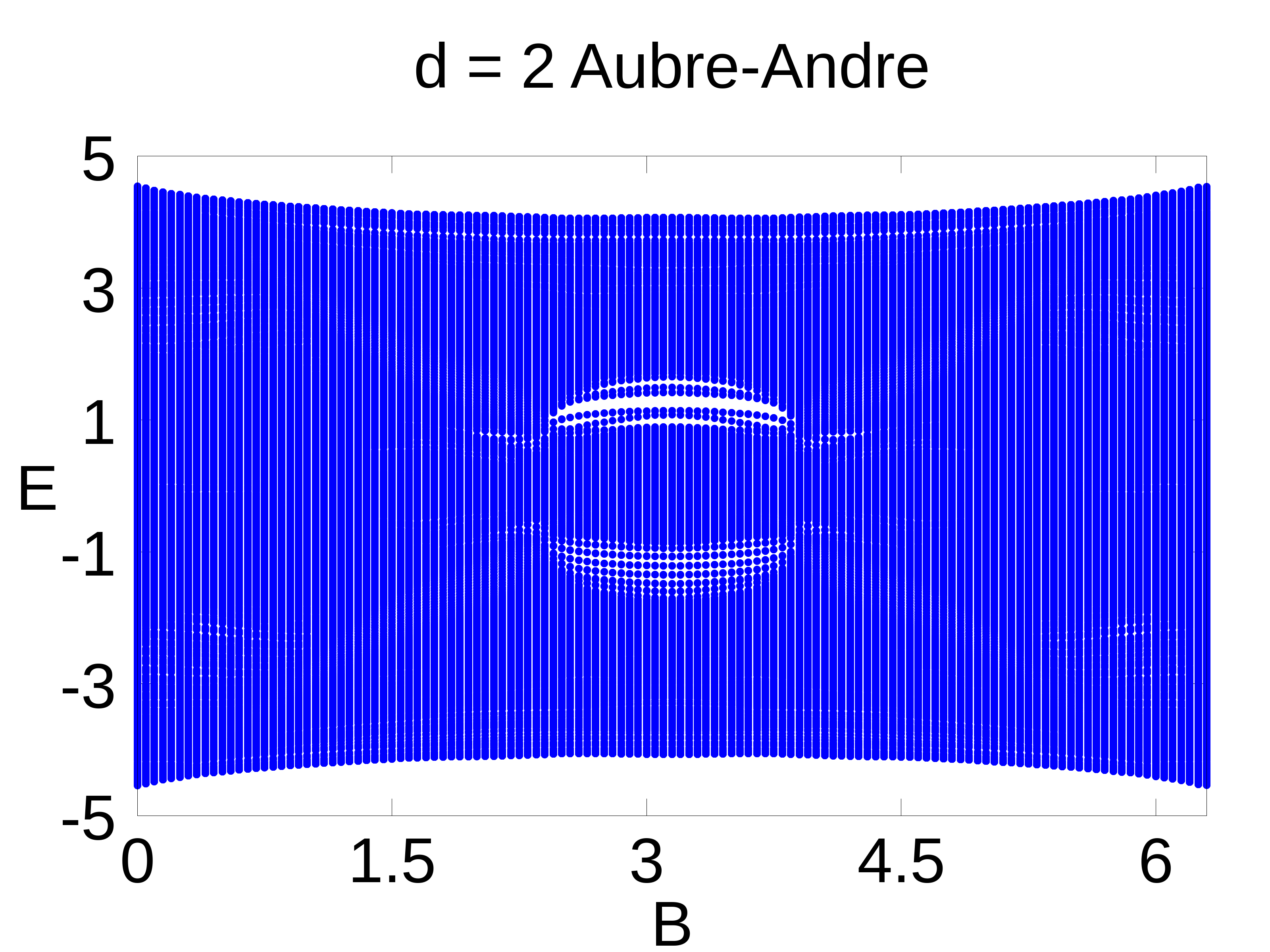}
			\subcaption{}
			\label{EBAD1}
		\end{subfigure}%
		\begin{subfigure}{.24\textwidth}
			\centering 
			\includegraphics[width=\linewidth]{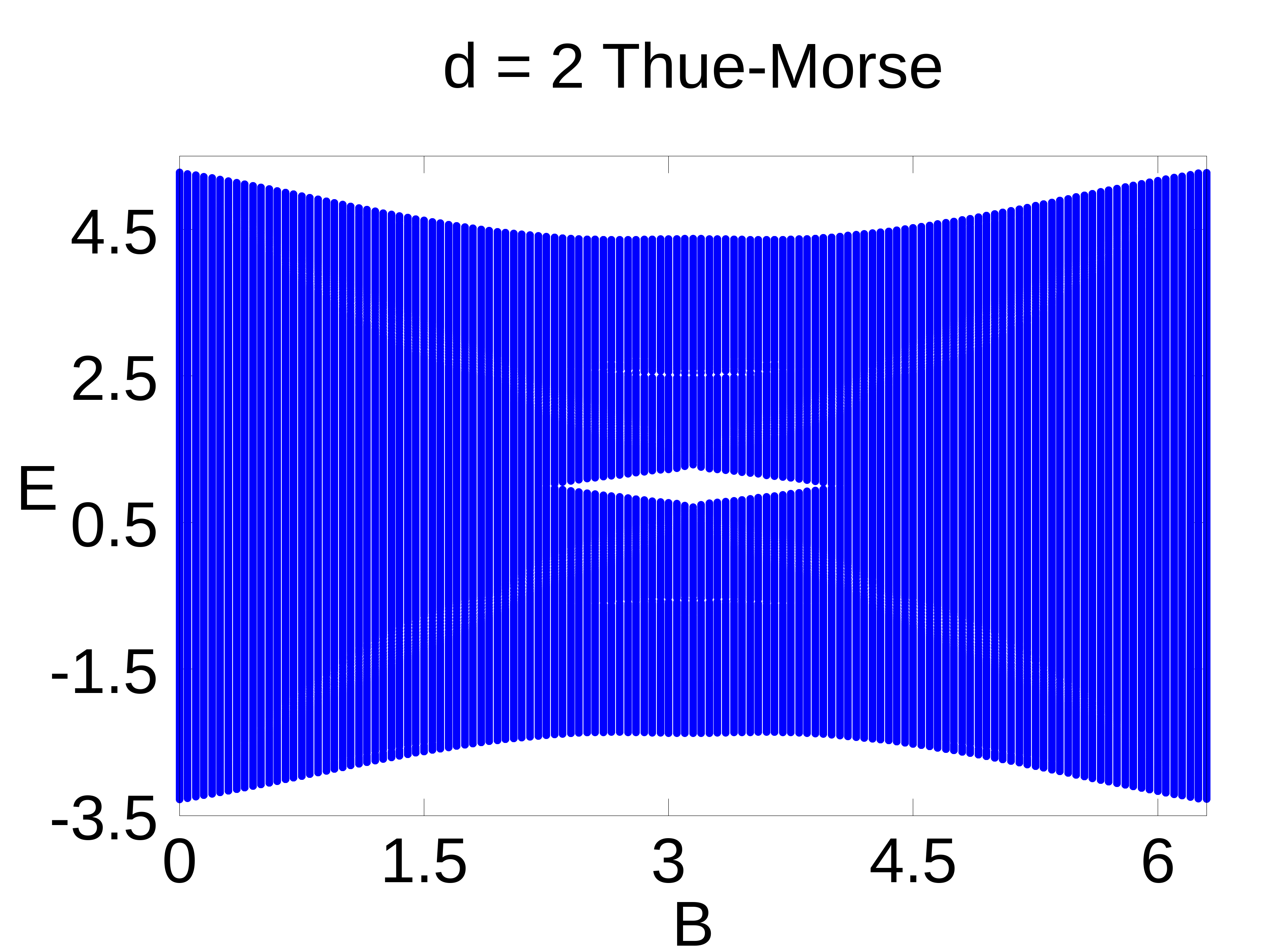}
			\subcaption{}
			\label{EBAD2}
		\end{subfigure}%
		\begin{subfigure}{.24\textwidth}
			\centering
			\includegraphics[width=\linewidth]{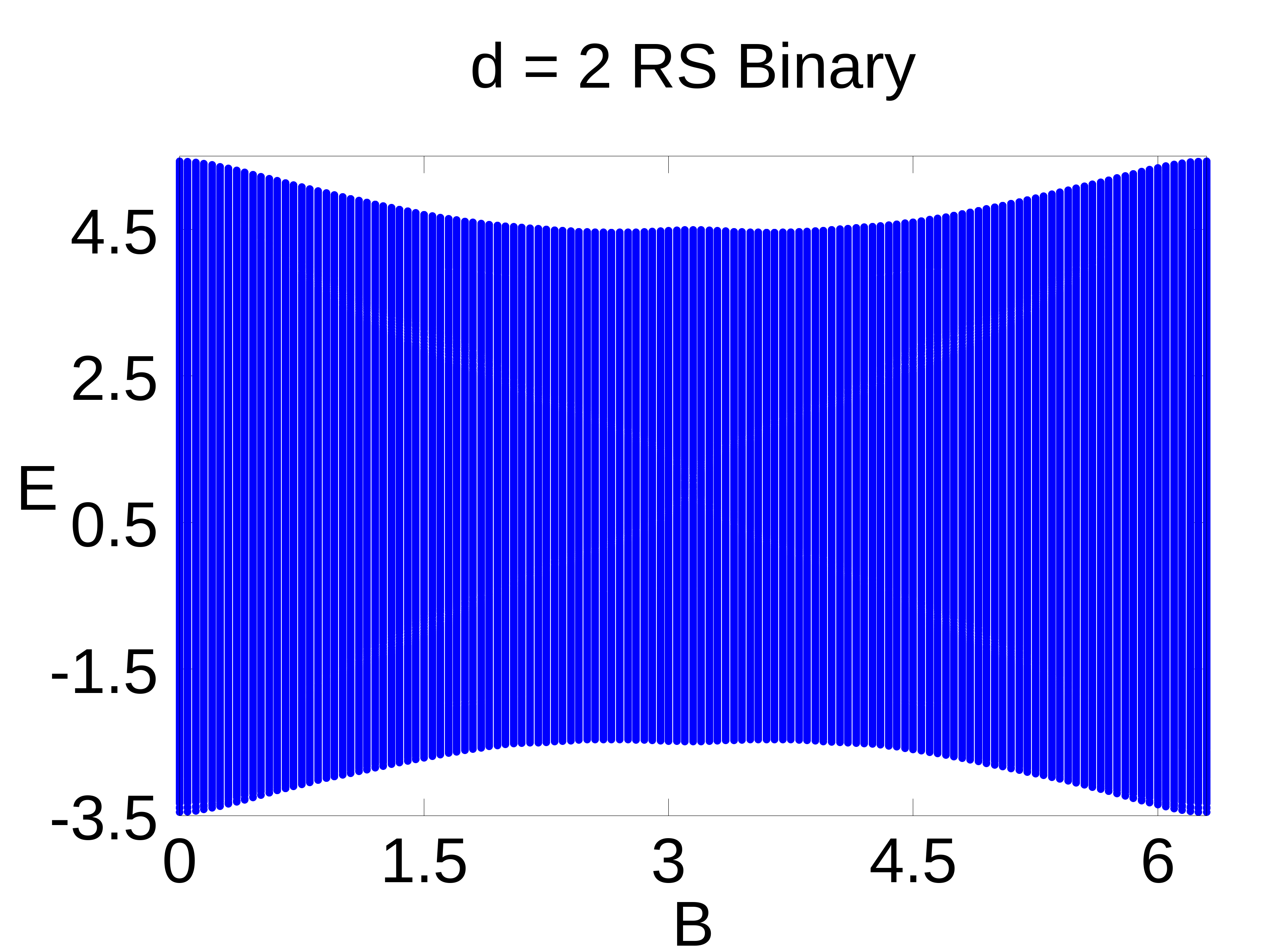}
			\subcaption{}
			\label{EBAD3}
		\end{subfigure}%
		\begin{subfigure}{.24\textwidth}
			\centering
			\includegraphics[width=\linewidth]{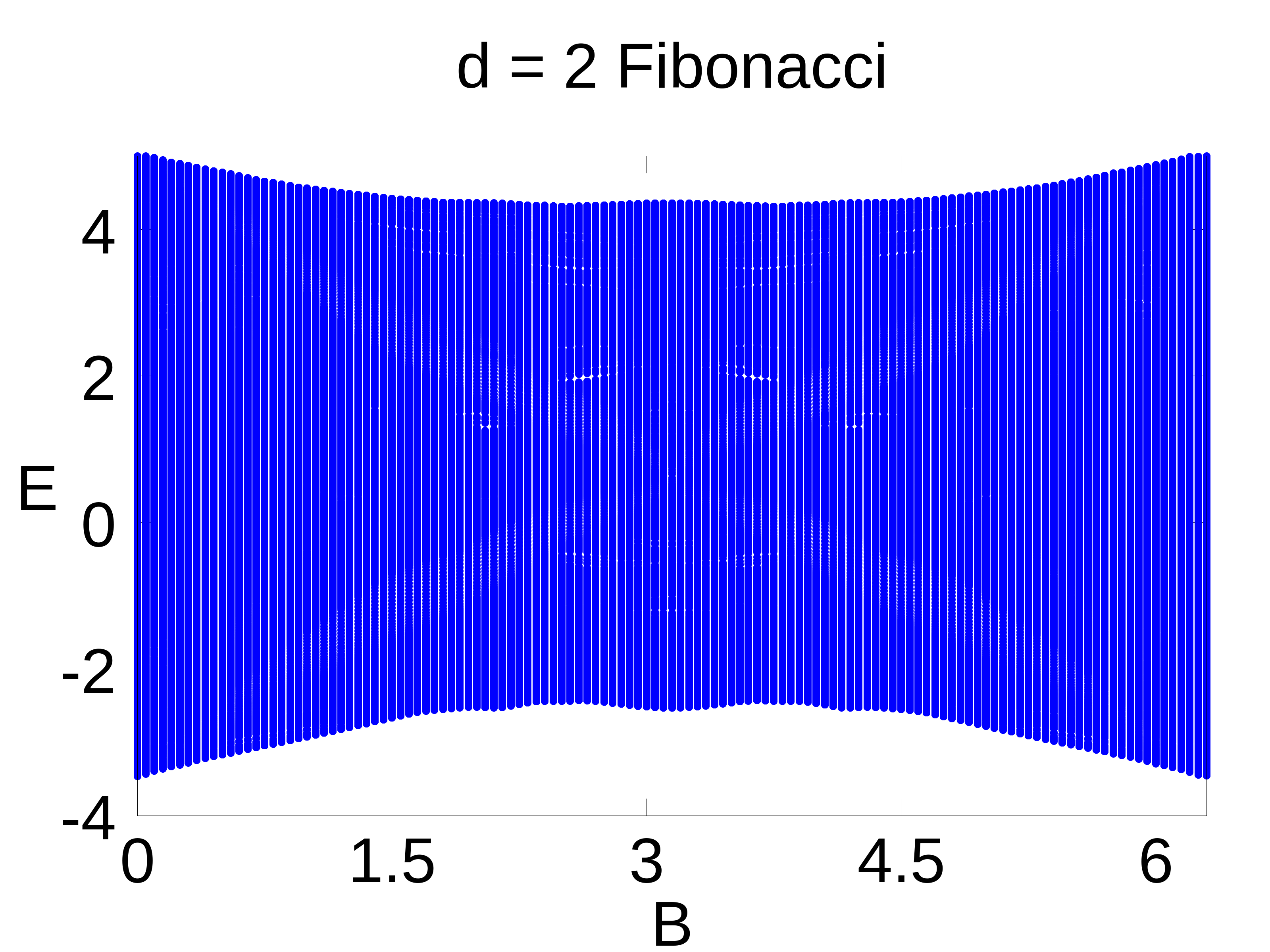}
			\subcaption{}
			\label{EBAD4}
		\end{subfigure}%

		\begin{subfigure}{.24\textwidth}
			\centering
			\includegraphics[width=\linewidth]{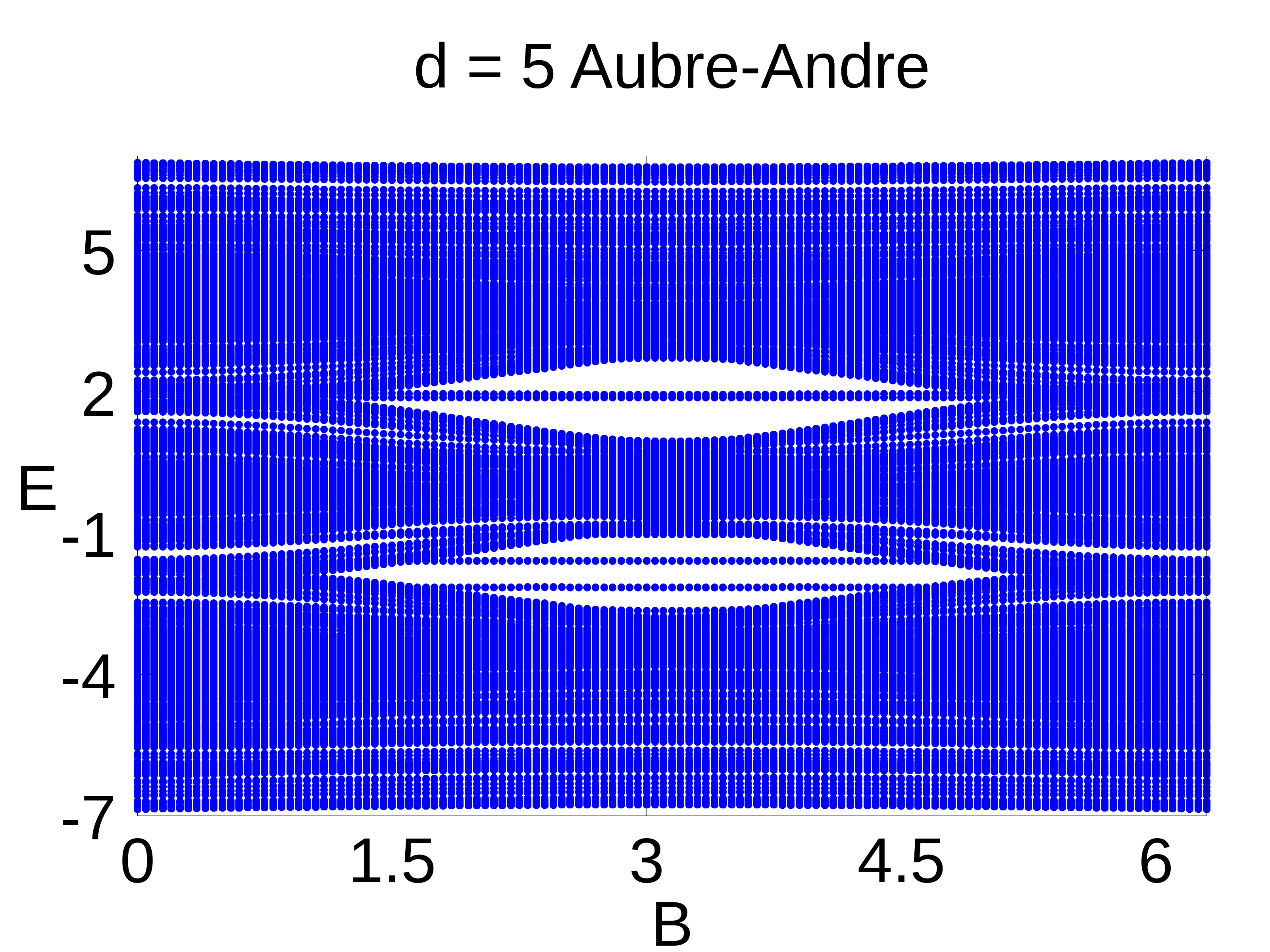}
			\subcaption{}
			\label{sq5}
		\end{subfigure}%
		\begin{subfigure}{.24\textwidth}
			\centering 
			\includegraphics[width=\linewidth]{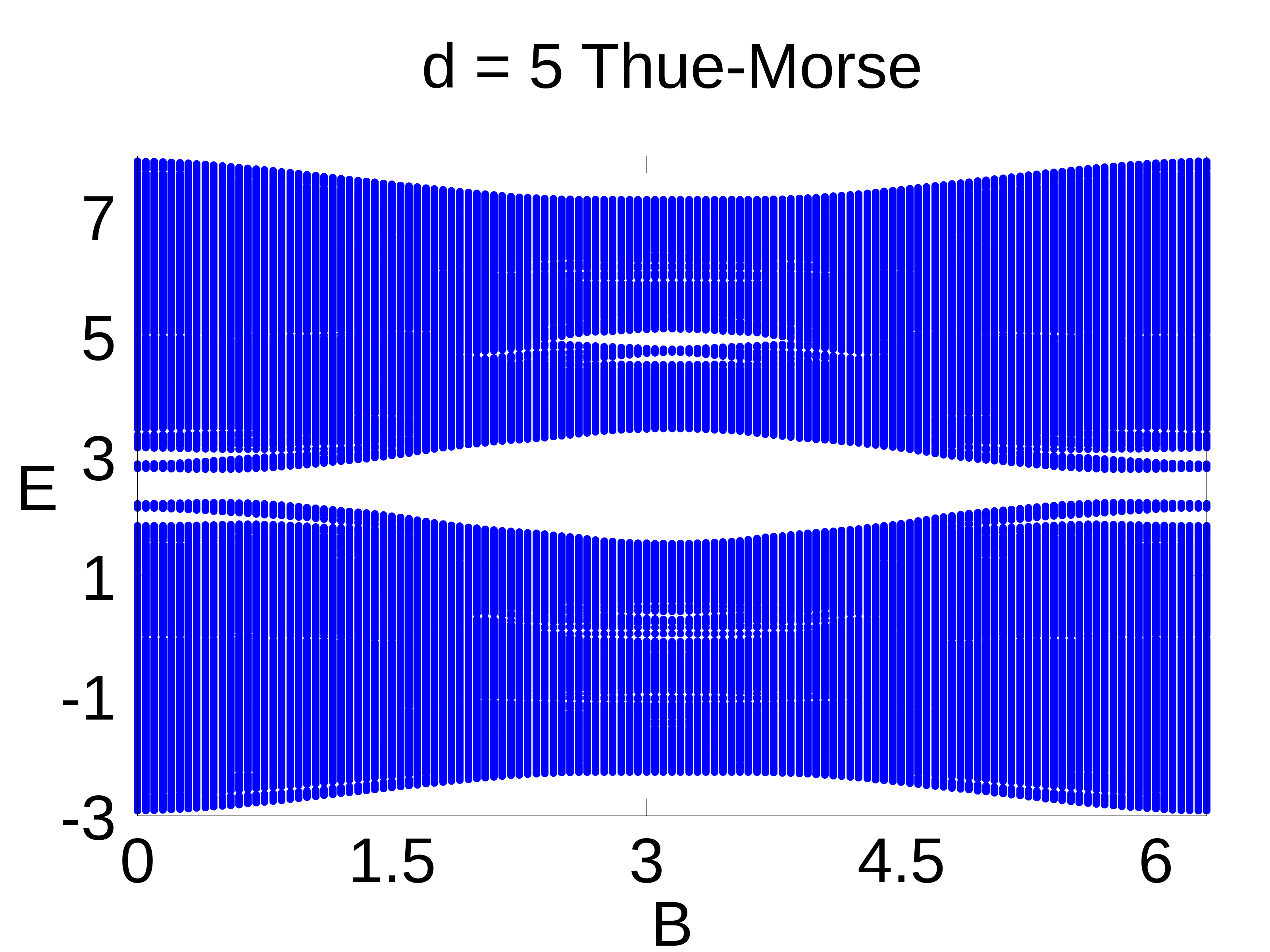}
			\subcaption{}
			\label{sq6}
		\end{subfigure}%
		\begin{subfigure}{.24\textwidth}
			\centering
			\includegraphics[width=\linewidth]{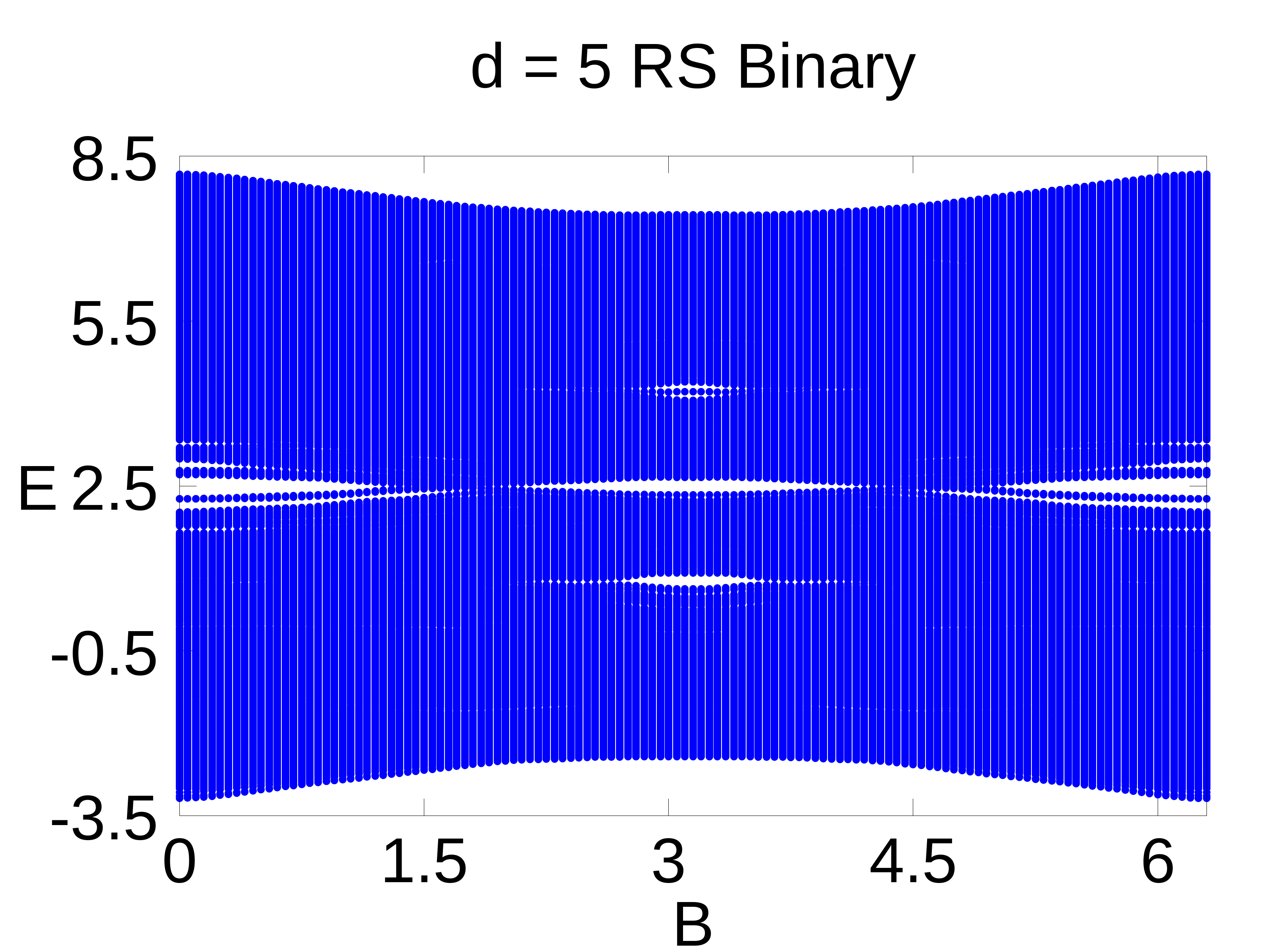}
			\subcaption{}
			\label{sq7}
		\end{subfigure}%
		\begin{subfigure}{.24\textwidth}
			\centering
			\includegraphics[width=\linewidth]{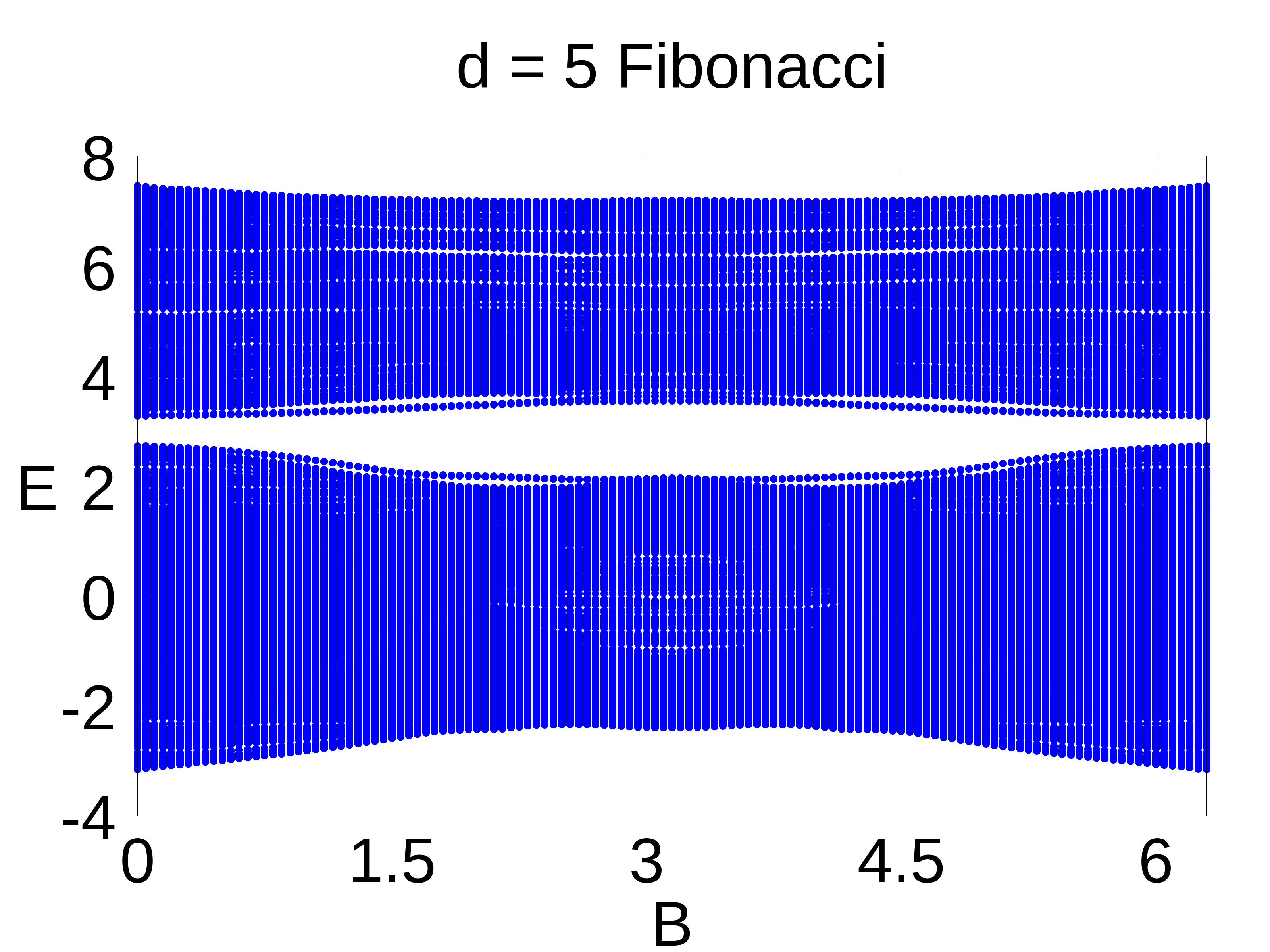}
			\subcaption{}
			\label{sq8}
		\end{subfigure}%

		\caption{The energy spectrum (E) is plotted as a function of the magnetic field (B) in a square lattice, illustrating the effects of various types of disorder for two values of their strength $d=2,5$. Each plot is labeled to represent a specific configuration of disorders and their strength.}
		\label{EBAD}
	\end{figure*}

	For AA disorder, if the disorder is weak, it has minimal impact on the energy spectrum, leaving the fractal features of the butterfly mostly unaffected. The localization effects only become significant at higher disorder strengths, but in the weak regime, the spectrum stays relatively unchanged. In contrast, TM disorder introduces a small gap at the center of the butterfly spectrum. However, this disorder does not disrupt the spectrum as significantly as RS disorder does. The overall energy structure remains largely unaffected, with only minor modifications around the gap.\\
	When disorder is introduced in the presence of interactions, we observe a remarkable revival of the Hofstadter butterfly for a critical value of the interaction strength. This revival is marked by the emergence of the largest wing, or gap, in the spectrum. However, the width of this wing has diminished compared to the clean system, and the subdominant wings appear faint but still discernible. At half-filling, the gap is fully closed in the presence of RS binary, indicating a strong impact of these quasi-periodic potential. For Fibonacci and AA disorders, the gap remains slightly open, reflecting their relatively weaker disorder effects. This phenomenon highlights a key interaction between localization effects due to disorder and interactions. These competing forces lead to a partial recovery of the Hofstadter butterfly spectrum. Interestingly, the better recovery is observed in the case of RS and Thue-Morse disorder, where the spectrum regains more of its original fractal structure despite the presence of disorder. Thus, we find clear evidence that the interplay between disorder and interaction effects can result in the partial revival of the butterfly spectrum, with varying degrees of recovery depending on the specific disorder.\\
	Figure \ref{EBAD} shows that how strengh of the disorders affacts the Hofstadter spectrum as a function of magnetic field. For small to moderate disorder strengths, the fractal features of the butterfly spectrum gradually fade, leading to a more uniform, structureless pattern. At strong disorder, a gap emerges at half filling as a result of localization effects, often referred to as the Anderson gap. In Figure \ref{EBAD}(a) showing that how the Hofstadter spectrum impact when the AA disorder strength increses. For AA strenth $d=2$, the fractal Hofstadter butterfly structure is severely suppressed and most of the extended bands and fine features from the clean limit are washed out. The energy spectrum undergoes a notable transformation characterized by the opening of symmetric gaps around $E=0$. When the AA disorder strength increses, the spectrum becomes progressively more fragmented and localized. At $d=5$, the Hofstadter spectrum shows clear disruption of its original fractal structure, with the central band splitting into a few distinct mini-bands and the emergence of pronounced energy gaps in Fig.\ref{EBAD}(e). The strong disorder dominates the system completely, leading to the formation of highly localized states confined within narrow, non-overlapping energy windows. The fractal nature of the Hofstadter butterfly is almost entirely obliterated, replaced by a set of well-separated flat bands with broad gaps between them, indicative of strong localization and suppression of magnetic field-induced interference effects.
	
When TM disorder is introduced, an aperiodic sequence replaces the periodic potential, adding a layer of complexity to the system. This disorder represents the strength of the aperiodic perturbation, which disrupts the regular lattice structure. In Fig. \ref{EBAD}(b), the spectrum begins to show initial distortions: the central gap around (E = 0) widens, and the fine structure of the energy levels becomes less distinct due to the weak disorder starting to scatter electrons. As (d) increases, the central gap expands further, and additional smaller gaps appear, reflecting the growing influence of the aperiodic potential. This progressive disruption breaks the symmetry and regularity of the original butterfly pattern, leading to a more fragmented energy spectrum. The underlying reason for this behavior is the competition between the magnetic field's quantization and the TM disorder's aperiodic scattering, which induces electron localization. For low disorder strength, energy gaps appear in the spectrum primarily around $B=\pi$. However, for higher disorder strength, gaps open across the entire range of the magnetic field. The gap remains more pronounced near $B=\pi$ even in the presence of strong disorder. This localization effect becomes more pronounced with increasing (d), transforming the fractal structure into a complex, gapped spectrum that reflects the intricate balance of quantum coherence and disorder.\\
In the case of RS disorder, at (d = 2), the fractal pattern of the Hofstadter butterfly completely vanishes, replaced by a smoother, more uniform distribution of energy states, indicating that the weak RS disorder immediately disrupts the delicate balance between magnetic flux and lattice periodicity in fig. \ref{EBAD}(c). For d=5 in fig.\ref{EBAD}(g), gaps start to open up at half filling, reflecting a stronger localization effect as the disorder intensifies. This transformation arises because the RS sequence, being quasi periodic, introduces irregular potential variations that interfere with the quantized Landau levels, leading to a breakdown of extended states. The progressive localization of electron wavefunctions, driven by this disorder, results in the observed gap formation and widening, ultimately reshaping the spectrum from a fractal to a gapped, localized state.\\
In the presence of Fibonacci disorder, the Hofstadter butterfly spectrum retains a faint fractal pattern at low disorder strength ($d=2$), with only a very small gap appearing near half-filling in fig.\ref{EBAD}(d). At strong disorder ($d=5$), the fractal structure is heavily degraded, with most fine features lost; a wide non uniform gap forms with magnetic field $B$, and the surrounding bands flatten further, indicating stronger localization. At very strong disorder, a large, dominant gap completely separates the spectrum into upper and lower energy bands, and the remaining bands are extremely flat, corresponding to highly localized eigenstates, a clear signature of strong localization driven by quasi-periodic disorder.
	\section{Interplay of Aubry-André Disorder with Other Quasi-Periodic Disorders}
	
The study of the Hofstadter butterfly under quasi-periodic disorders, such as AA disorder combined with TM, RS binary, and Fibonacci sequences, hinges on the interpolating potential that governs the system's energy spectrum. The potential is defined as	
\begin{eqnarray}
	V_{i}= \lambda(1-X)  cos(2 \pi \alpha i)+\lambda_p X V_{p}(i)
\end{eqnarray}	
where $\lambda$ and $\lambda_{p}$ denote the disorder strengths of the AA and additional quasi-periodic potentials, respectively. The parameter $X \in [0,1]$ controls the relative contribution of the two disorder types. We consider $X = 0.2, 0.4, 0.6, 0.8$ to study the gradual interplay between the disorders. Here, $V_p(i)$ represents the additional quasi-periodic disorder, with $p = 0, 1, 2$ corresponding to TM, RS, and Fibonacci sequences, respectively, and $X$ (ranging from 0 to 1) acts as an interpolation parameter that modulates the transition between the two disorder types.\\
In figure (\ref{IEB}(a-d) shows the energy spectrum interpolation between AA($\lambda=5$) and TM ($\lambda_0=5$) as a function of magnetic flux. The evolution of the energy spectra with increasing $X$ reveals a non-monotonic dependence of the central gap on disorder. At low $X$ (0.2), the AA potential establishes two symmetric gap near $B=\pi$ within a fractal spectrum. As $X$ increases, TM disorder disrupts the fractal symmetry, initially reducing the gap ($X=0.4$) before reinstating it at higher $X$ (0.6-0.8) through localization. This non-monotonic evolution underscores the competition between quasiperiodicity-induced bandgaps and disorder-driven localization in strong-potential regimes. The $X=0.8$ spectrum, with its wide central gap, denotes the onset of a localized phase dominated by TM disorder. This behavior highlights the competition between quasiperiodicity-induced bandgaps and disorder-driven localization, where the gap first closes due to loss of AA coherence and later reopens under TM-dominated localization. In Fig. \ref{IEB} (e-h), the energy spectra illustrate the interpolation between AA and Fibonacci disorders as a function of magnetic flux $ B $, with AA disorder ($ \lambda = 5 $) and Fibonacci strength ($ \lambda_2 = 5 $). At $X=0.2$, the AA disorder dominates, resulting in two broader central gaps and a complete loss of the butterfly spectrum's fractal structure. As $ X $ increases, these gaps begin to narrow, and at $ X = 0.8 $, the lower gap closes while the upper gap persists, highlighting the overriding influence of AA disorder due to the relatively weak contribution of the Fibonacci disorder. Details of the interpolation between the AA quasiperiodic potential and the RS binary disorder are provided in the Appendix.\\ 
	\begin{figure*}[!htb]
		\centering	
		\begin{subfigure}{.24\textwidth}
			\centering
			\includegraphics[width=\linewidth]{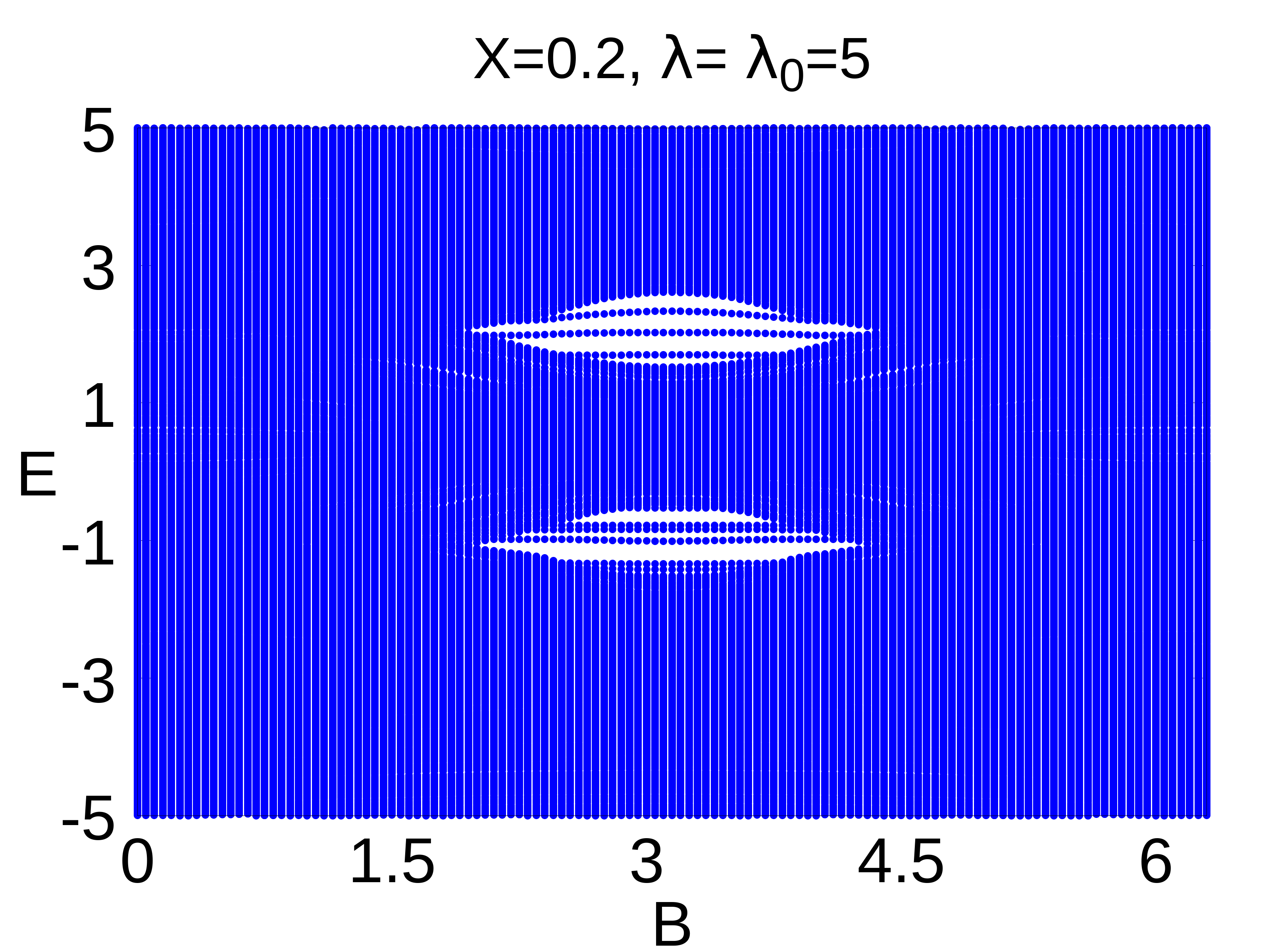}
			\subcaption{}
			\label{IEB5}
		\end{subfigure}%
		\begin{subfigure}{.24\textwidth}
			\centering 
			\includegraphics[width=\linewidth]{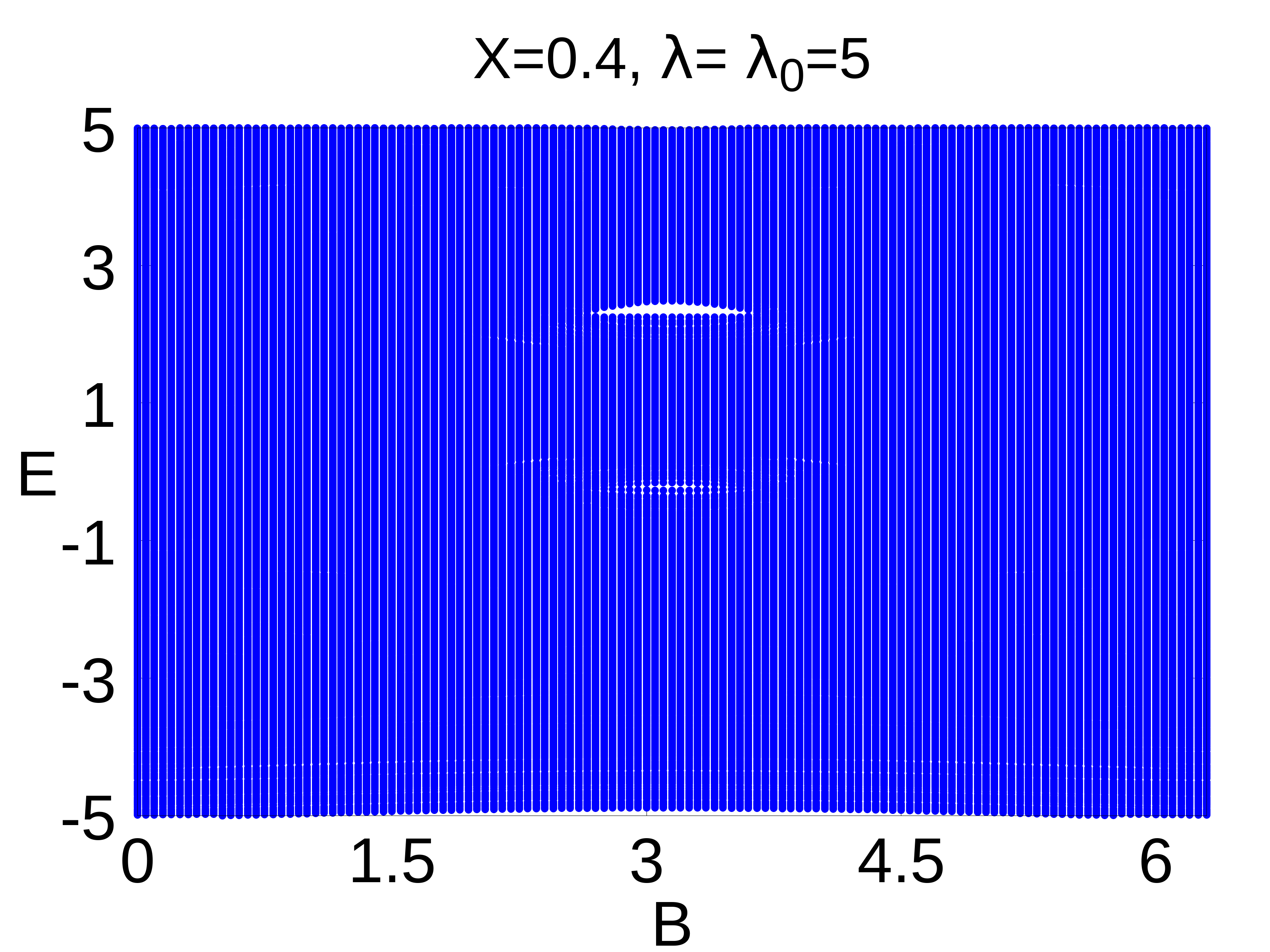}
			\subcaption{}
			\label{IEB6}
		\end{subfigure}%
		\begin{subfigure}{.24\textwidth}
			\centering
			\includegraphics[width=\linewidth]{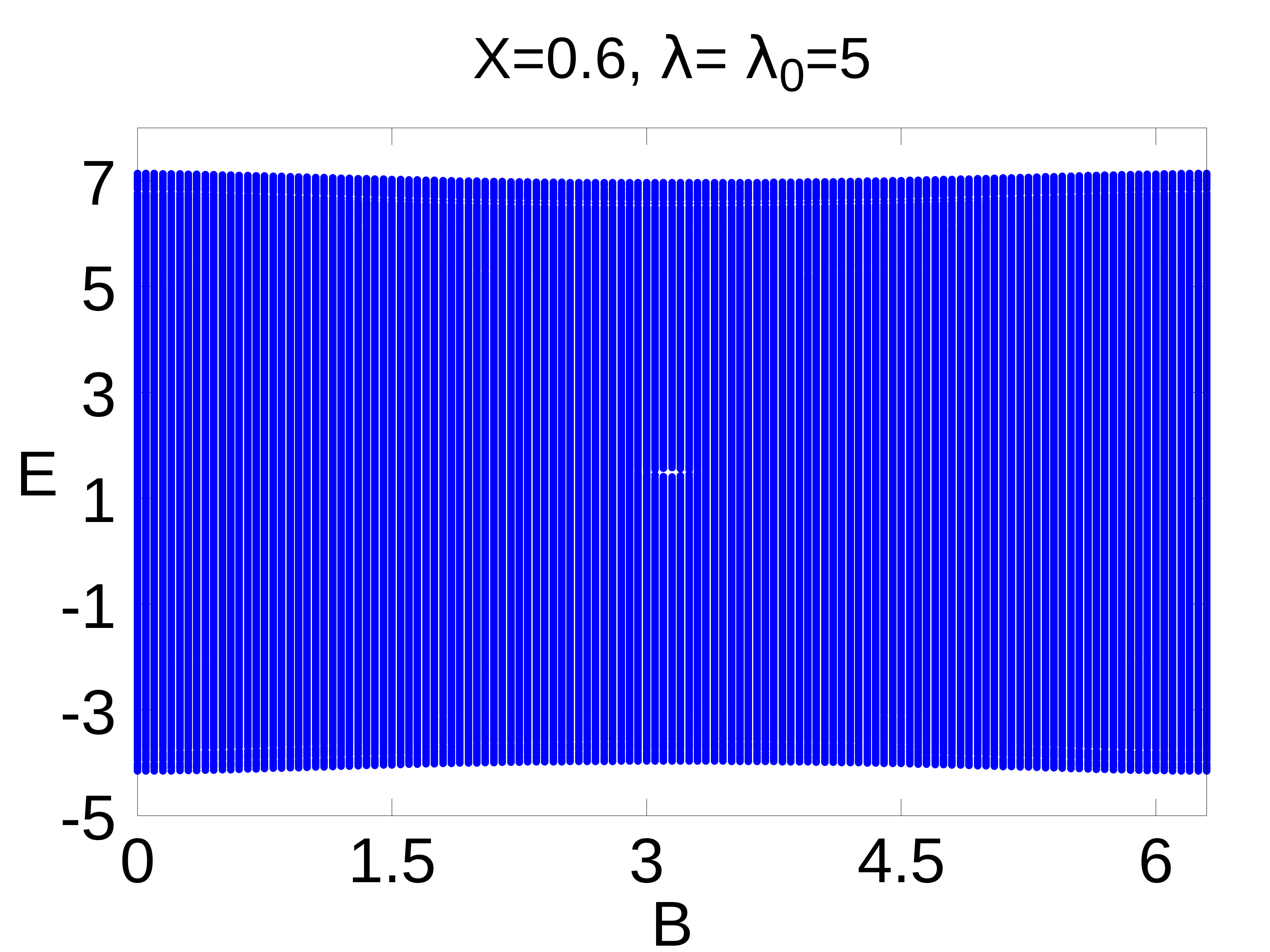}
			\subcaption{}
			\label{IEB7}
		\end{subfigure}%
		\begin{subfigure}{.24\textwidth}
			\centering
			\includegraphics[width=\linewidth]{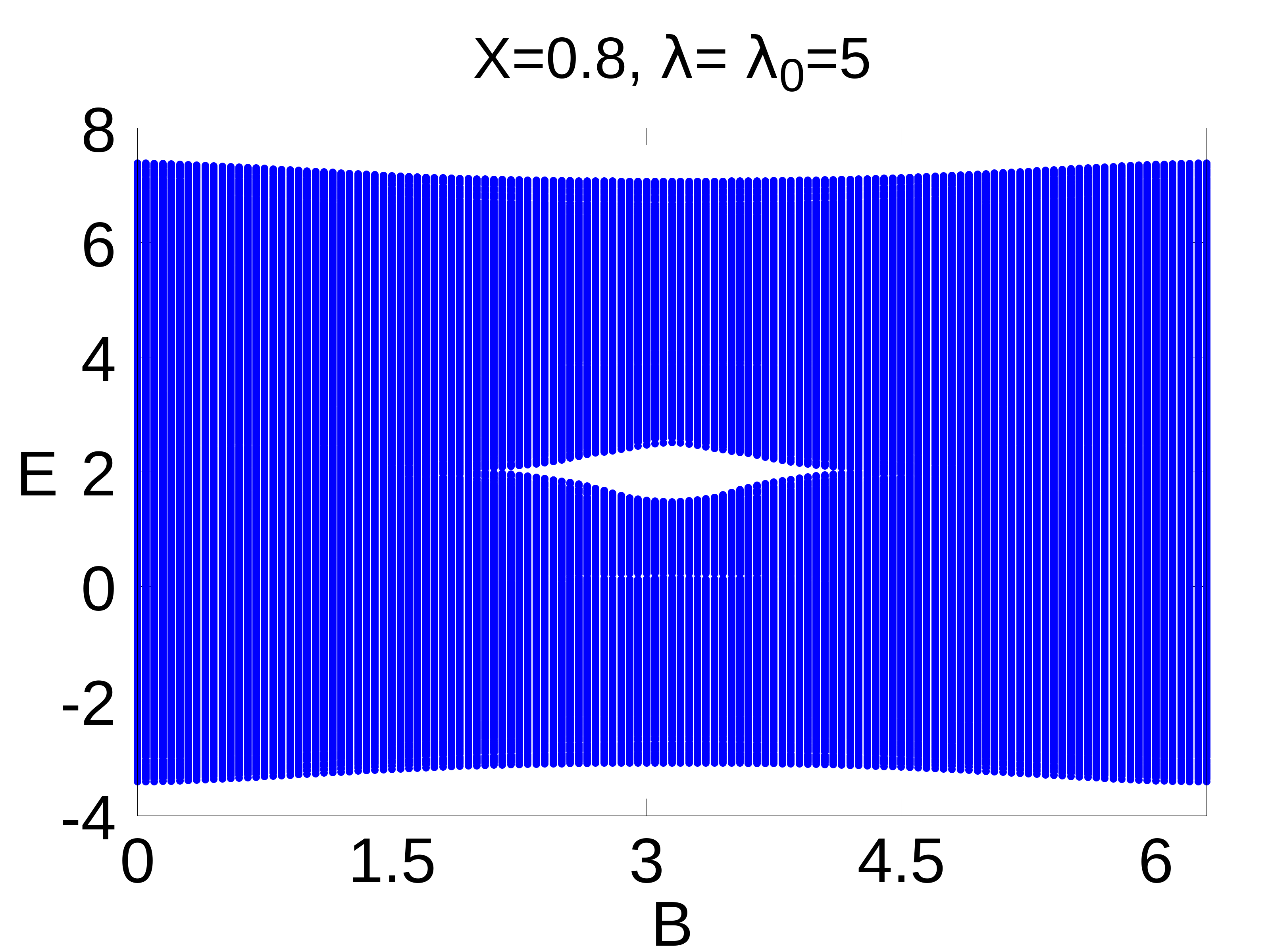}
			\subcaption{}
			\label{IEB8}
		\end{subfigure}%
		
\begin{subfigure}{.24\textwidth}
	\centering
	\includegraphics[width=\linewidth]{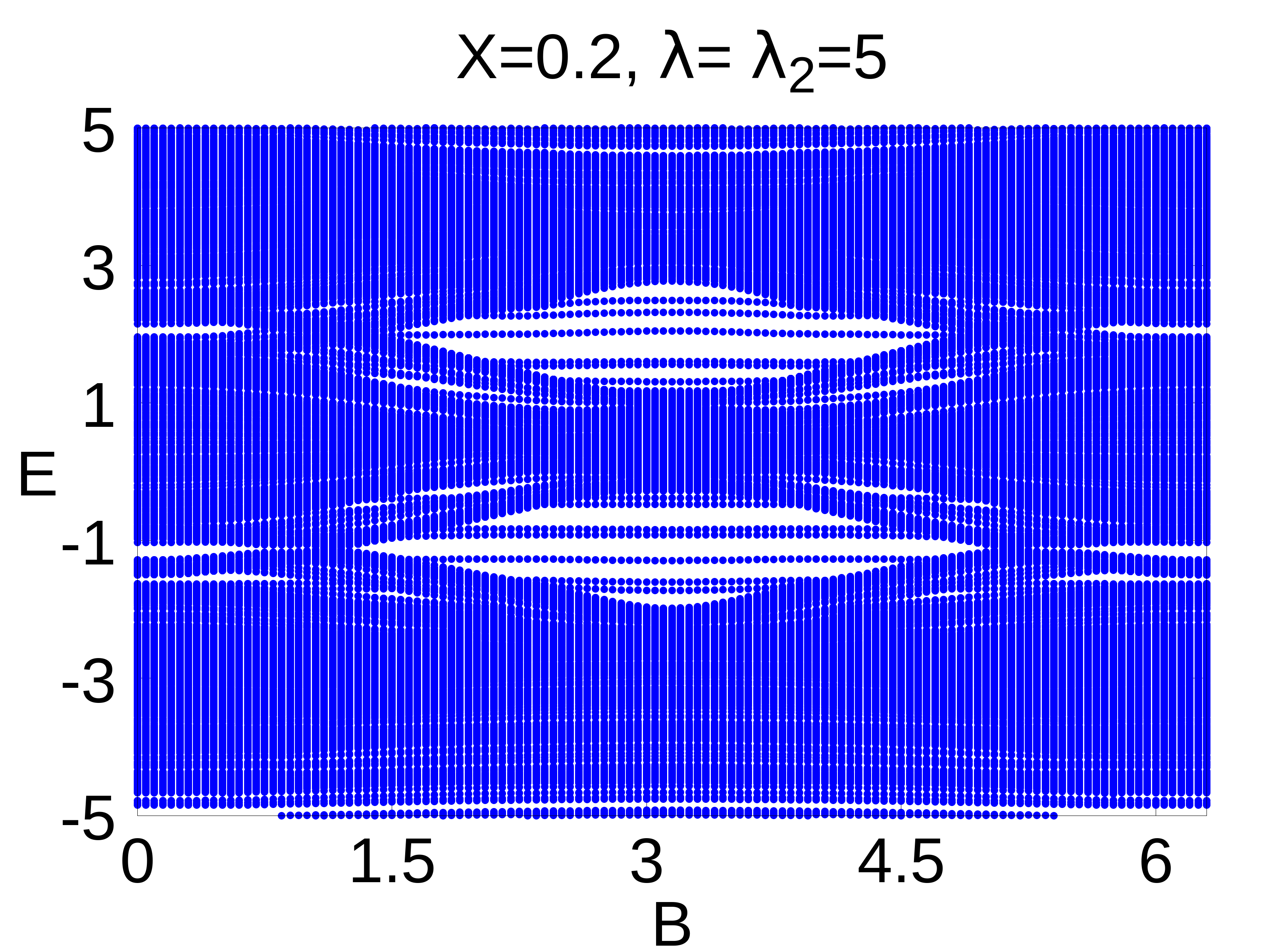}
	\subcaption{}
	\label{IEB21}
\end{subfigure}%
\begin{subfigure}{.24\textwidth}
	\centering 
	\includegraphics[width=\linewidth]{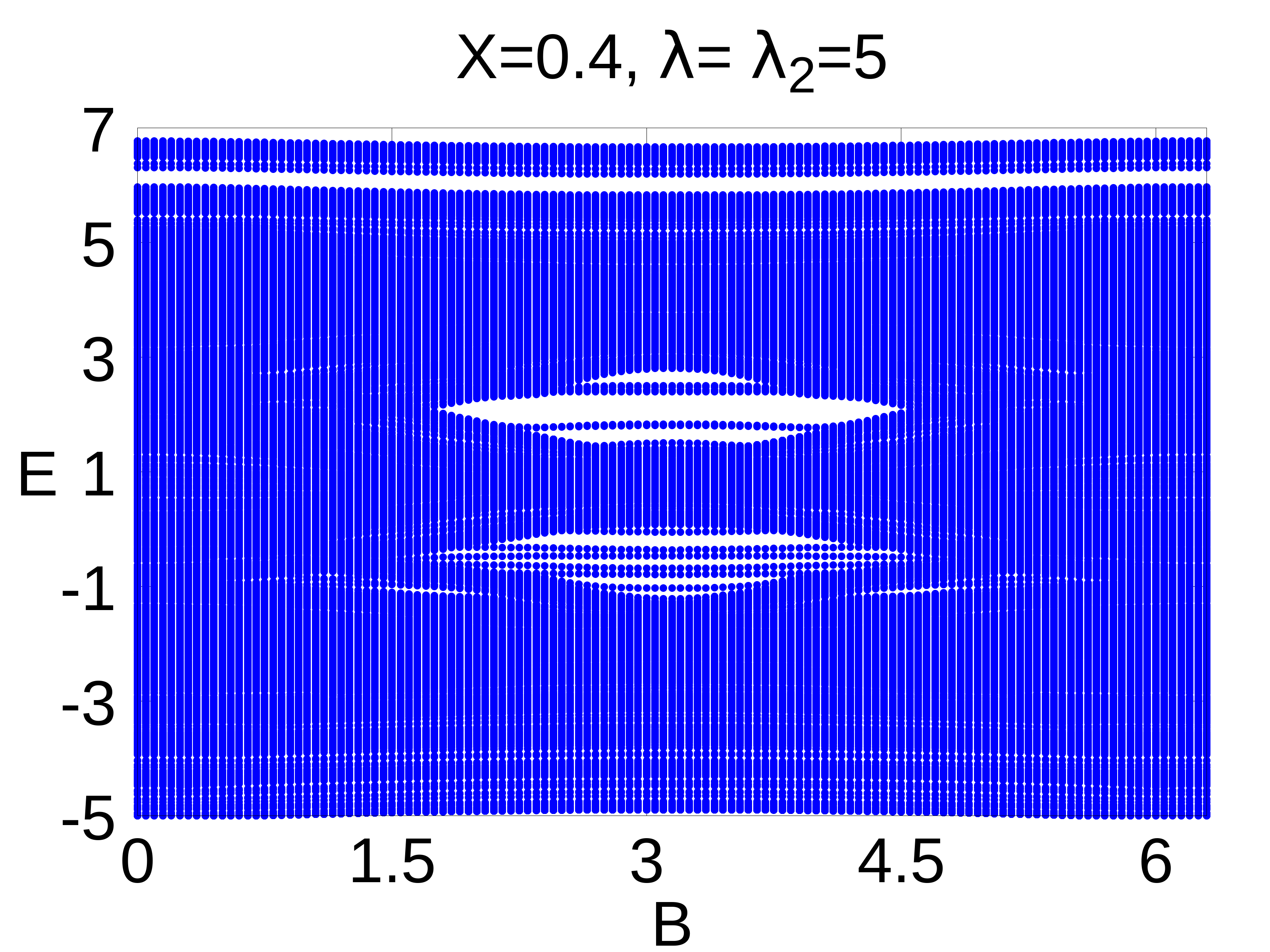}
	\subcaption{}
	\label{IEB22}
\end{subfigure}%
\begin{subfigure}{.24\textwidth}
	\centering
	\includegraphics[width=\linewidth]{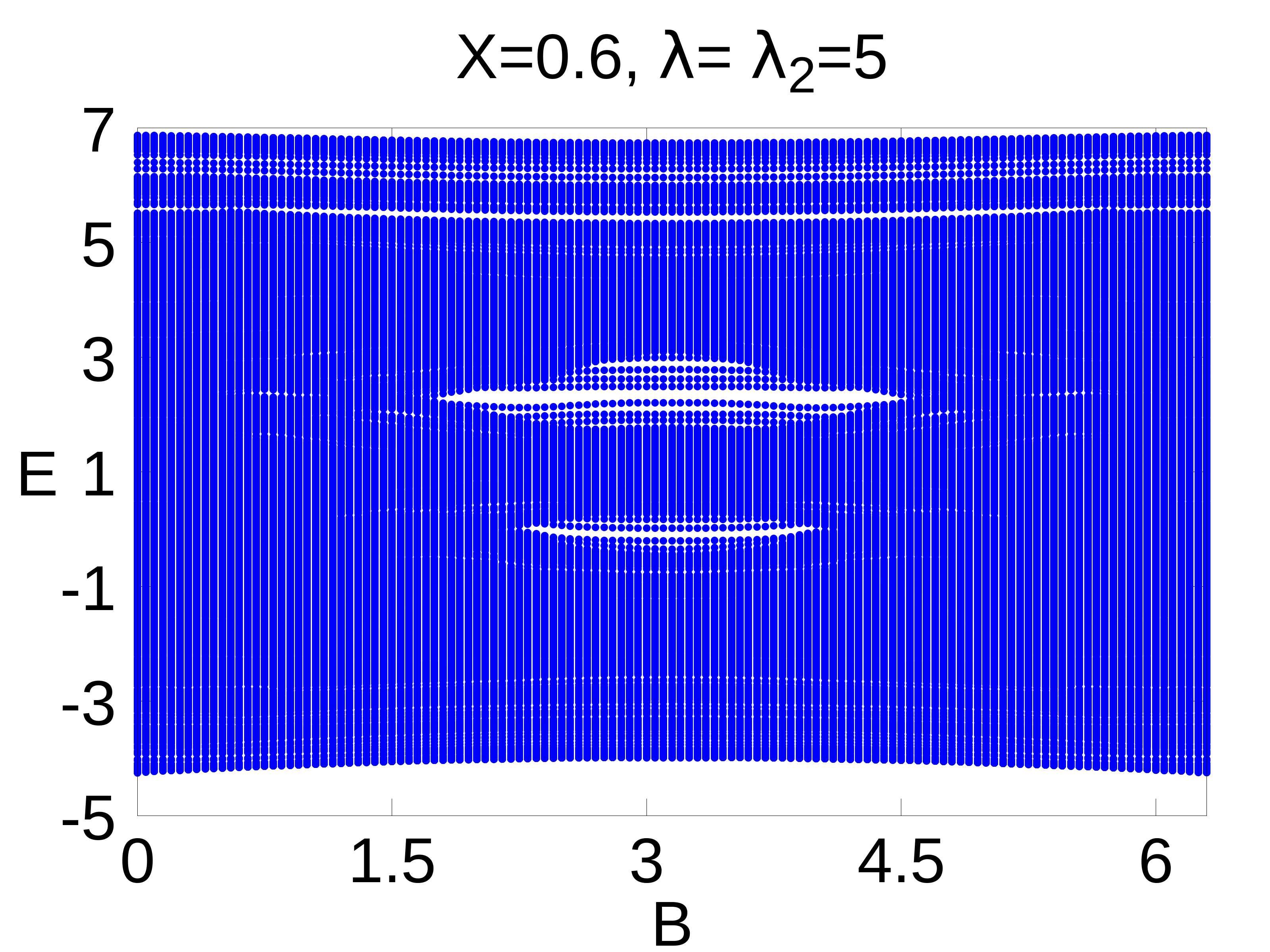}
	\subcaption{}
	\label{IEB23}
\end{subfigure}%
\begin{subfigure}{.24\textwidth}
	\centering
	\includegraphics[width=\linewidth]{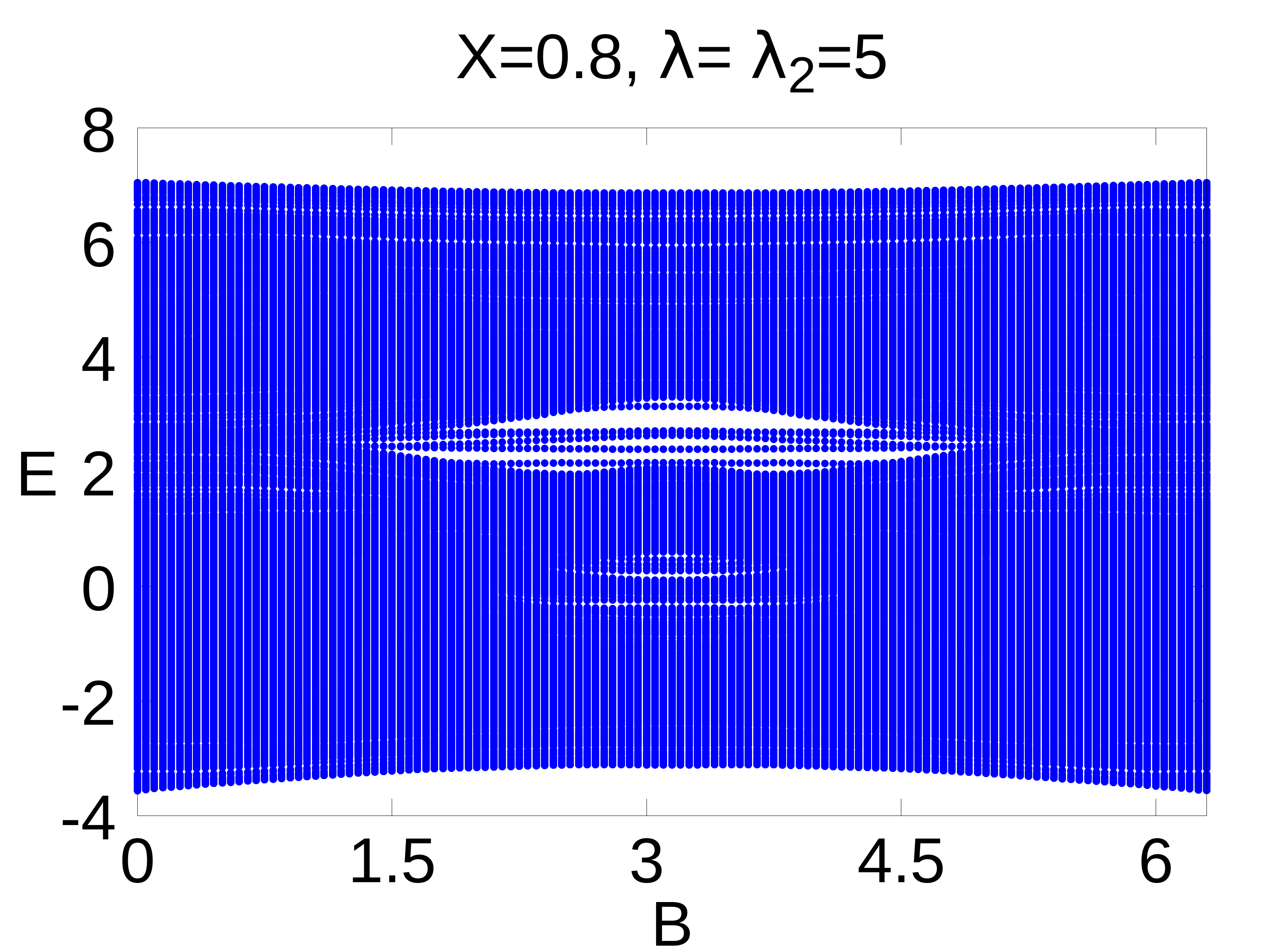}
	\subcaption{}
	\label{IEB24}
\end{subfigure}%
		
		\caption{Energy spectrum $E$ as a function of the magnetic field $B$ for a square lattice, showing the combined effects of the Aubry--André potential and other quasiperiodic disorders. The upper panel (a--d) illustrates the competition between AA and Thue--Morse disorders: the system evolves from the AA-dominated regime, characterized by two symmetric gaps, to the TM-dominated regime, where a single central gap appears. The lower panel displays the corresponding interpolation between the AA and Fibonacci disorders.
		    }
		\label{IEB}
	\end{figure*}	
	\section{Effect of interaction and disorder on entanglement}

\begin{figure*}[!htb]
		\centering
		\begin{subfigure}{.24\textwidth}
			\centering
			\includegraphics[width=\linewidth]{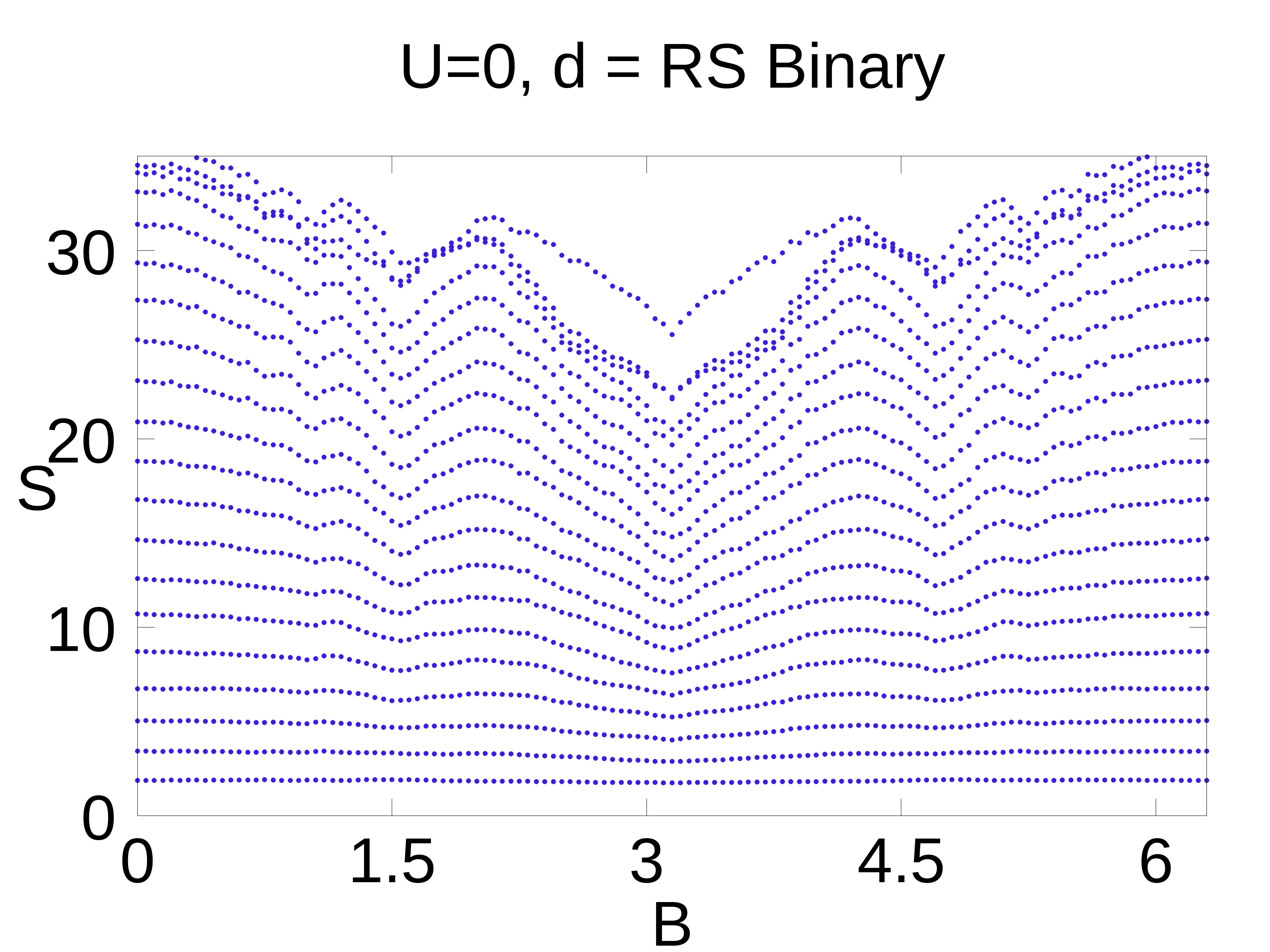}
			\subcaption{}
			\label{sq1}
		\end{subfigure}%
		\begin{subfigure}{.24\textwidth}
			\centering 
			\includegraphics[width=\linewidth]{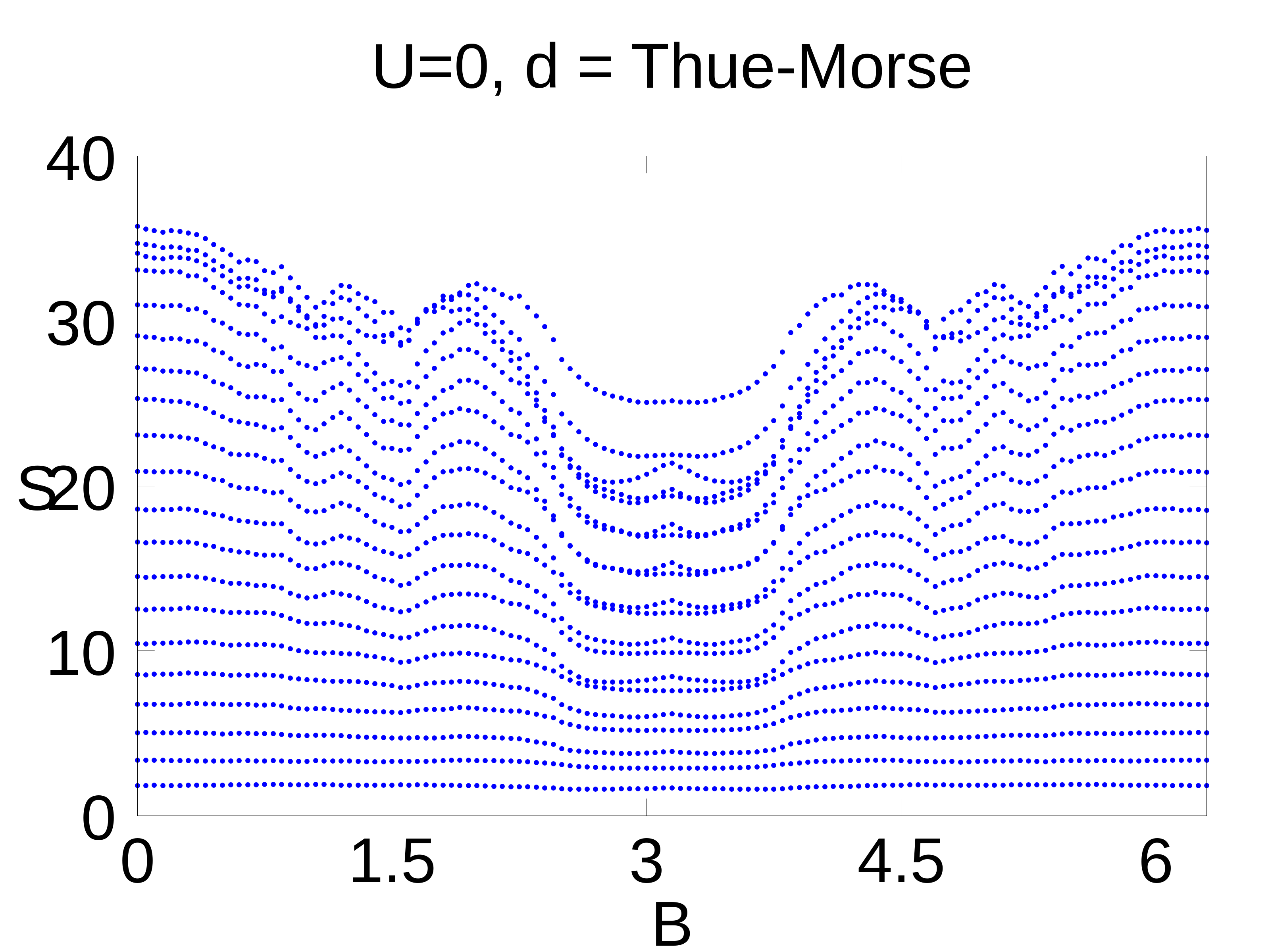}
			\subcaption{}
			\label{sq2}
		\end{subfigure}%
		\begin{subfigure}{.24\textwidth}
			\centering
			\includegraphics[width=\linewidth]{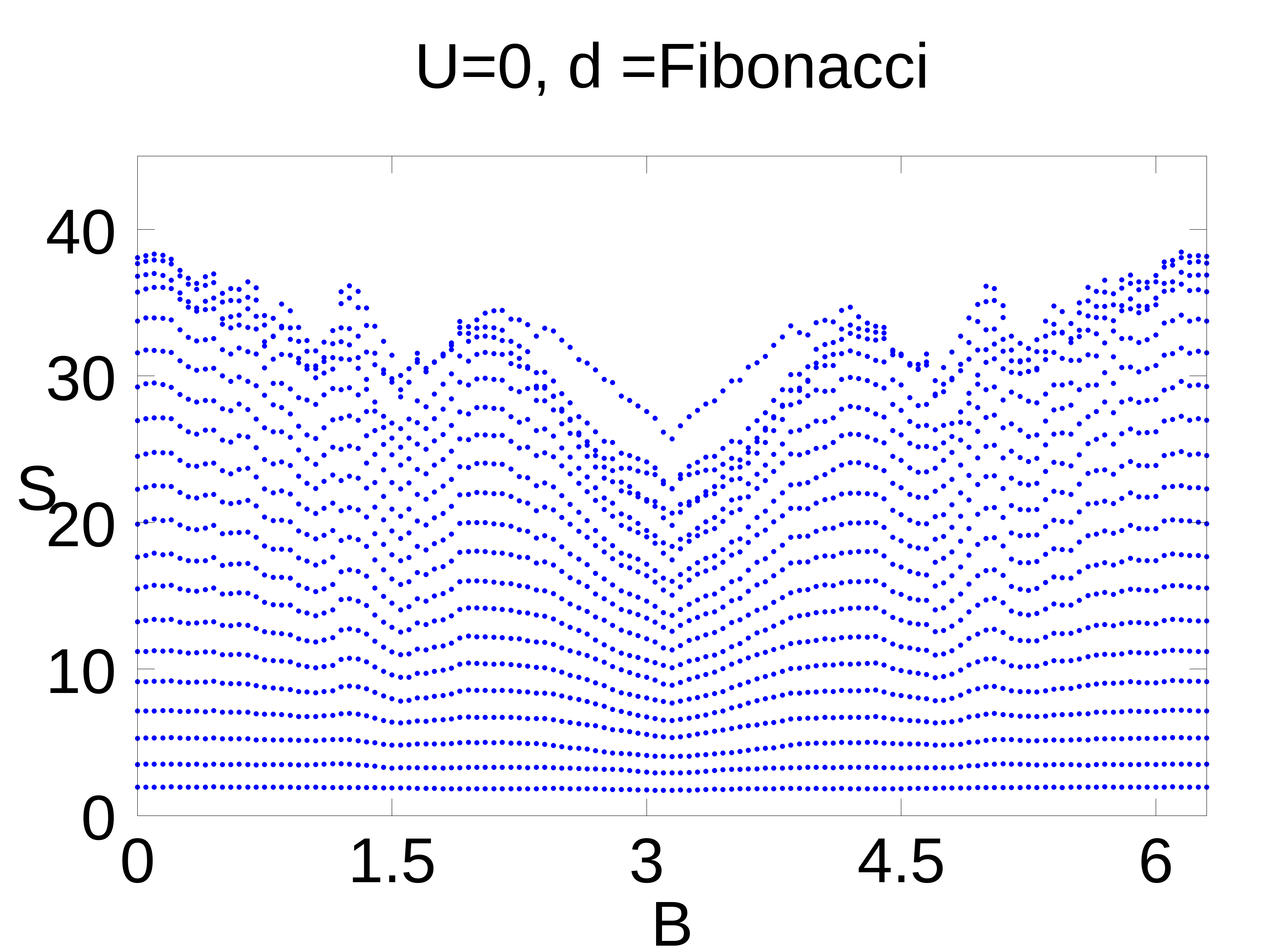}
			\subcaption{}
			\label{sq3}
		\end{subfigure}%
		\begin{subfigure}{.24\textwidth}
			\centering
			\includegraphics[width=\linewidth]{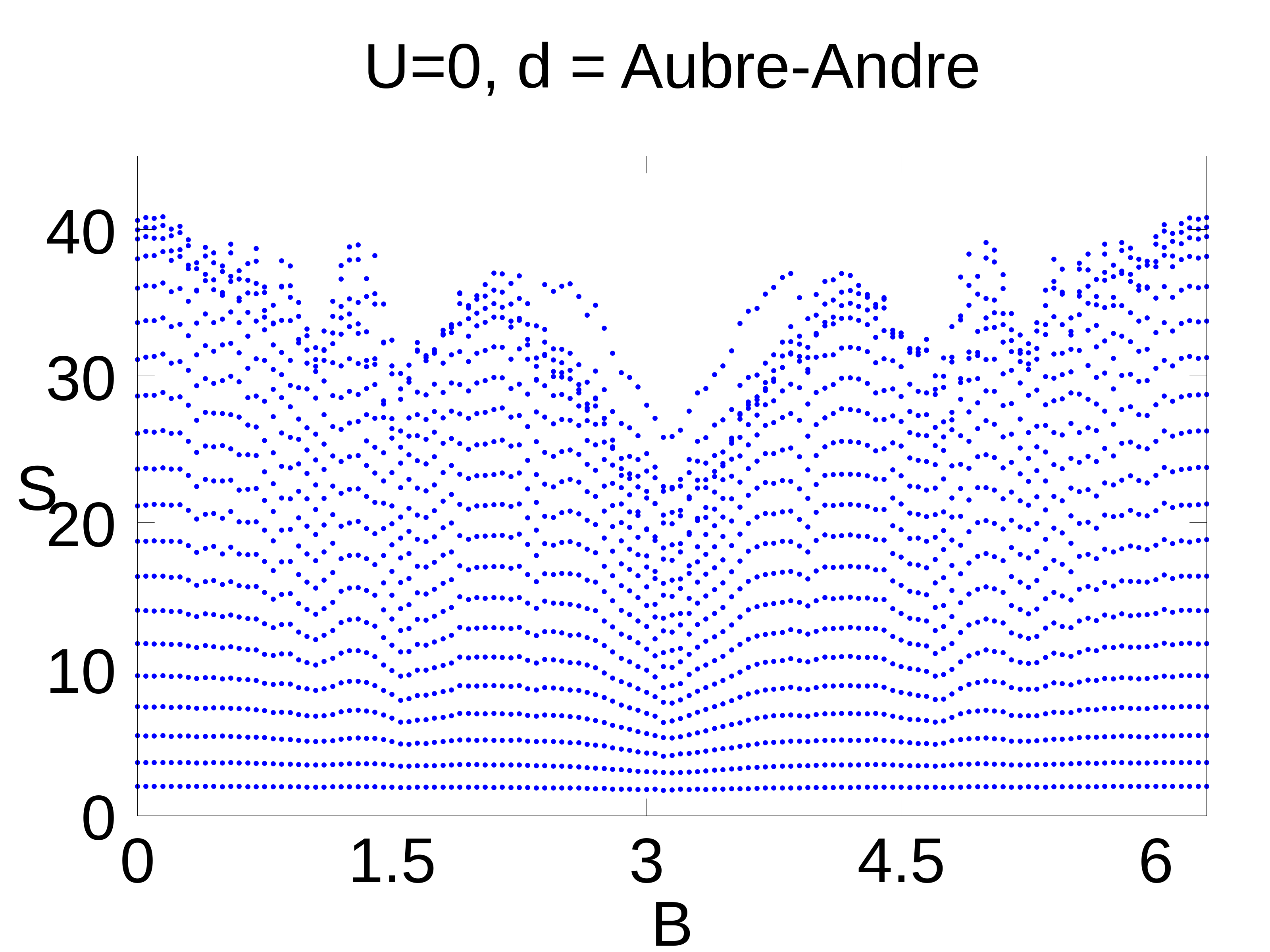}
			\subcaption{}
			\label{sq4}
		\end{subfigure}%

		\begin{subfigure}{.24\textwidth}
			\centering
			\includegraphics[width=\linewidth]{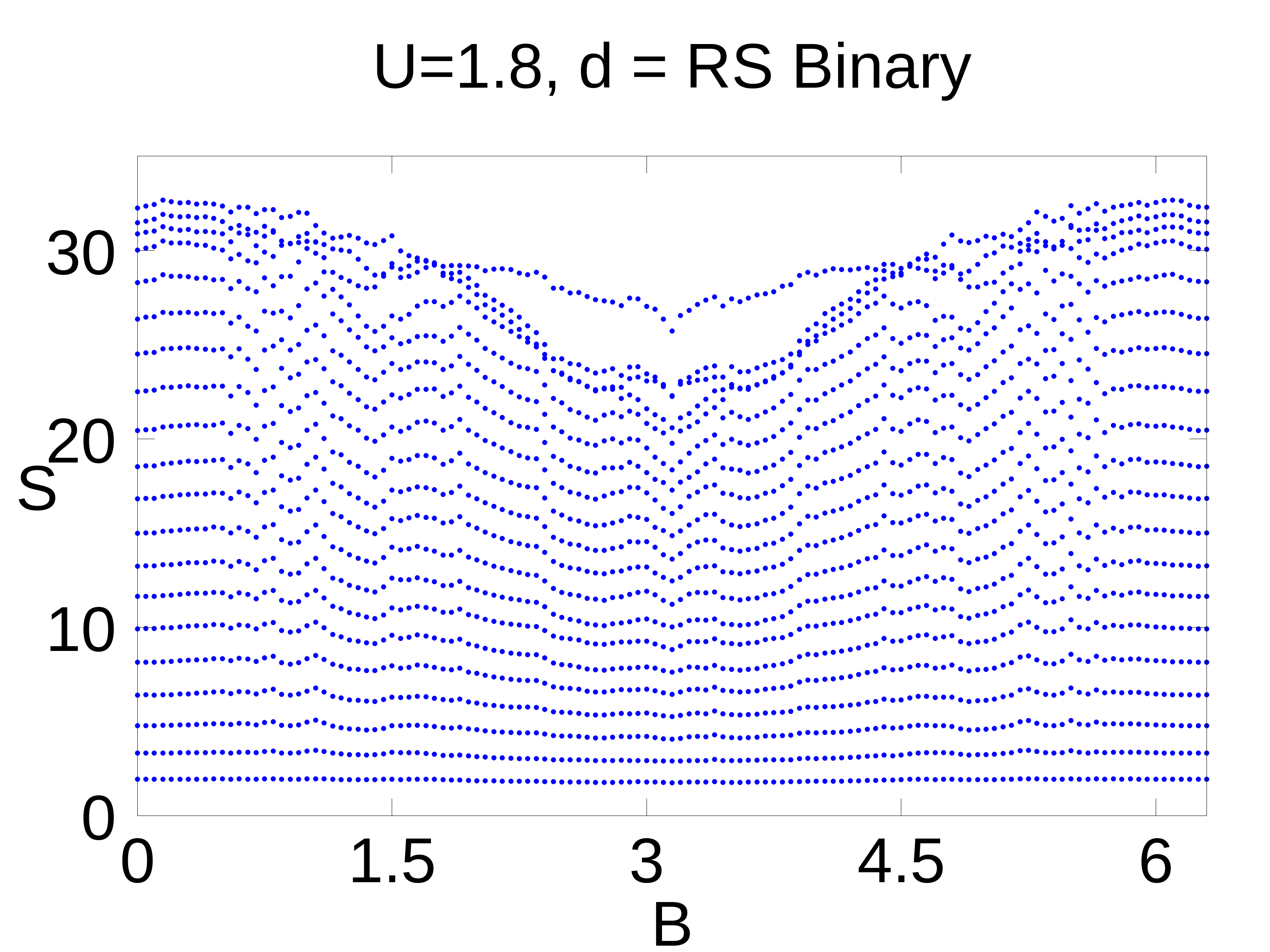}
			\subcaption{}
			\label{sq5}
		\end{subfigure}%
		\begin{subfigure}{.24\textwidth}
			\centering 
			\includegraphics[width=\linewidth]{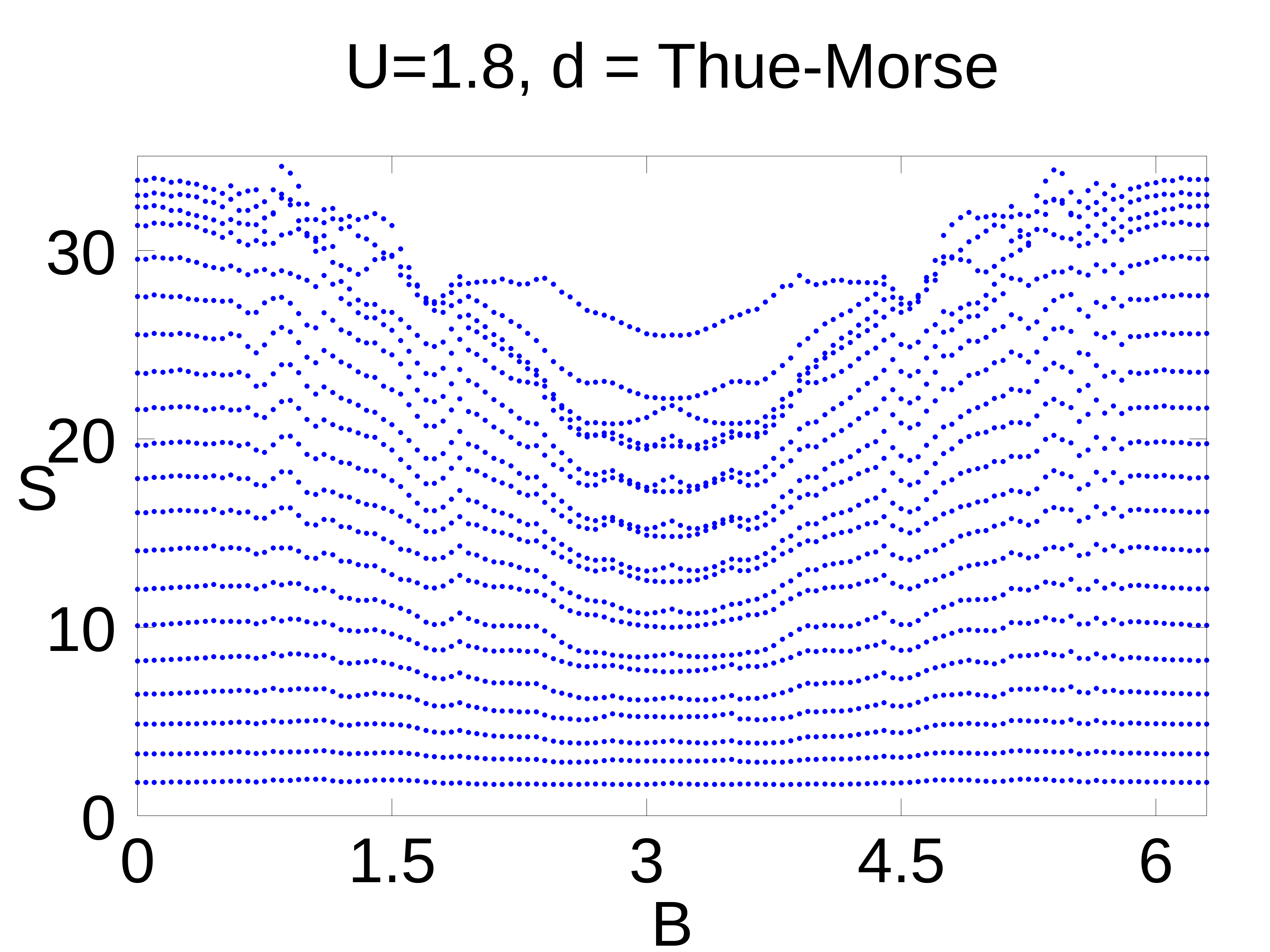}
			\subcaption{}
			\label{sq6}
		\end{subfigure}%
		\begin{subfigure}{.24\textwidth}
			\centering
			\includegraphics[width=\linewidth]{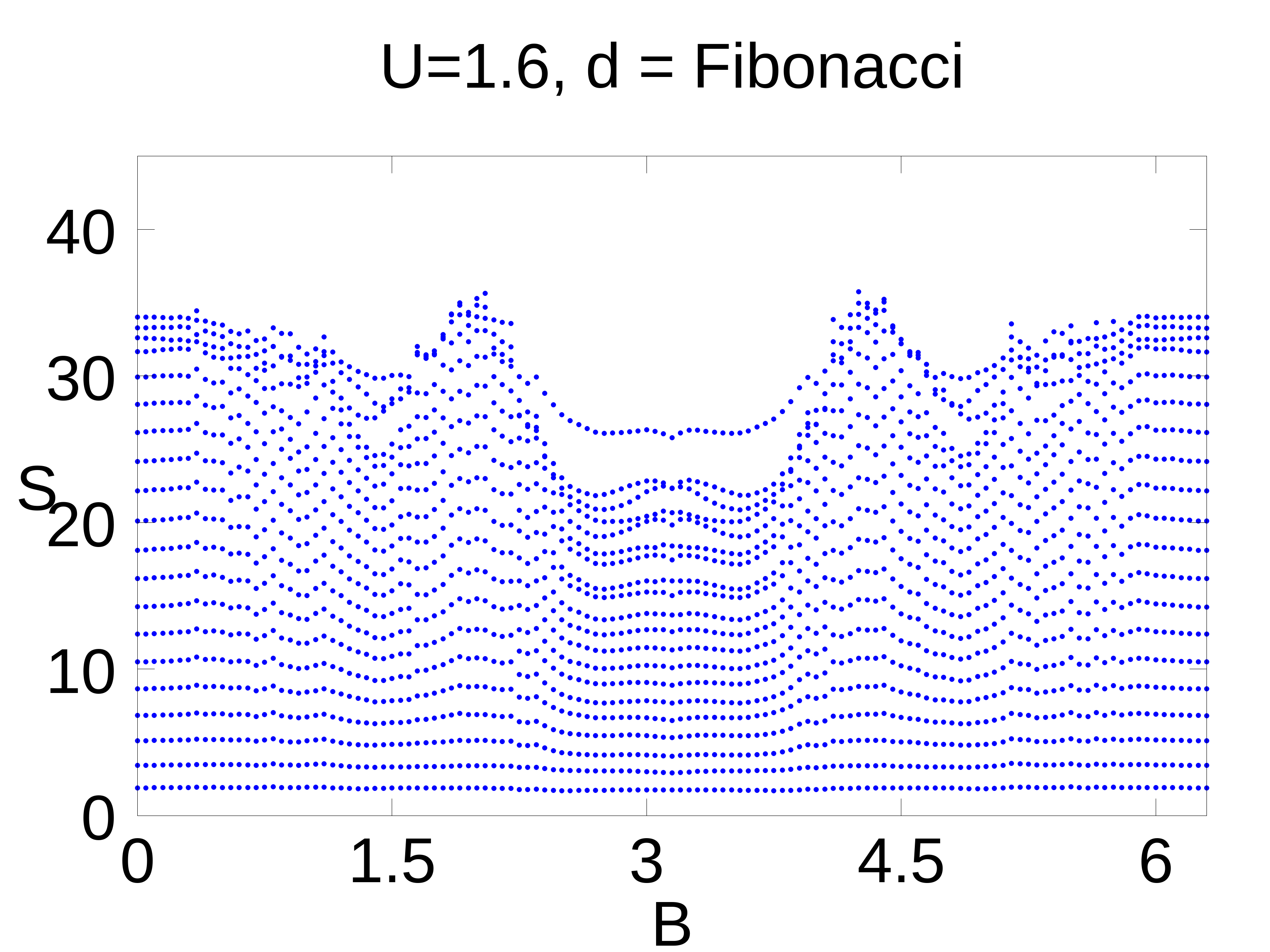}
			\subcaption{}
			\label{sq7}
		\end{subfigure}%
		\begin{subfigure}{.24\textwidth}
			\centering
			\includegraphics[width=\linewidth]{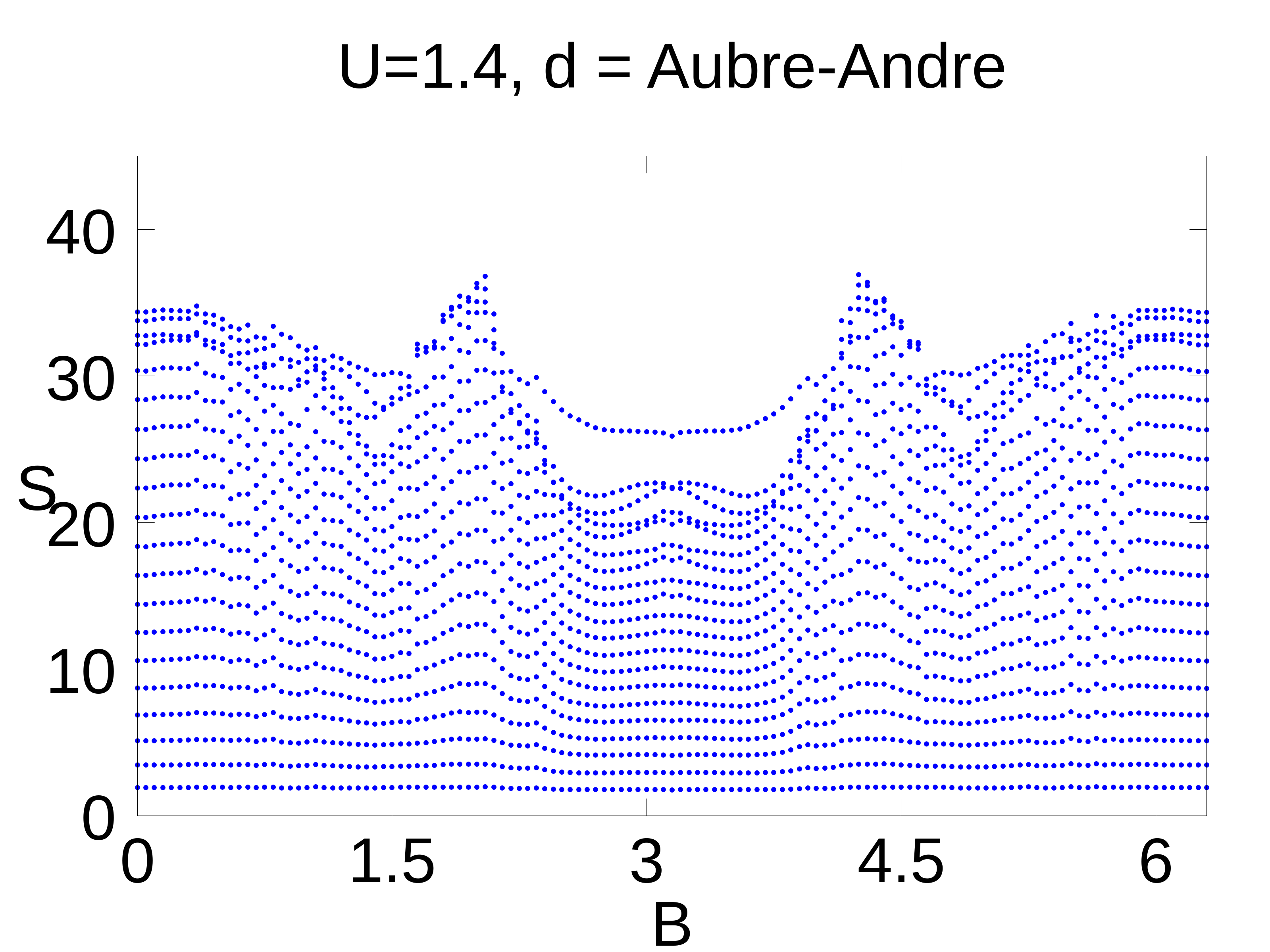}
			\subcaption{}
			\label{sq8}
		\end{subfigure}%
		
		\caption{ The etanglement entropy (S) is plotted as a function of the magnetic field (B) in a square lattice, illustrating the effects of various types of disorder and interaction. For a fixed value of the magnetic field, moving vertically corresponds to entanglement entropy values for different subsystem sizes, ranging from $4 \times 4$ up to $L/2 \times L/2$. The nearly equally spaced curves generally reflect an area-law scaling, although noticeable deviations occur, as discussed in the main text.}
		\label{ASB}
	\end{figure*}

	In this discussion, we explore the effects of interaction and various types of disorder on entanglement. Specifically, we have analyzed the entanglement entropy for different subsystem sizes and examined the correlation matrix spectrum for the largest subsystem, whose dimension is half of the full system. Figure \ref{ASB} presents the entanglement entropy under different conditions (as described earlier, involving interactions and disorders). In each plot, the horizontal axis represents the magnetic field, while the vertical axis denotes the entanglement entropy for various block sizes. For a fixed magnetic field, for example $B = 0$, the first blue point on the vertical line corresponds to the entanglement entropy of a $4 \times 4$ block, the next point represents a $5 \times 5$ block, and so on. The nearly uniform 
	spacing of these entropy values at a given magnetic field reflects the validity 
	of the area law. For all disorder types, the area law is well satisfied for 
	smaller subsystem sizes over the entire range of magnetic fields. It also holds 
	accurately in both the low- and high-field regimes, while deviations appear at 
	intermediate field strengths for larger subsystems. In the clean system, the 
	area law shows noticeable breakdown near $B = \pi$, whereas the inclusion of 
	disorder and interactions improves compliance with area-law behavior.
	In Figure \ref{ASB} among all disorders, RS disorder provides a better agreement with the area law of entanglement entropy compared to others. Additionally, when both disorder and interaction are present, the entanglement entropy becomes more evenly spaced for smaller subsystems. 	However, a departure from the area law is observed for larger subsystems. 	Interestingly, the entanglement entropy exhibits a local minimum at $B=\pi$ in the presence of disorder but shows a local maximum when interactions are included.
	
	\begin{figure*}[!htb]
	\centering
	\begin{subfigure}{.24\textwidth}
		\centering
		\includegraphics[width=\linewidth]{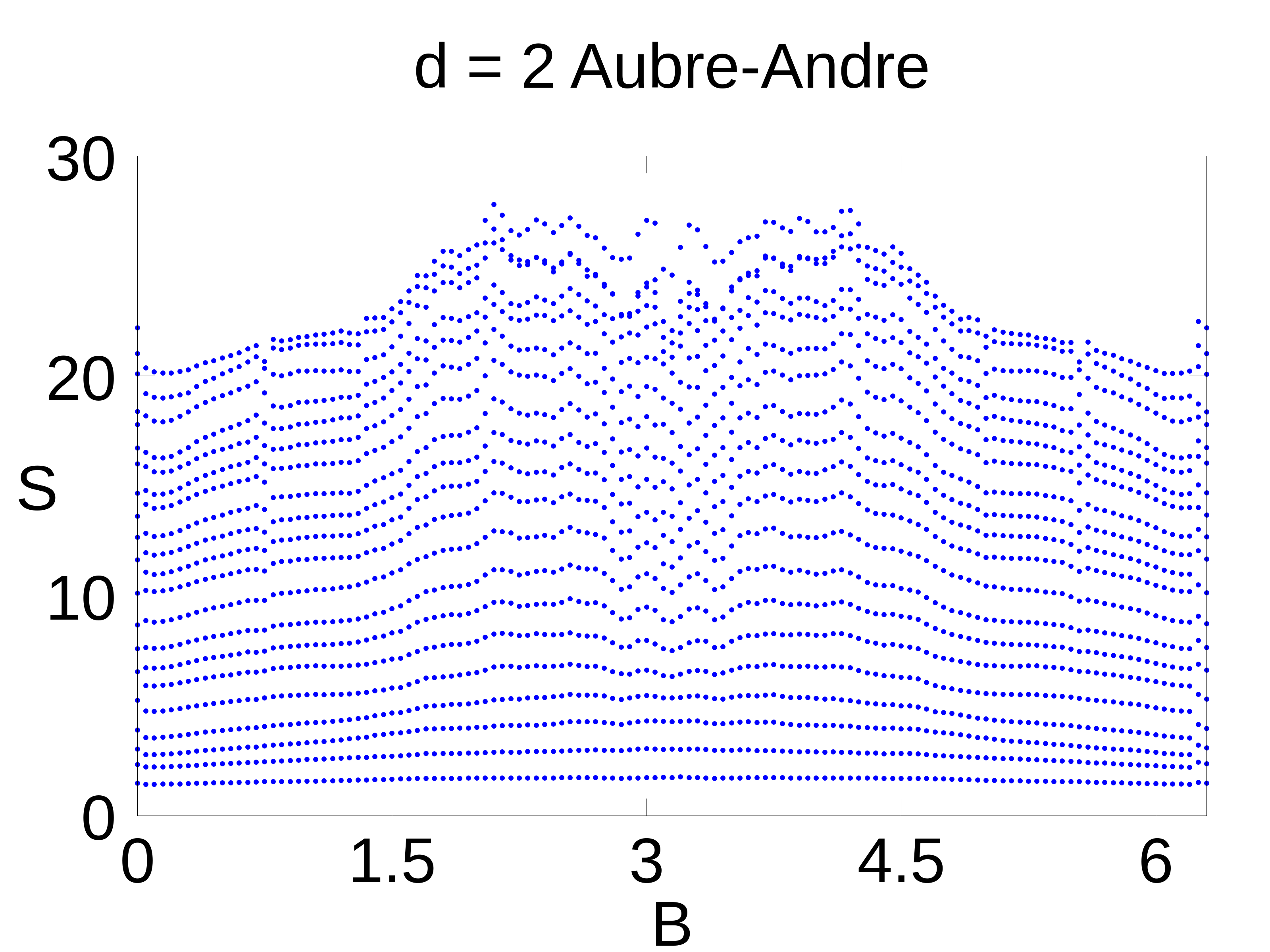}
		\subcaption{}
		\label{ARSB1}
	\end{subfigure}%
	\begin{subfigure}{.24\textwidth}
		\centering 
		\includegraphics[width=\linewidth]{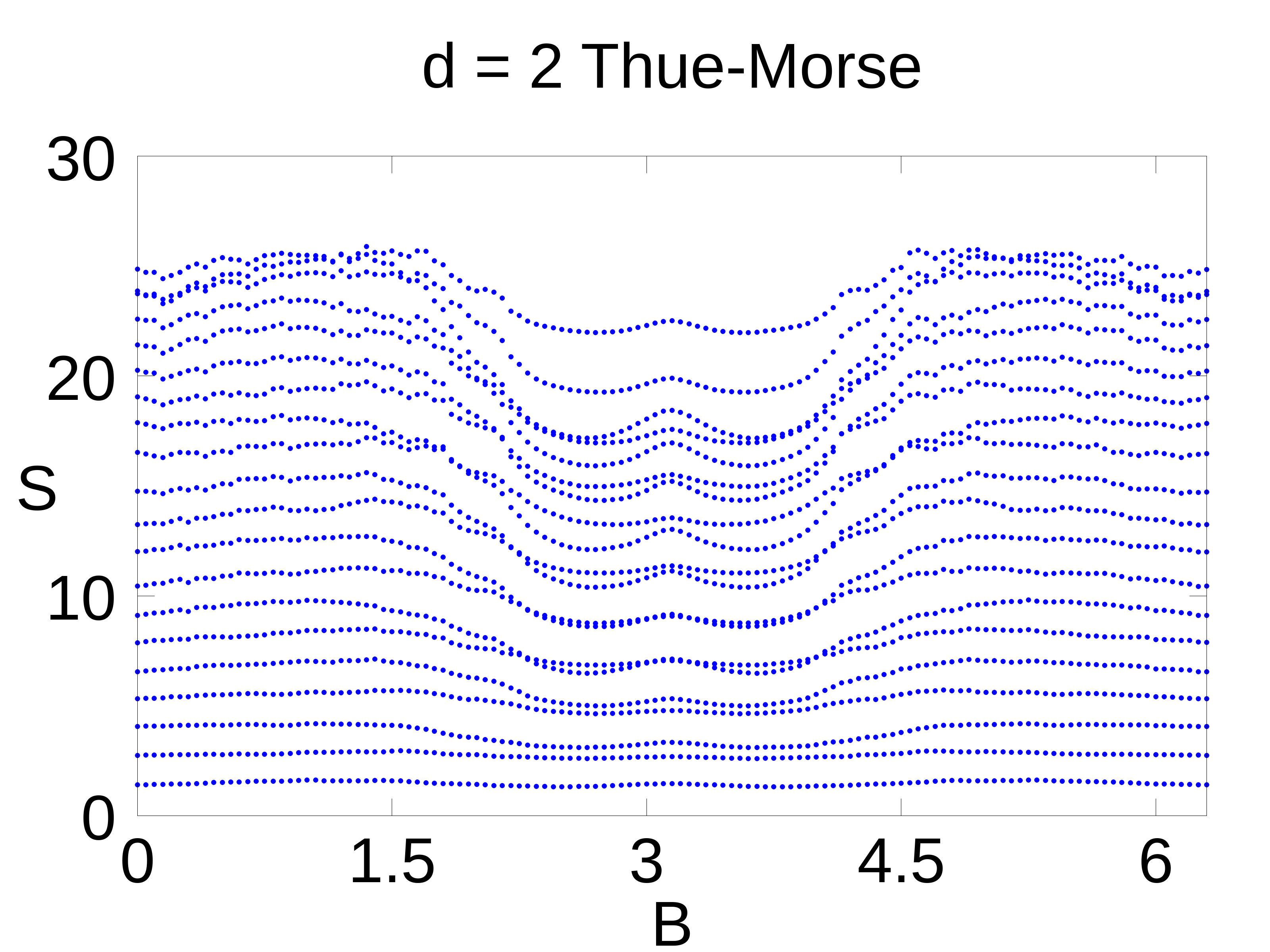}
		\subcaption{}
		\label{ARSB2}
	\end{subfigure}%
	\begin{subfigure}{.24\textwidth}
		\centering
		\includegraphics[width=\linewidth]{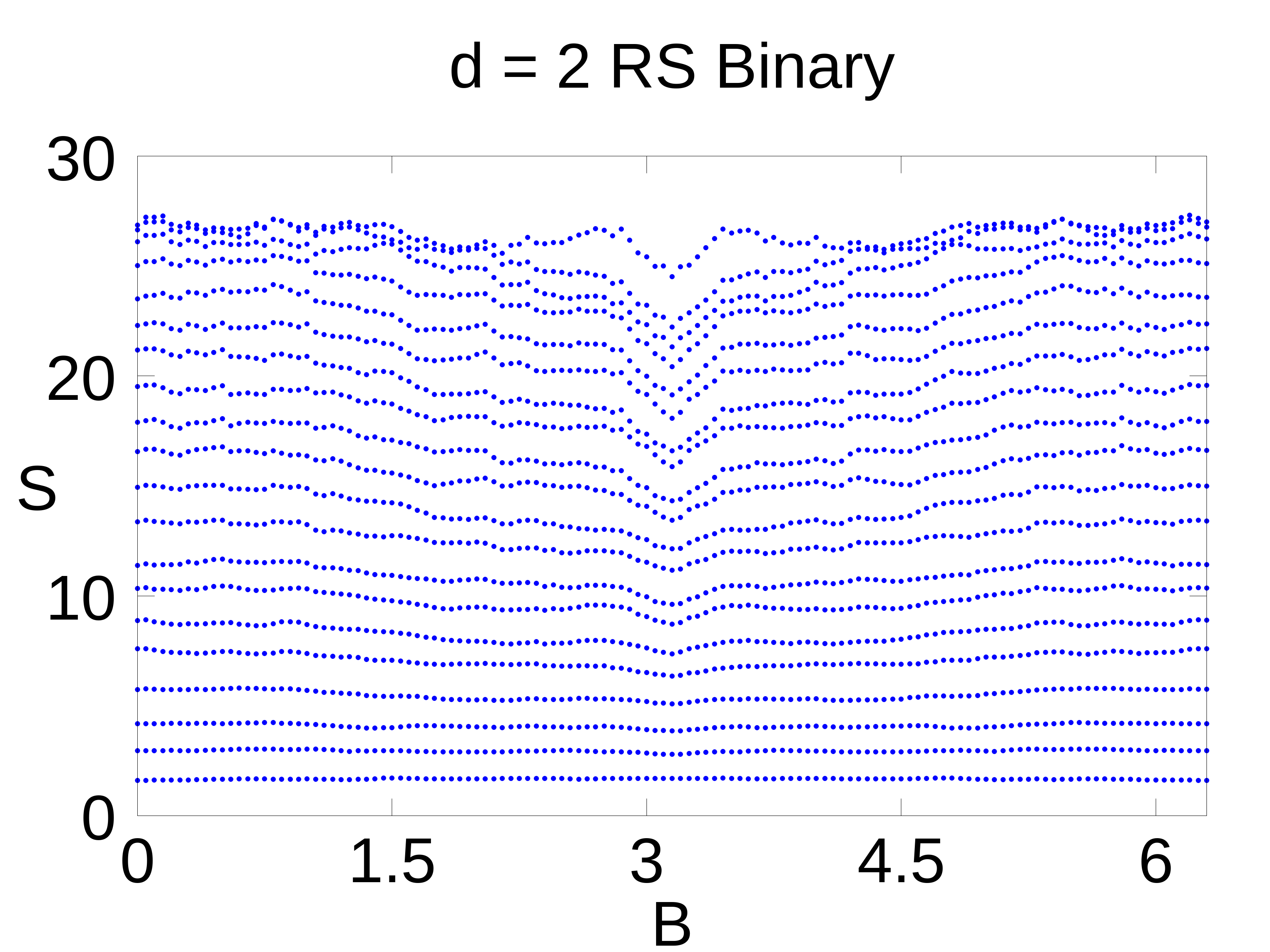}
		\subcaption{}
		\label{ARSB3}
	\end{subfigure}%
	\begin{subfigure}{.24\textwidth}
		\centering
		\includegraphics[width=\linewidth]{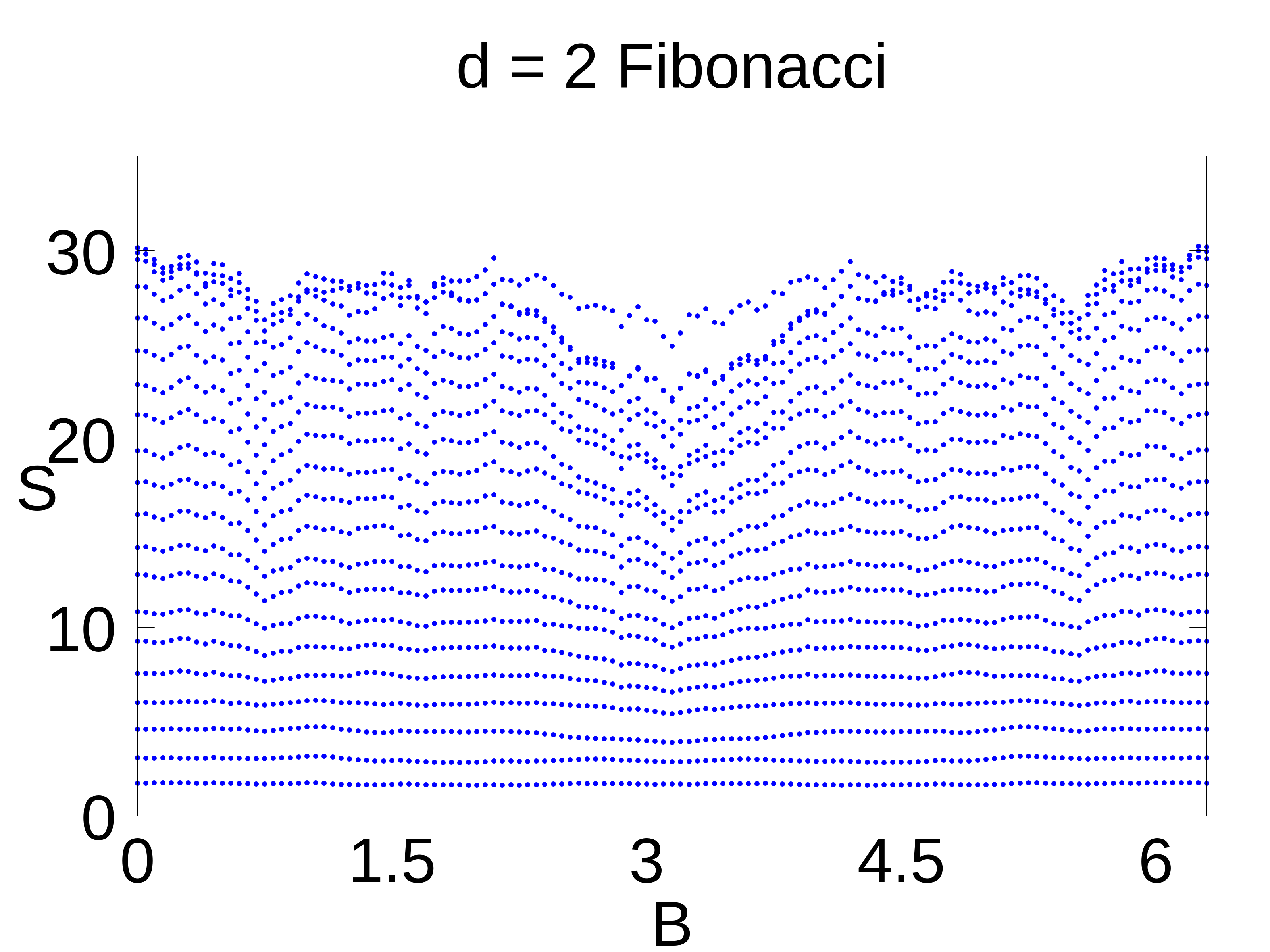}
		\subcaption{}
		\label{ARSB4}
	\end{subfigure}%

	\begin{subfigure}{.24\textwidth}
		\centering
		\includegraphics[width=\linewidth]{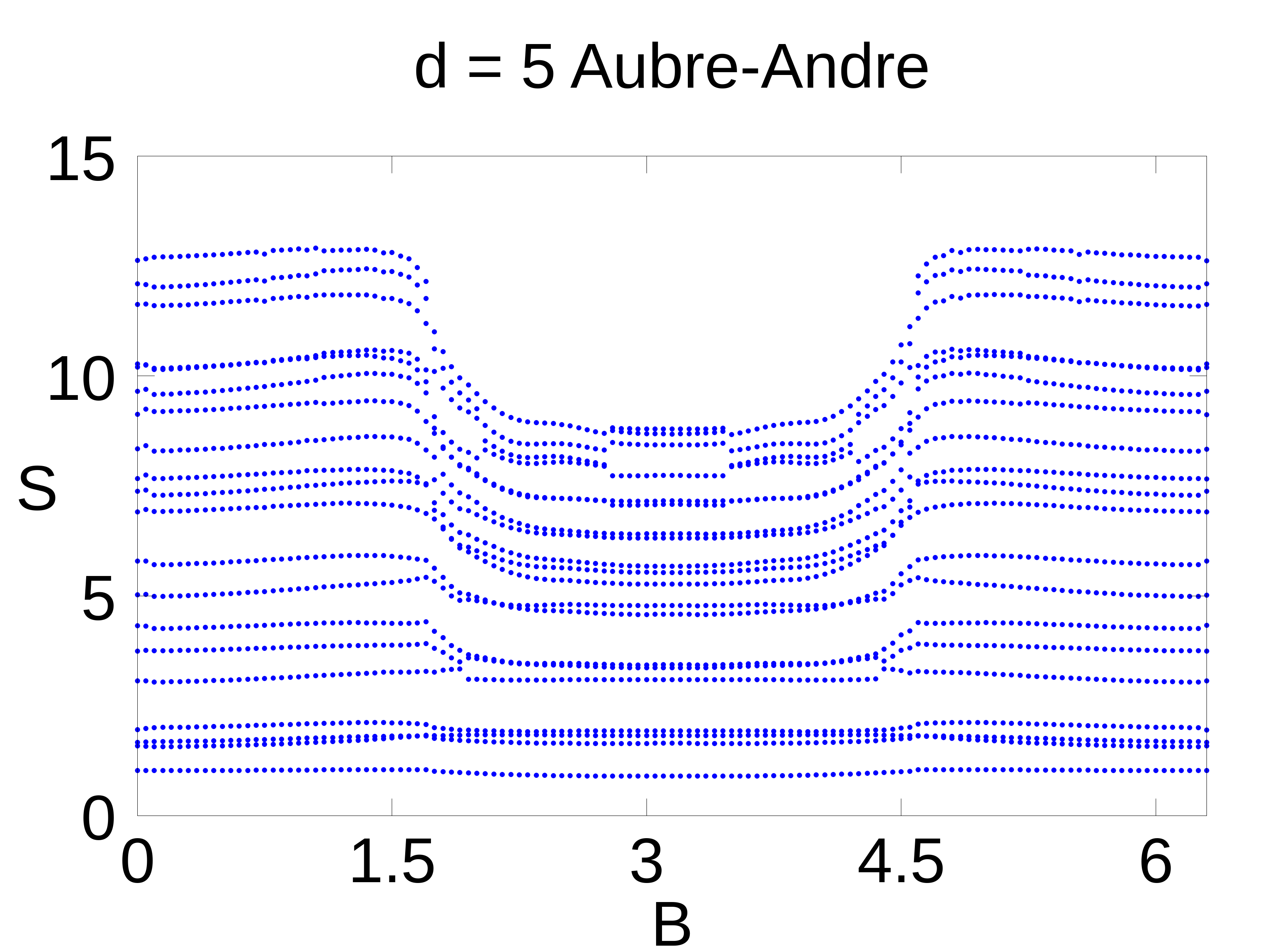}
		\subcaption{}
		\label{ARSB5}
	\end{subfigure}%
	\begin{subfigure}{.24\textwidth}
		\centering 
		\includegraphics[width=\linewidth]{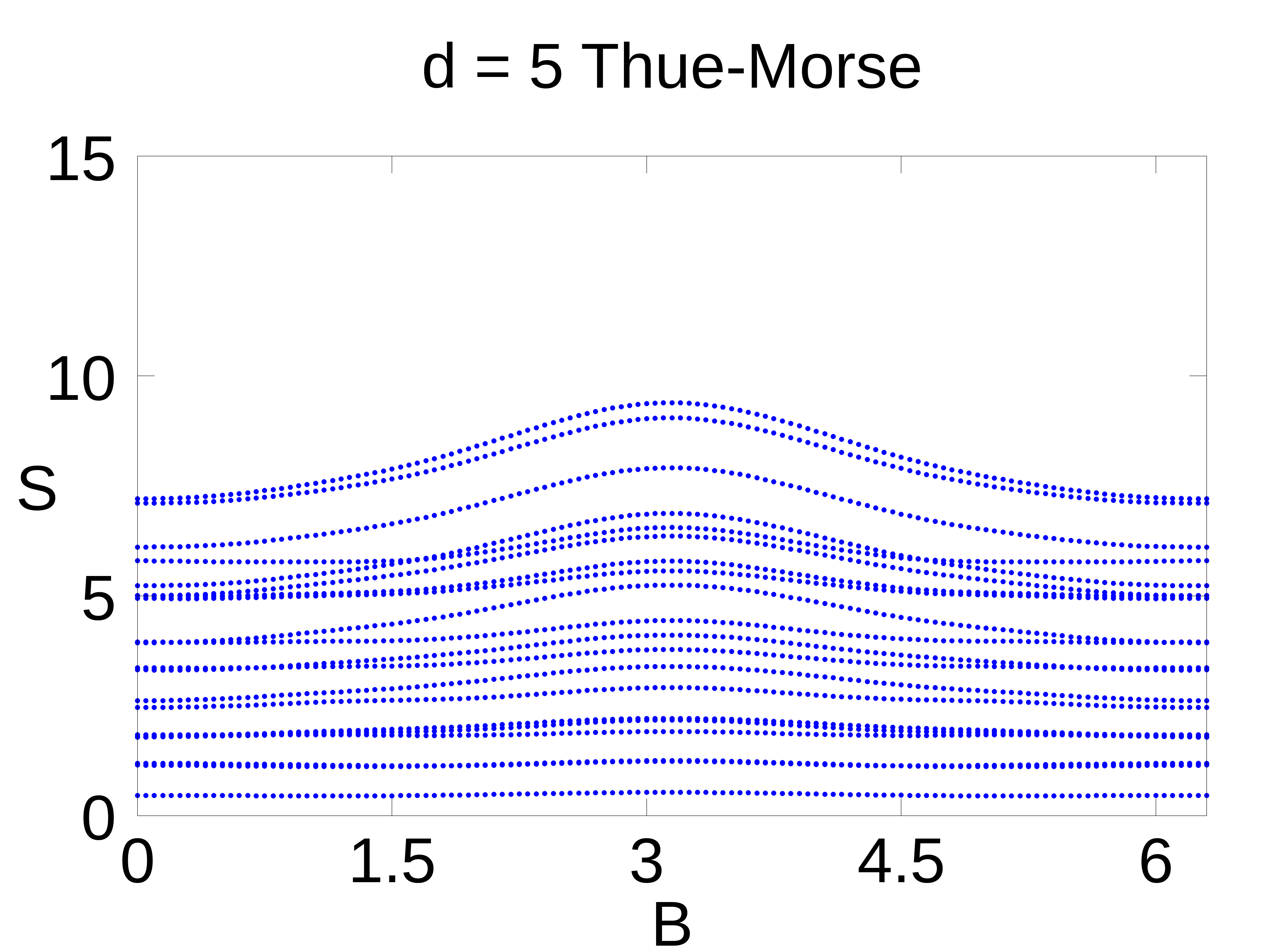}
		\subcaption{}
		\label{ARSB6}
	\end{subfigure}%
	\begin{subfigure}{.24\textwidth}
		\centering
		\includegraphics[width=\linewidth]{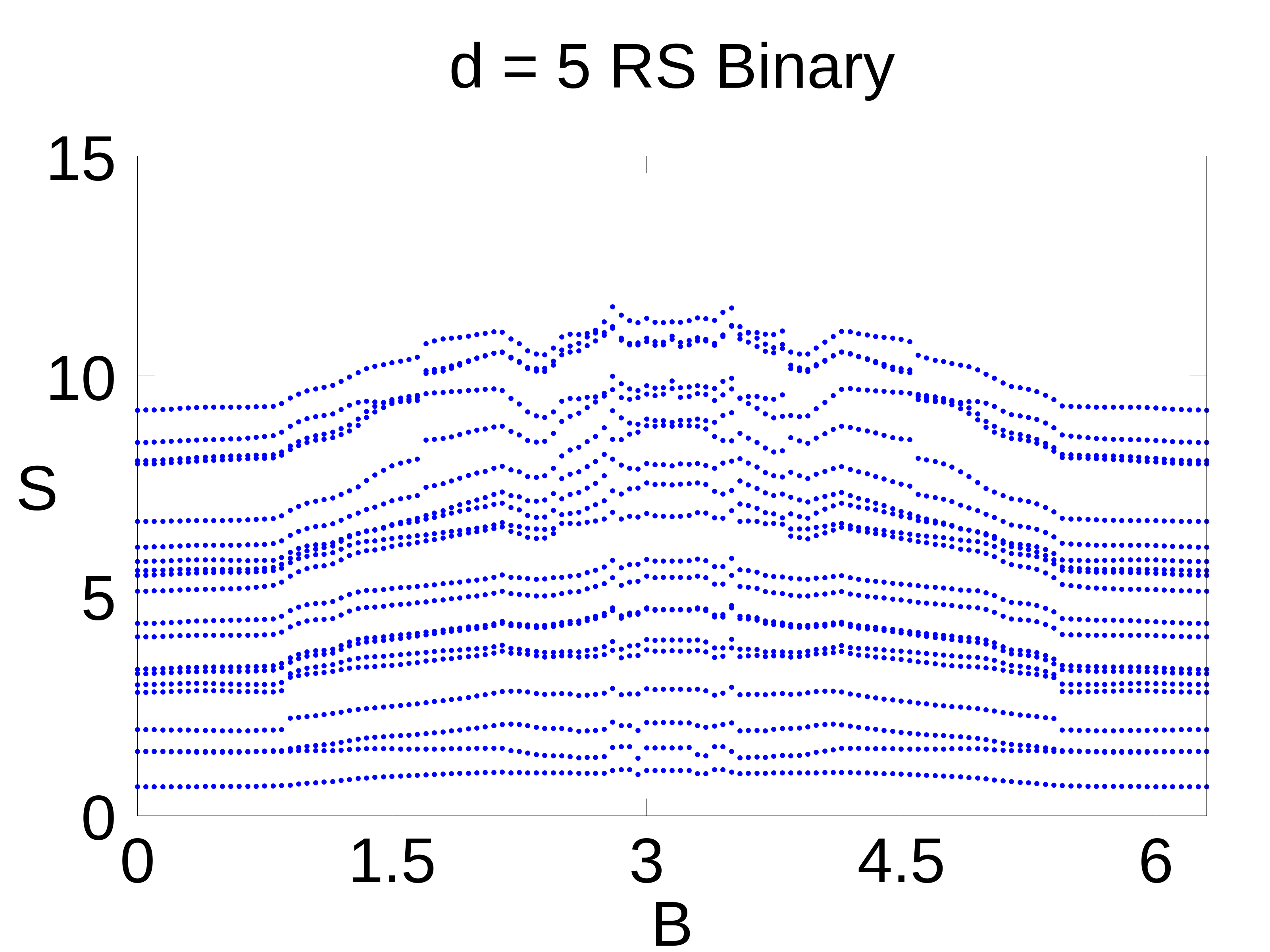}
		\subcaption{}
		\label{ARSB7}
	\end{subfigure}%
	\begin{subfigure}{.24\textwidth}
		\centering
		\includegraphics[width=\linewidth]{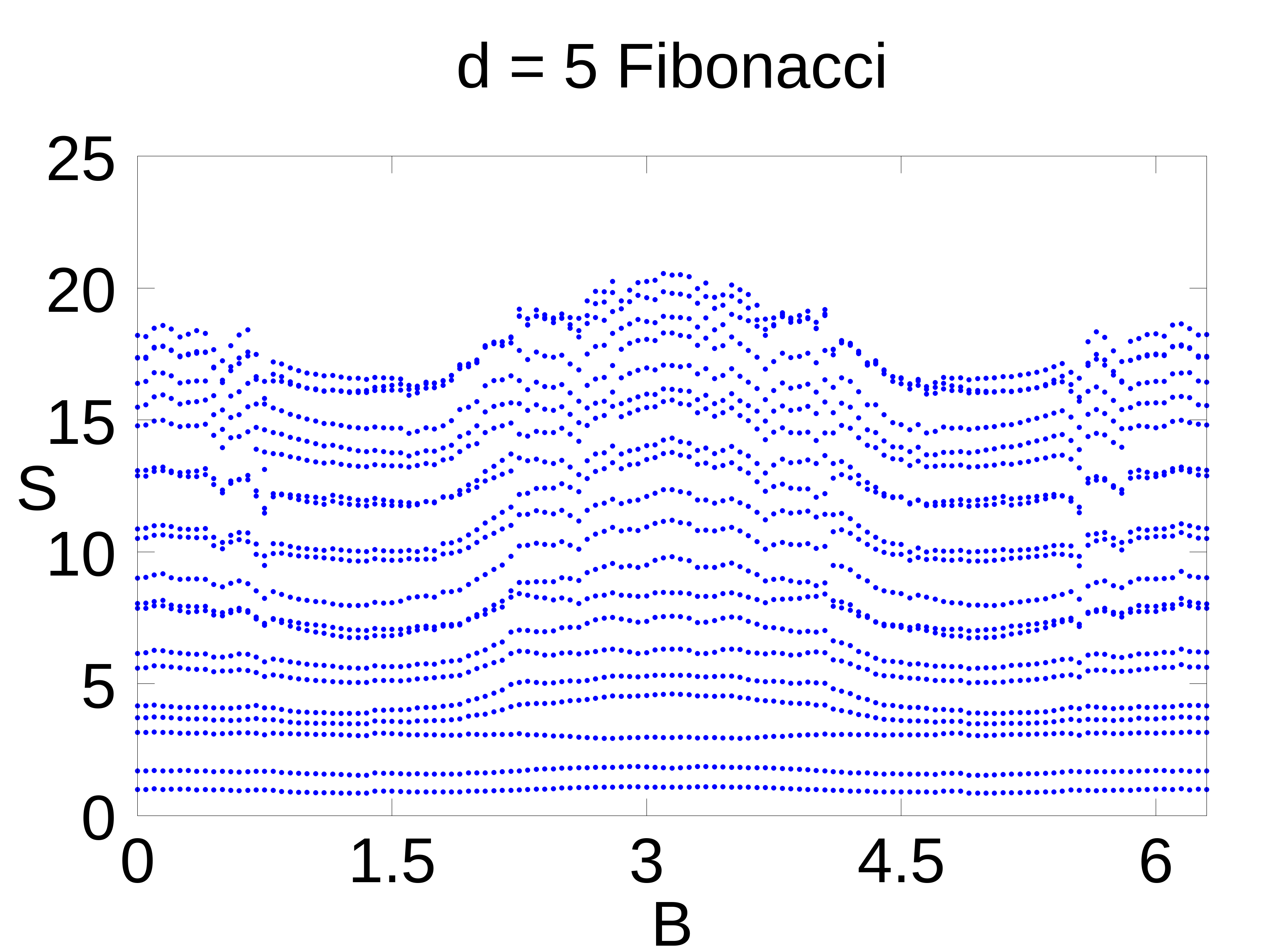}
		\subcaption{}
		\label{ARSB8}
	\end{subfigure}%

	\caption{Entanglement entropy plotted as a function of the magnetic field $B$ for a square lattice, illustrating the effects of different quasiperiodic disorders at two 
	disorder strengths. The upper panel corresponds to the weak-disorder case $d = 2$ for all disorder types, while the lower panel shows the results for the strong-disorder regime.}
	\label{ARSB}
\end{figure*}

In Fig. \ref{ARSB}, the entanglement entropy (EE) is shown as a function of the magnetic field ($B$) for two different disorder strengths, $d = 2$ and $d = 5$. Figures \ref{ARSB}(a–d) illustrate the entropy behavior at $d = 2$ for all types of disorders.   As the strength of the disorder increases, the entanglement entropy decreases. At low and high magnetic fields, the entropy becomes slightly more stable, adhering to the area law, which indicates that it scales with the boundary area of the subsystem. However, at intermediate magnetic fields, the entanglement entropy shows significant variation and deviates from the area law. For $d=5$, AA disorder strength, an interesting U-shaped minimum emerges in the entropy profile at intermediate magnetic field values ($ B $), reflecting a complex response to the heightened disorder that modulates the system's quantum correlations. The TM disorder ($d=2$), the EE is elevated at both small and large magnetic fields, displaying a pronounced minima at intermediate $B$, while adhering more closely to the area law for smaller system sizes, indicative of localized states. As the disorder strength increases ($d=5$), the overall EE magnitude diminishes, signaling stronger localization. Notably, the minima at intermediate $B$ evolves into a maxima, reflecting a shift in the entanglement structure due to enhanced disorder. For the strong disorder, the EE becomes more stable across $B$, suggesting that the system approaches a highly localized regime where entanglement fluctuations are suppressed. This progression underscores the competition between disorder-induced localization and magnetic field effects, with TM disorder ultimately driving the system toward a more homogeneous entanglement distribution.\\ 
Figure \ref{ARSB}(c) shows the variation of EE for RS binary disorder. At low disorder $(d=2)$, the EE follows the area law closely, with smooth behavior and minimal fluctuationsfor smaller subsystem size. For $(d=5)$, the EE exhibits stronger fluctuations, especially around intermediate values of $B$, indicating deviations from the area law. In Fig. \ref{ARSB}(d), the effect of Fibonacci disorder on the EE as a function of magnetic field is qualitatively similar to that of the TM and RS disorders. At weak disorder, the EE exhibits a minimum at intermediate magnetic fields and for larger subsystem sizes. As the disorder strength increases, the EE decreases overall, and the minimum evolves into a maximum at intermediate fields for larger subsystems. In general, increasing the strength of any type of disorder reduces the EE, consistent with enhanced localization of the system. For strong AA disorder, the EE retains a minimum-like structure at intermediate fields, whereas for Fibonacci, TM, and RS disorders it instead develops a maximum. In all cases, strong disorder leads to a violation of the area law.\\
Figure \ref{IASB} shows the EE of intropolation of AA disorder with other quasi periodic disorders, which we explain above section. Fig. \ref{IASB}(a-d) shows the EE interpolation between AA($\lambda=5$) and TM ($\lambda_0=5$) as a function of magnetic flux. For lower $X$ (e.g., $X=0.2$), the system is dominated by the AA potential, exhibiting stronger localization and lower EE with distinct dips near $B=\pi$, consistent with the area law. As $X$ increases (e.g., From $X=0.6$ to $X=0.8$), TM disorder dominates, leading to more extended states and enhance the EE. These results highlight how the nature of disorder governs entanglement characteristics and localization behavior in the system. The transition in EE behavior across values of $X$ captures how the interplay between two quasiperiodic potentials governs the entanglement structure of the system. The EE becomes nearly independent of the magnetic flux and violates the area law, showing overlap across different system sizes further confirming the dominance of localized states.\\
In fig \ref{IASB}(e-h) show the EE, intropolation between AA ($\lambda=5$) and Fibonacci ($\lambda_2=5$) disorder as a function of magnetic flux. Across small values of the interpolation parameter $X$, the EE behavior is largely governed by the AA potential.    Interestingly, at  $X=0.2$, the EE exhibits a minimum near $B=\pi$ for larger subsystem sizes. As the interpolation parameter $X$ increases, the influence of the Fibonacci disorder becomes dominant, and this minimum evolves into a maximum. It shows the transition between AA to Fibonacci disoders.  This indicates stronger localization, as a majority of the system’s states become highly localized due to the enhanced potential strength. Consequently, the entanglement entropy further deviates from the area law, reinforcing the suppression of extended states in the strongly disordered regime.\\

\begin{figure*}[!htb]
	\centering

	\begin{subfigure}{.24\textwidth}
		\centering
		\includegraphics[width=\linewidth]{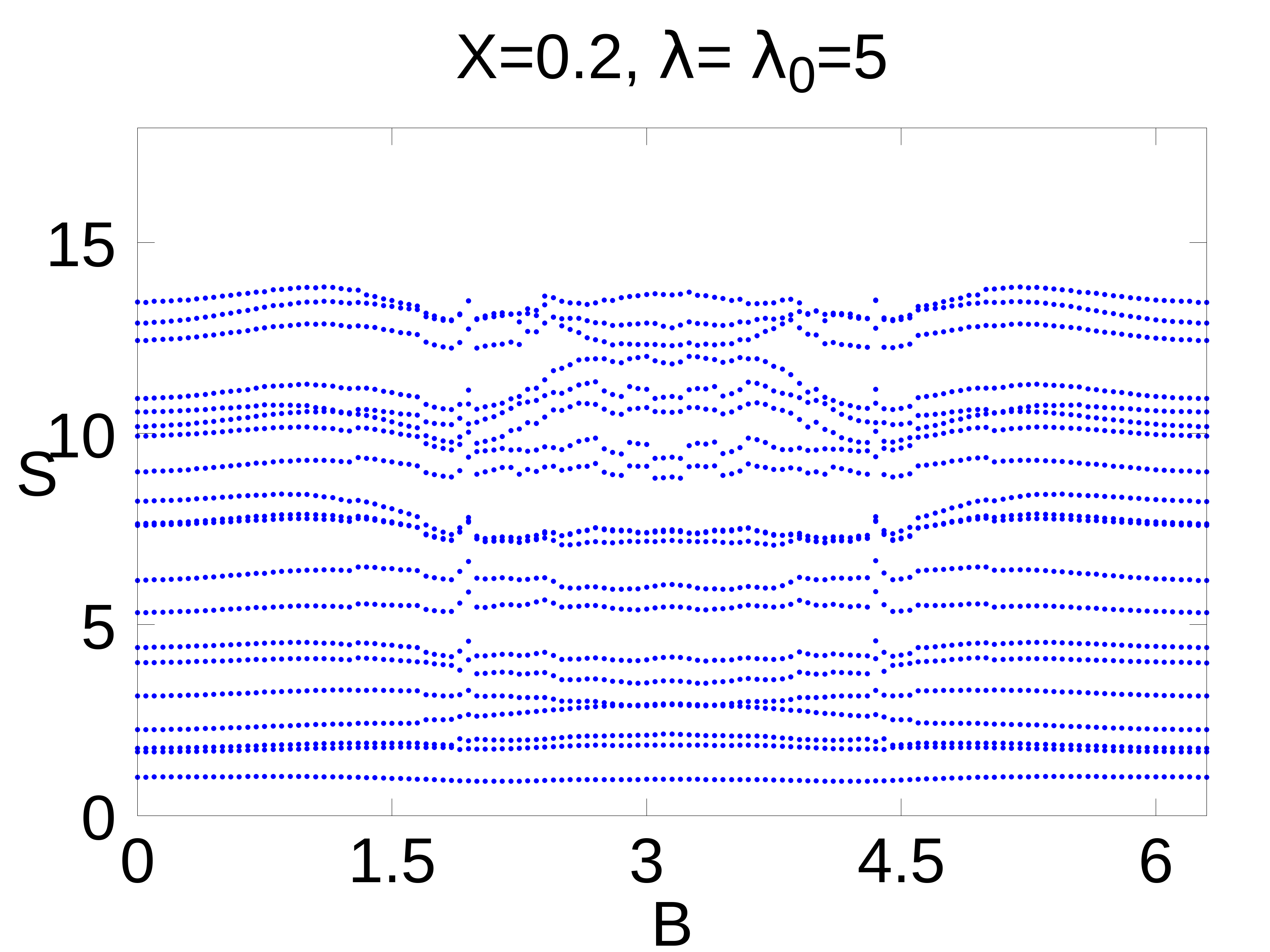}
		\subcaption{}
		\label{sq5}
	\end{subfigure}%
	\begin{subfigure}{.24\textwidth}
		\centering 
		\includegraphics[width=\linewidth]{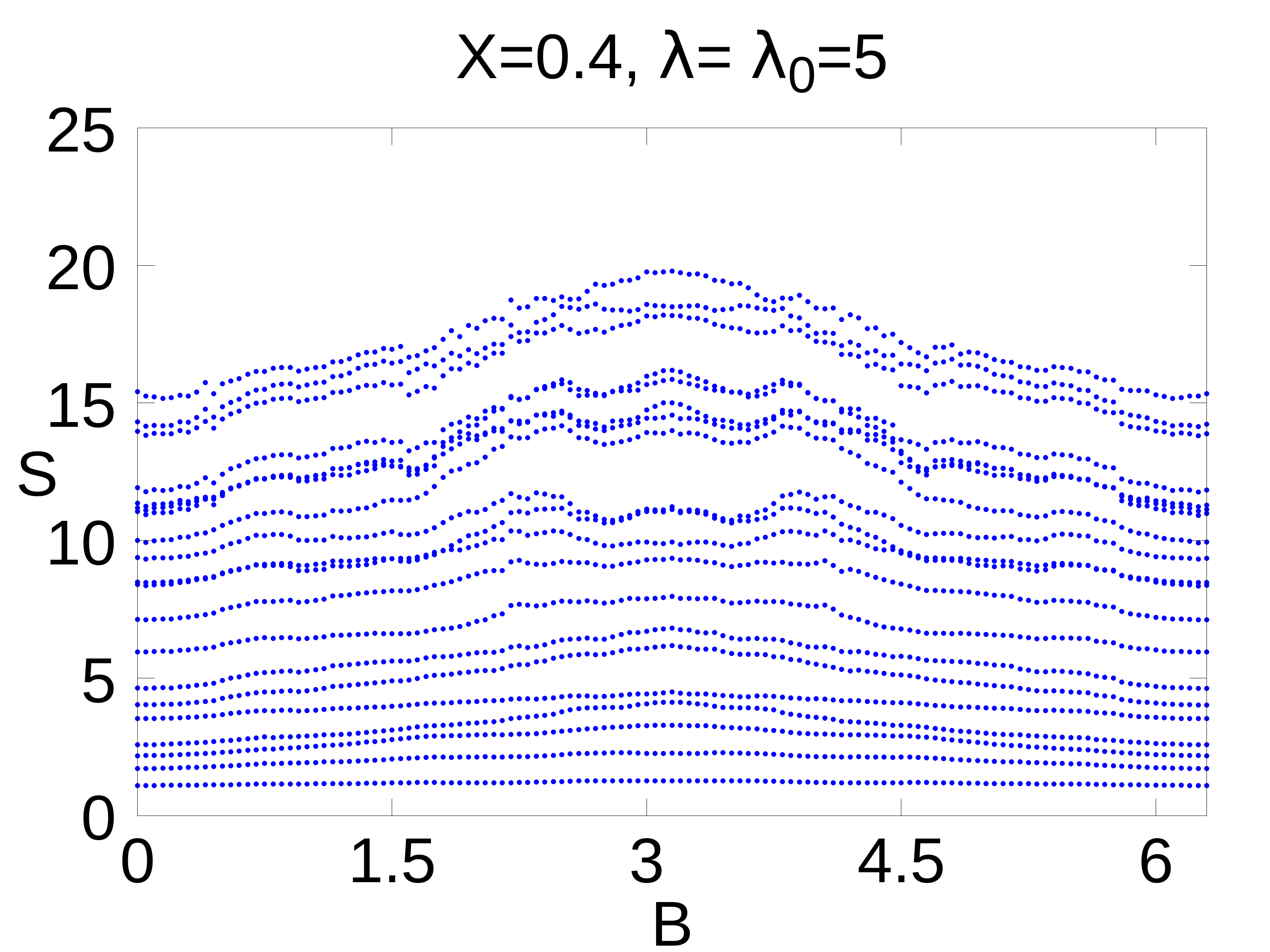}
		\subcaption{}
		\label{sq6}
	\end{subfigure}%
	\begin{subfigure}{.24\textwidth}
		\centering
		\includegraphics[width=\linewidth]{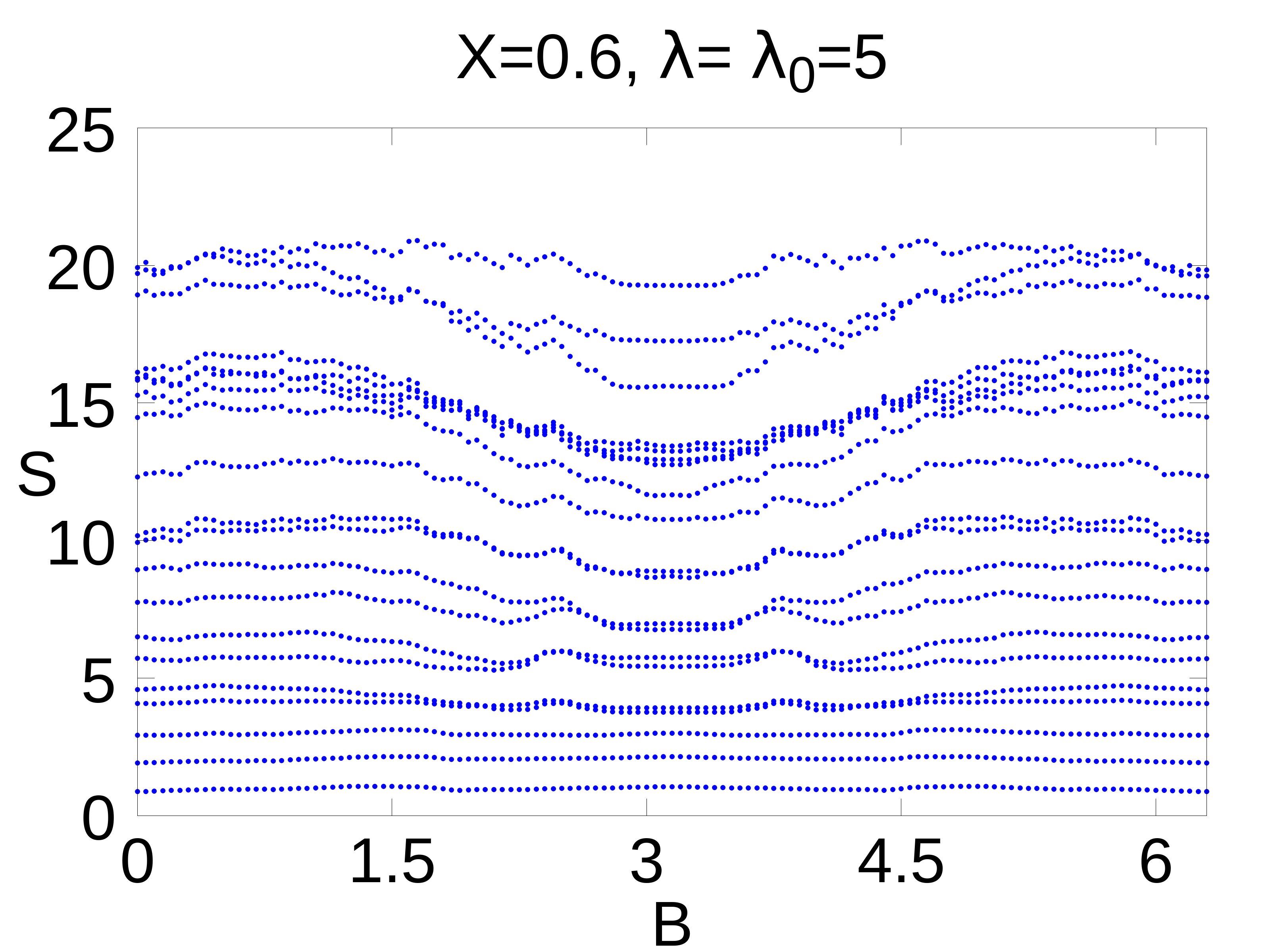}
		\subcaption{}
		\label{sq7}
	\end{subfigure}%
	\begin{subfigure}{.24\textwidth}
		\centering
		\includegraphics[width=\linewidth]{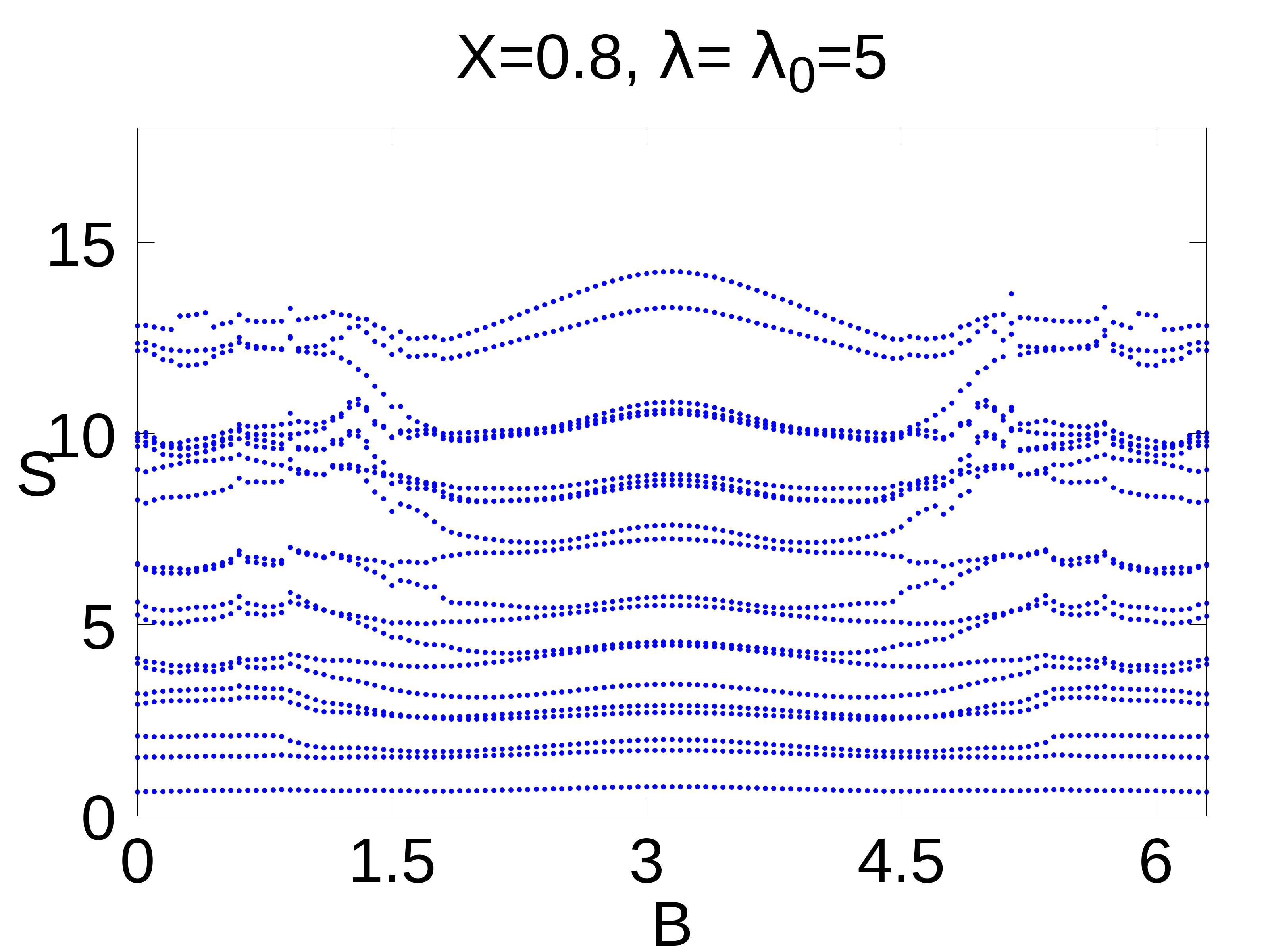}
		\subcaption{}
		\label{sq8}
	\end{subfigure}%

\begin{subfigure}{.24\textwidth}
	\centering
	\includegraphics[width=\linewidth]{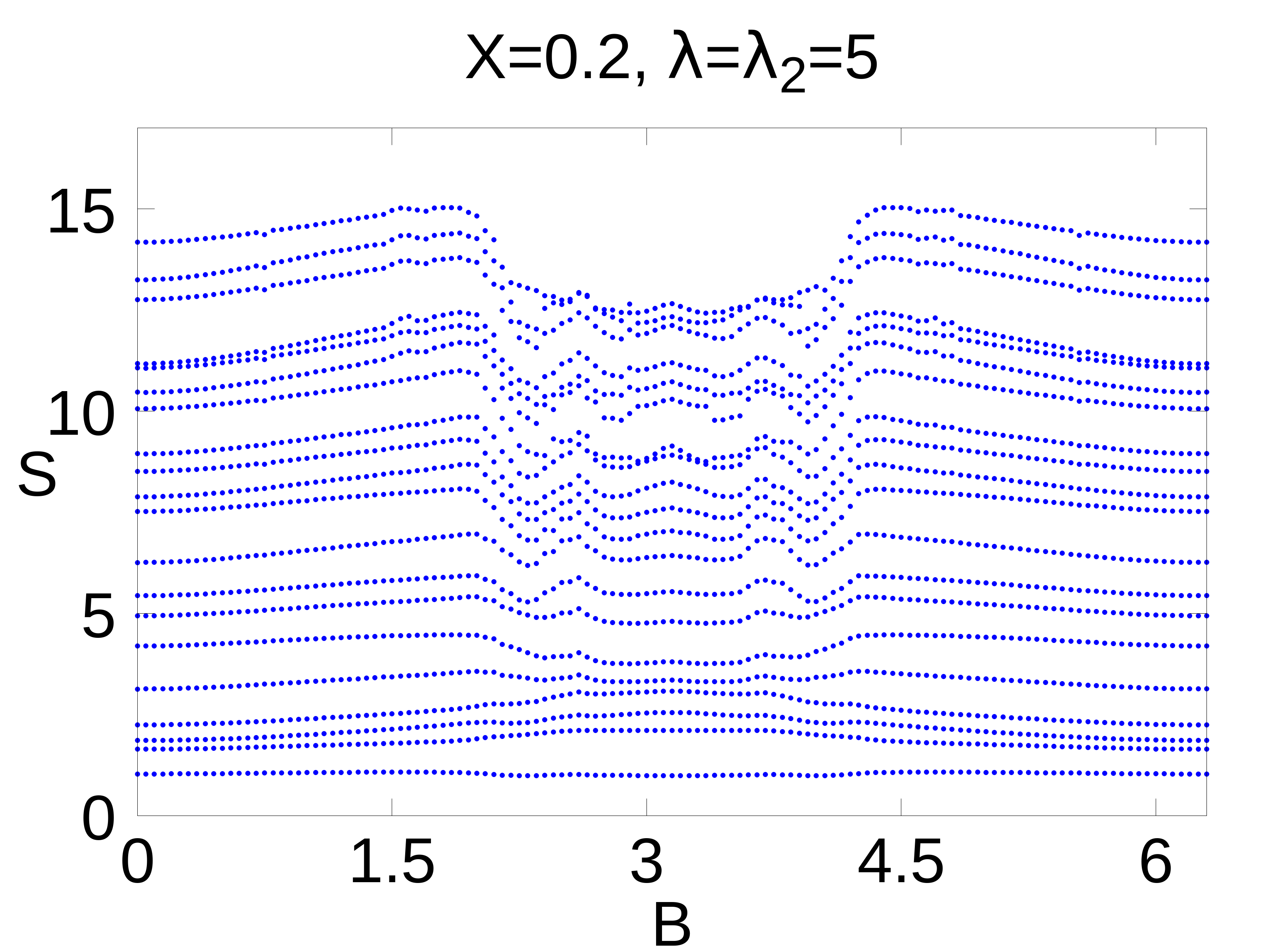}
	\subcaption{}
	\label{sq5}
\end{subfigure}%
\begin{subfigure}{.24\textwidth}
	\centering 
	\includegraphics[width=\linewidth]{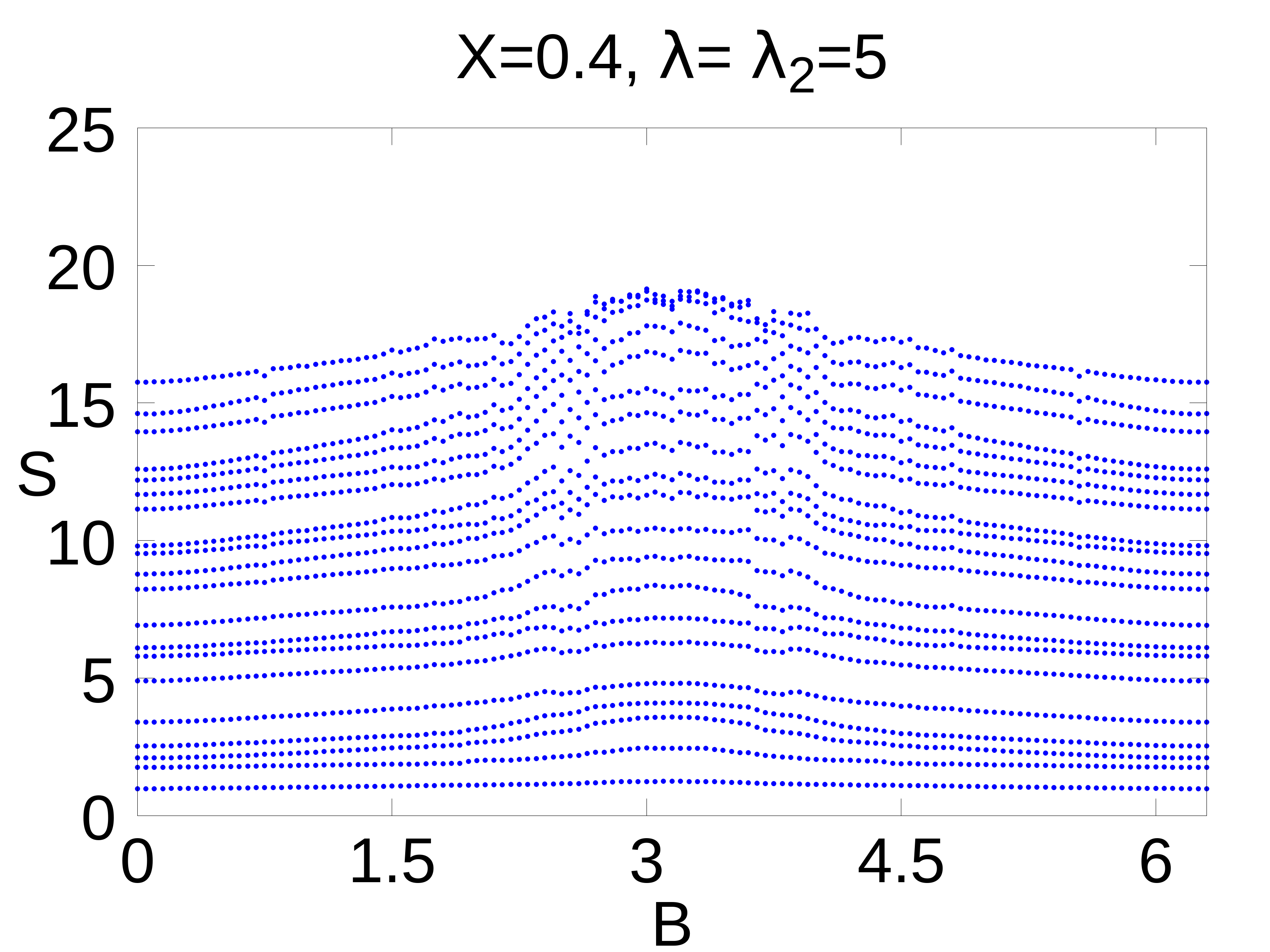}
	\subcaption{}
	\label{sq6}
\end{subfigure}%
\begin{subfigure}{.24\textwidth}
	\centering
	\includegraphics[width=\linewidth]{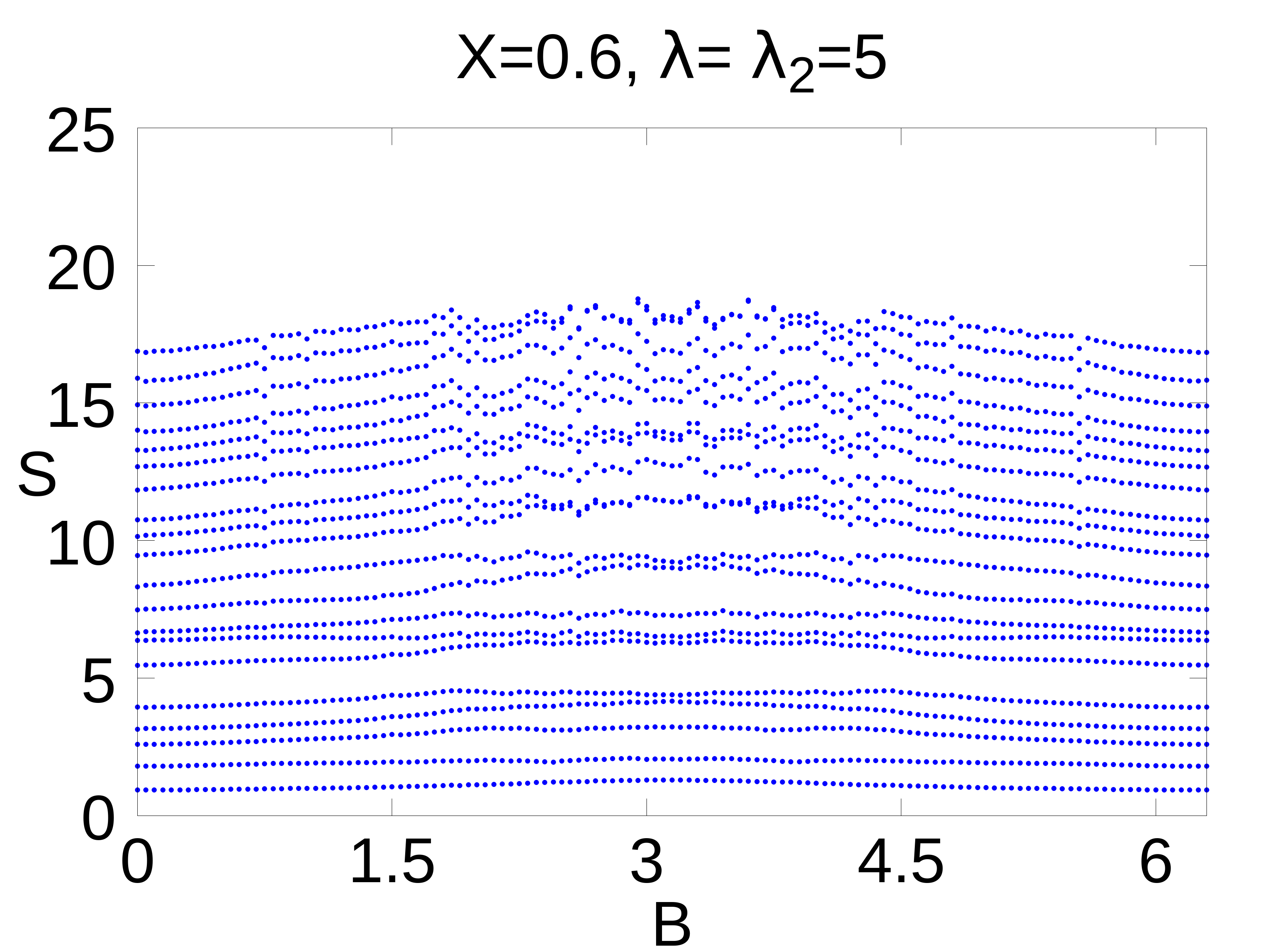}
	\subcaption{}
	\label{sq7}
\end{subfigure}%
\begin{subfigure}{.24\textwidth}
	\centering
	\includegraphics[width=\linewidth]{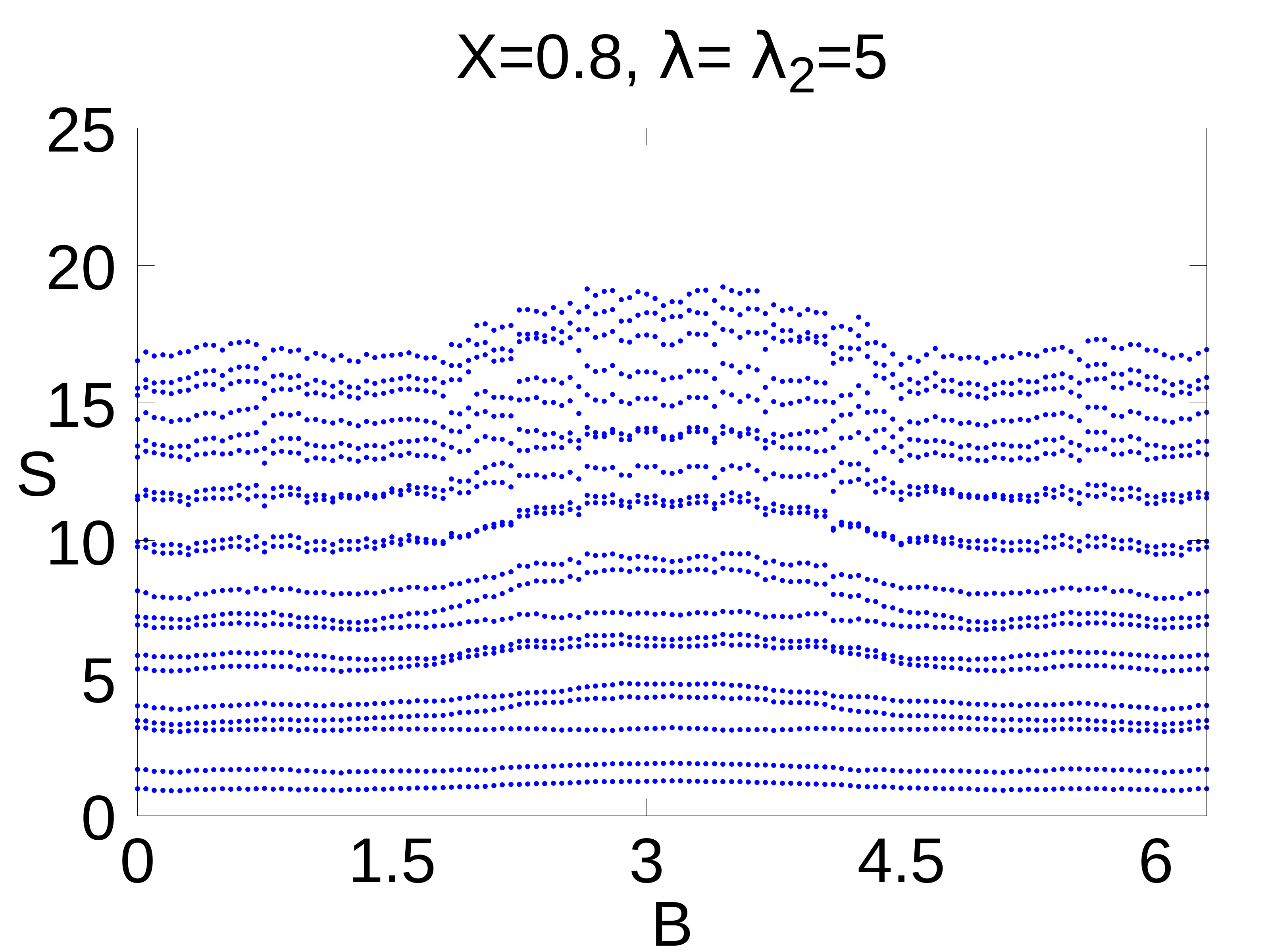}
	\subcaption{}
	\label{sq8}
\end{subfigure}%
	
	\caption{Entanglement entropy $S$ as a function of the magnetic field $B$ for a square 
		lattice, illustrating the combined effects of the Aubry--André potential with 
		different quasiperiodic disorders. The upper panel shows the interpolation 
		between AA and Thue--Morse disorders, while the lower panel presents the 
		corresponding results for the AA–Fibonacci case.
	}
	\label{IASB}
\end{figure*}

	\begin{figure*}[!htb]
		\centering
		\begin{subfigure}{.24\textwidth}
			\centering
			\includegraphics[width=\linewidth]{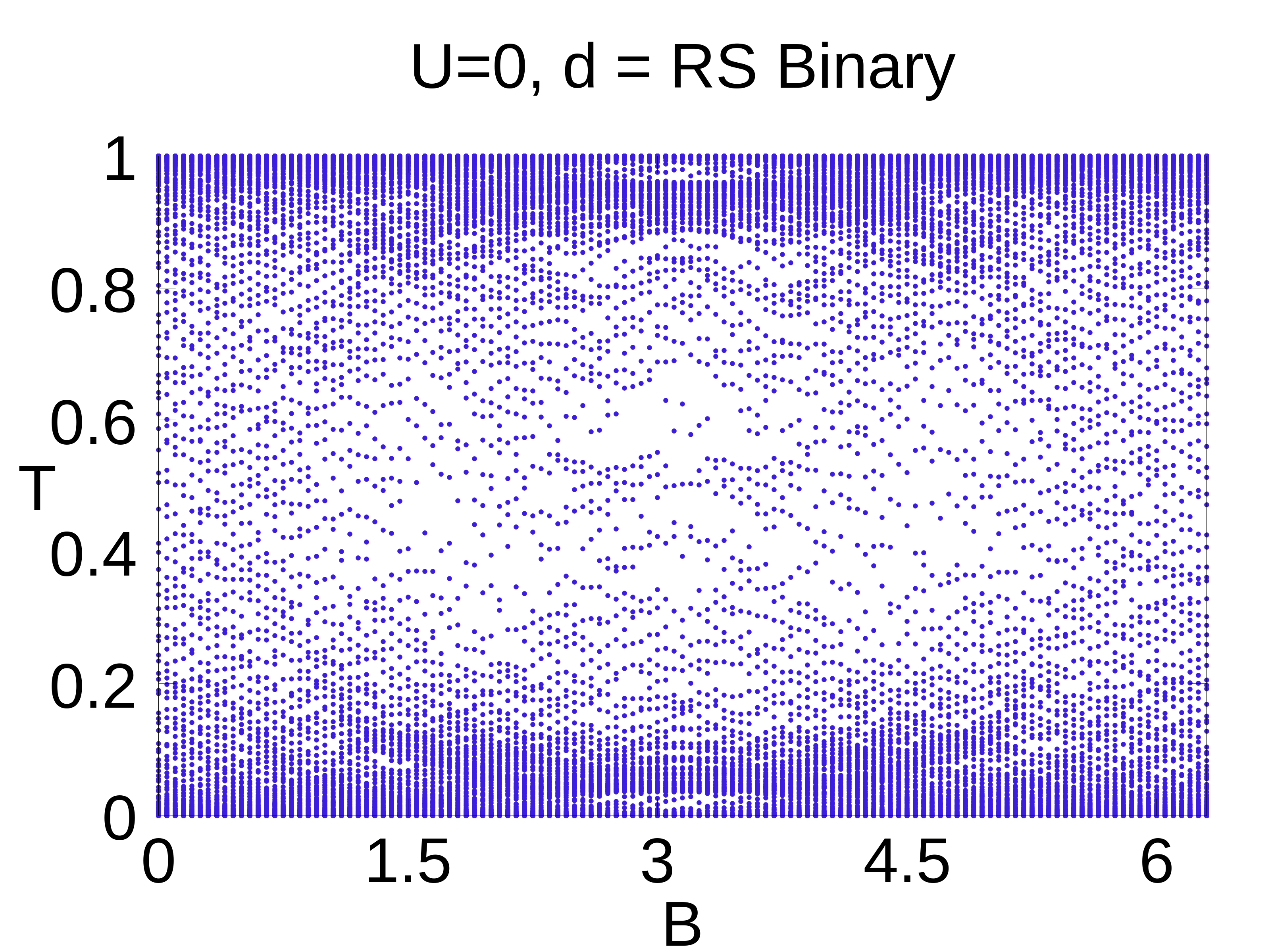}
			\subcaption{}
			\label{sq1}
		\end{subfigure}%
		\begin{subfigure}{.24\textwidth}
			\centering 
			\includegraphics[width=\linewidth]{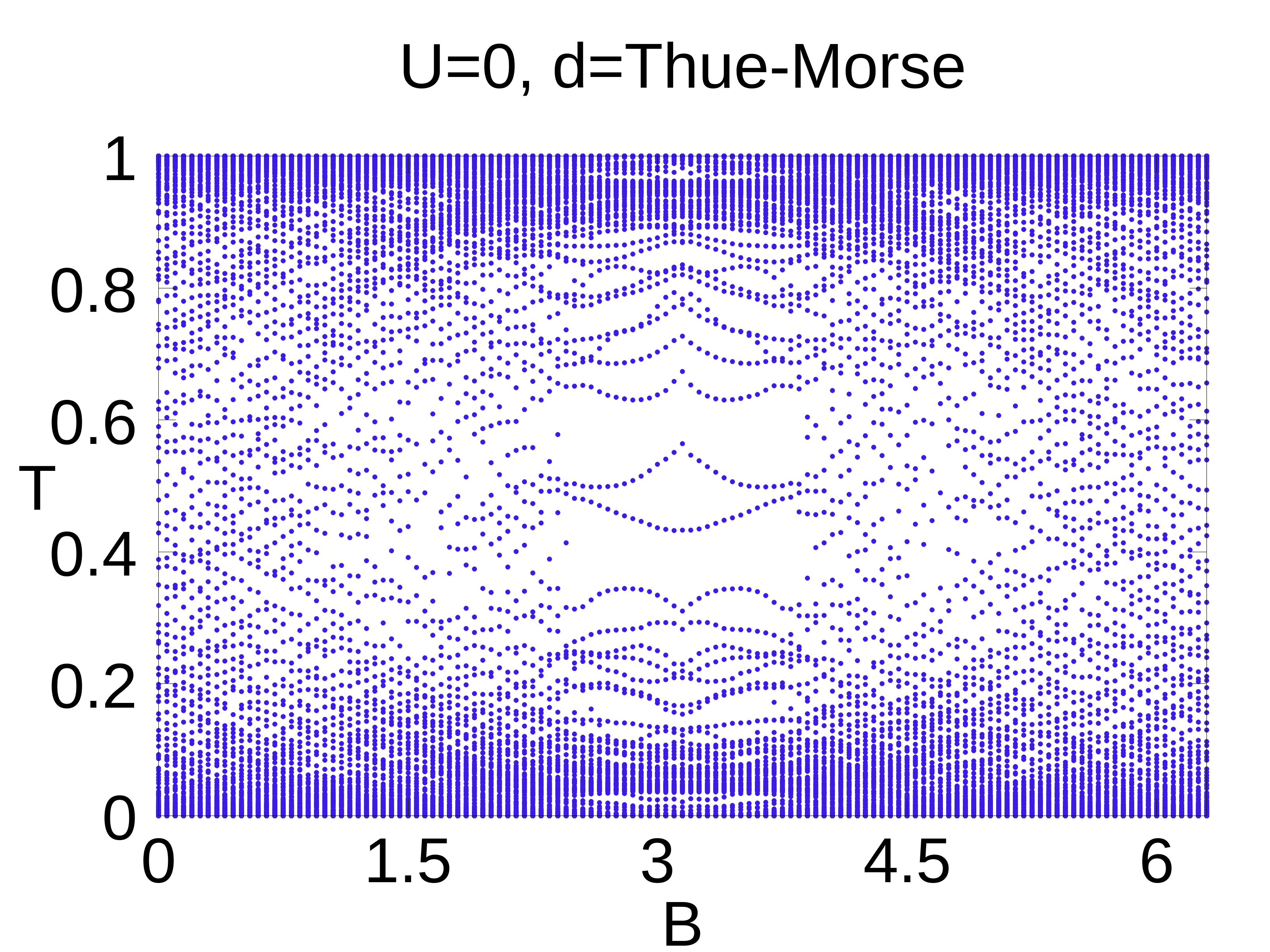}
			\subcaption{}
			\label{sq2}
		\end{subfigure}%
		\begin{subfigure}{.24\textwidth}
			\centering
			\includegraphics[width=\linewidth]{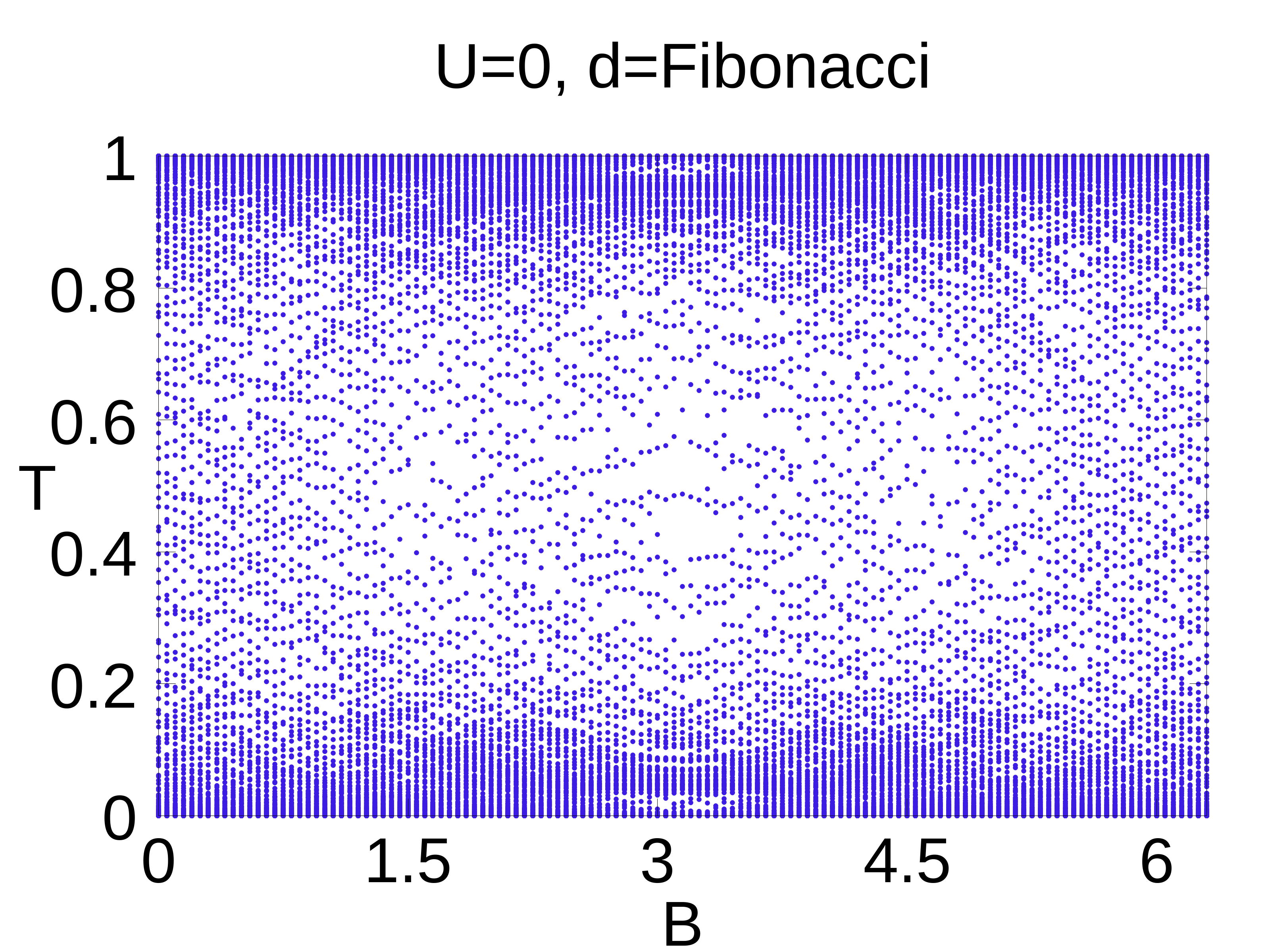}
			\subcaption{}
			\label{sq3}
		\end{subfigure}%
		\begin{subfigure}{.24\textwidth}
			\centering
			\includegraphics[width=\linewidth]{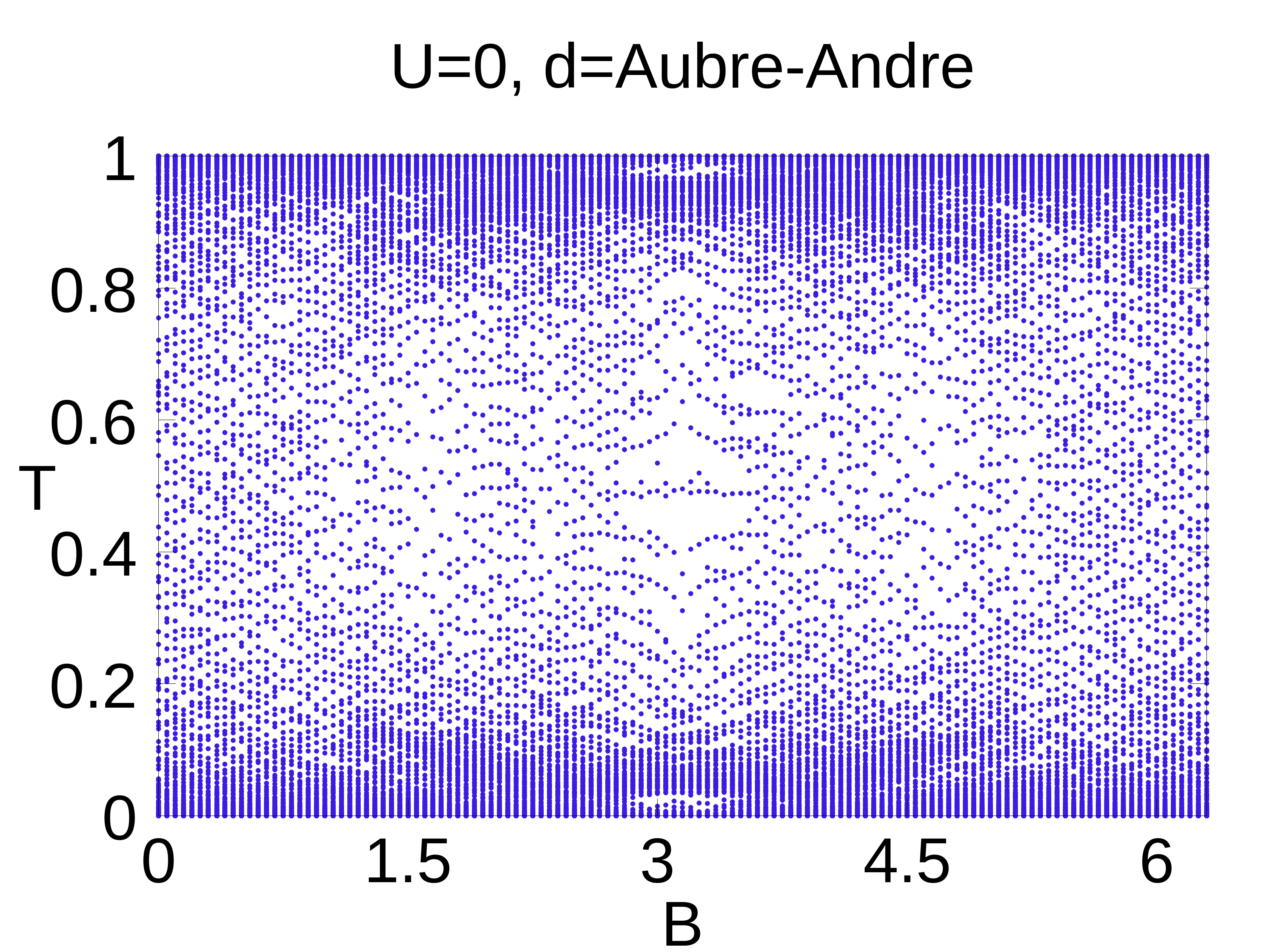}
			\subcaption{}
			\label{sq4}
		\end{subfigure}%

		\begin{subfigure}{.24\textwidth}
			\centering
			\includegraphics[width=\linewidth]{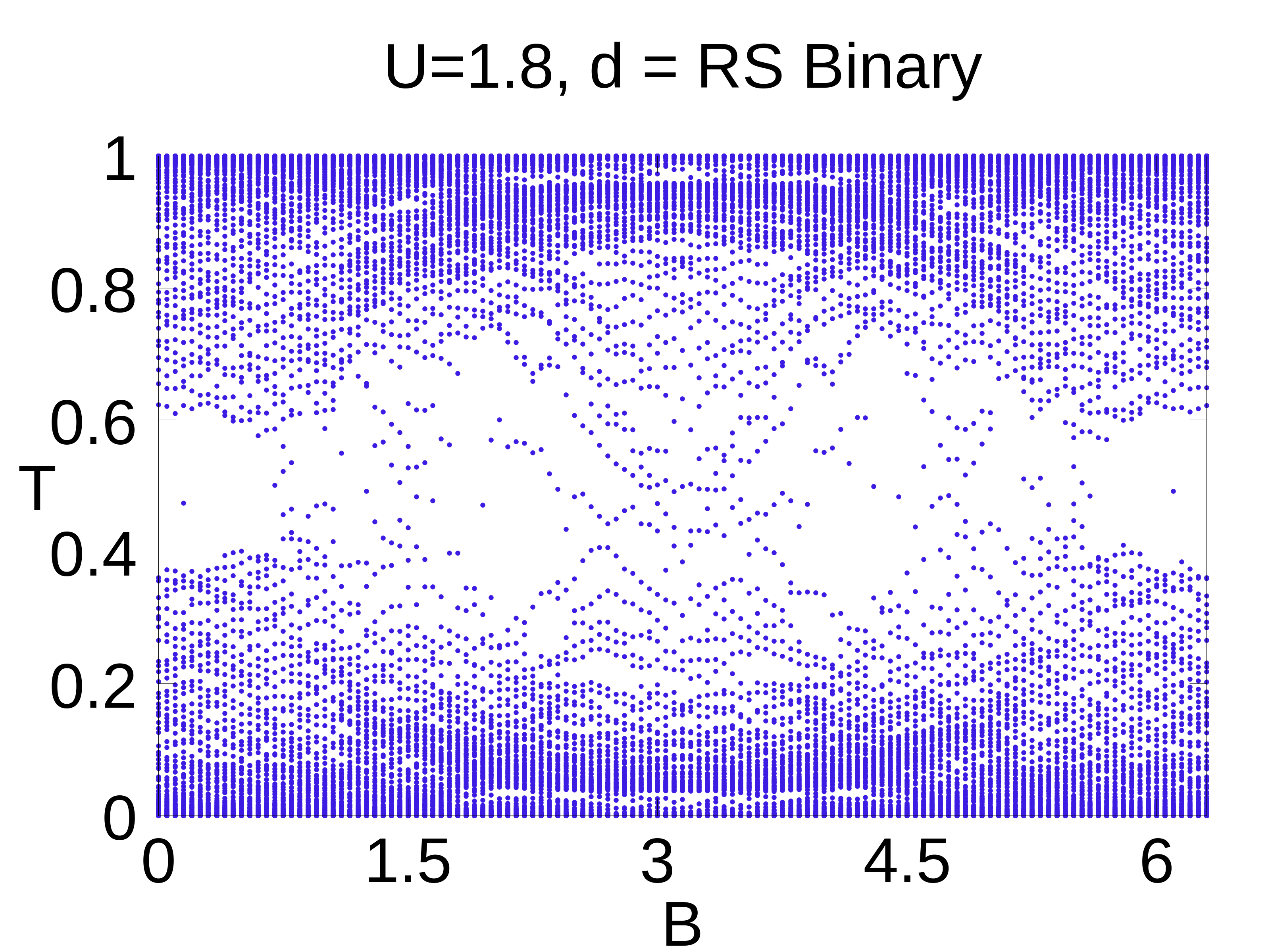}
			\subcaption{}
			\label{sq5}
		\end{subfigure}%
		\begin{subfigure}{.24\textwidth}
			\centering 
			\includegraphics[width=\linewidth]{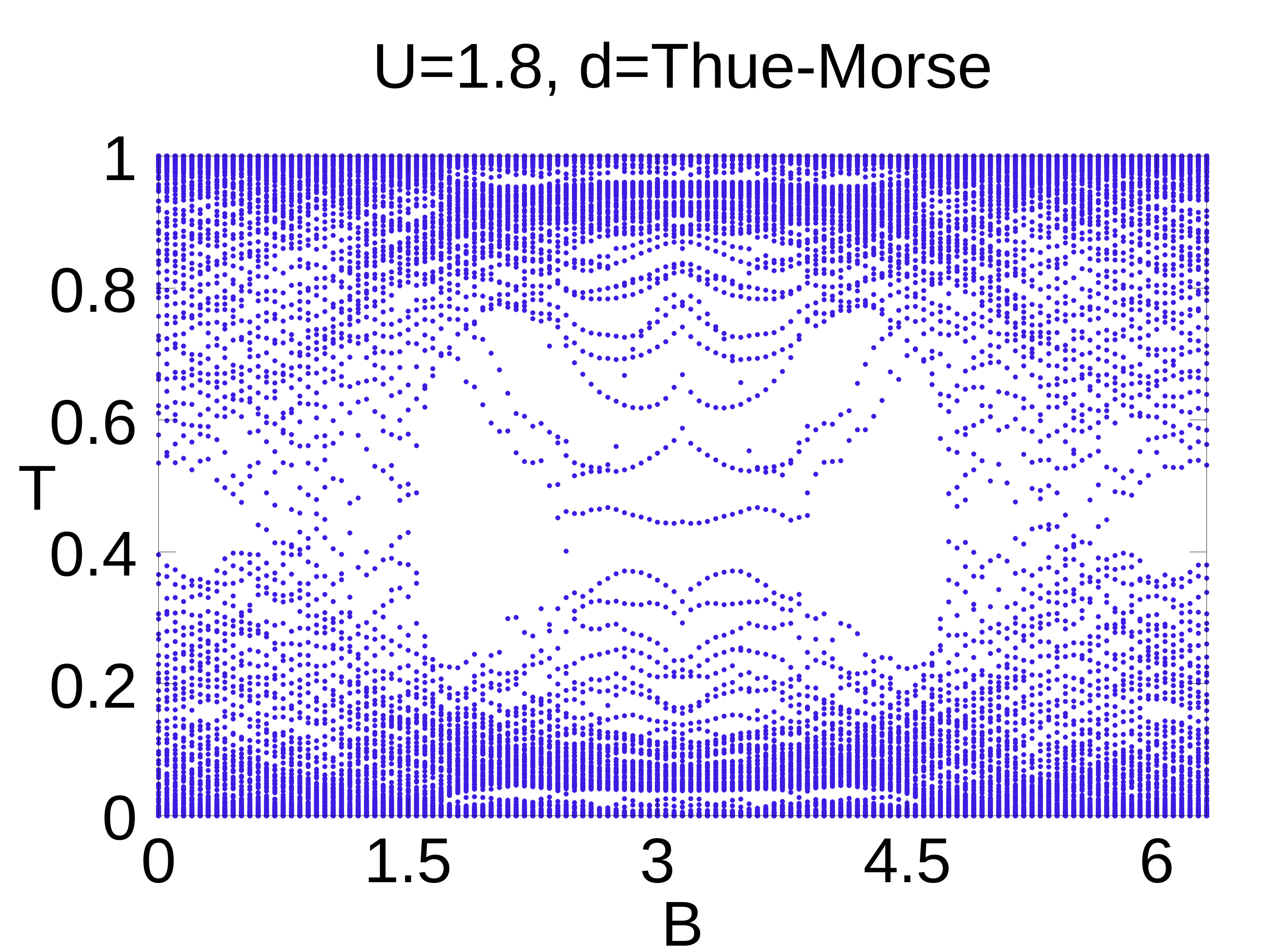}
			\subcaption{}
			\label{sq6}
		\end{subfigure}%
		\begin{subfigure}{.24\textwidth}
			\centering
			\includegraphics[width=\linewidth]{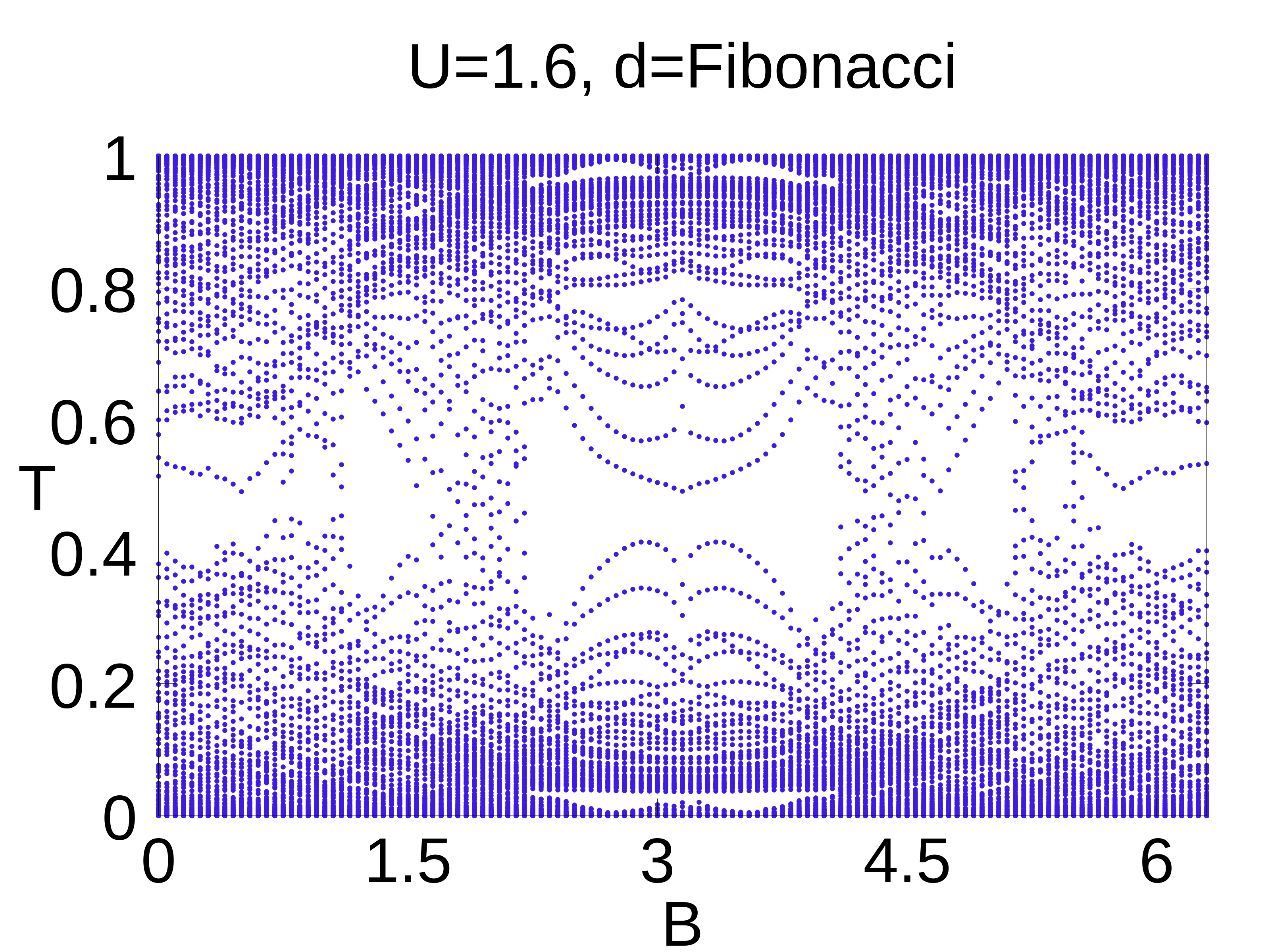}
			\subcaption{}
			\label{sq7}
		\end{subfigure}%
		\begin{subfigure}{.24\textwidth}
			\centering
			\includegraphics[width=\linewidth]{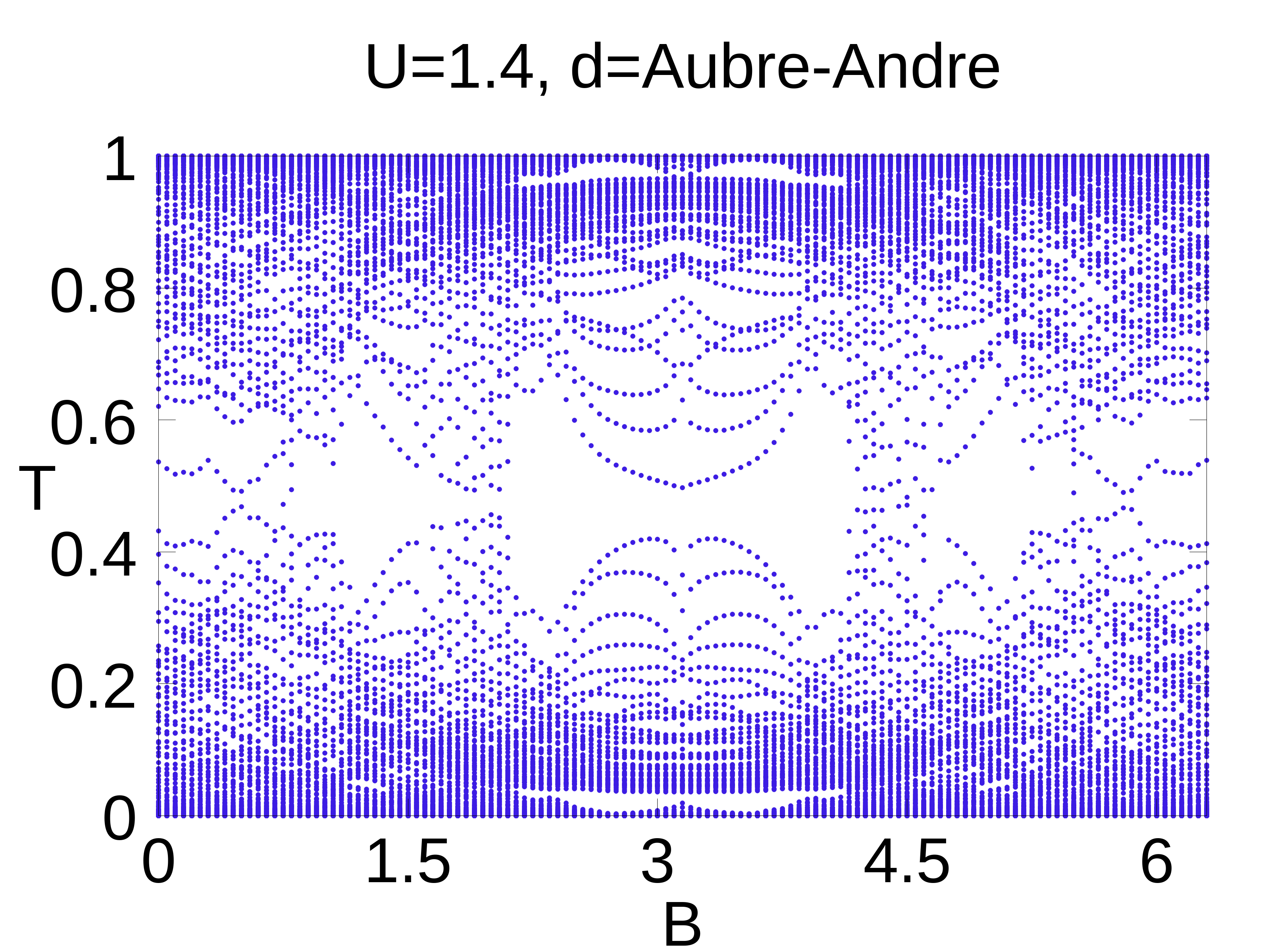}
			\subcaption{}
			\label{sq8}
		\end{subfigure}%

		\caption{The correlation spectrum (T) is plotted for the largest subsystem under various disorder schemes in the square lattice. The headers of each plot indicate the specific combination of disorders and interactions. In the uniform case, the spectrum exhibits reflection symmetry \cite{Mandal}. However, interactions open up a gap, while disorder generally smears the spectrum, eliminating any distinct features.}
		\label{ATB}
	\end{figure*}
\begin{figure*}[!htb]
		\centering
		\begin{subfigure}{.24\textwidth}
			\centering
			\includegraphics[width=\linewidth]{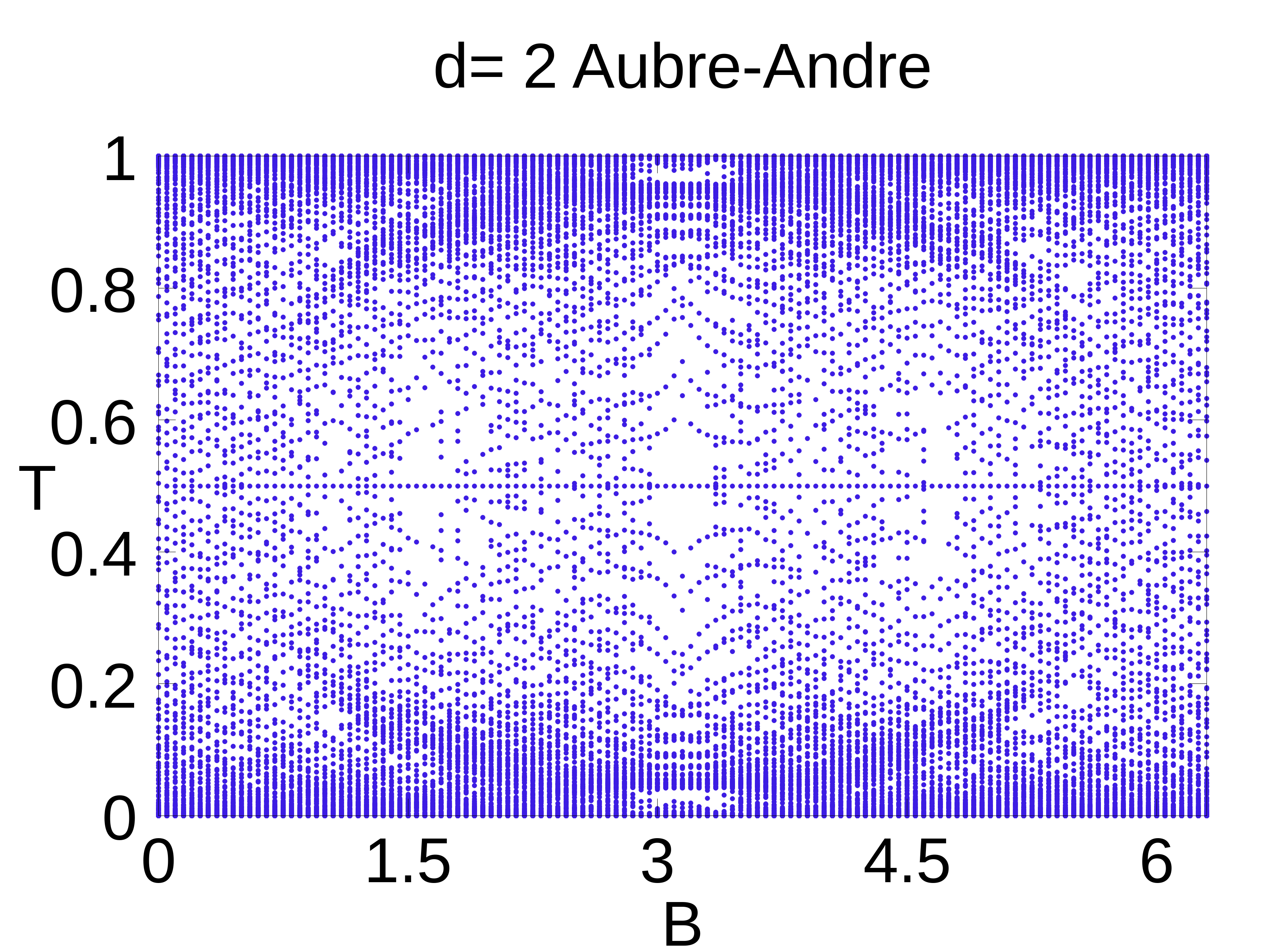}
			\subcaption{}
			\label{sq1}
		\end{subfigure}%
		\begin{subfigure}{.24\textwidth}
			\centering 
			\includegraphics[width=\linewidth]{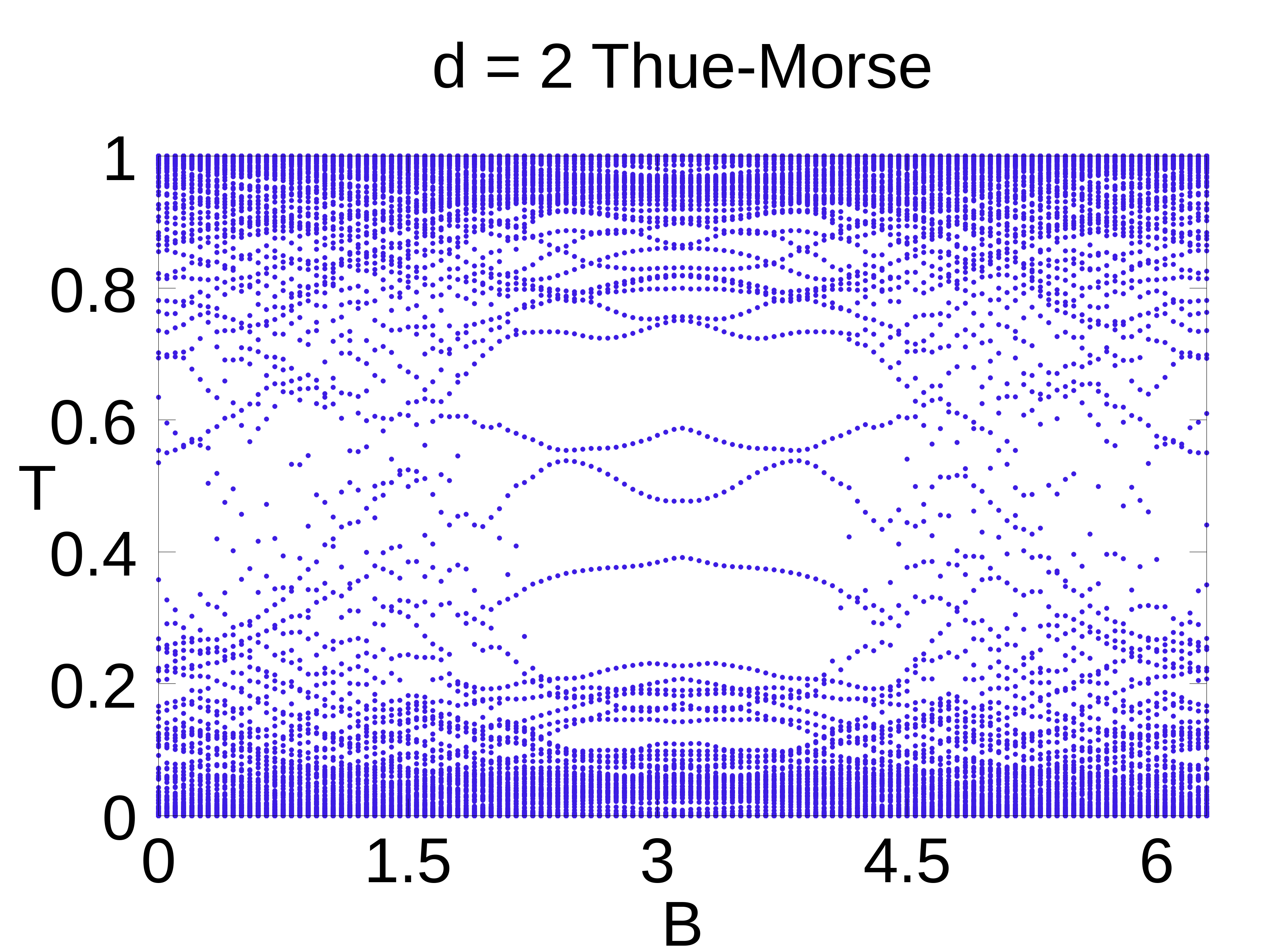}
			\subcaption{}
			\label{sq2}
		\end{subfigure}%
		\begin{subfigure}{.24\textwidth}
			\centering
			\includegraphics[width=\linewidth]{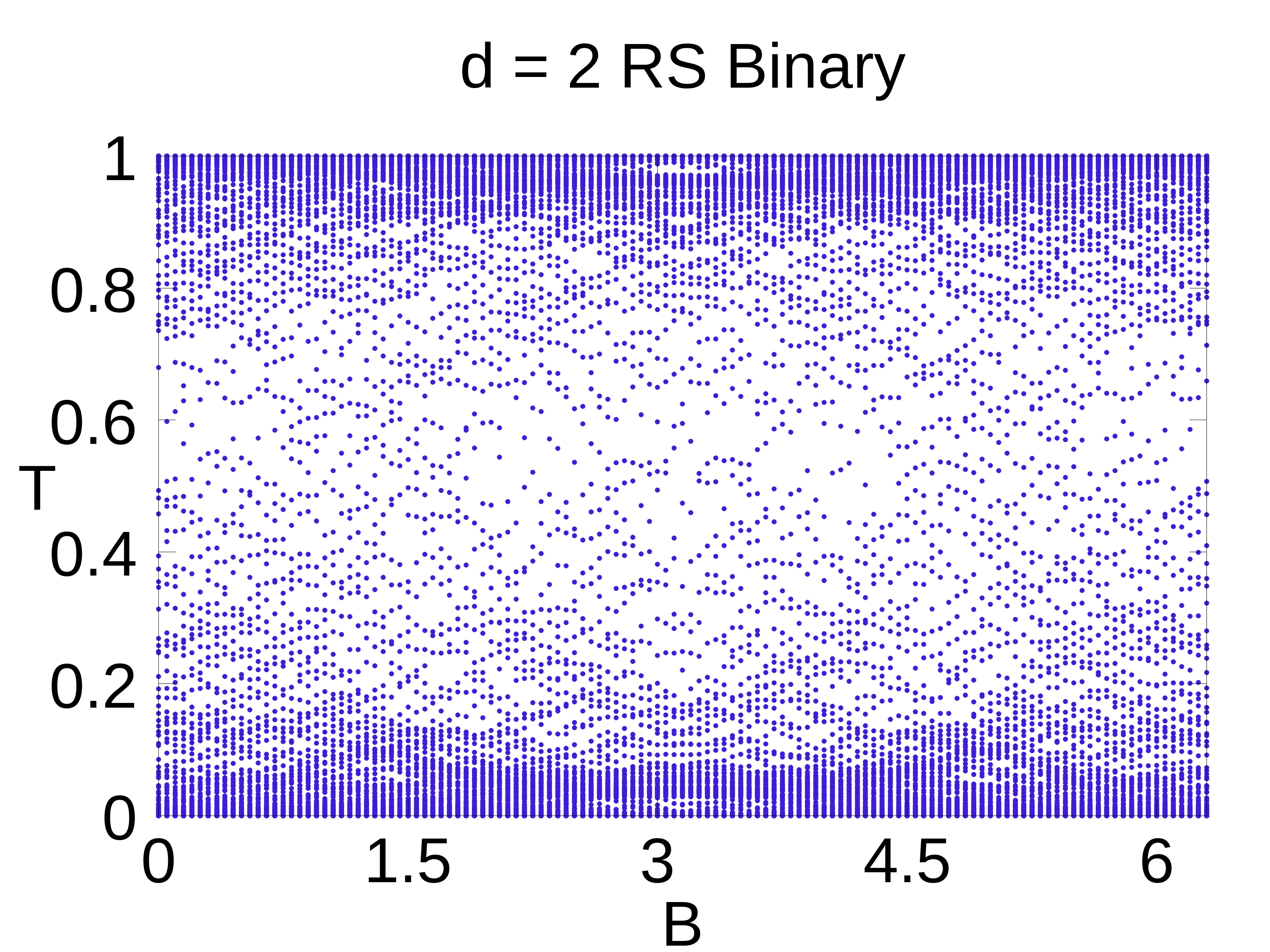}
			\subcaption{}
			\label{sq3}
		\end{subfigure}%
		\begin{subfigure}{.24\textwidth}
			\centering
			\includegraphics[width=\linewidth]{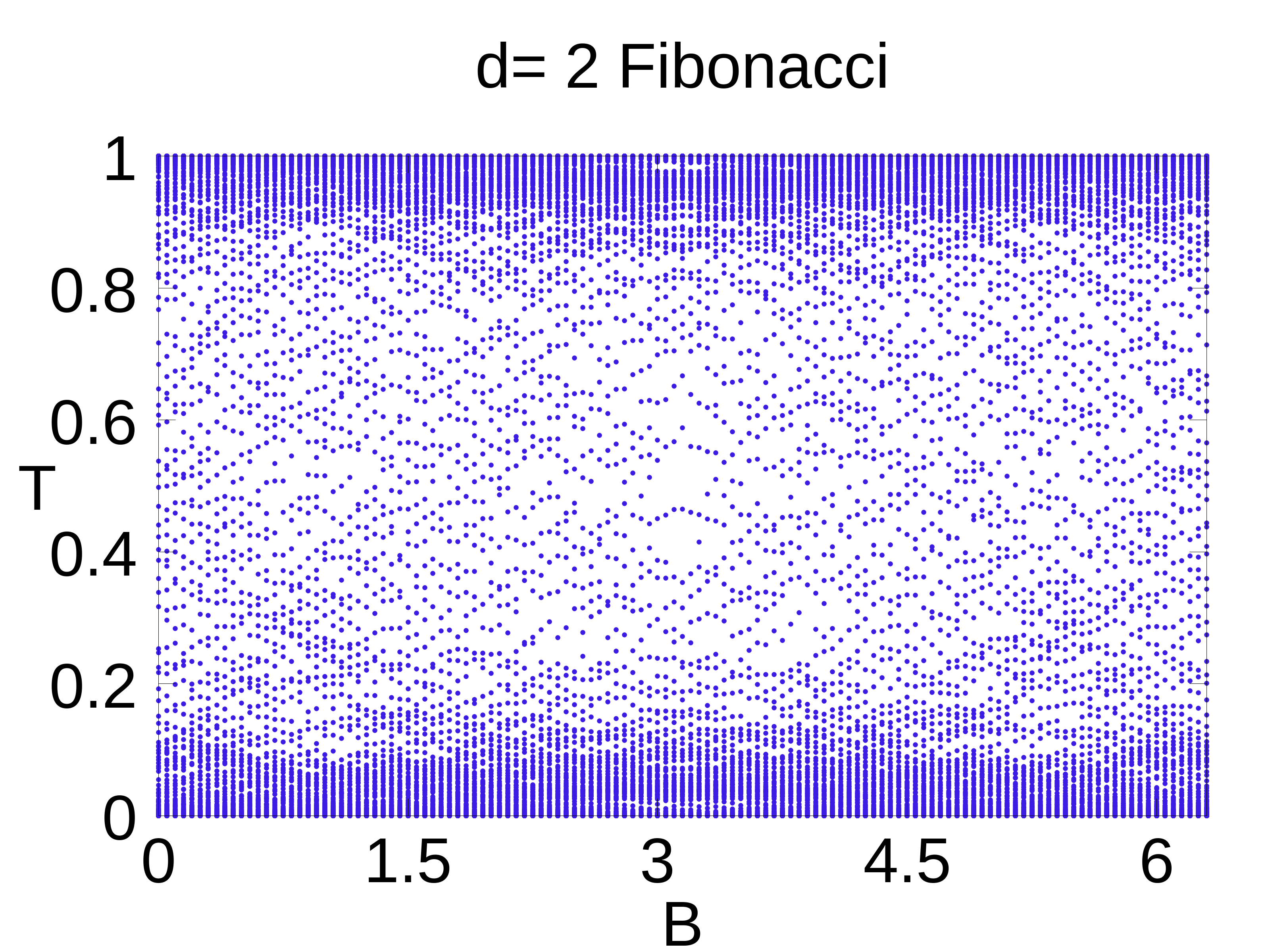}
			\subcaption{}
			\label{sq4}
		\end{subfigure}%

		\begin{subfigure}{.24\textwidth}
			\centering
			\includegraphics[width=\linewidth]{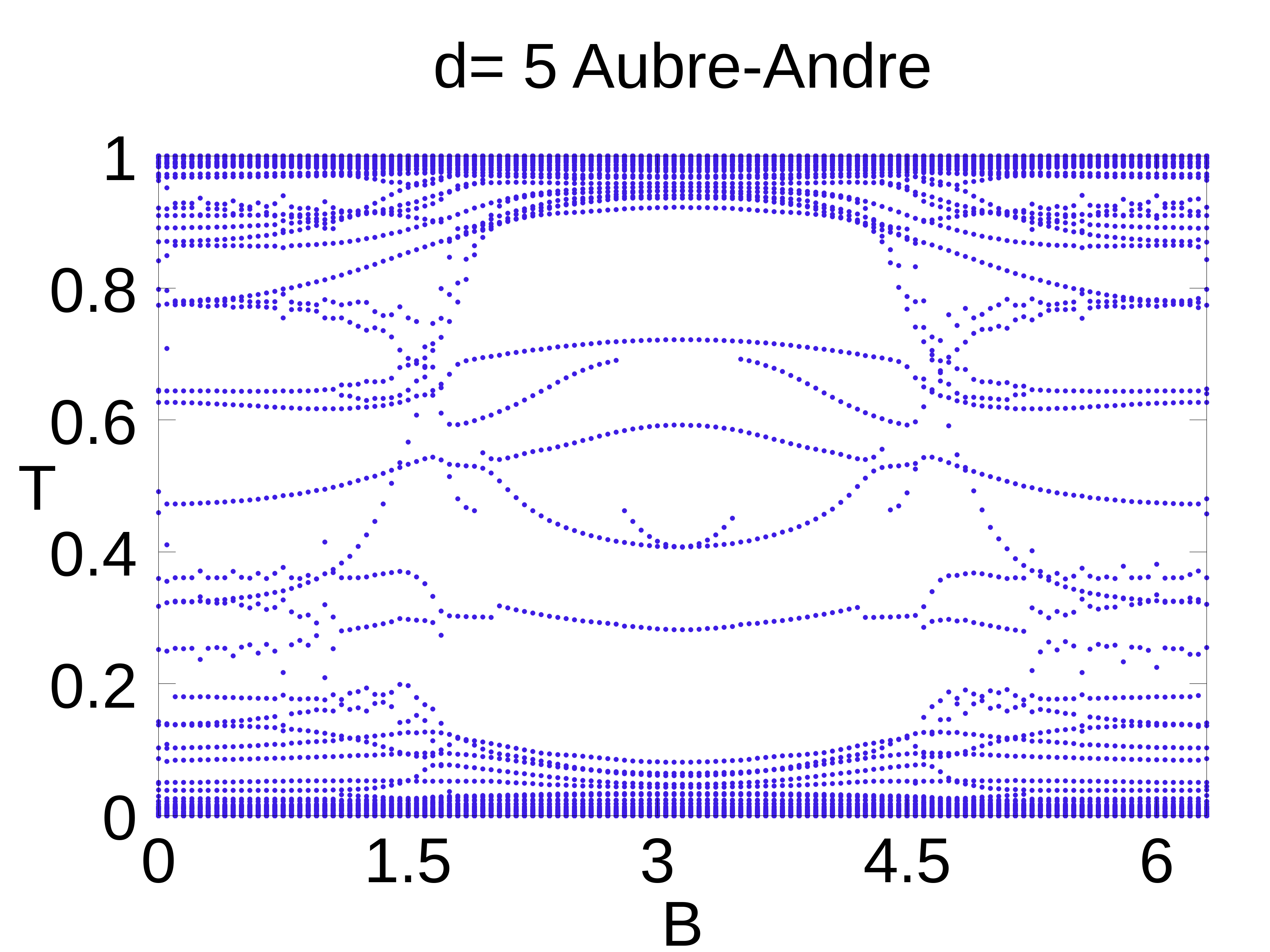}
			\subcaption{}
			\label{sq5}
		\end{subfigure}%
		\begin{subfigure}{.24\textwidth}
			\centering 
			\includegraphics[width=\linewidth]{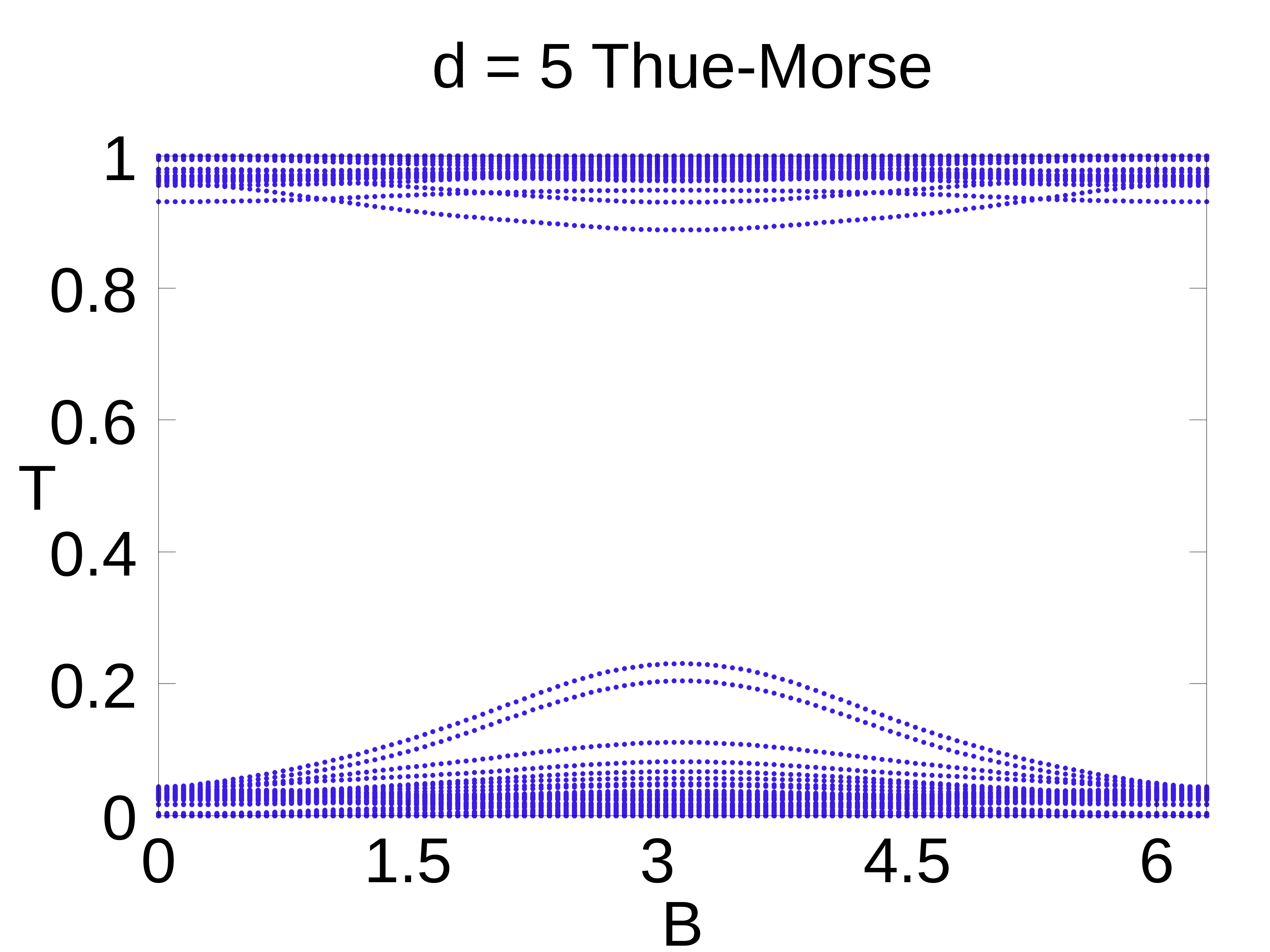}
			\subcaption{}
			\label{sq6}
		\end{subfigure}%
		\begin{subfigure}{.24\textwidth}
			\centering
			\includegraphics[width=\linewidth]{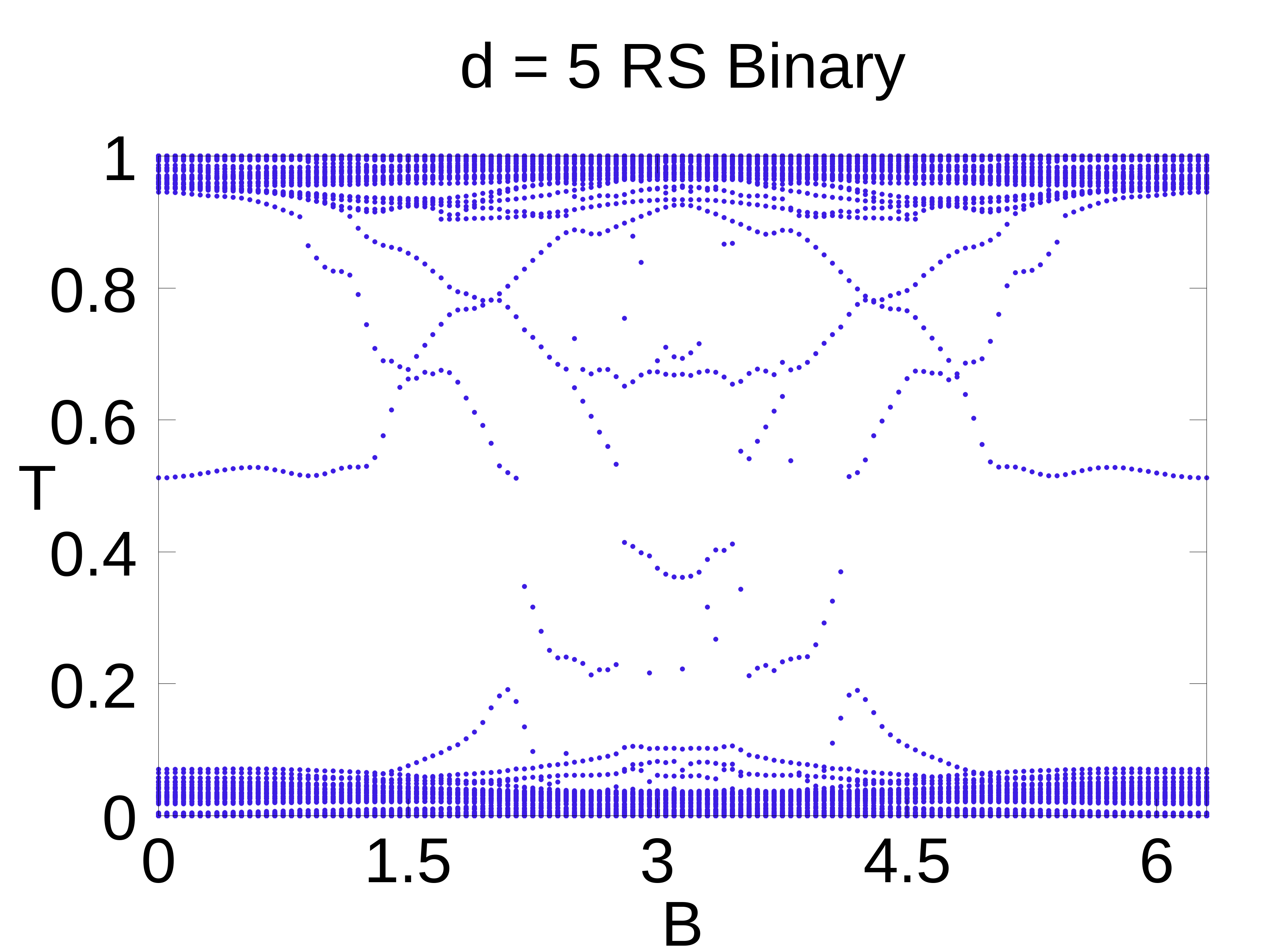}
			\subcaption{}
			\label{sq7}
		\end{subfigure}%
		\begin{subfigure}{.24\textwidth}
			\centering
			\includegraphics[width=\linewidth]{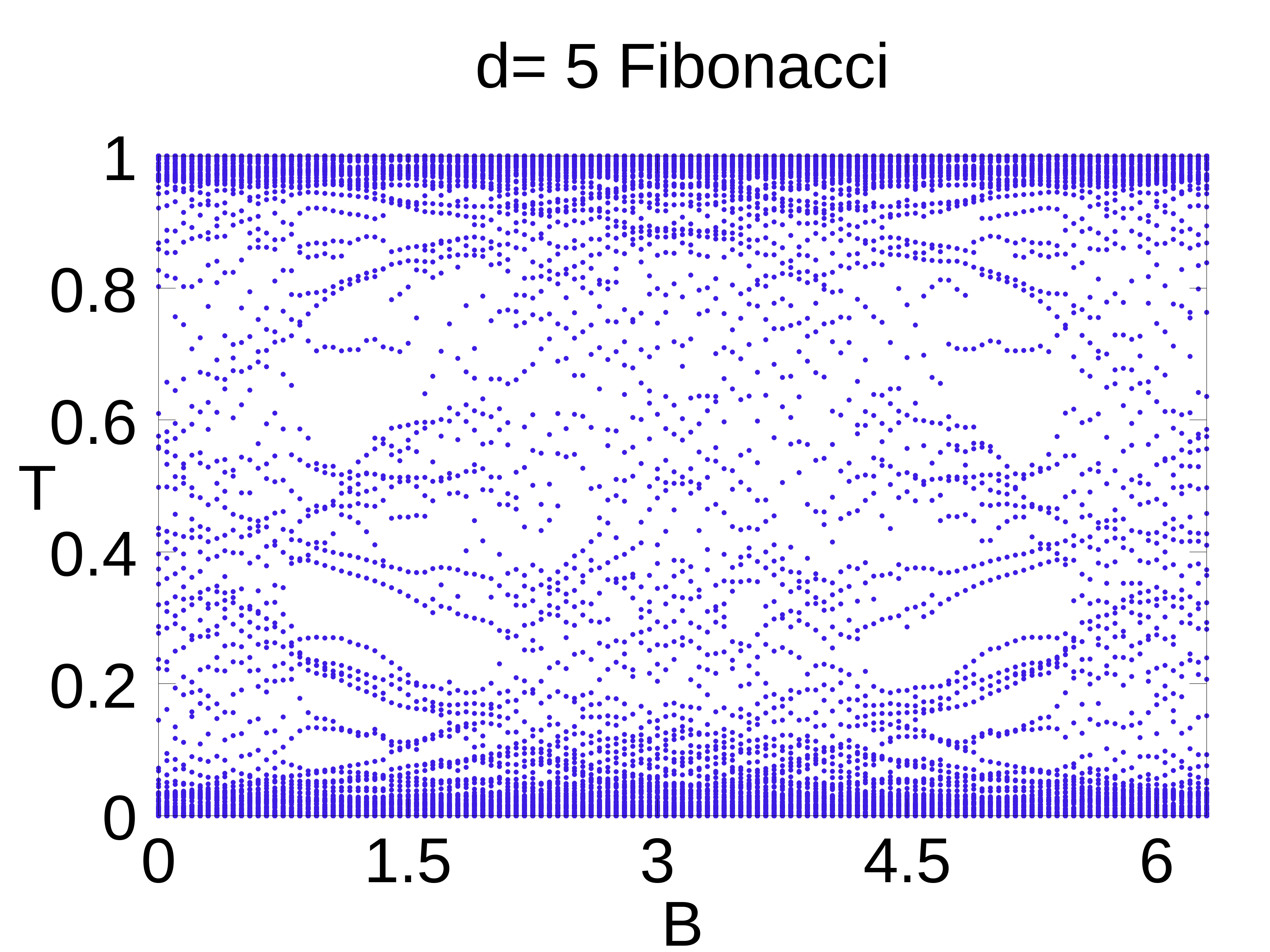}
			\subcaption{}
			\label{sq8}
		\end{subfigure}%

		\caption{Entanglement spectrum $T$ plotted as a function of the magnetic field $B$ for a square lattice, showing the effects of all quasiperiodic disorders at different strengths. The upper panel corresponds to the weak-disorder case ($d = 2$), while the lower panel displays the results for the stronger disorder ($d = 5$), where a clear gap develops in the correlation spectrum. }
		\label{DTB}
	\end{figure*}
\begin{figure*}[!htb]
	\centering
	
	\begin{subfigure}{.24\textwidth}
		\centering
		\includegraphics[width=\linewidth]{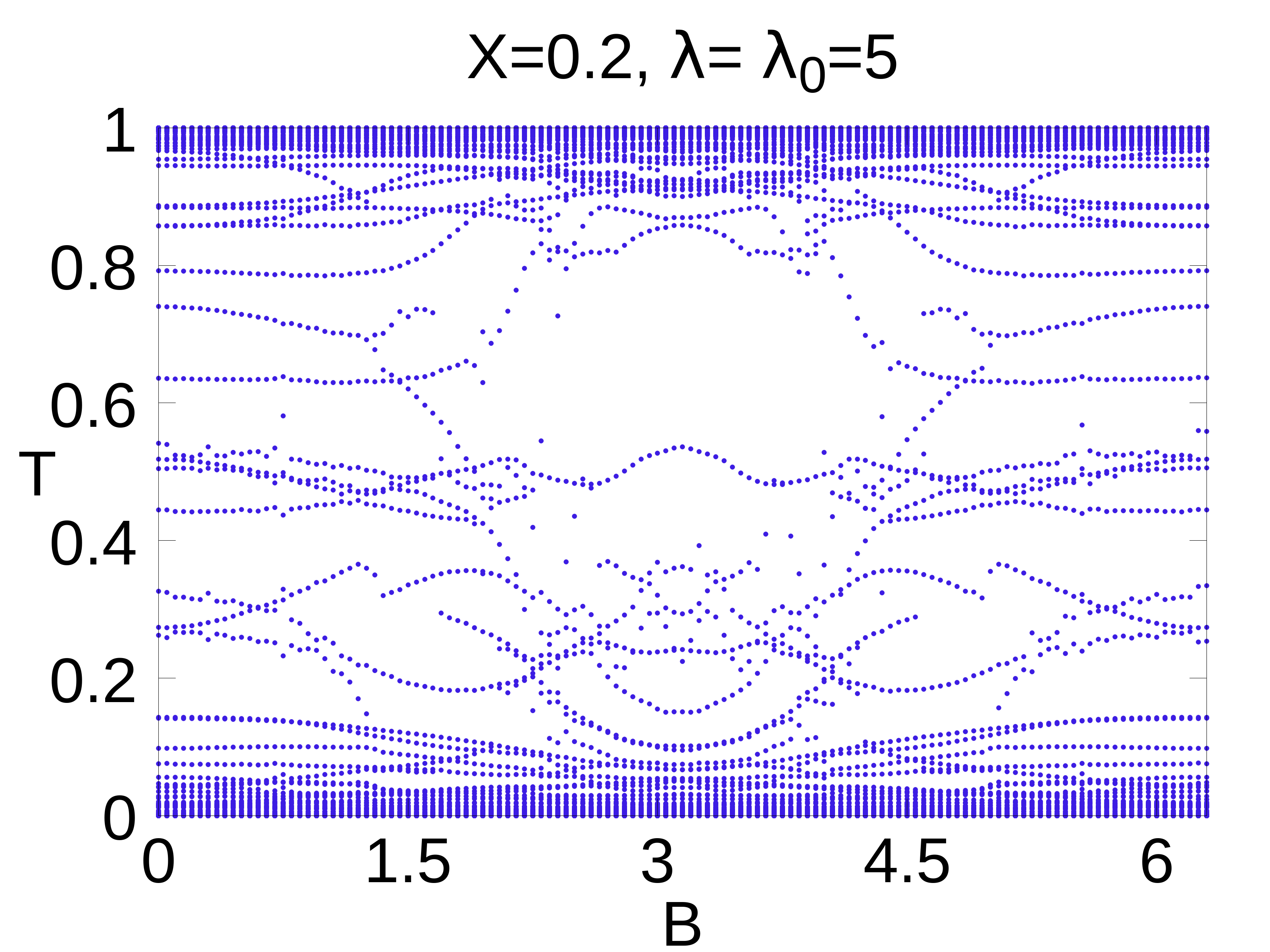}
		\subcaption{}
		\label{sq5}
	\end{subfigure}%
	\begin{subfigure}{.24\textwidth}
		\centering 
		\includegraphics[width=\linewidth]{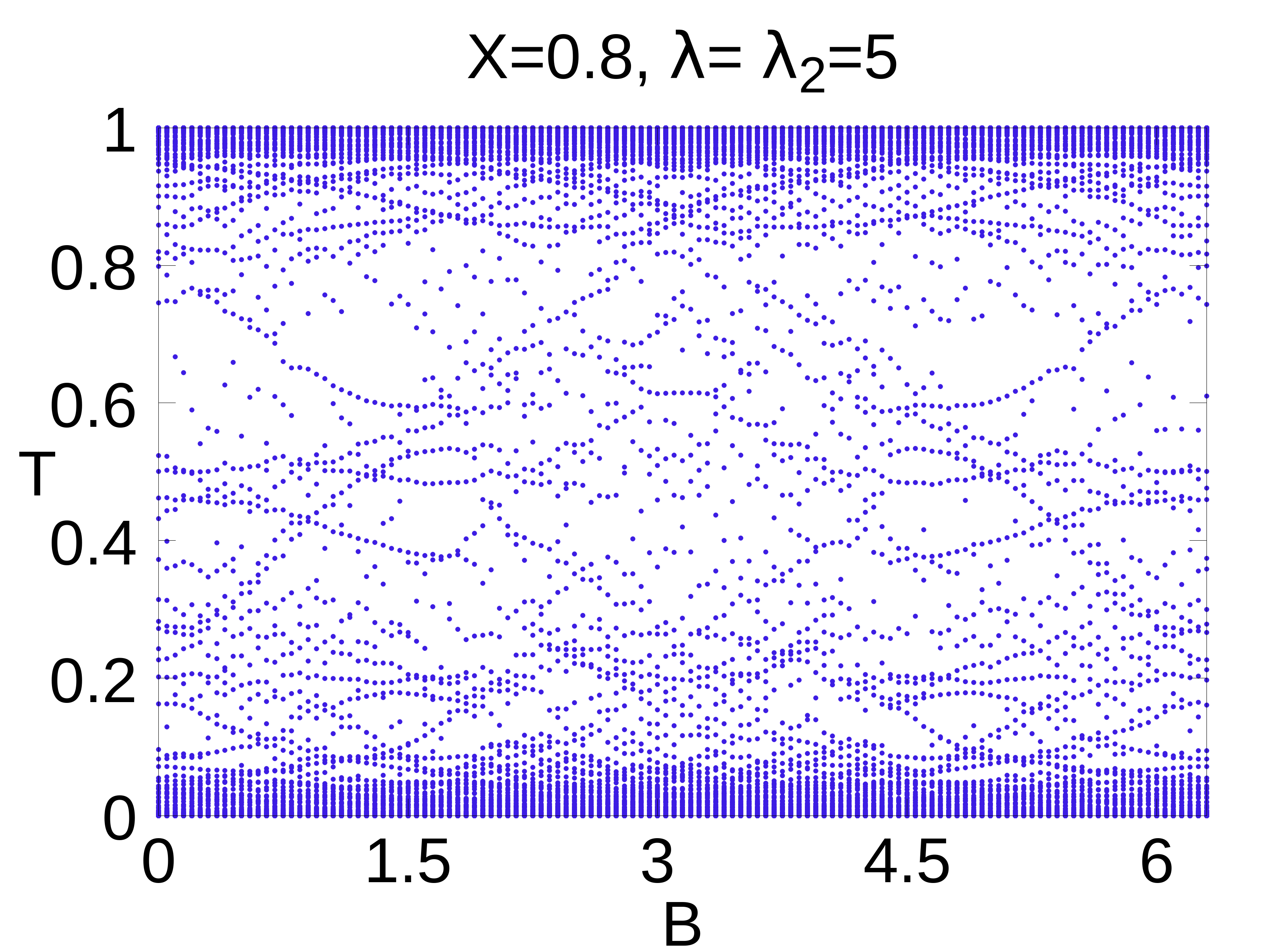}
		\subcaption{}
		\label{sq6}
	\end{subfigure}%
	\begin{subfigure}{.24\textwidth}
		\centering
		\includegraphics[width=\linewidth]{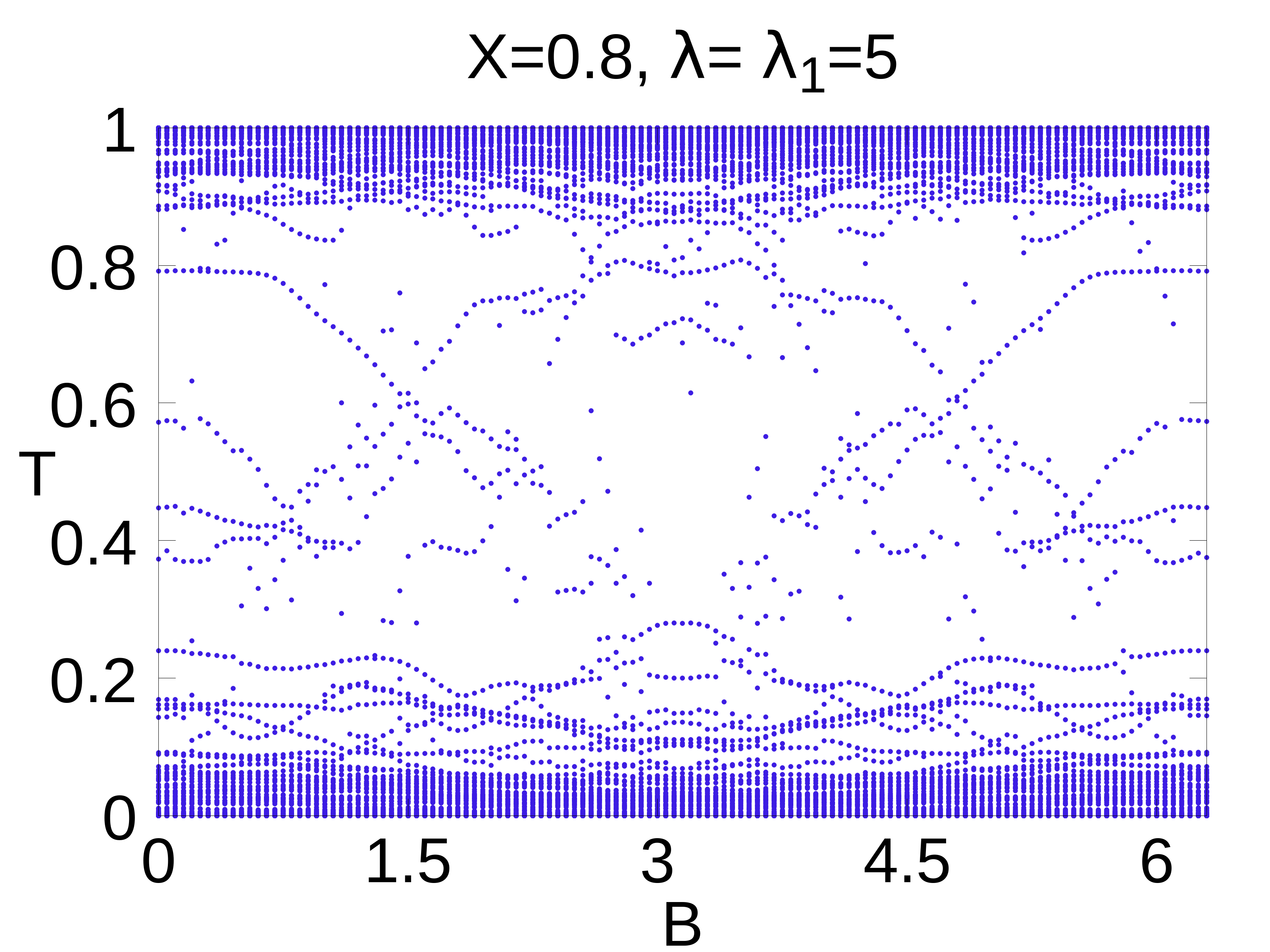}
		\subcaption{}
		\label{sq7}
	\end{subfigure}%
	\begin{subfigure}{.24\textwidth}
		\centering
		\includegraphics[width=\linewidth]{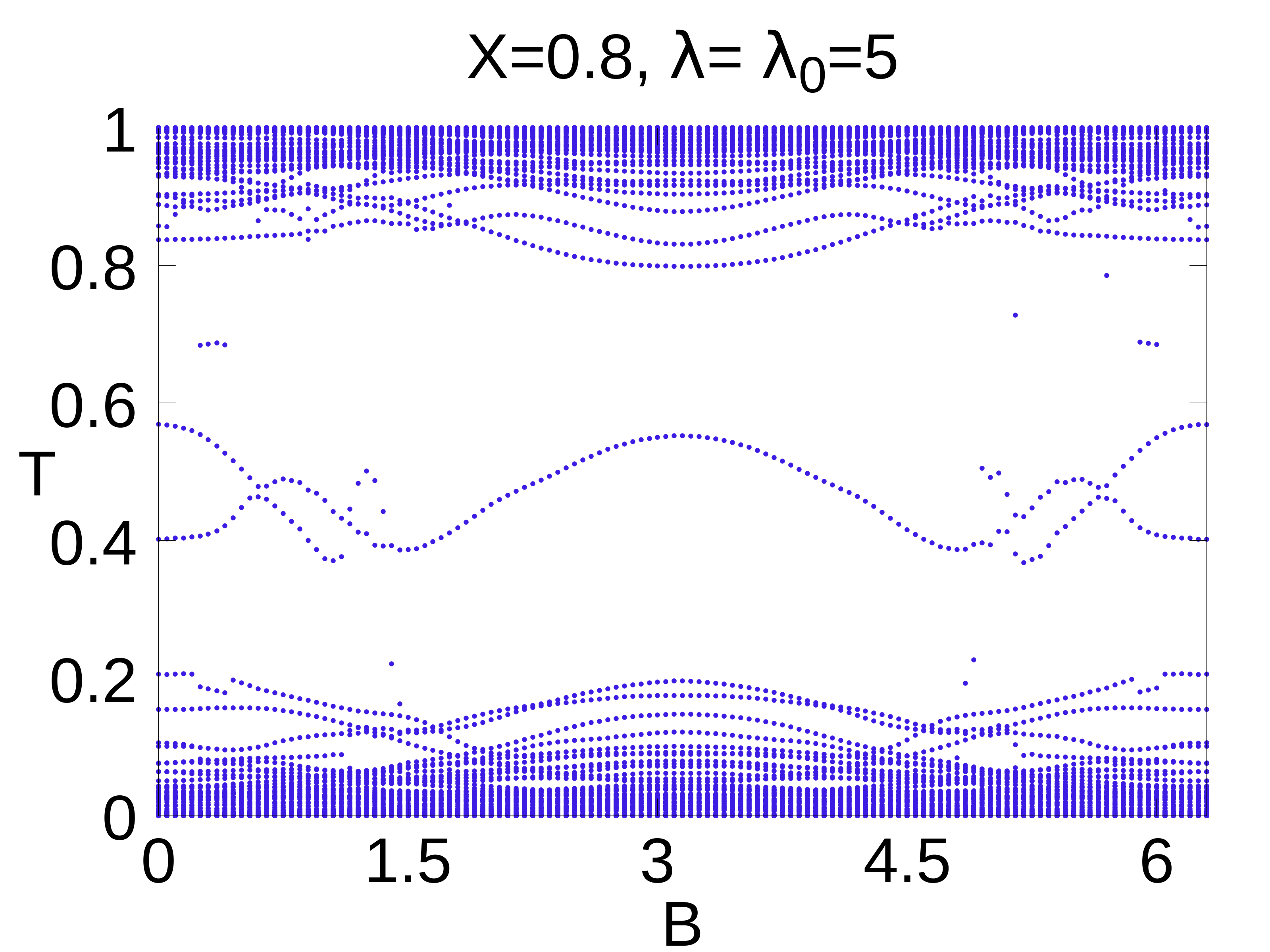}
		\subcaption{}
		\label{sq8}
	\end{subfigure}%

	\caption{Entanglement spectrum $T$ plotted as a function of the magnetic field $B$ for a square lattice, showing the combined effects of the Aubry--André potential with 
	other quasiperiodic disorders. The panels illustrate the crossover from an AA-dominated regime to regimes where the Fibonacci, Rudin--Shapiro, and 
		Thue--Morse components become dominant, respectively.}
	\label{ATTB}
\end{figure*}

	 In the pure case, the correlation spectrum exhibits reflection symmetry and fine structures reminiscent of the Hofstadter spectrum. All types of disorders break both the symmetry and the fine structure of the spectrum, as shown in Fig. \ref{ATB}(a–d). For TM disorder, the correlation spectrum retains some fractal features, with a gap appearing near $B=\pi$. It is well known that interactions create a gap in the spectrum. When both disorder and interactions are present, the gap at half filling is slightly reduced, and the edge gap of the spectrum is also closed which is open due to interaction \cite{Mandal}.\\
	In Fig. \ref{DTB}, we show the entanglement spectrum as a function of magnetic flux $B$ for different disorder strengths and types. For weak and moderate disorder, the spectrum exhibits multiple bands and fine structures that vary with $B$. As the disorder strength increases, a noticeable gap opens in the spectrum, with the fine features gradually disappearing. In particular, for strong TM disorder (e.g., $d=5$), the spectrum becomes almost flat apart from a few isolated bands, indicating strong localization and the suppression of $B$-dependent structure. In these cases, the gap is wide and the correlation spectrum loses its intricate patterns, reflecting a highly localized phase.  Figure \ref{ATTB}(a), \ref{ATTB}(b),\ref{ATTB}(c) and \ref{ATTB}(d) show the entanglement spectrum as a functiom of $B$ for the interpolation between dominating  AA with TM and AA with dominating component of Fibonacci, RS, and TM disorders, respectively. \\

\section{Inverse Participate Ratio (IPR)}	

The inverse participation ratio (IPR) is a useful quantity to analyze the degree of localization of a quantum state in a lattice system. For a normalized single-particle wavefunction expressed as
\begin{equation}
	|\Psi\rangle = \sum_i a_i\, c_i^\dagger |0\rangle,
\end{equation}
where $c_i^\dagger$ is the creation operator at site $i$ and $a_i$ represents the complex amplitude at that site, the IPR is defined as
\begin{equation}
	I = \sum_i |a_i|^4.
\end{equation}

For a completely delocalized state with uniform amplitude $a_i \approx 1/\sqrt{N}$ over $N$ sites, the IPR scales as $I \approx 1/N \to 0$ in the thermodynamic limit. In contrast, for a state fully localized at a single site, $a_j = 1$, $a_{i \ne j} = 0$, one obtains $I = 1$. Hence, $I \sim 1$ indicates strong localization, while $I \sim 0$ characterizes extended states. In condensed matter systems, these correspond to insulating and metallic phases, respectively.
The normalized participation ratio (NPR) behaves inversely to the IPR; smaller NPR values signify localization, whereas larger values correspond to extended states.

\begin{figure*}[!htb]
	\centering
	\begin{subfigure}{.25\textwidth}
		\centering
		\includegraphics[width=\linewidth]{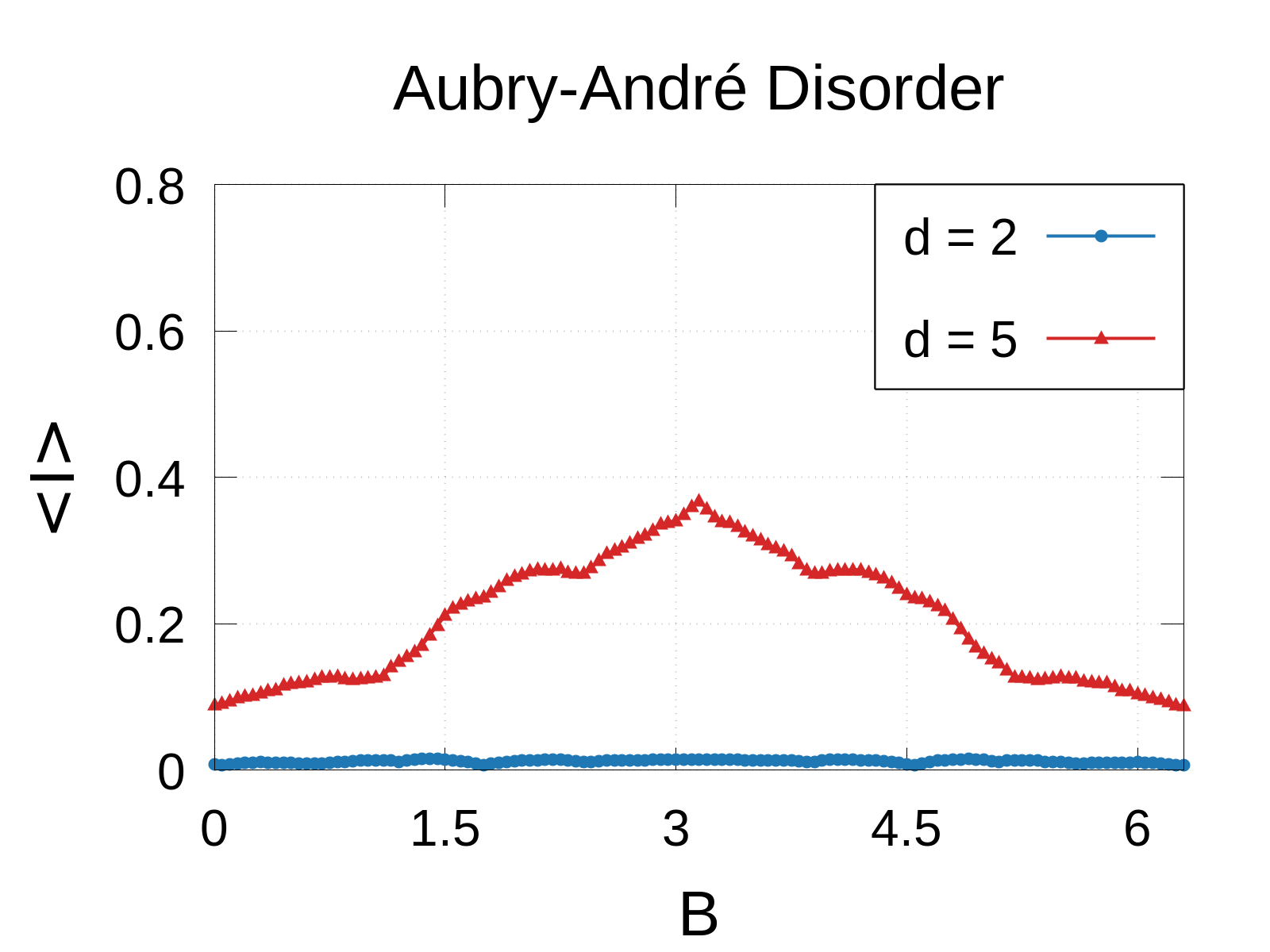}
		\subcaption{}
		\label{sq1}
	\end{subfigure}%
	\begin{subfigure}{.25\textwidth}
		\centering 
		\includegraphics[width=\linewidth]{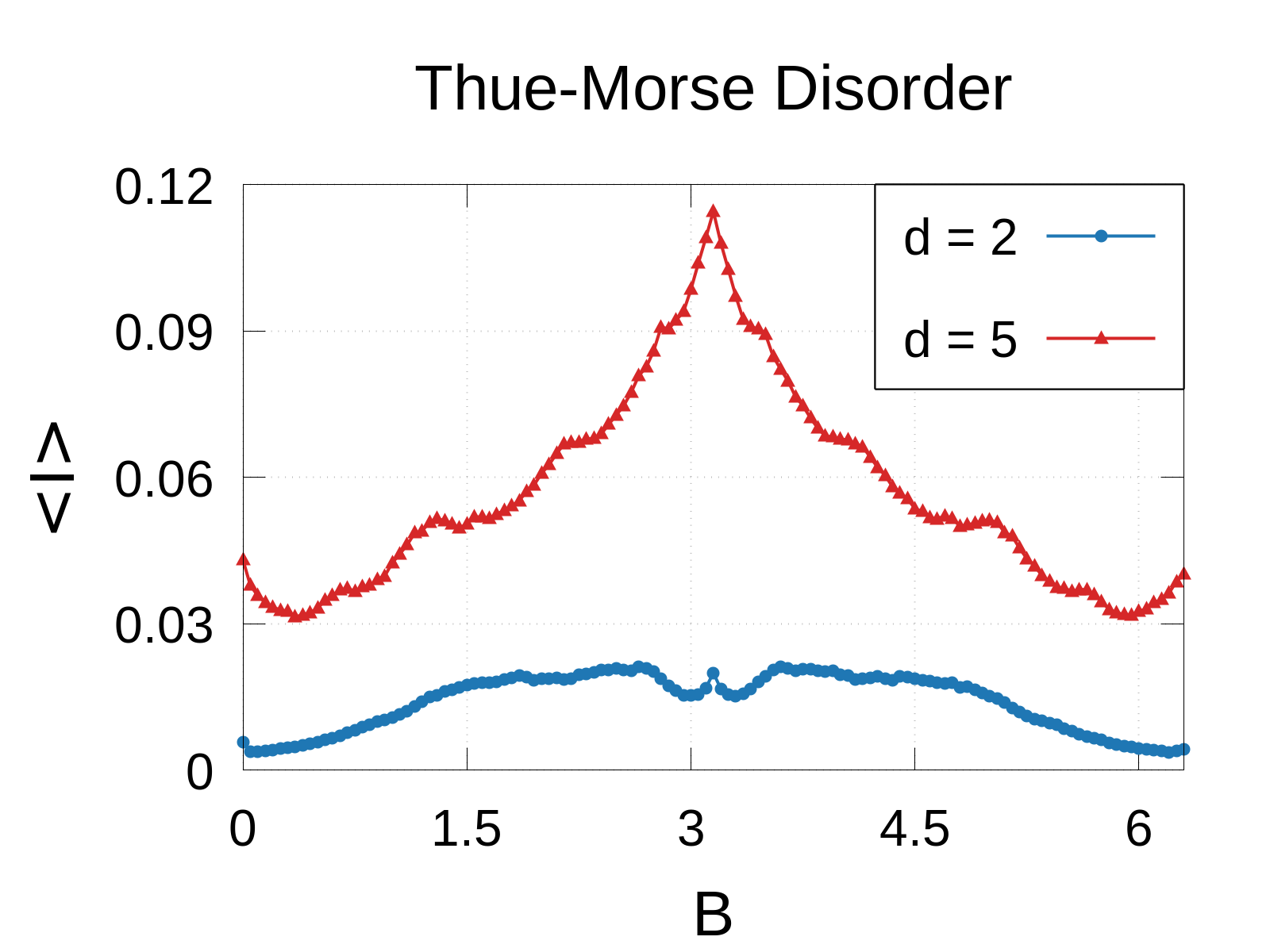}
		\subcaption{}
		\label{sq2}
	\end{subfigure}%
	\begin{subfigure}{.25\textwidth}
		\centering
		\includegraphics[width=\linewidth]{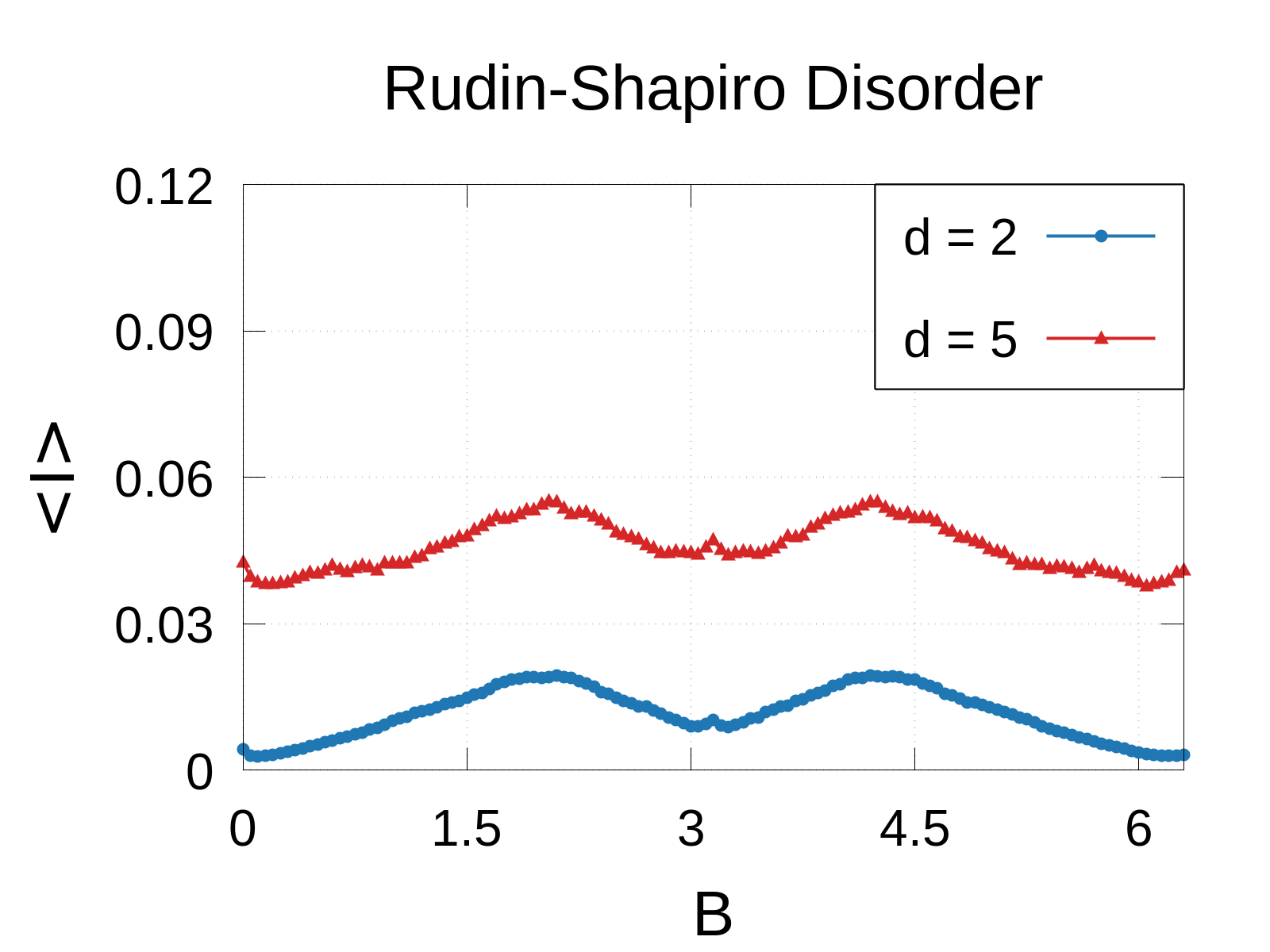}
		\subcaption{}
		\label{sq3}
	\end{subfigure}%
	\begin{subfigure}{.25\textwidth}
		\centering
		\includegraphics[width=\linewidth]{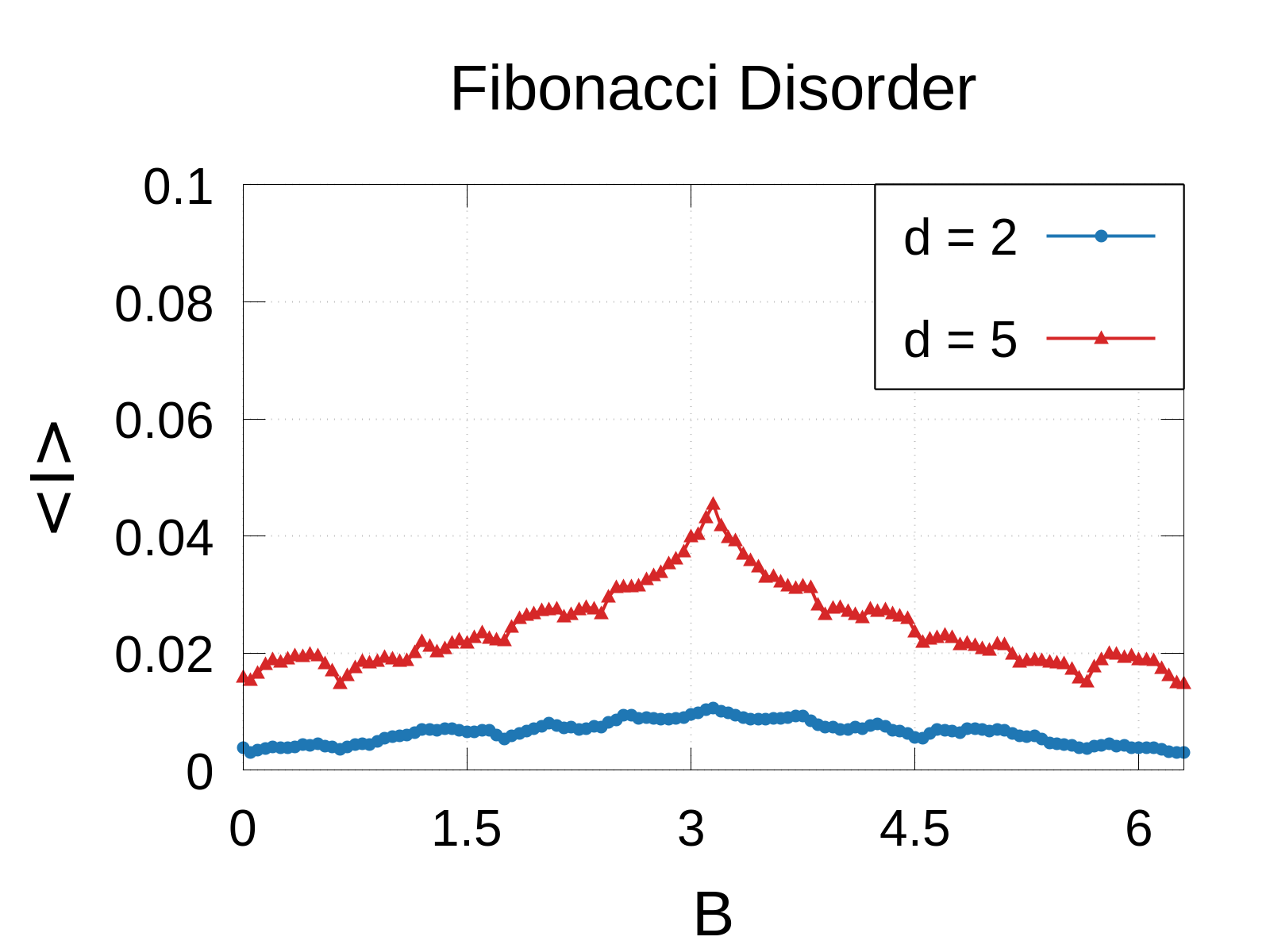}
		\subcaption{}
		\label{sq4}
	\end{subfigure}%

	\begin{subfigure}{.25\textwidth}
		\centering
		\includegraphics[width=\linewidth]{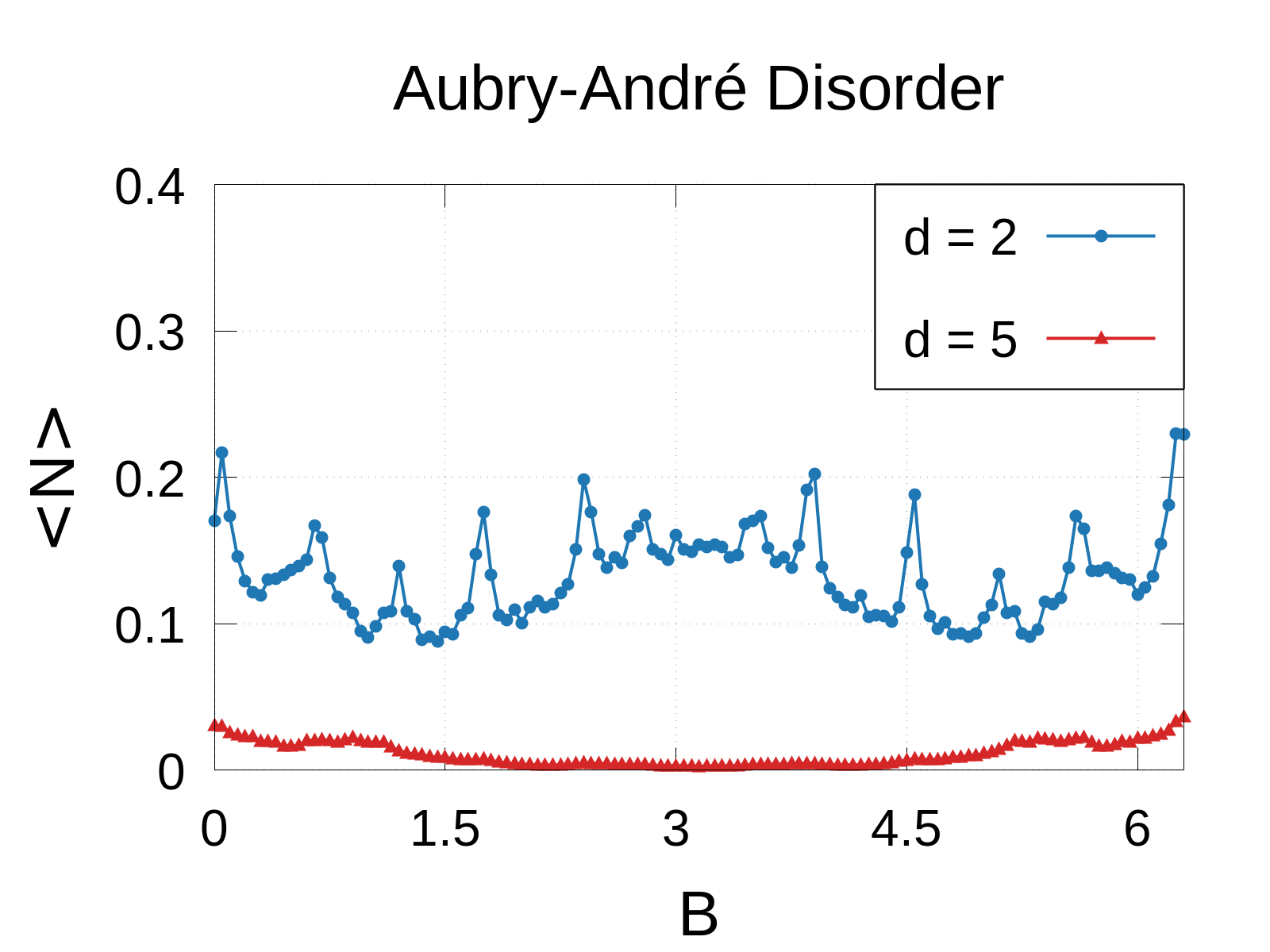}
		\subcaption{}
		\label{sq1}
	\end{subfigure}%
	\begin{subfigure}{.25\textwidth}
		\centering 
		\includegraphics[width=\linewidth]{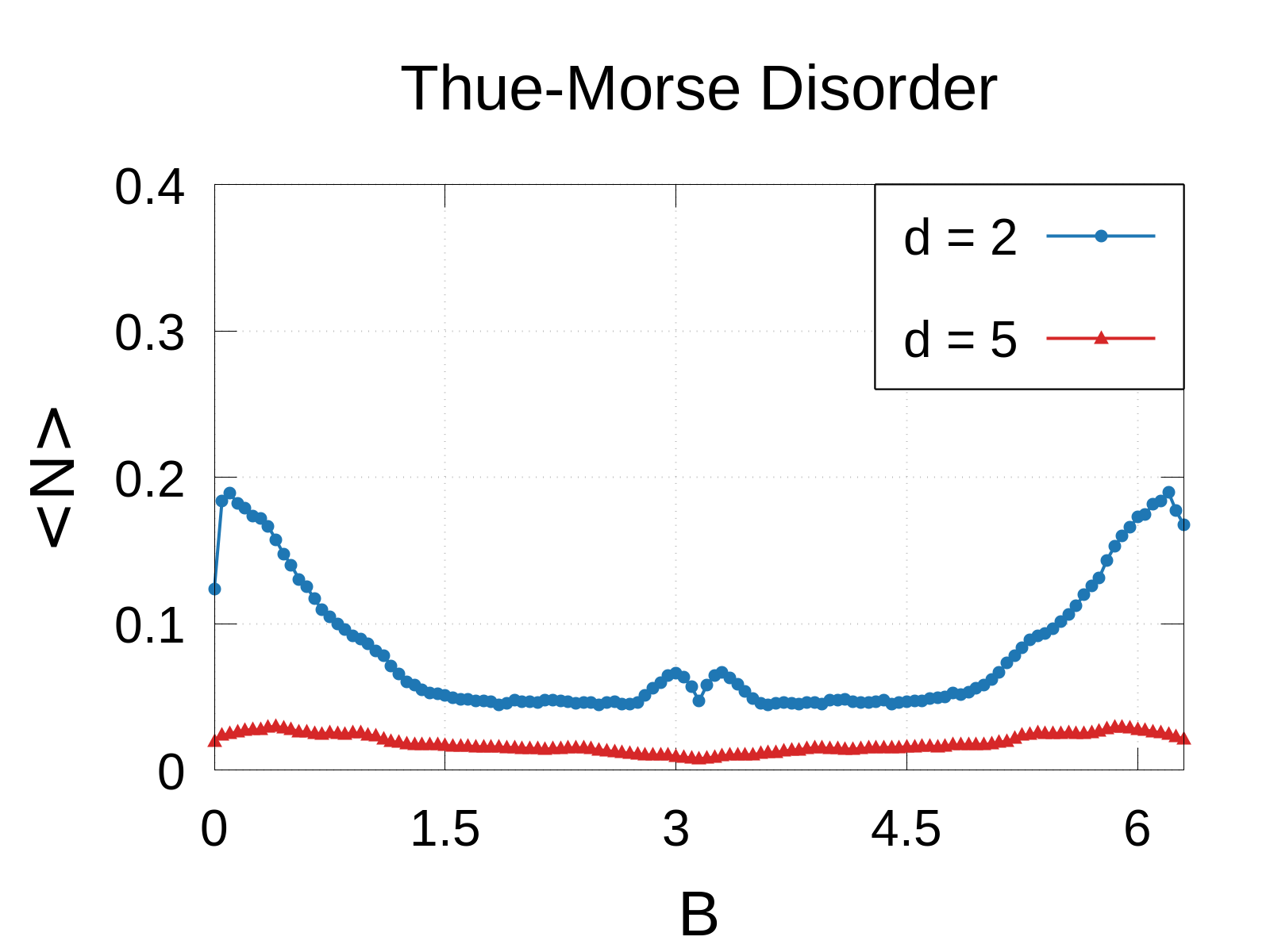}
		\subcaption{}
		\label{sq2}
	\end{subfigure}%
	\begin{subfigure}{.25\textwidth}
		\centering
		\includegraphics[width=\linewidth]{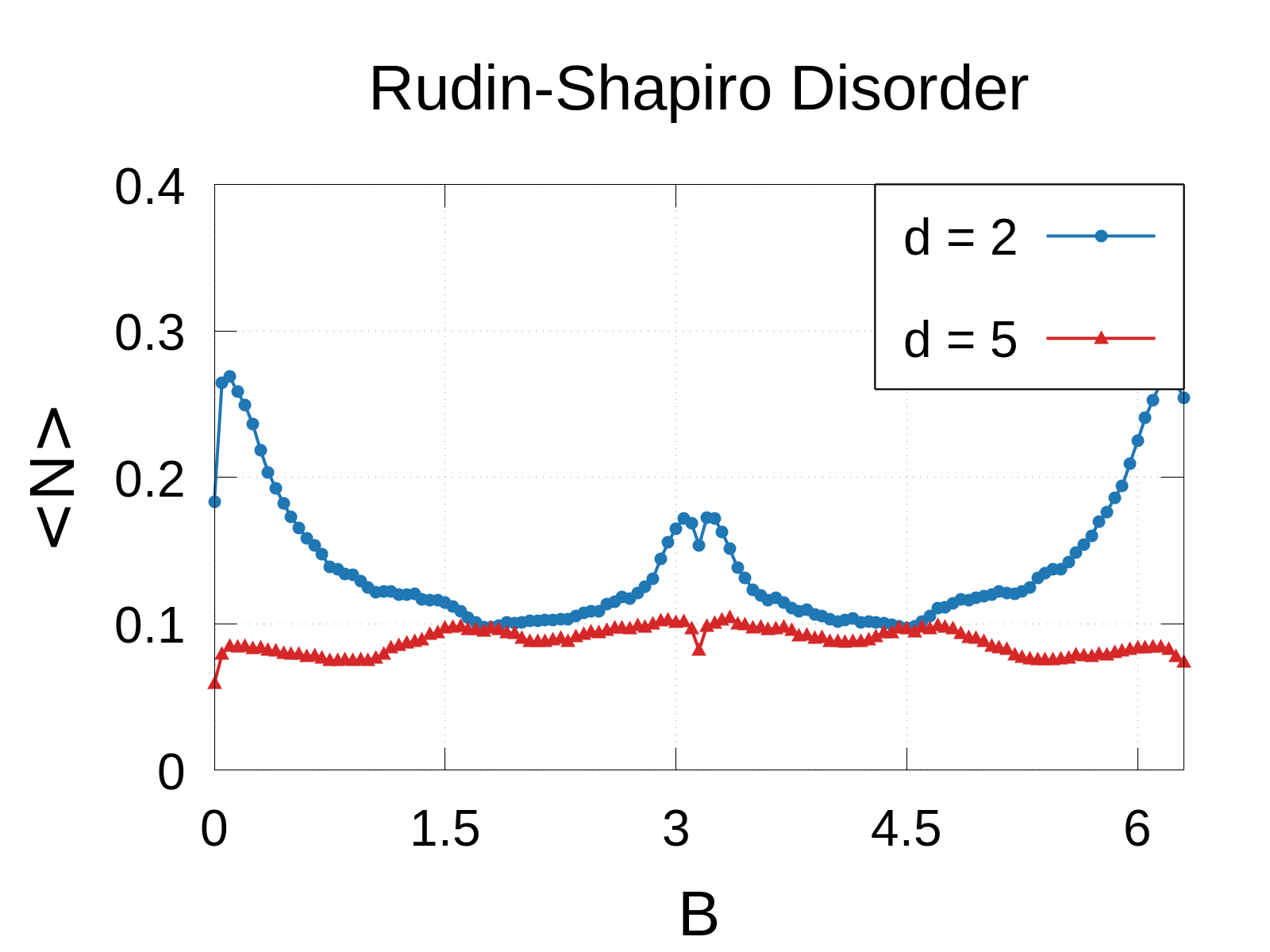}
		\subcaption{}
		\label{sq3}
	\end{subfigure}%
	\begin{subfigure}{.25\textwidth}
		\centering
		\includegraphics[width=\linewidth]{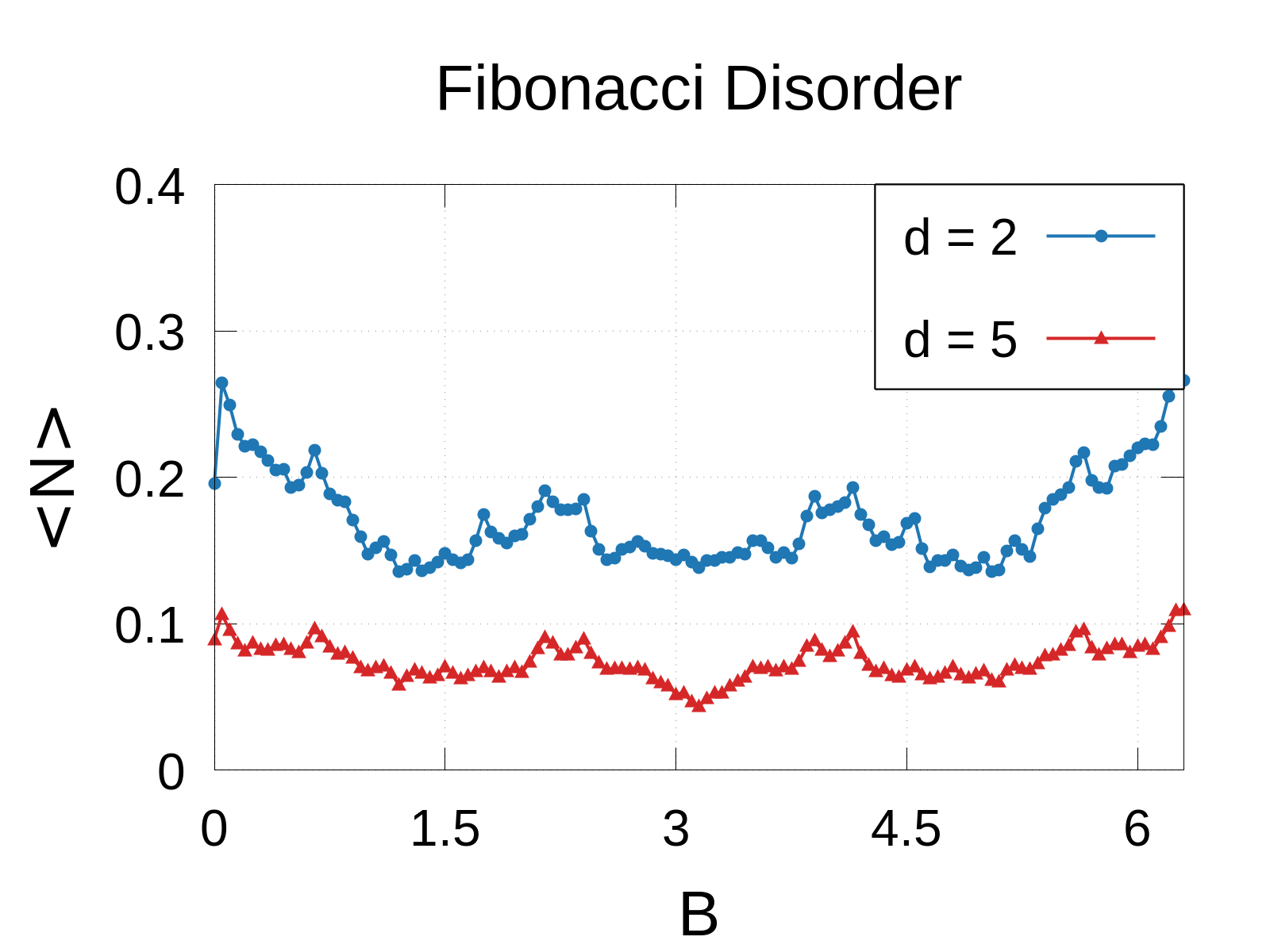}
		\subcaption{}
		\label{sq4}
	\end{subfigure}%
	
	\caption{Average IPR $\langle I \rangle$ (upper panel) and average NPR $\langle N \rangle$ (lower panel) as functions of the magnetic field $B$ for different quasiperiodic disorders. Results are shown for two disorder strengths, $d = 2$ and $d = 5$.}
	\label{AIP40}
\end{figure*}

Figure \ref{AIP40}(a–d) shows the variation of the avrage IPR as a function of the magnetic field $B$ for lattice size 40 and different  quasi-periodic potential. From the figures, it is clearly seen that as the disorder strength increases, the value of the IPR also increases. Among all four types of disorders, the average IPR value is highest for the AA disorder, indicating that the maximum number of states are localized in this case. For all disorders, the average IPR remain symmetric around  $B=\pi$. For small values of the disorder strength, the average IPR $<I>$ is minimal, while for larger disorder values, it reaches a maximum near $B=\pi$ for AA,TM and Fibonacci. In Fig. Figure~\ref{AIP40}(e-h) illustrates the average NPR ($<N>$) as functions of the magnetic field $B$ for a lattice size of 40 under various quasiperiodic disorders. As the NPR is inversely related to the IPR, higher values correspond to delocalized states at weak AA disorder, whereas lower values indicate localization at strong disorder. For larger system sizes and stronger disorder strengths, the IPR approaches unity while the NPR tends toward zero, indicating that almost all states become localized. 

\vspace{0.2cm}


\begin{figure*}[!htb]
	\centering
	\begin{subfigure}{.30\textwidth}
		\centering
		\includegraphics[width=\linewidth]{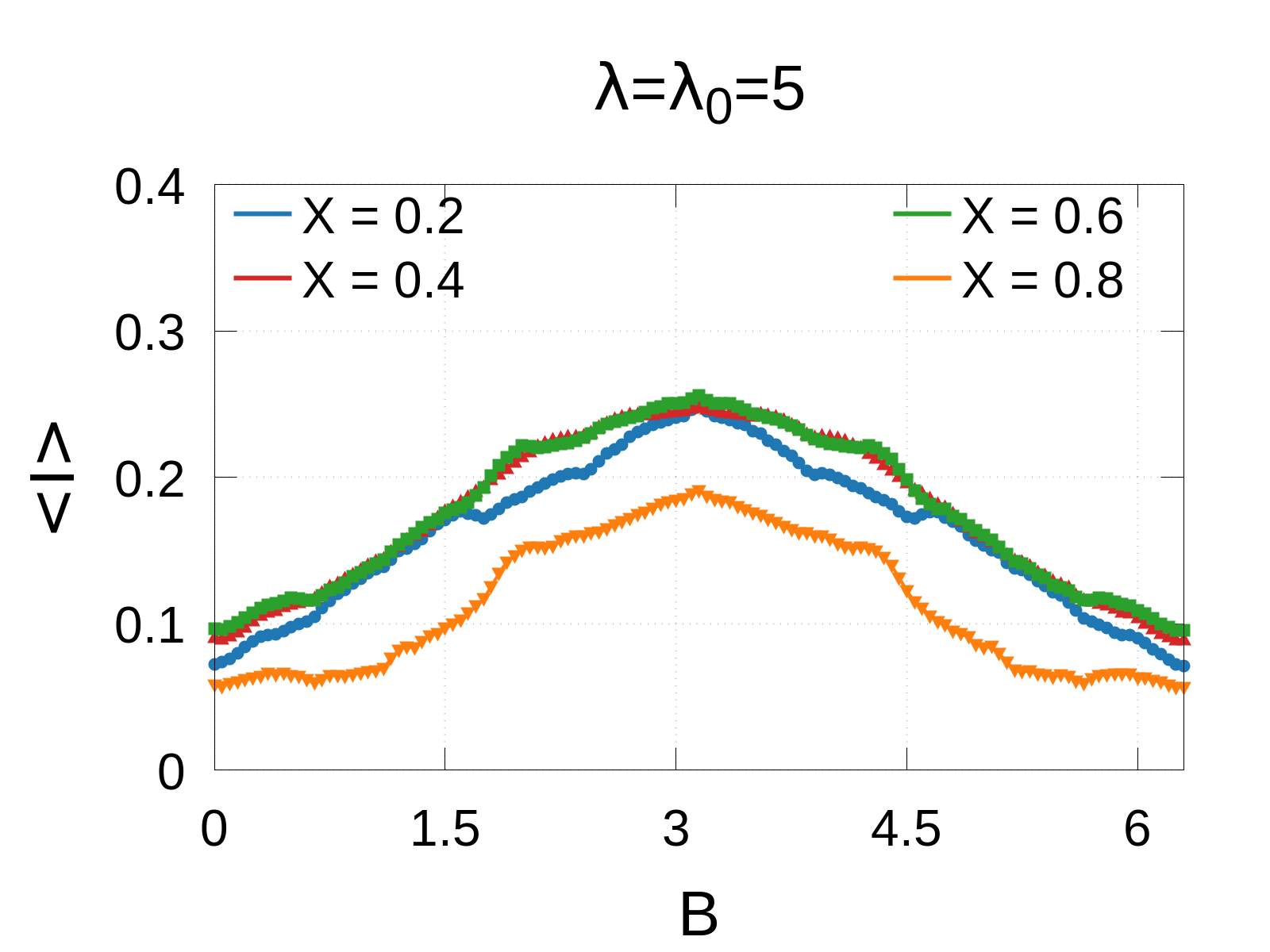}
		\subcaption{}
		\label{sq1}
	\end{subfigure}%
	\begin{subfigure}{.30\textwidth}
		\centering 
		\includegraphics[width=\linewidth]{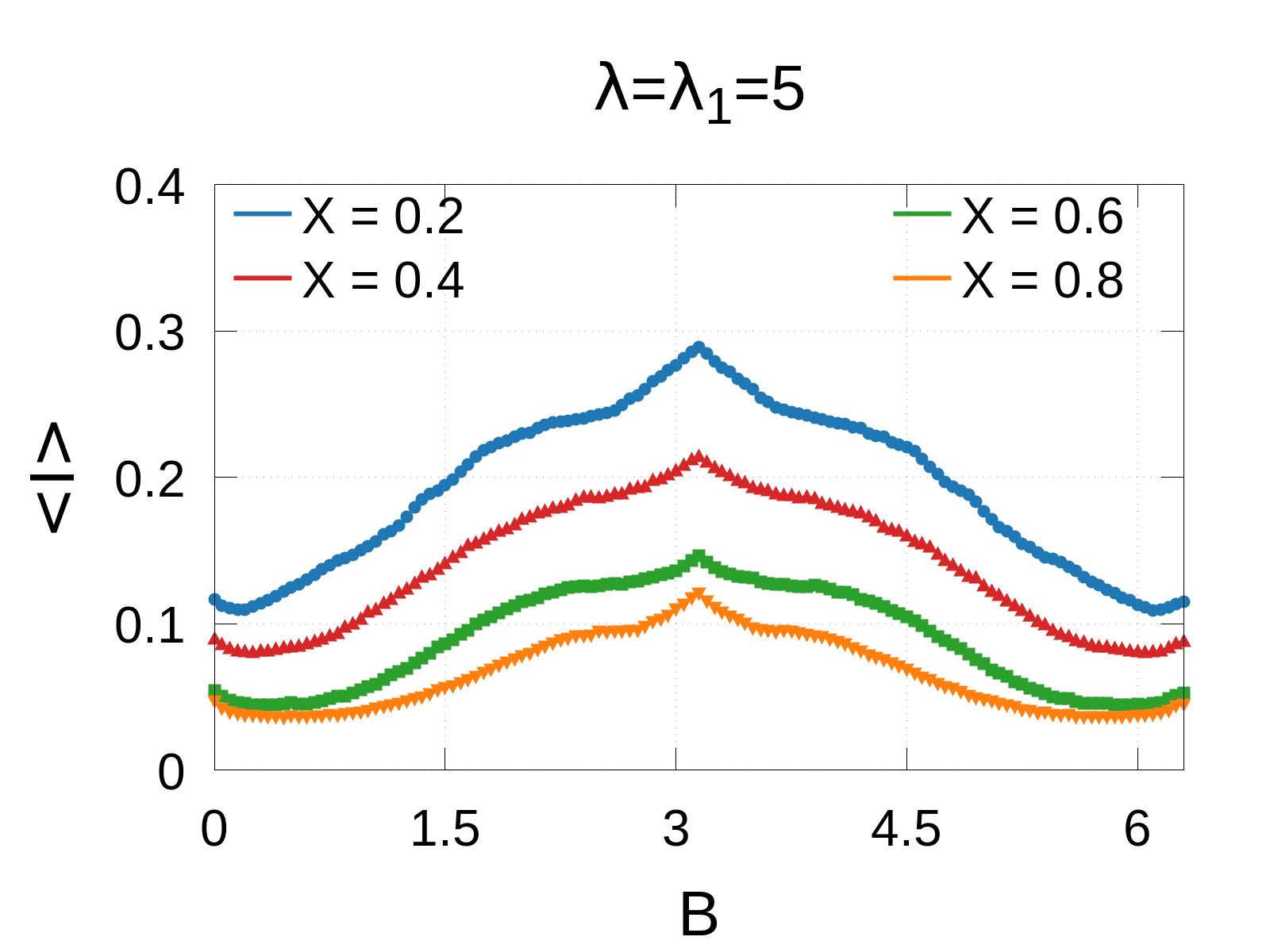}
		\subcaption{}
		\label{sq2}
	\end{subfigure}%
	\begin{subfigure}{.30\textwidth}
		\centering
		\includegraphics[width=\linewidth]{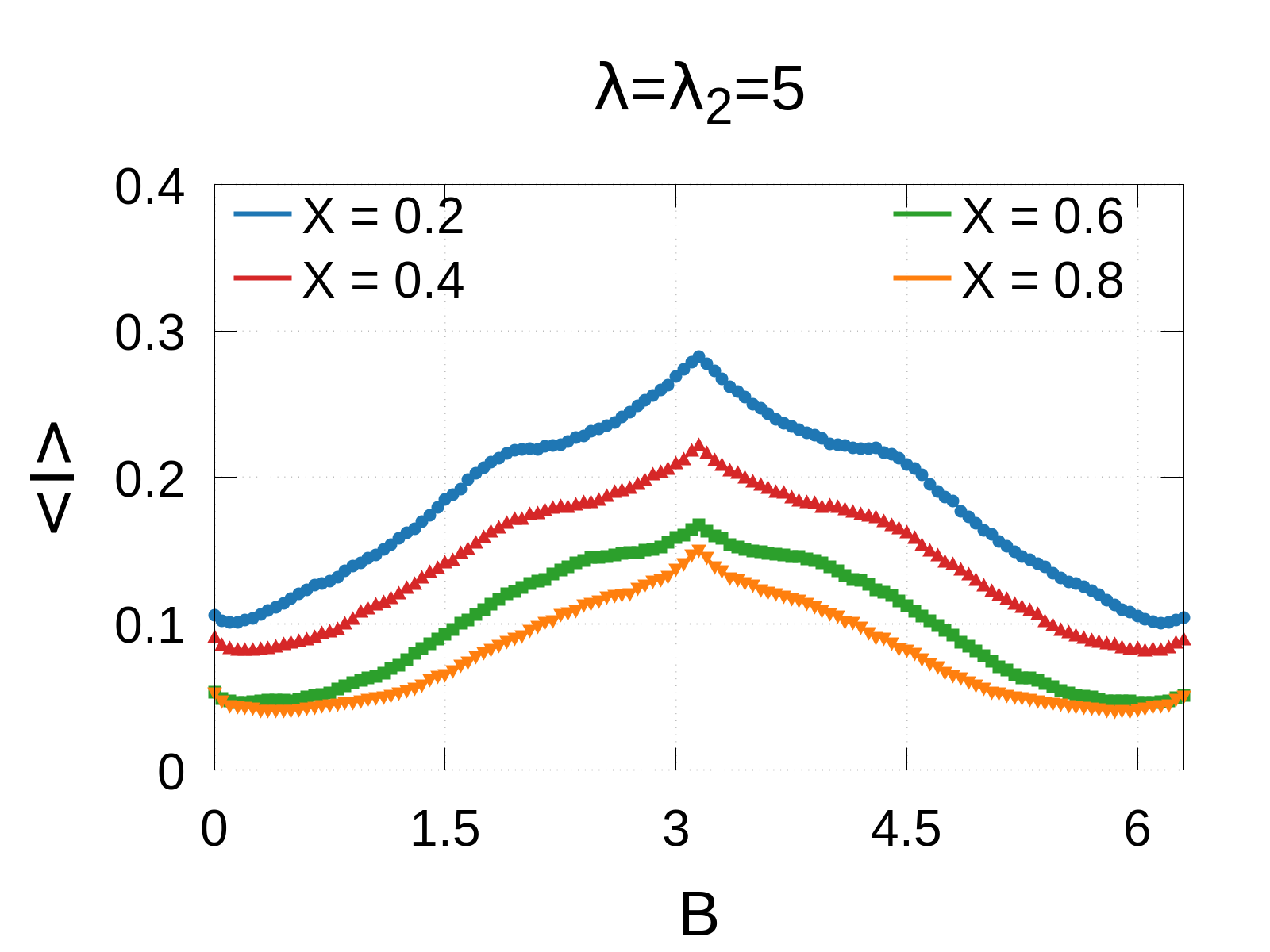}
		\subcaption{}
		\label{sq3}
	\end{subfigure}%

	\begin{subfigure}{.30\textwidth}
	\centering
	\includegraphics[width=\linewidth]{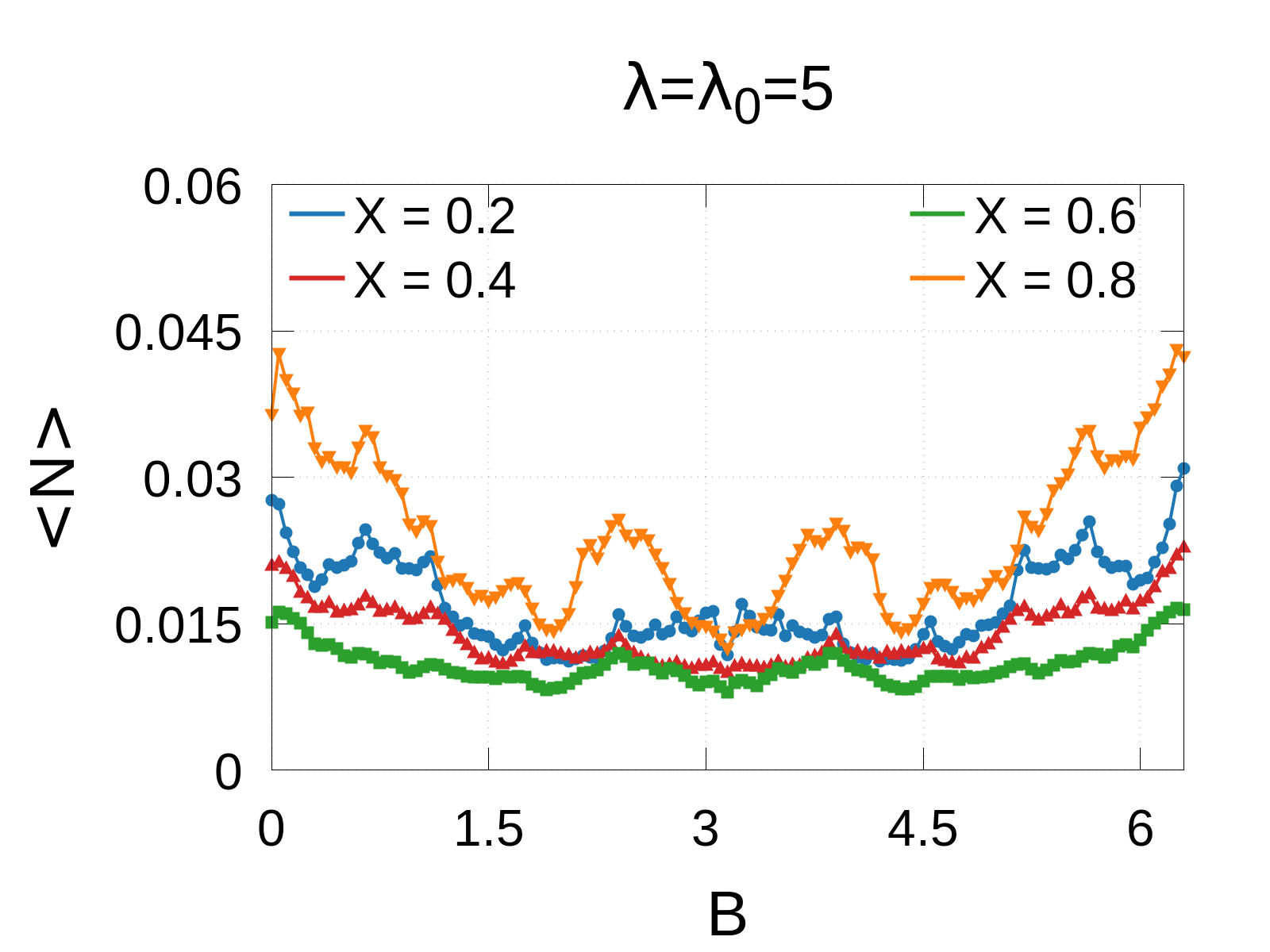}
	\subcaption{}
	\label{sq1}
\end{subfigure}%
\begin{subfigure}{.30\textwidth}
	\centering 
	\includegraphics[width=\linewidth]{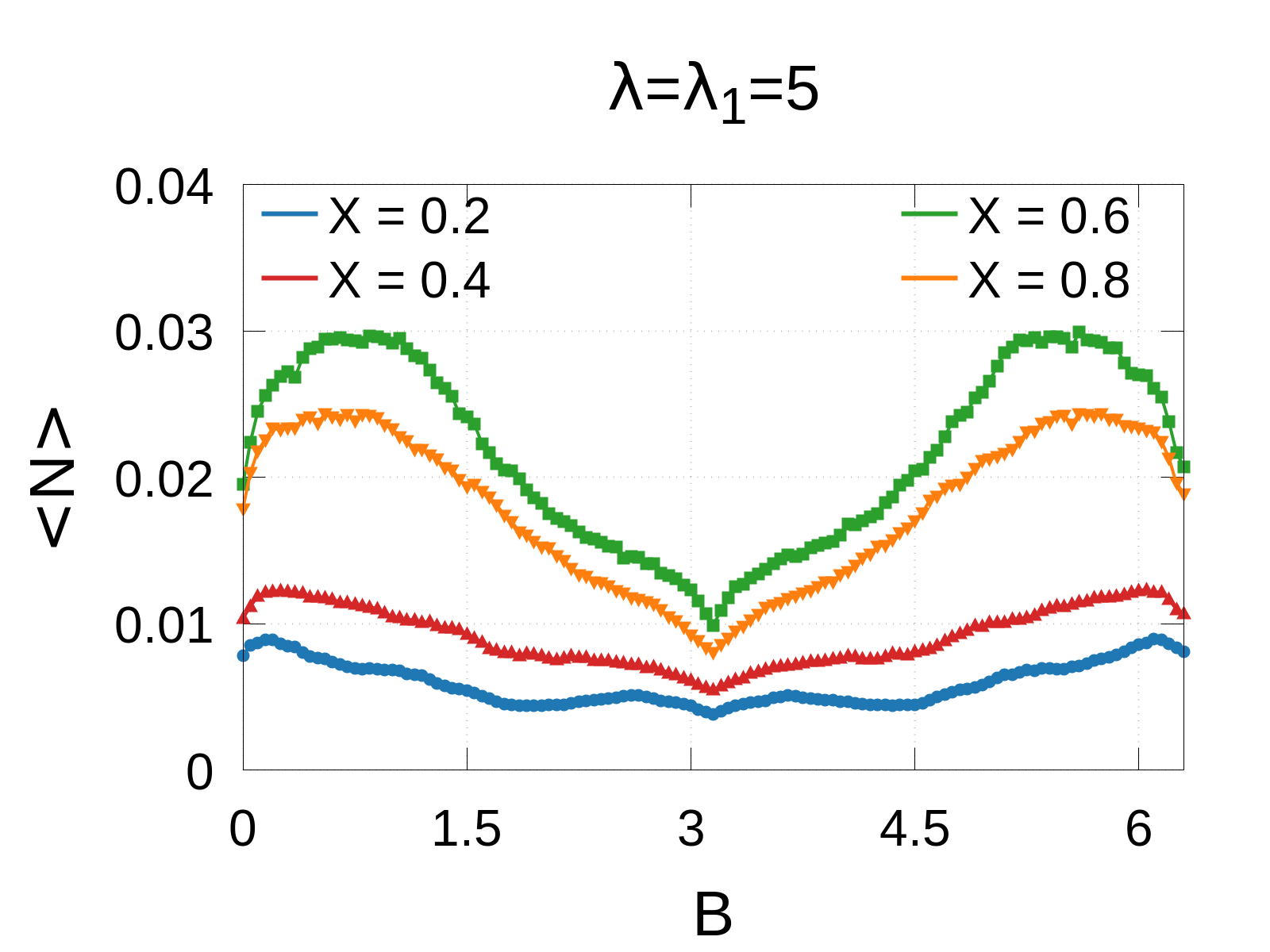}
	\subcaption{}
	\label{sq2}
\end{subfigure}%
\begin{subfigure}{.30\textwidth}
	\centering
	\includegraphics[width=\linewidth]{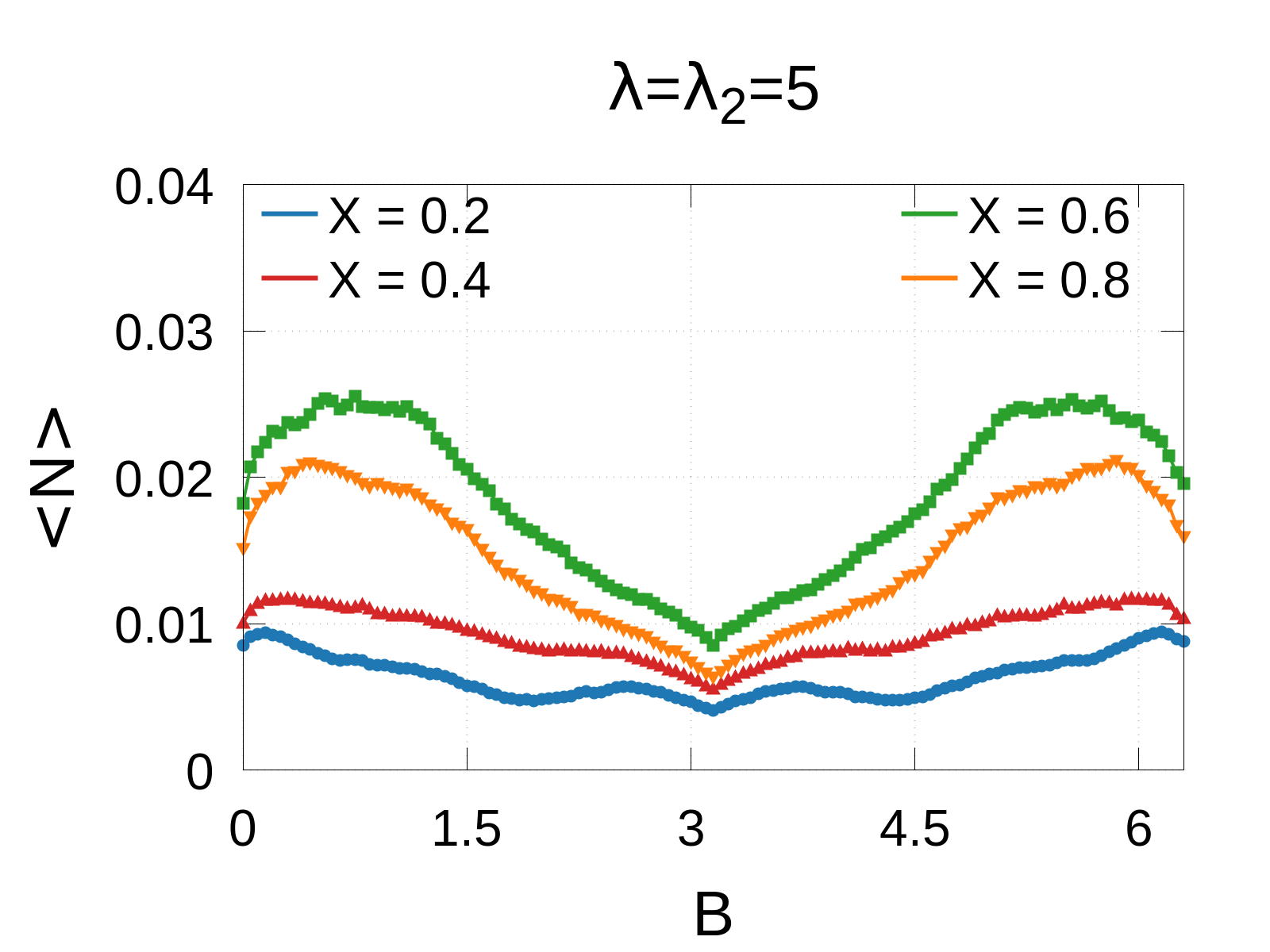}
	\subcaption{}
	\label{sq3}
\end{subfigure}%
	
	\caption{Average IPR $\langle I \rangle$ (upper panel) and average NPR $\langle N \rangle$ (lower panel) as functions of the magnetic field $B$ for the interpolation between the Aubry--André potential and other quasiperiodic disorders. The left column shows the interpolation between AA and Thue--Morse, the middle column corresponds to AA and Rudin--Shapiro, and the right column presents the AA–Fibonacci interpolation.}
	\label{IPR40}
\end{figure*}

The average IPR $\langle I \rangle$ and NPR $\langle N \rangle$ as functions of 
the magnetic flux for the system interpolating between the AA and TM disorders 
are shown in Figs.~\ref{IPR40}(a) and \ref{IPR40}(d), respectively. For small values of the interpolation parameter, the system is dominated by the AA disorder, exhibiting relatively high IPR values. Interestingly, for intermediate interpolation parameters ($X=0.4$ and $0.6$), where the effects of AA and TM disorders mix, the average IPR value increases. However, for larger interpolation parameters $X=0.8$, where the TM disorder dominates, $<I>$ decreases compared to the AA regime. As shown in Fig. \ref{IPR40}(d), for small values of the interpolation parameter, the NPR $<N>$ remains low, reflecting strong localization arising from the dominant AA disorder. The IPR reaches its maximum and the NPR attains a minimum at $B=\pi$, demonstrating the symmetric nature of the system.\\


Figures \ref{IPR40}(b-c) present the IPR as a function of magnetic flux $B$ for systems with interpolated disorder between AA and RS disorders (Fig. \ref{IPR40}(b)), and between AA and Fibonacci disorders (Fig. \ref{IPR40}(c)), respectively. Similar to previous observations, we find that for small values of the interpolation parameter $X$, the AA disorder dominates, resulting in higher IPR values that indicate stronger localization of states. In particular, the combination of AA and RS disorder produces a significant localization effect at low $X$. Interestingly, the strongest localization (i.e., the highest IPR values) is observed when AA is mixed with RS disorder. This is because the RS disorder, by itself, is relatively strong in inducing localization. Even when its strength is increased, its overall impact on localization remains limited. Therefore, the localization in this case is primarily driven by the AA component, especially at lower $X$. \\
The NPR as a function of magnetic flux $B$ for systems with interpolation between AA and RS disorders and between AA and Fibonacci disorders is shown in Figs. \ref{IPR40}(e) and \ref{IPR40}(f), respectively. The system exhibits lower NPR values for the interpolation between AA and RS disorders, whereas the interpolation between AA and Fibonacci disorders results in comparatively higher NPR values. This suggests that the combination of AA and RS disorders enhances localization within the system.


\section{Conclusions and Outlook}
Our work investigates the effects of various quasi-periodic disorders, as well as their interpolation with the AA disorder, on the Hofstadter butterfly in a two-dimensional square lattice. We have shown that for small disorder strengths, the Hofstadter butterfly becomes slightly smeared, while for large quasi-periodic potentials, the butterfly structure disappears completely and multiple gaps emerge in the spectrum. The AA quasiperiodic disorder is stronger compared to the TM, RS, and Fibonacci disorders, and for larger strengths it creates multiple gaps in the energy spectrum. In the strong-disorder regime, the energy spectrum undergoes substantial broadening and develops several distinct gaps around half-filling. We also have seen the interpolation between AA and other quasi-periodic disorders on energy spectrum, entangment entropy and IPR. At small interpolation parameters, the AA quasiperiodic disorder dominates, producing gaps immediately above and below half-filling. In contrast, for large interpolation parameters, the competing quasiperiodic disorders become stronger and generate a pronounced gap at half-filling. In the intermediate regime, these spectral gaps progressively diminish, resulting in a partial collapse of the Hofstadter butterfly structure.\\
The influence of different quasiperiodic disorders, together with interaction effects, is clearly reflected in the entanglement properties. For all the quasiperiodic disorders, the entanglement entropy follows the area law in the low- and high-field regimes. At intermediate magnetic fields, however, noticeable deviations occur for larger subsystem sizes. Increasing the overall disorder strength leads to a reduction in entanglement entropy and a clear departure from the area-law behavior. In the case of AA disorder, a stronger potential suppresses the peak of the entanglement entropy, which eventually turns into a pronounced minimum. For the other quasiperiodic disorders, however, the opposite trend is observed: the entropy minimum evolves into a maximum at intermediate magnetic fields. We have analyzed how the entanglement properties interpolate between the AA disorder and the other quasiperiodic disorders. \\
To understand the localization–delocalization behavior of the system, we studied the avrage IPR and NPR. With increasing disorder strength, the IPR systematically increases and the NPR correspondingly decreases, signaling enhanced localization. Notably, the AA disorder yields the largest IPR among all the quasiperiodic potentials studied. Under strong disorder, the IPR attains its maximum at $B=\pi$ for every disorder except the RS case. We investigated how the localization–delocalization behavior evolves as the AA disorder is interpolated with the other quasiperiodic potentials. 

\section*{Acknowledgements}
We acknowledge the use of SAMKHYA, the High Performance Computing Facility at the 
Institute of Physics, Bhubaneswar. We also thank Prof.~Julian Vidal and Prof.~ Arunava Chakrabarti 
for valuable discussions.

\section*{Appendix}
In this appendix, we briefly discuss the effects of the RS disorder and the 
interpolation between the AA and RS potentials on the energy spectrum and the 
entanglement entropy. The results for the average IPR and NPR, evaluated for a system size of $40$, are also presented here for all disorder types.\\

The energy spectra in figure \ref{IEB}(a-d) reveal a non-monotonic evolution as the disorder transitions from AA-dominated ($\lambda=5, X=0.2$) to RS-dominated ($\lambda_1=5, X=0.8$). At $X=0.2$, the spectrum retains a blur Hofstadter butterfly structure with two ditinct gaps. For intermediate mixing ($X=0.4-0.6$), increasing RS disorder first narrows ($X=0.4$) and then eliminates ($X=0.6$) the gap while fragmenting the fractal pattern. At $ X = 0.8 $, the RS effect becomes significant, influencing the spectral characteristics. 
\begin{figure*}
	\begin{subfigure}{.24\textwidth}
		\centering
		\includegraphics[width=\linewidth]{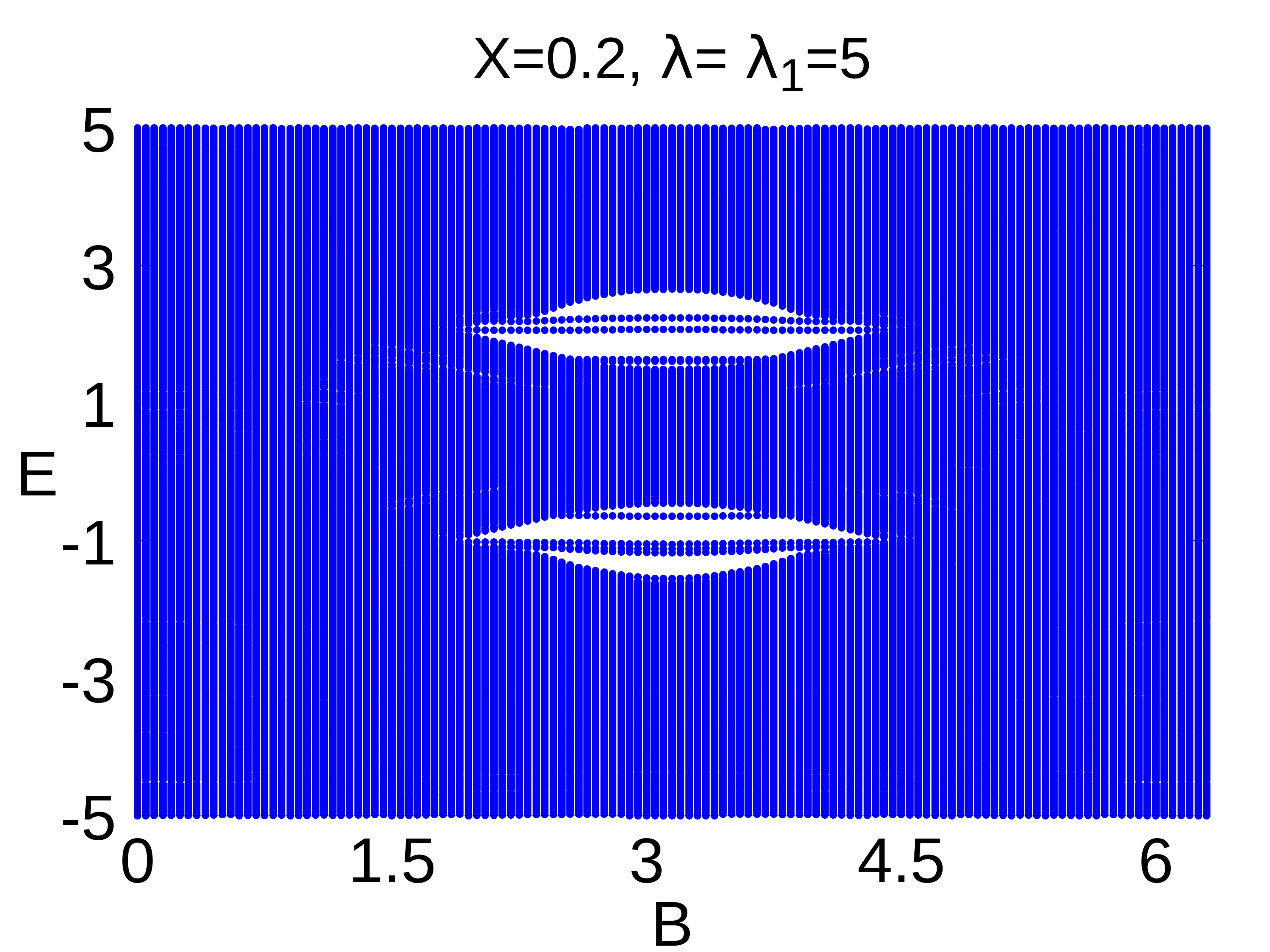}
		\subcaption{}
		\label{IEB13}
	\end{subfigure}%
	\begin{subfigure}{.24\textwidth}
		\centering 
		\includegraphics[width=\linewidth]{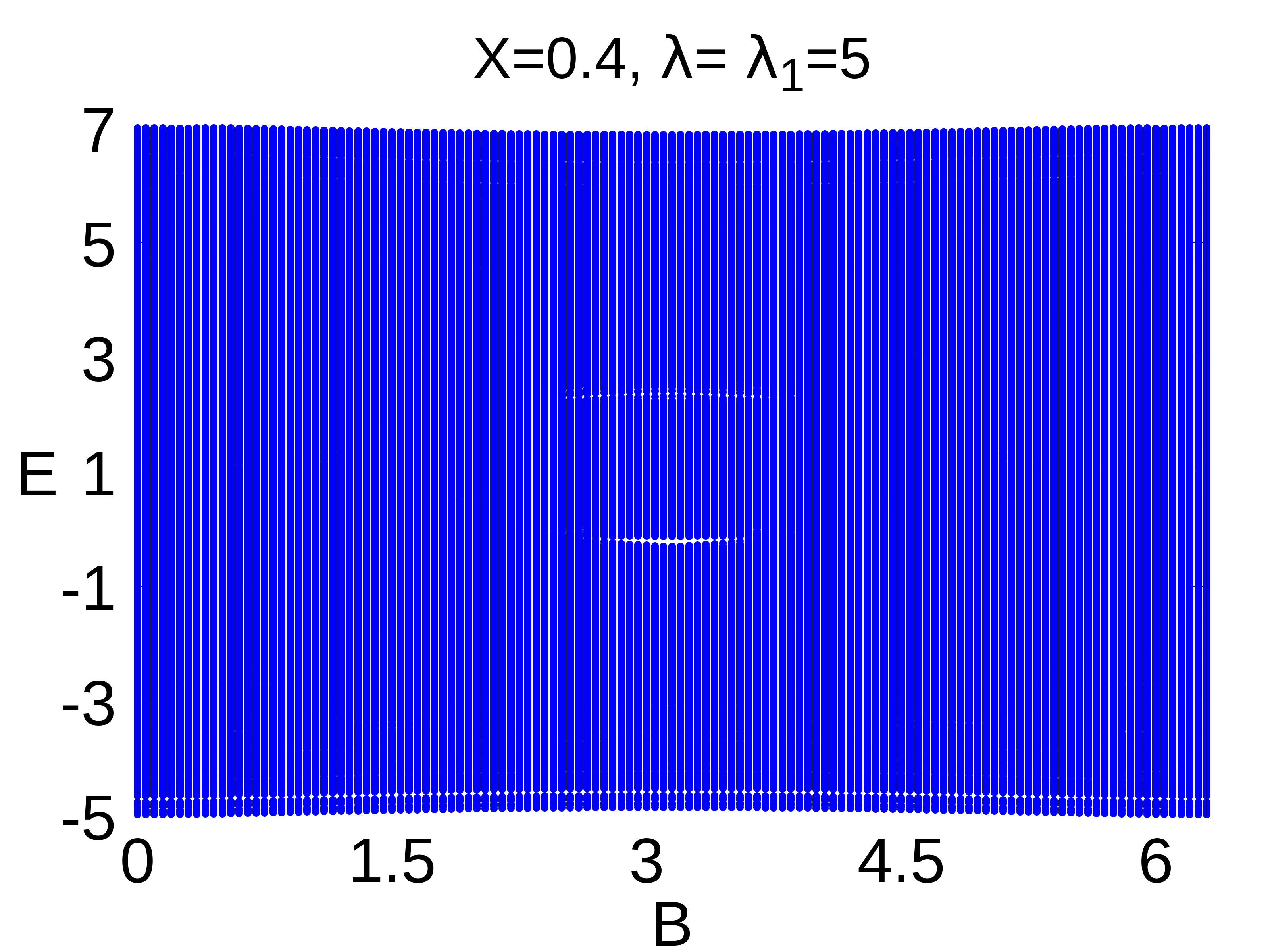}
		\subcaption{}
		\label{IEB14}
	\end{subfigure}%
	\begin{subfigure}{.24\textwidth}
		\centering
		\includegraphics[width=\linewidth]{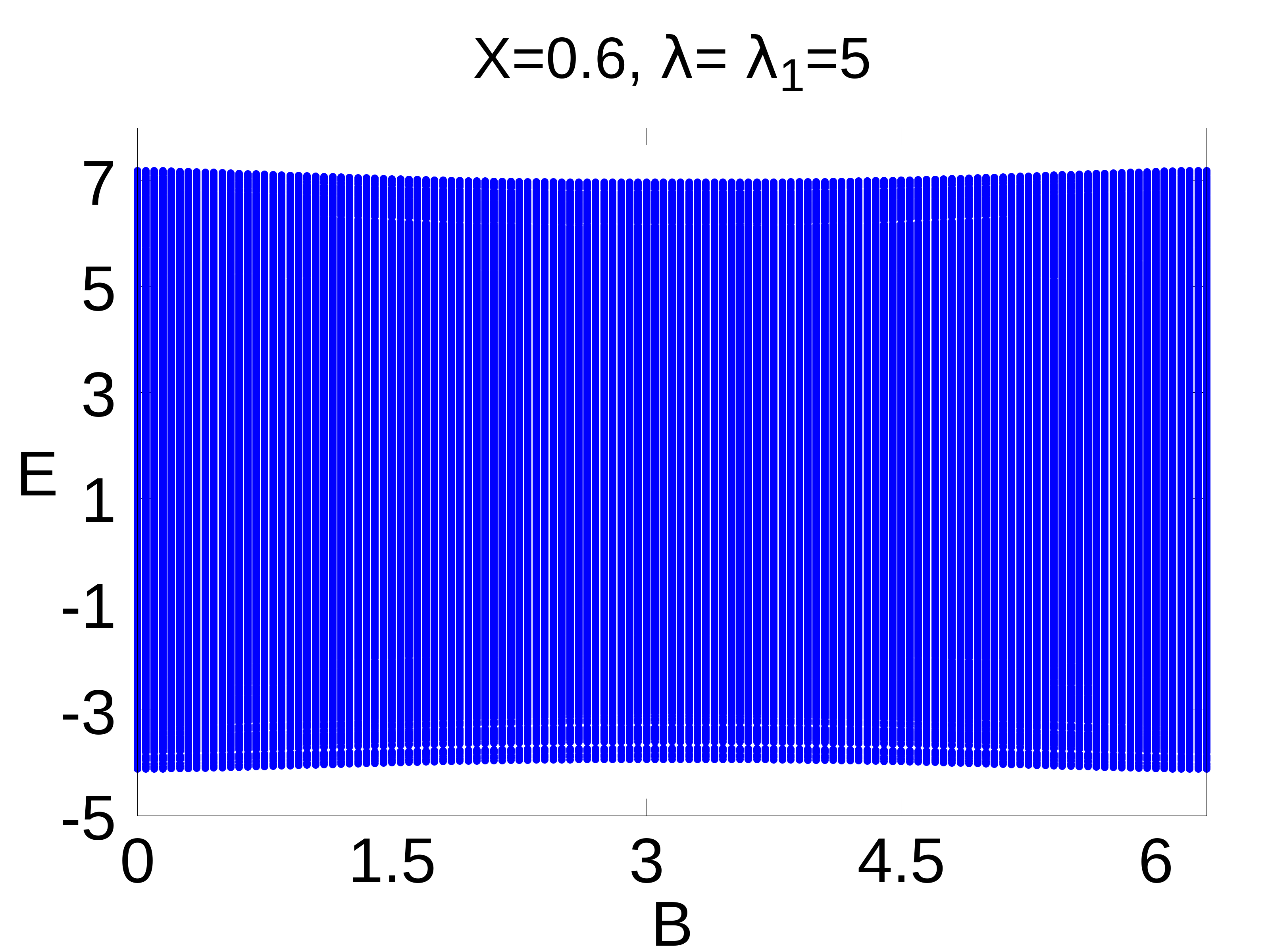}
		\subcaption{}
		\label{IEB15}
	\end{subfigure}%
	\begin{subfigure}{.24\textwidth}
		\centering
		\includegraphics[width=\linewidth]{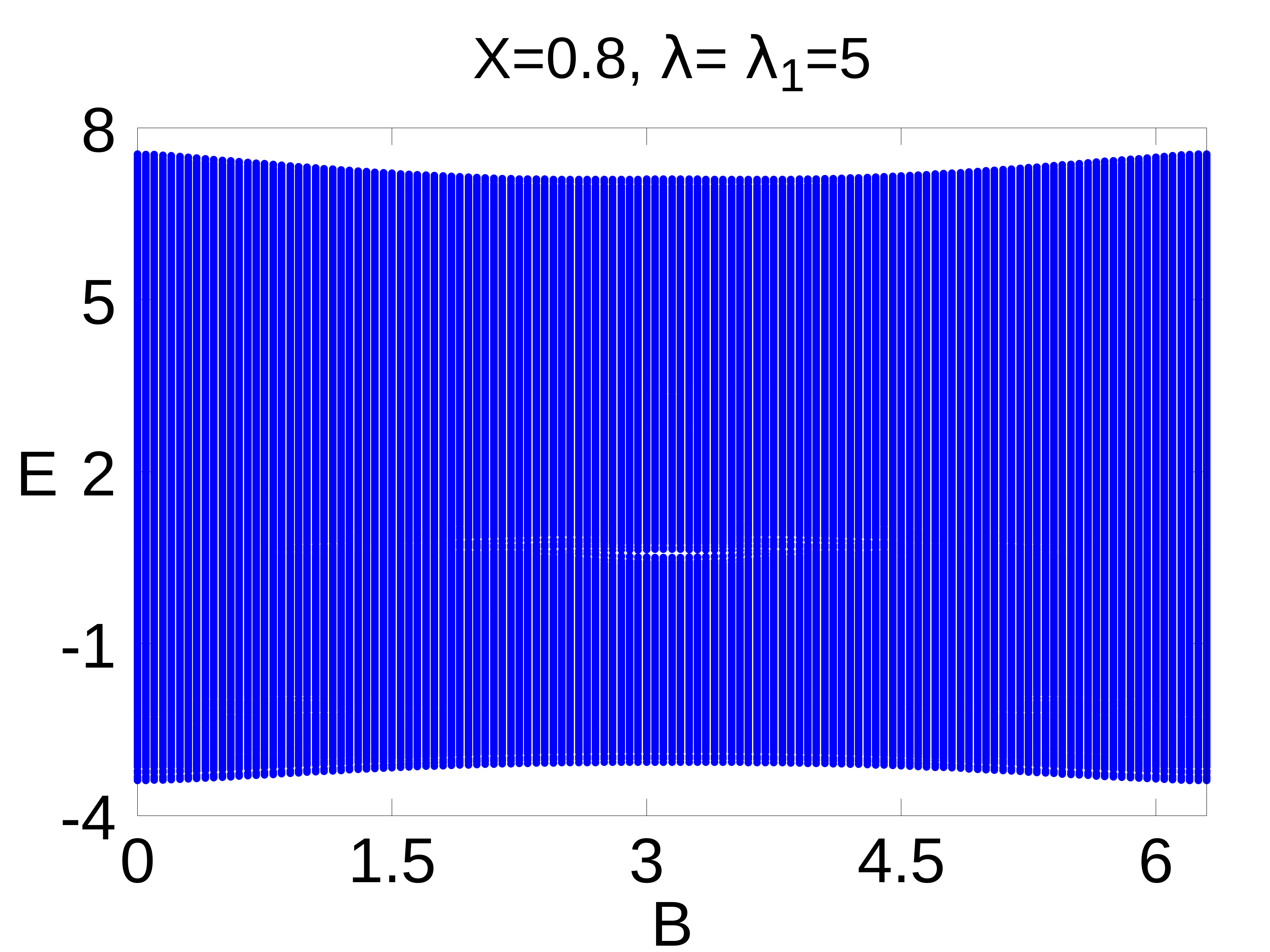}
		\subcaption{}
		\label{IEB16}
	\end{subfigure}%
	
	\centering
	\caption{The energy spectrum (E) is plotted as a function of the magnetic field (B) in a square lattice, illustrating the combined effects of Aubre-Andre and Rudin-Saphiro disorder. }
	\label{IEB}
\end{figure*}
Fig. \ref{IASB} (a-d) shows the EE interpolation between AA($\lambda=5$) and RS binary ($\lambda_2=5$) as a function of magnetic flux. At lower values of the interpolation parameter $X$ (e.g., $X=0.2$), the AA potential dominates the system. In this strong disorder regime, the EE is relatively low and departure the area law, indicating stronger localization of states. As $X$ increases (e.g., From $X=0.6$ to $X=0.8$), the contribution from RS binary disorder becomes more significant. This leads to a noticeable enhancement in EE, reflecting the transition towards more extended states and weaker localization. This transition is characterized by increased EE across a wide range of magnetic flux values.  In this strongly disordered regime, the entanglement entropy no longer follows the area-law behavior, as the entanglement spreads beyond isolated subsystems, and overlaps between subsystem boundaries become prominent, further reflecting the suppression of extended states.

\begin{figure*}
	\begin{subfigure}{.24\textwidth}
		\centering
		\includegraphics[width=\linewidth]{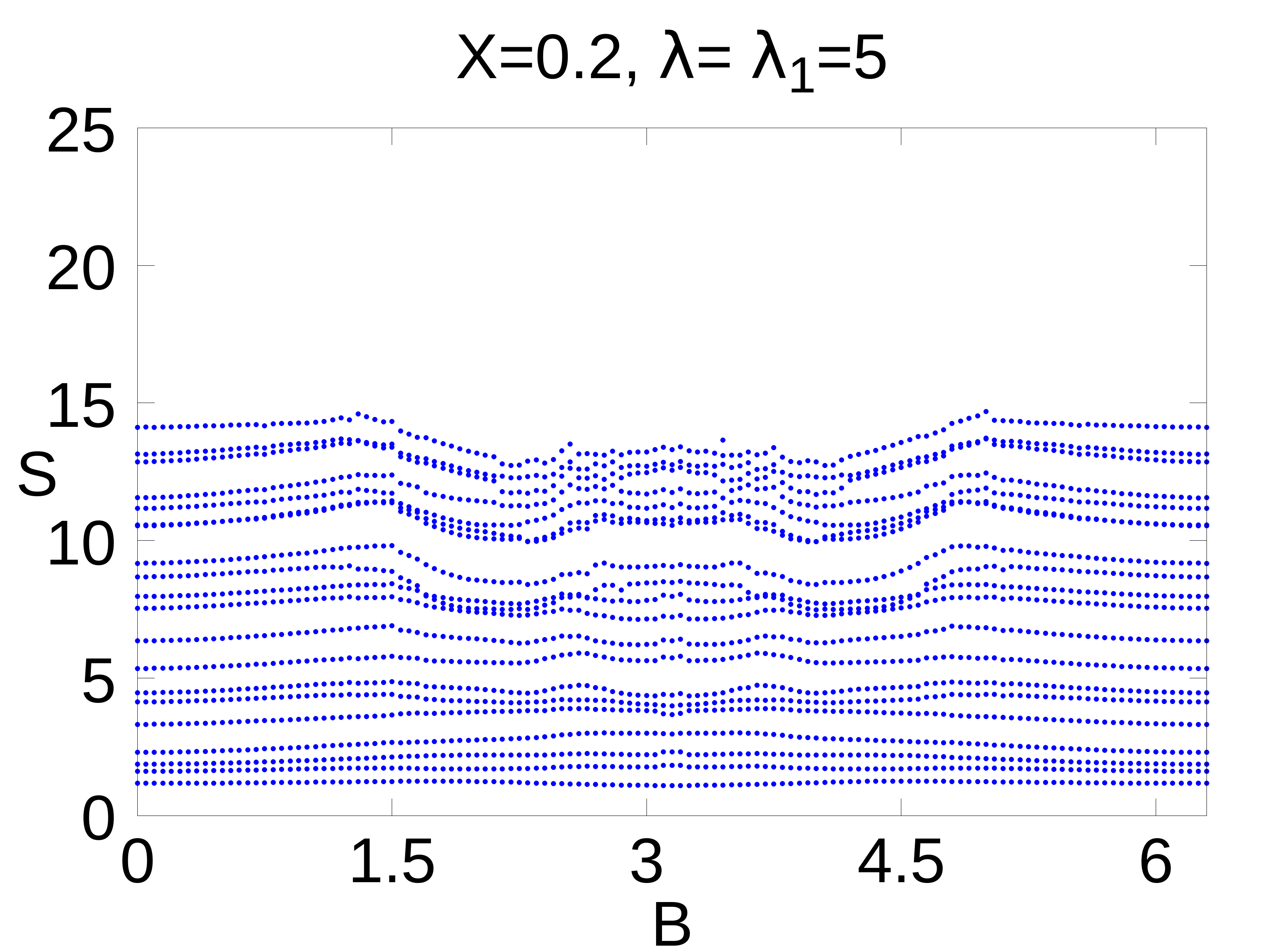}
		\subcaption{}
		\label{sq5}
	\end{subfigure}%
	\begin{subfigure}{.24\textwidth}
		\centering 
		\includegraphics[width=\linewidth]{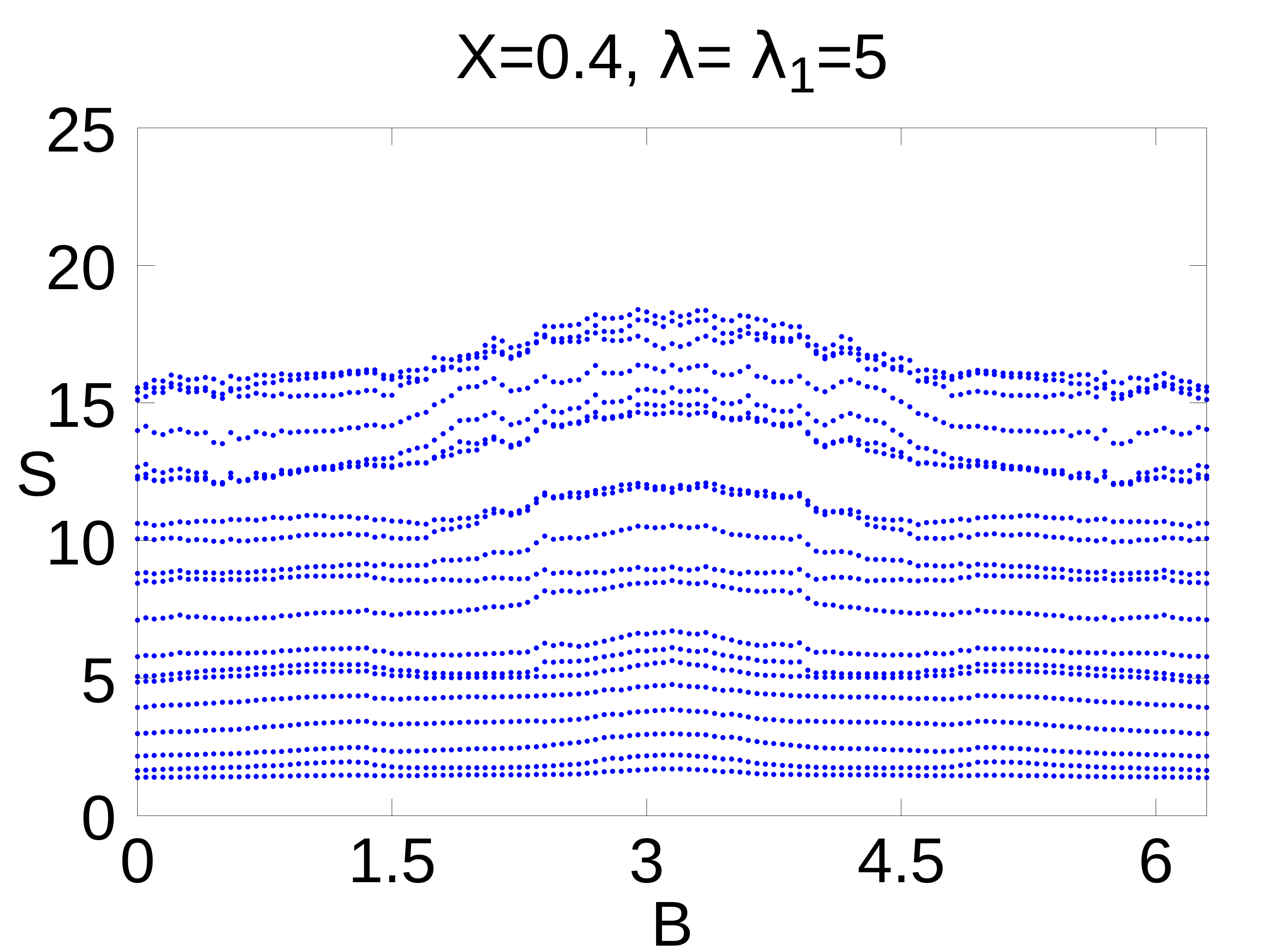}
		\subcaption{}
		\label{sq6}
	\end{subfigure}%
	\begin{subfigure}{.24\textwidth}
		\centering
		\includegraphics[width=\linewidth]{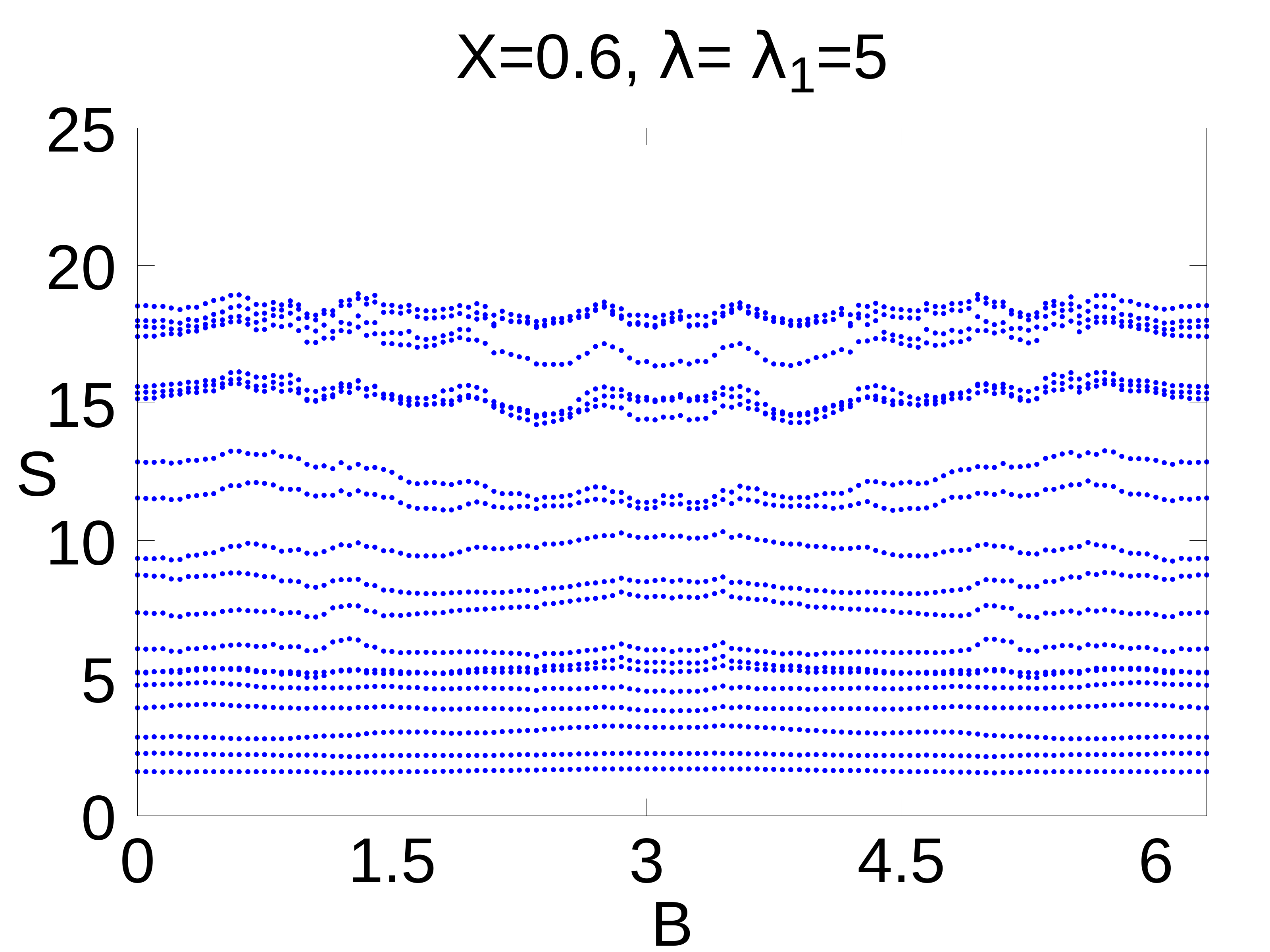}
		\subcaption{}
		\label{sq7}
	\end{subfigure}%
	\begin{subfigure}{.24\textwidth}
		\centering
		\includegraphics[width=\linewidth]{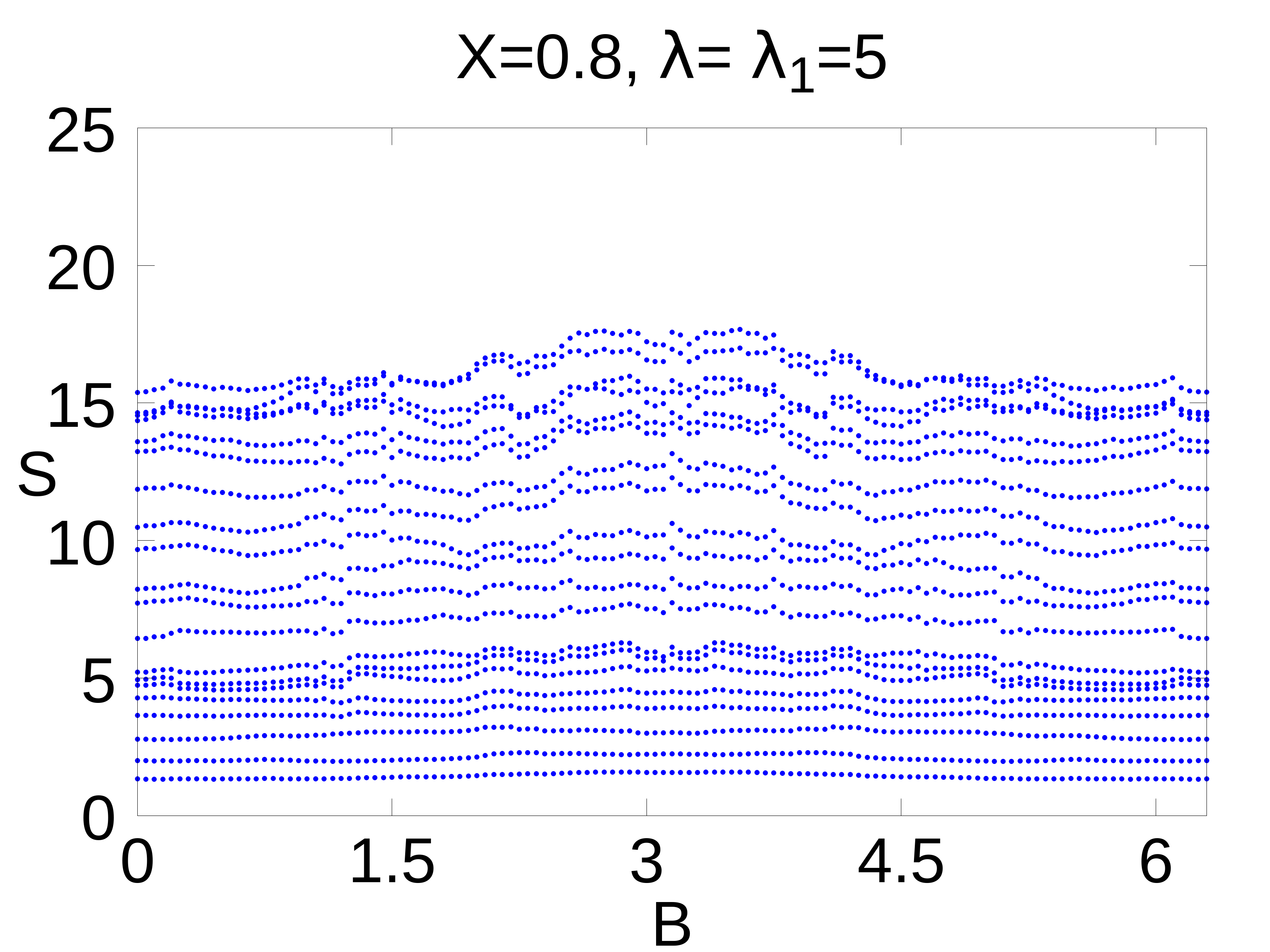}
		\subcaption{}
		\label{sq8}
	\end{subfigure}%

\caption{The entanglement entropy (S) is plotted as a function of the magnetic field (B) in a square lattice, illustrating the combined effects of Aubre-Andre and RS binary disorder}
	\label{IASB}
\end{figure*}

\begin{figure*}

\begin{subfigure}{.24\textwidth}
	\centering
	\includegraphics[width=\linewidth]{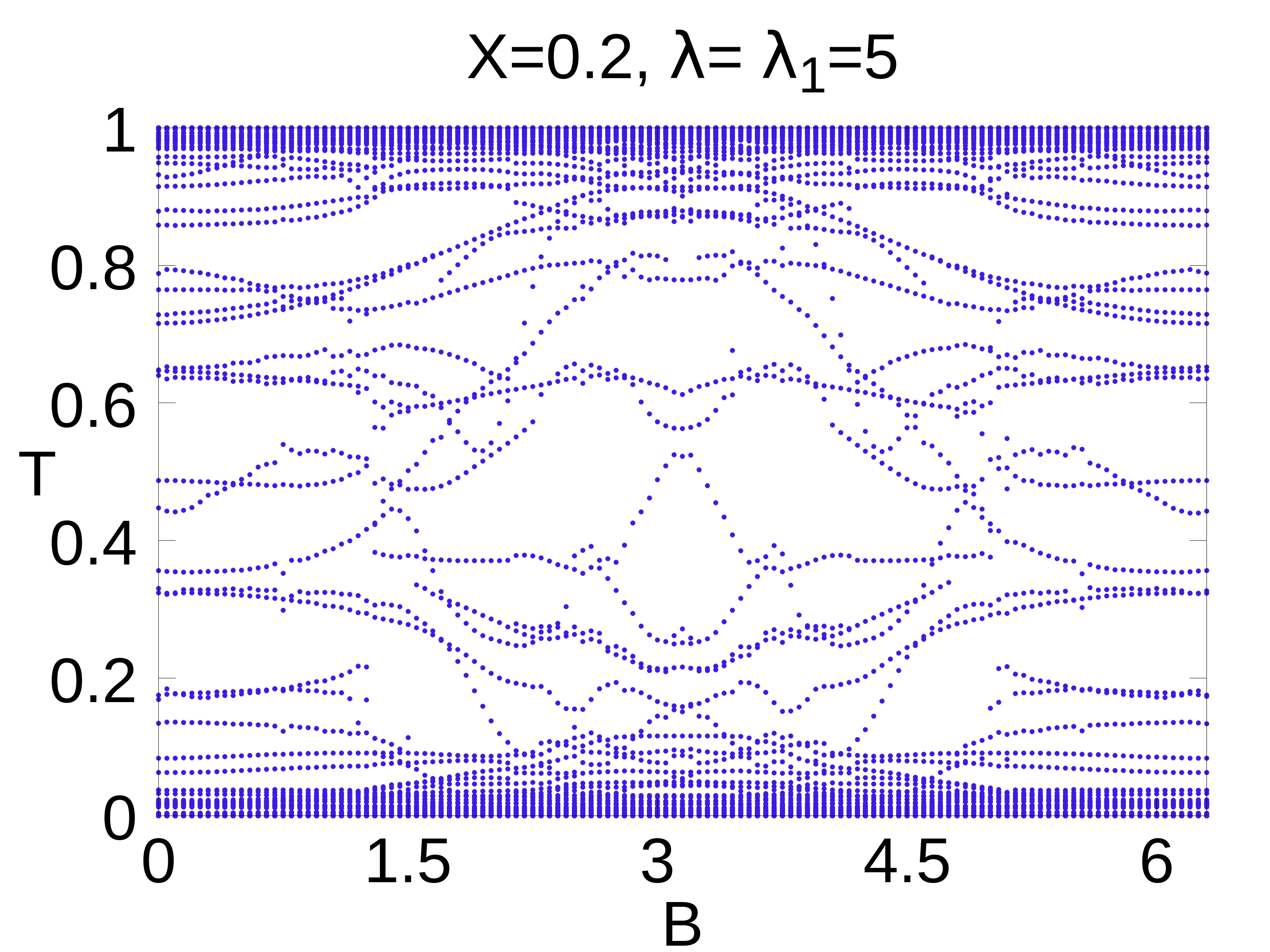}
	\subcaption{}
	\label{sq5}
\end{subfigure}%
\begin{subfigure}{.24\textwidth}
	\centering 
	\includegraphics[width=\linewidth]{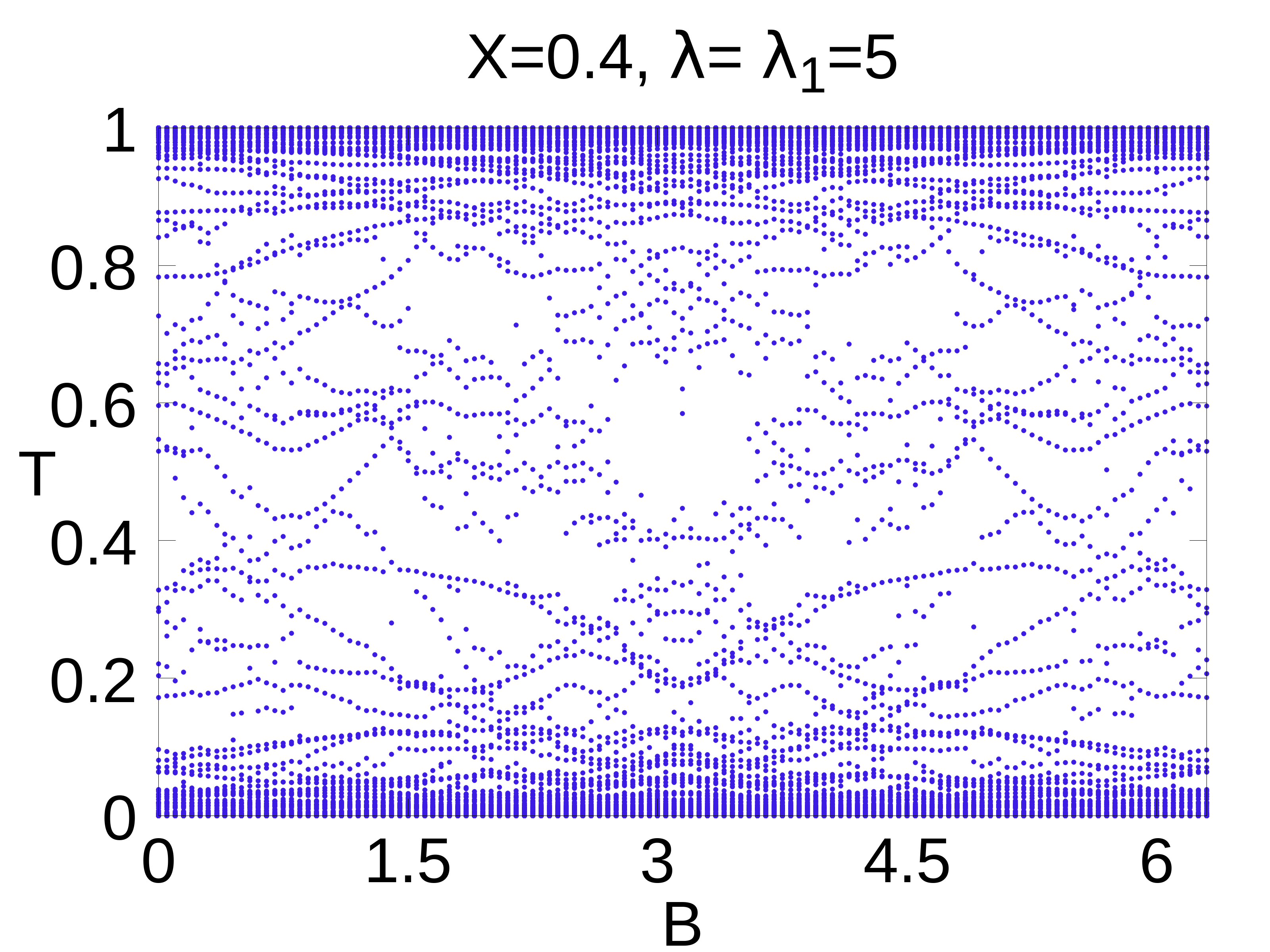}
	\subcaption{}
	\label{sq6}
\end{subfigure}%
\begin{subfigure}{.24\textwidth}
	\centering
	\includegraphics[width=\linewidth]{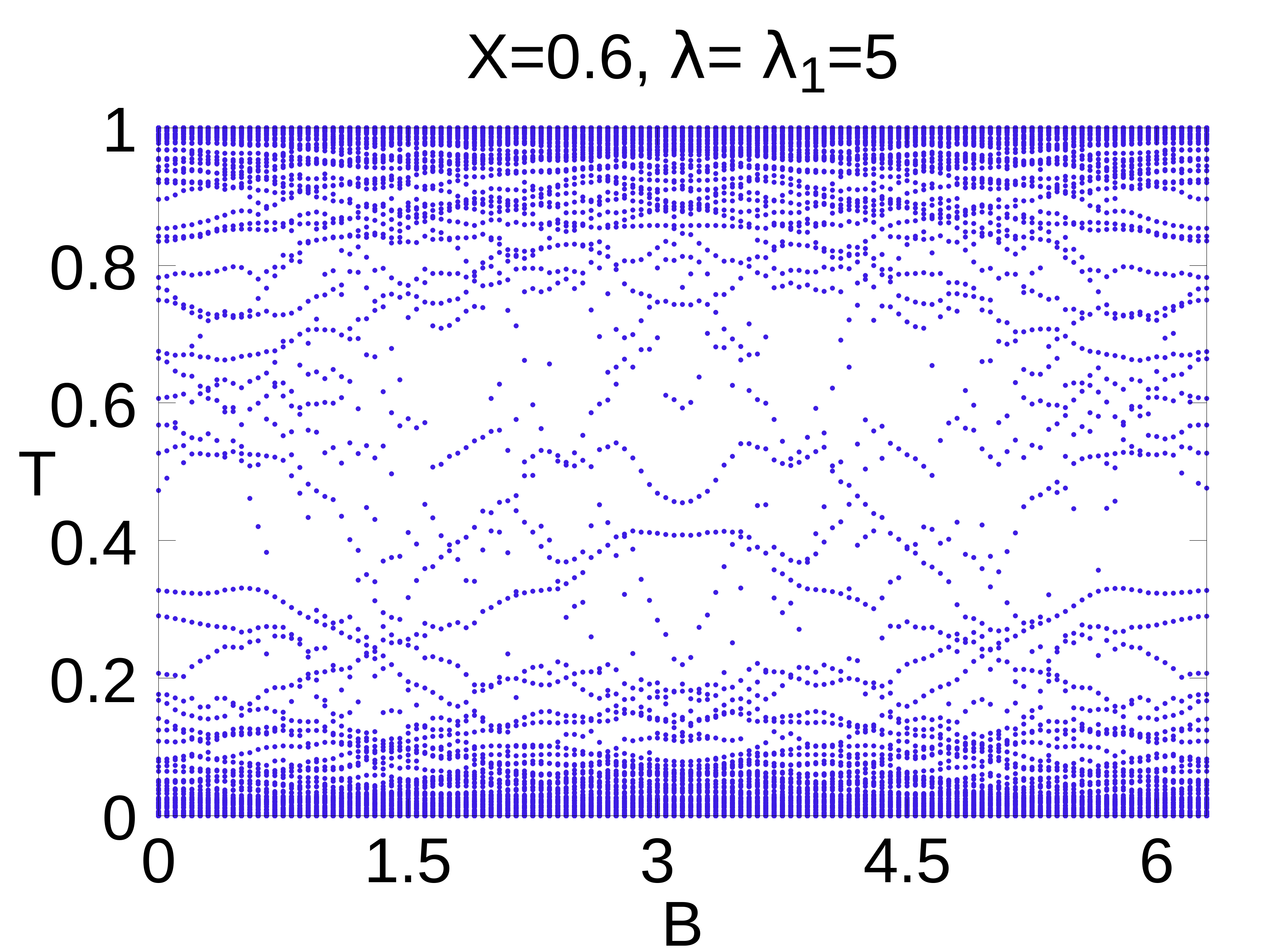}
	\subcaption{}
	\label{sq7}
\end{subfigure}%
\begin{subfigure}{.24\textwidth}
	\centering
	\includegraphics[width=\linewidth]{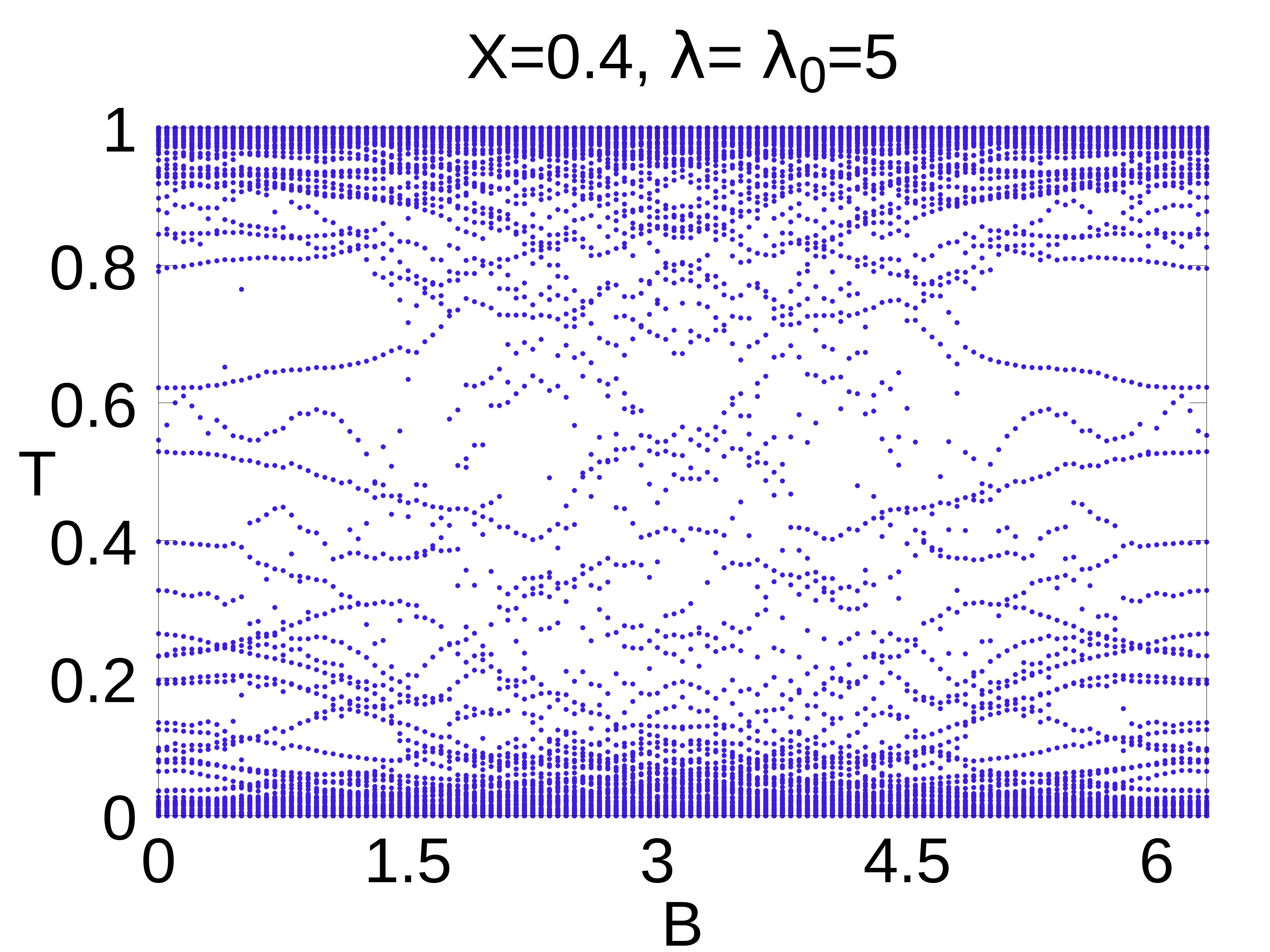}
	\subcaption{}
	\label{sq8}
\end{subfigure}%

\begin{subfigure}{.24\textwidth}
	\centering
	\includegraphics[width=\linewidth]{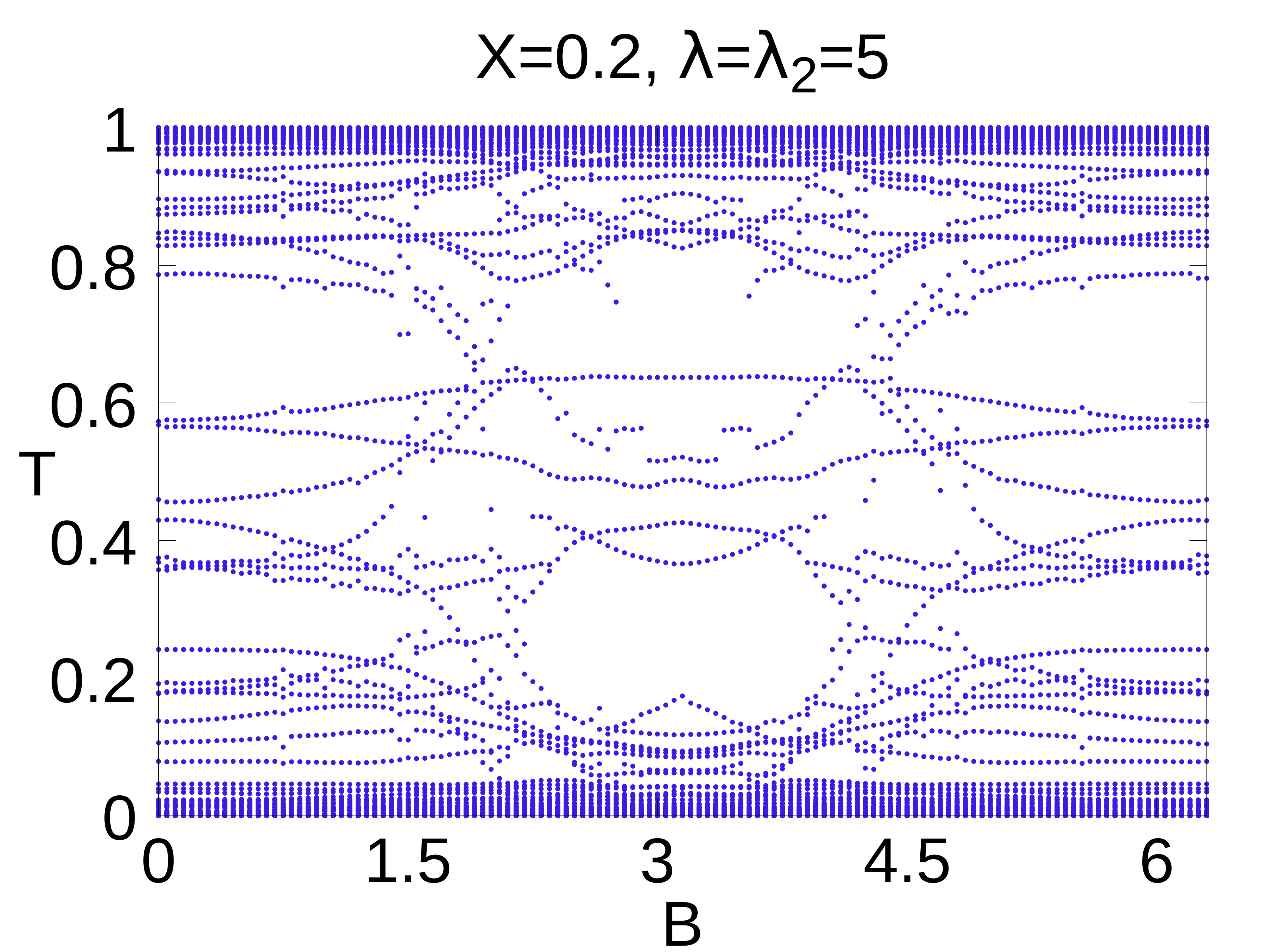}
	\subcaption{}
	\label{sq5}
\end{subfigure}%
\begin{subfigure}{.24\textwidth}
	\centering 
	\includegraphics[width=\linewidth]{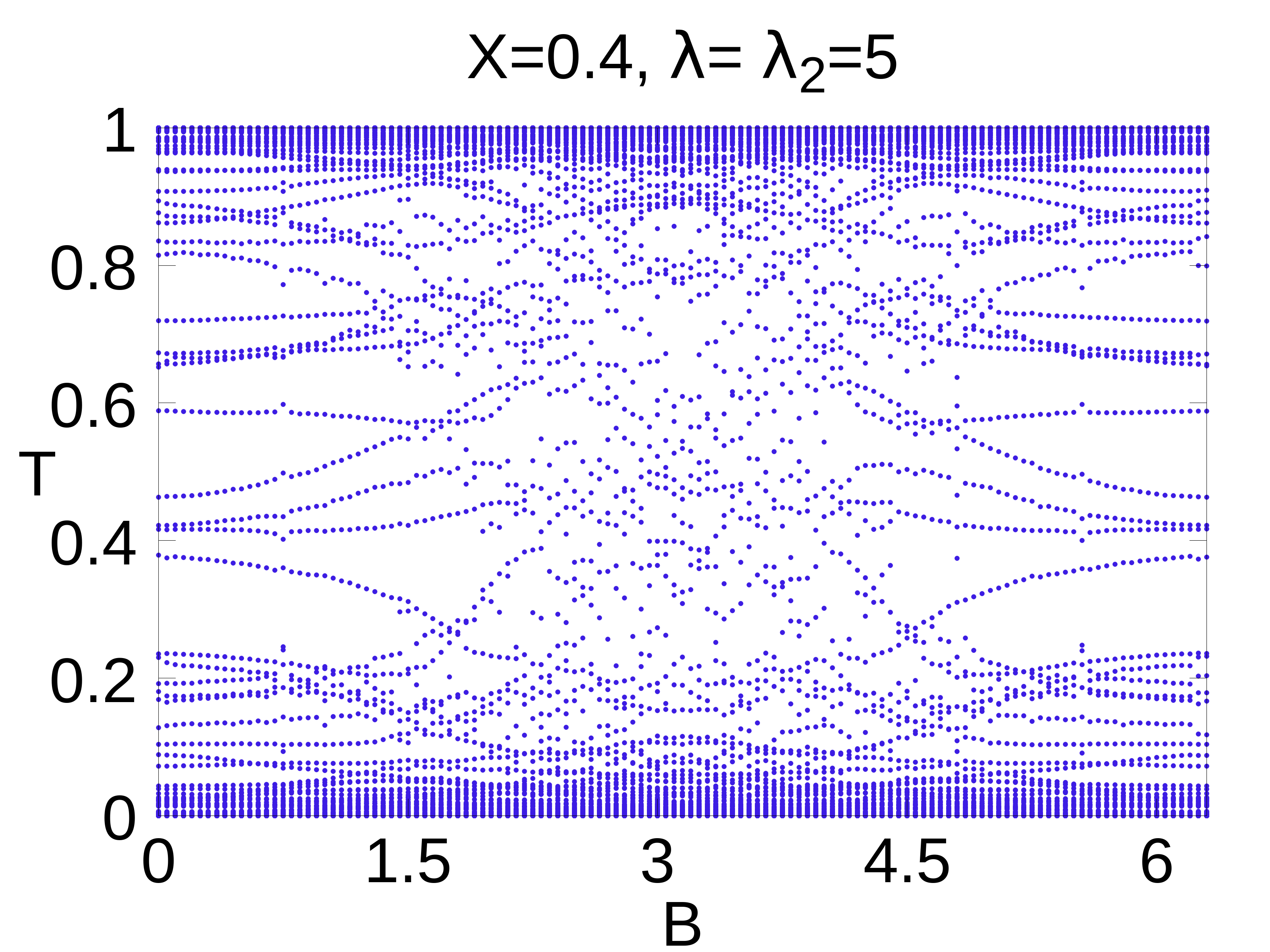}
	\subcaption{}
	\label{sq6}
\end{subfigure}%
\begin{subfigure}{.24\textwidth}
	\centering
	\includegraphics[width=\linewidth]{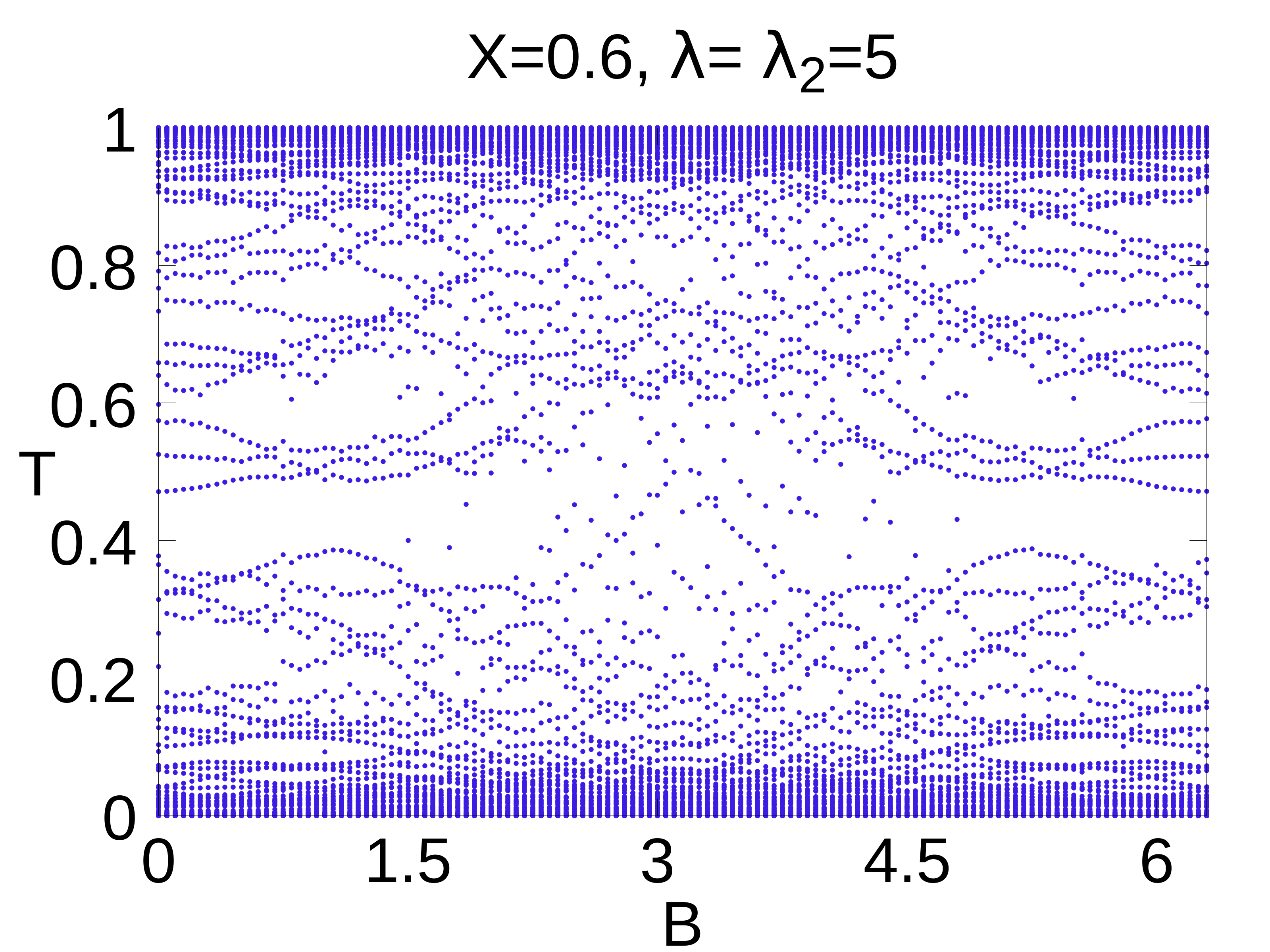}
	\subcaption{}
	\label{sq7}
\end{subfigure}%
\begin{subfigure}{.24\textwidth}
	\centering
	\includegraphics[width=\linewidth]{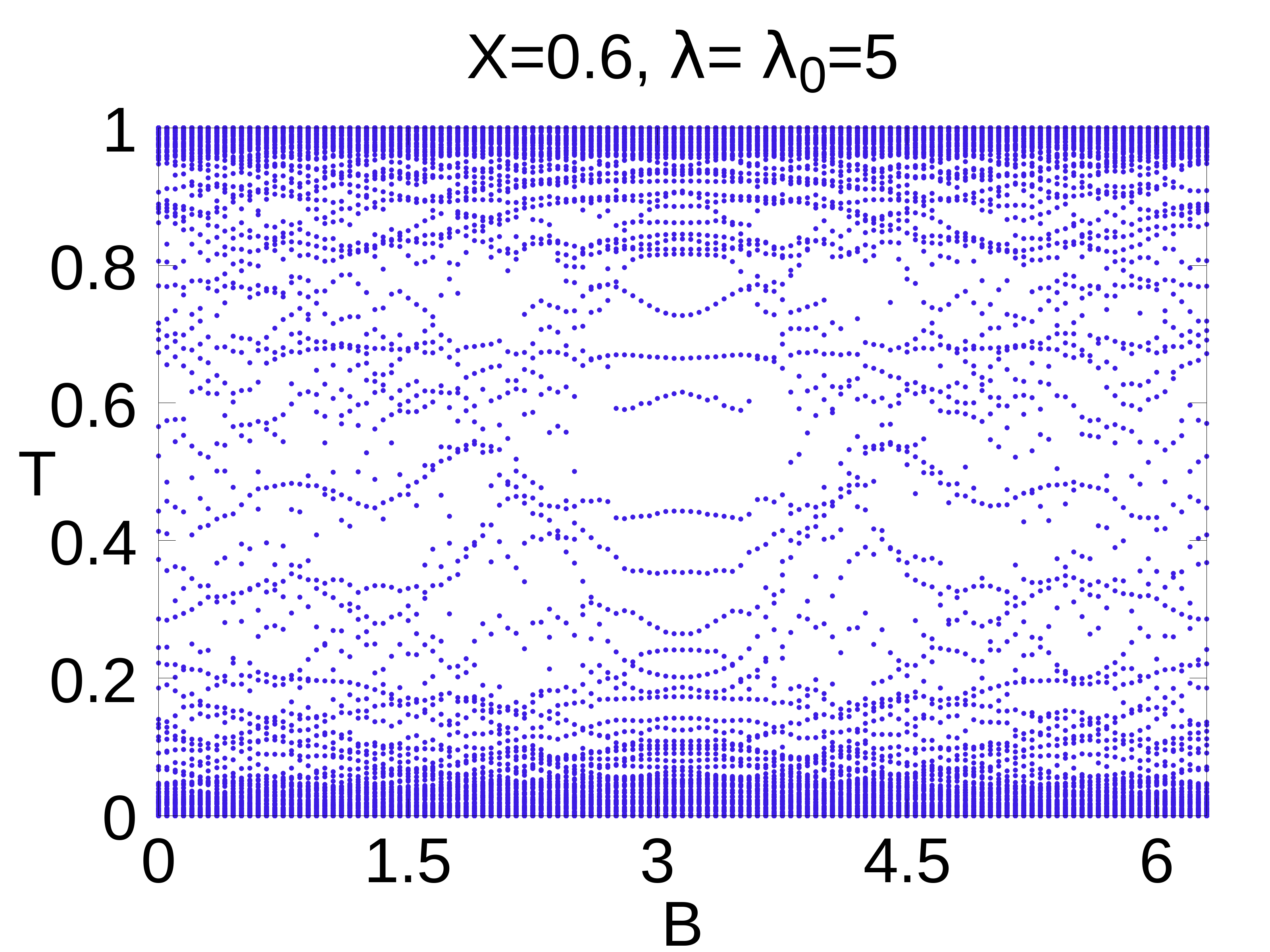}
	\subcaption{}
	\label{sq8}
\end{subfigure}%
	\caption{Correlation spectrum as a function of the magnetic field $B$ for the interpolation between the AA potential and different quasiperiodic disorders. Here, $\lambda$ denotes the strength of the AA component, while $\lambda_{0}$, $\lambda_{1}$, and $\lambda_{2}$ correspond to the TM, RS, and Fibonacci disorders, respectively. The parameter $X$ represents the interpolation factor between the AA and the respective quasiperiodic potentials.}
\label{CS40}
\end{figure*}

\begin{figure*}[]
	\centering
	\begin{subfigure}{.25\textwidth}
		\centering
		\includegraphics[width=\linewidth]{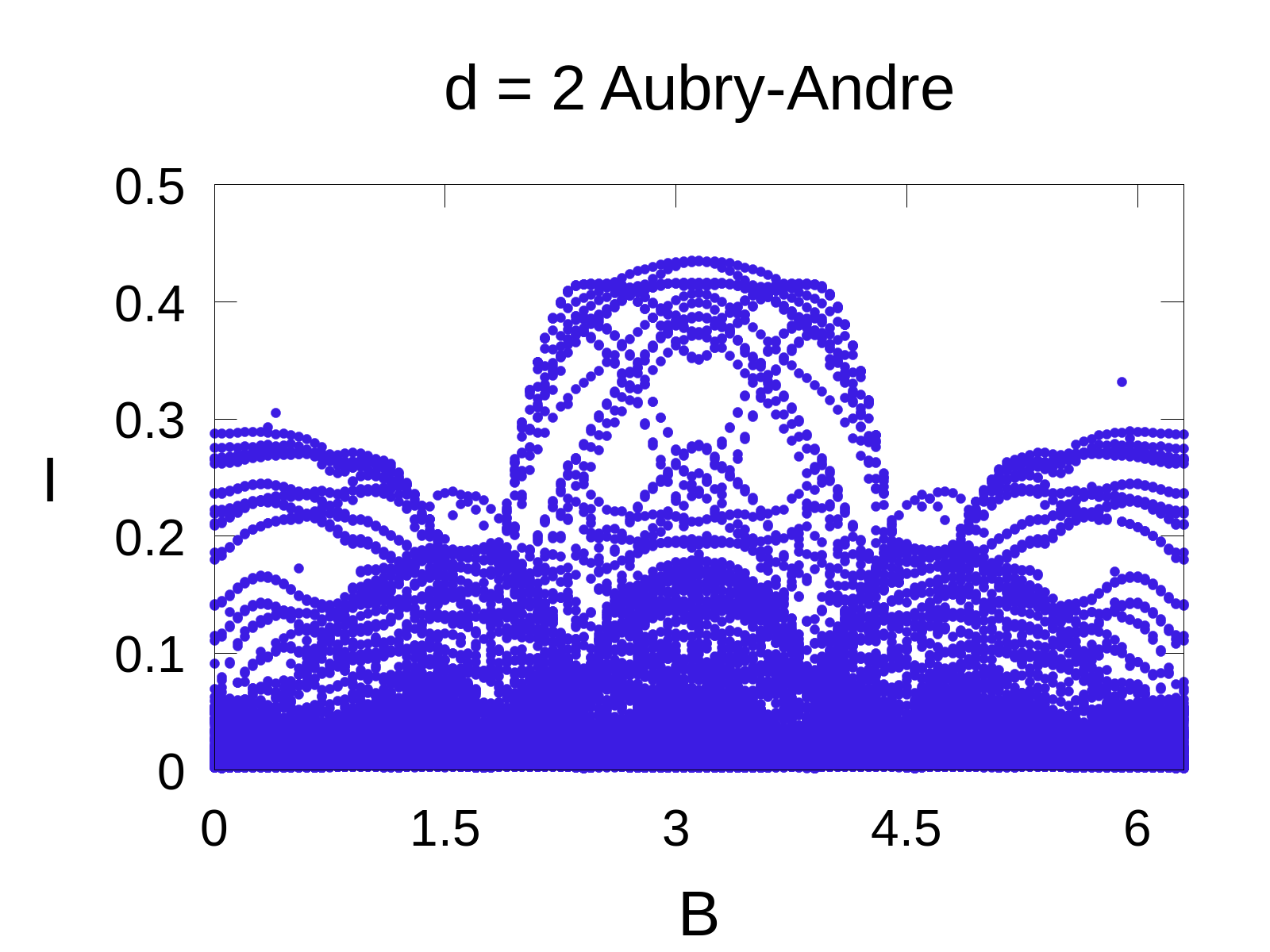}
		\subcaption{}
		\label{sq1}
	\end{subfigure}%
	\begin{subfigure}{.25\textwidth}
		\centering 
		\includegraphics[width=\linewidth]{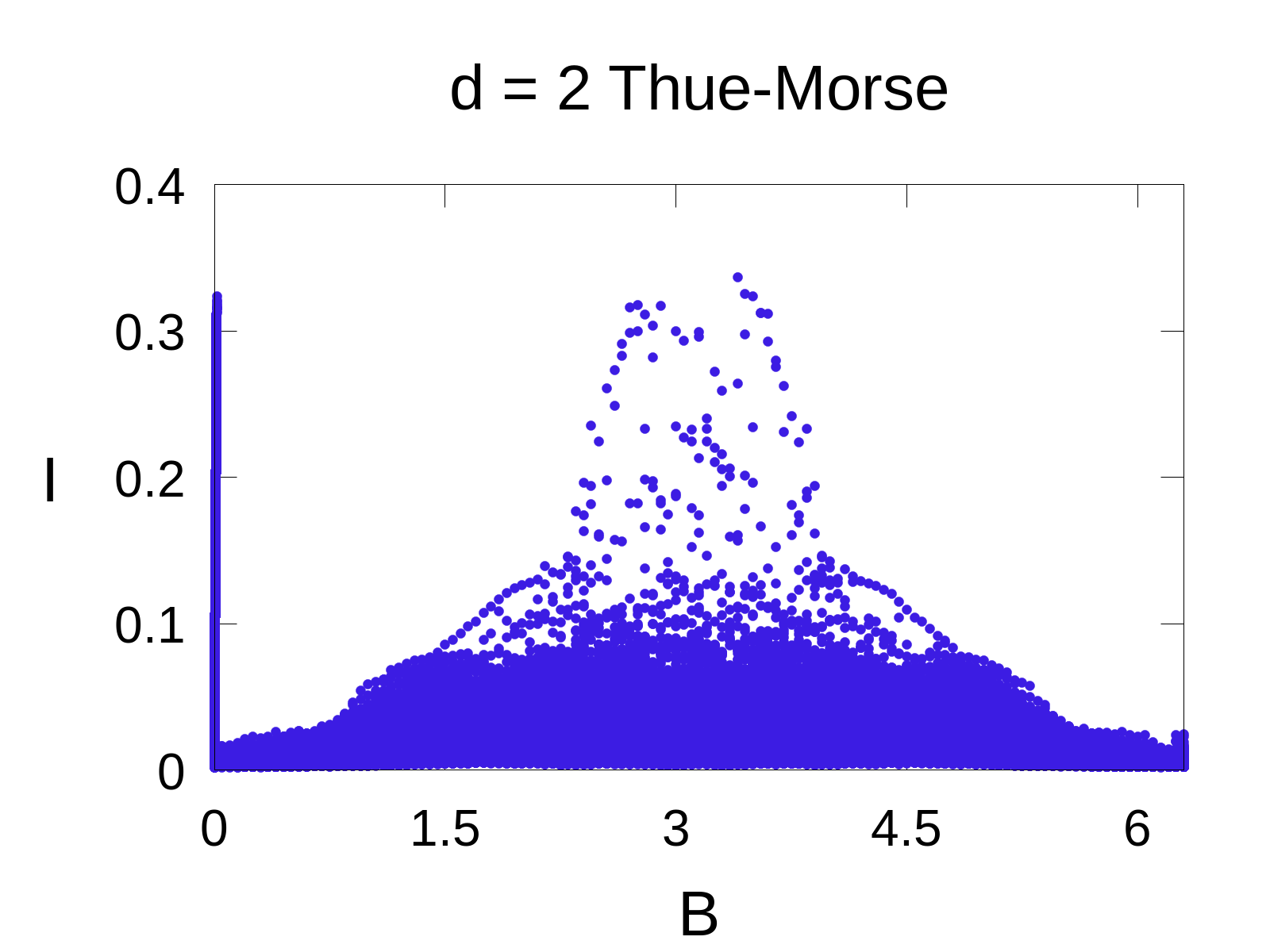}
		\subcaption{}
		\label{sq2}
	\end{subfigure}%
	\begin{subfigure}{.25\textwidth}
		\centering
		\includegraphics[width=\linewidth]{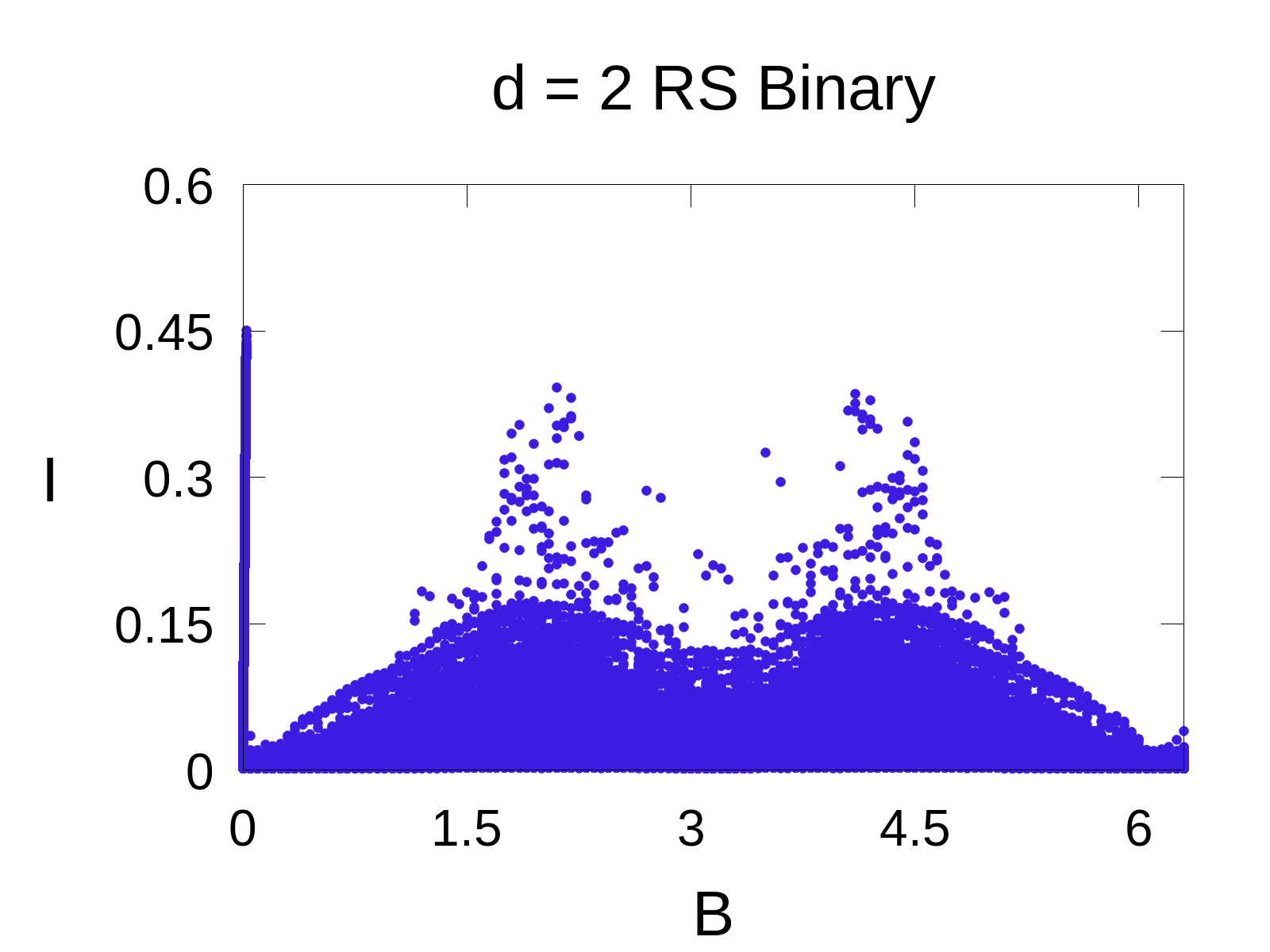}
		\subcaption{}
		\label{sq3}
	\end{subfigure}%
	\begin{subfigure}{.25\textwidth}
		\centering
		\includegraphics[width=\linewidth]{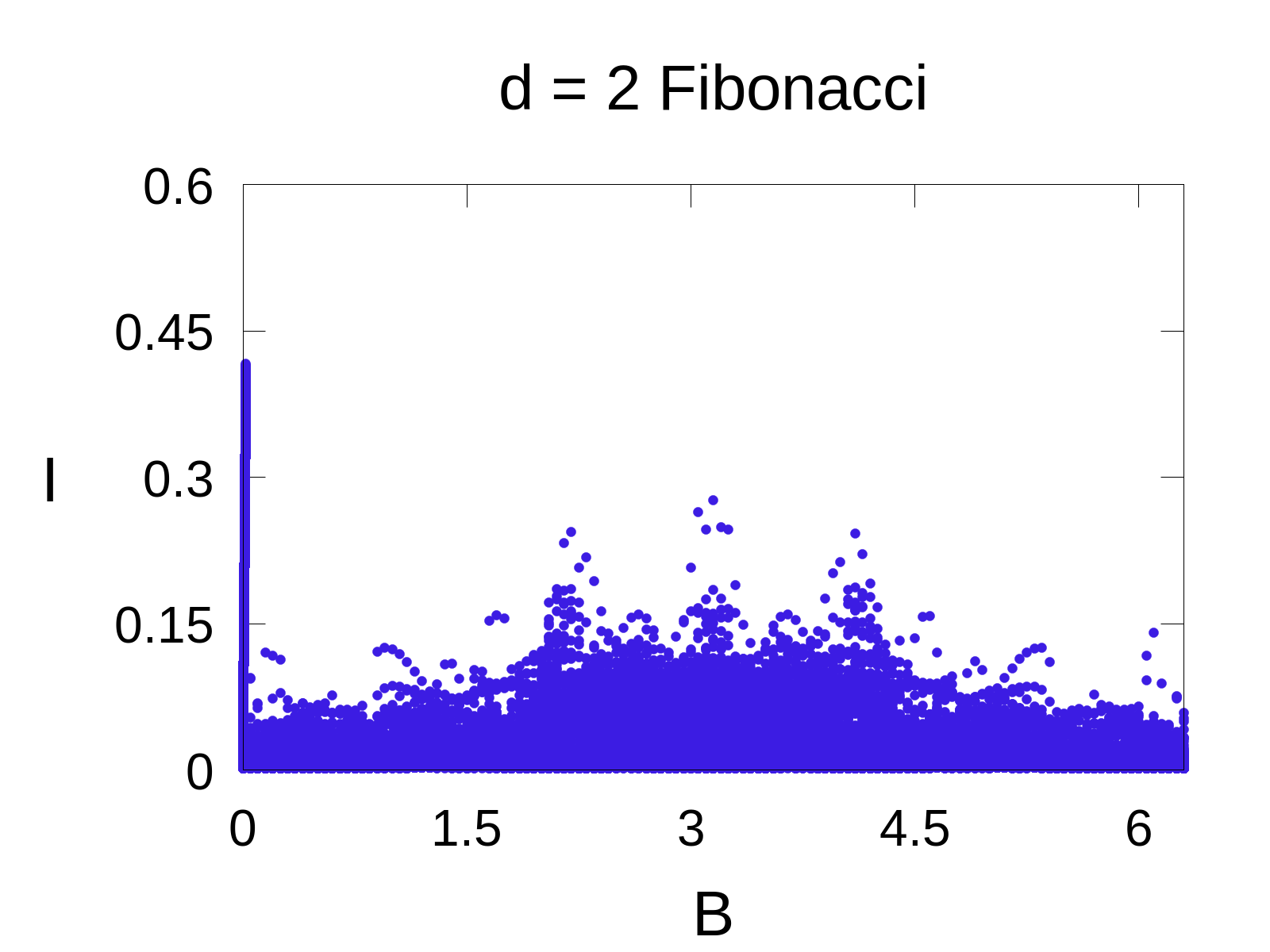}
		\subcaption{}
		\label{sq4}
	\end{subfigure}%

	\begin{subfigure}{.25\textwidth}
		\centering
		\includegraphics[width=\linewidth]{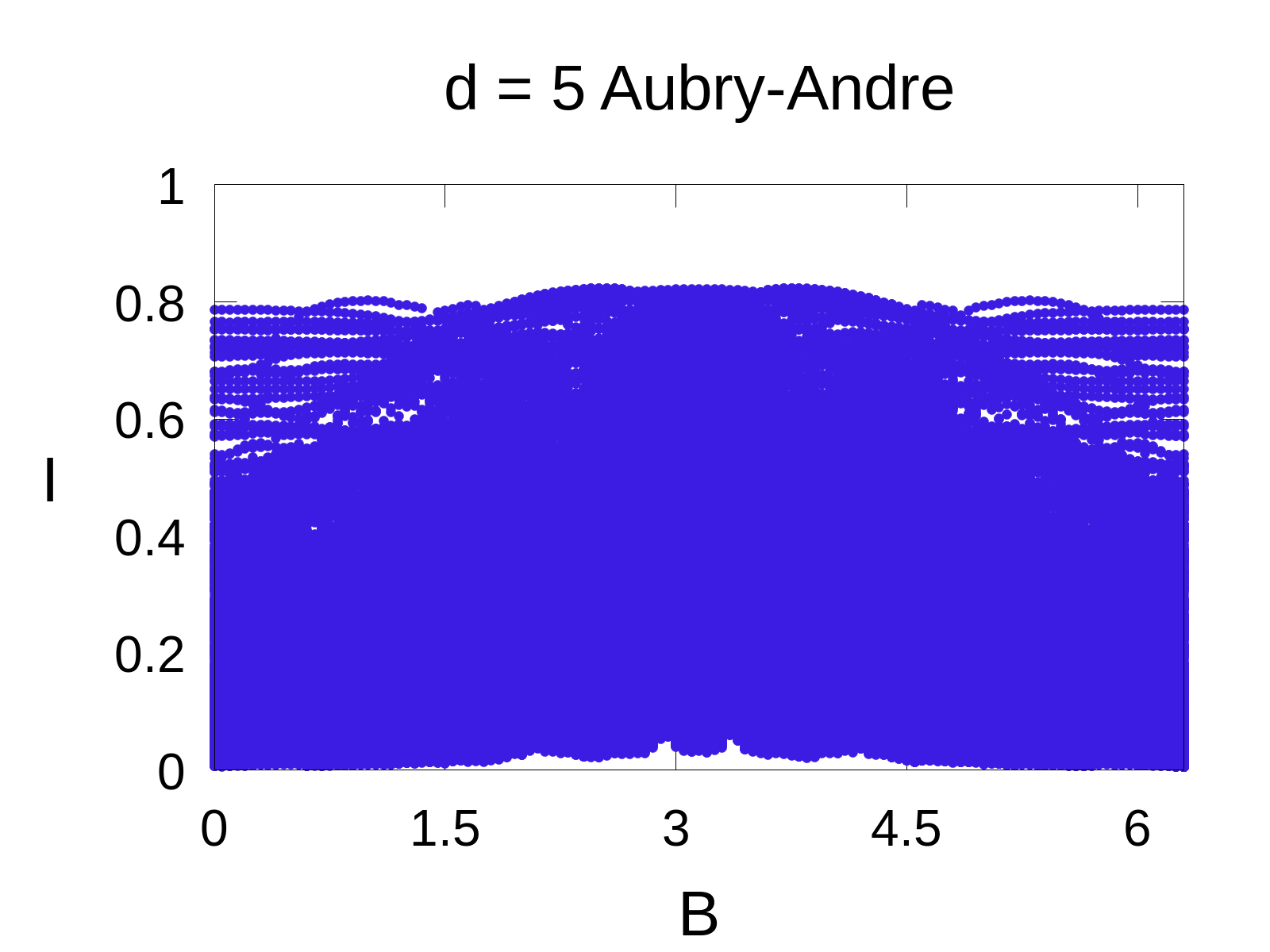}
		\subcaption{}
		\label{sq1}
	\end{subfigure}%
	\begin{subfigure}{.25\textwidth}
		\centering 
		\includegraphics[width=\linewidth]{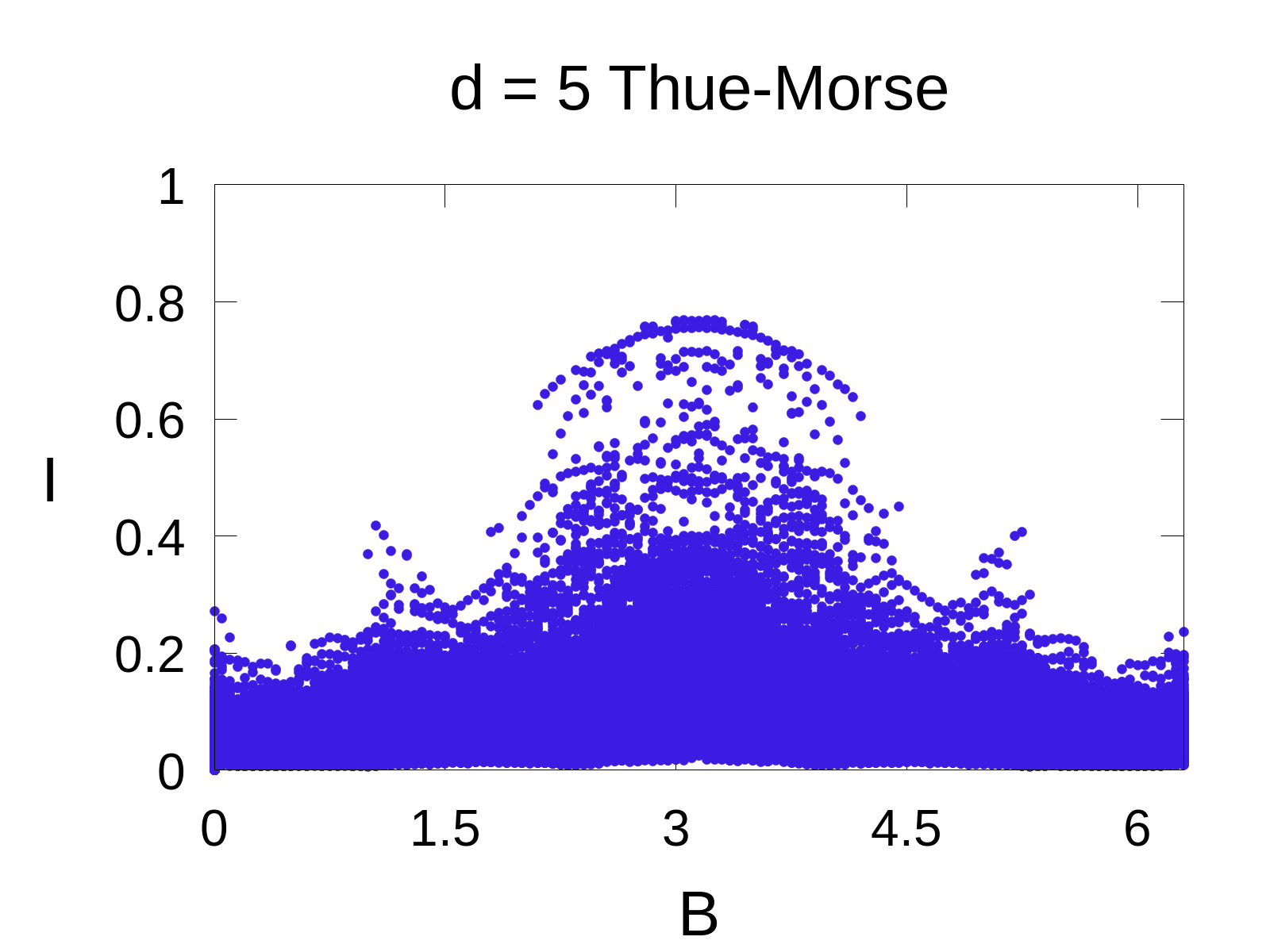}
		\subcaption{}
		\label{sq2}
	\end{subfigure}%
	\begin{subfigure}{.25\textwidth}
		\centering
		\includegraphics[width=\linewidth]{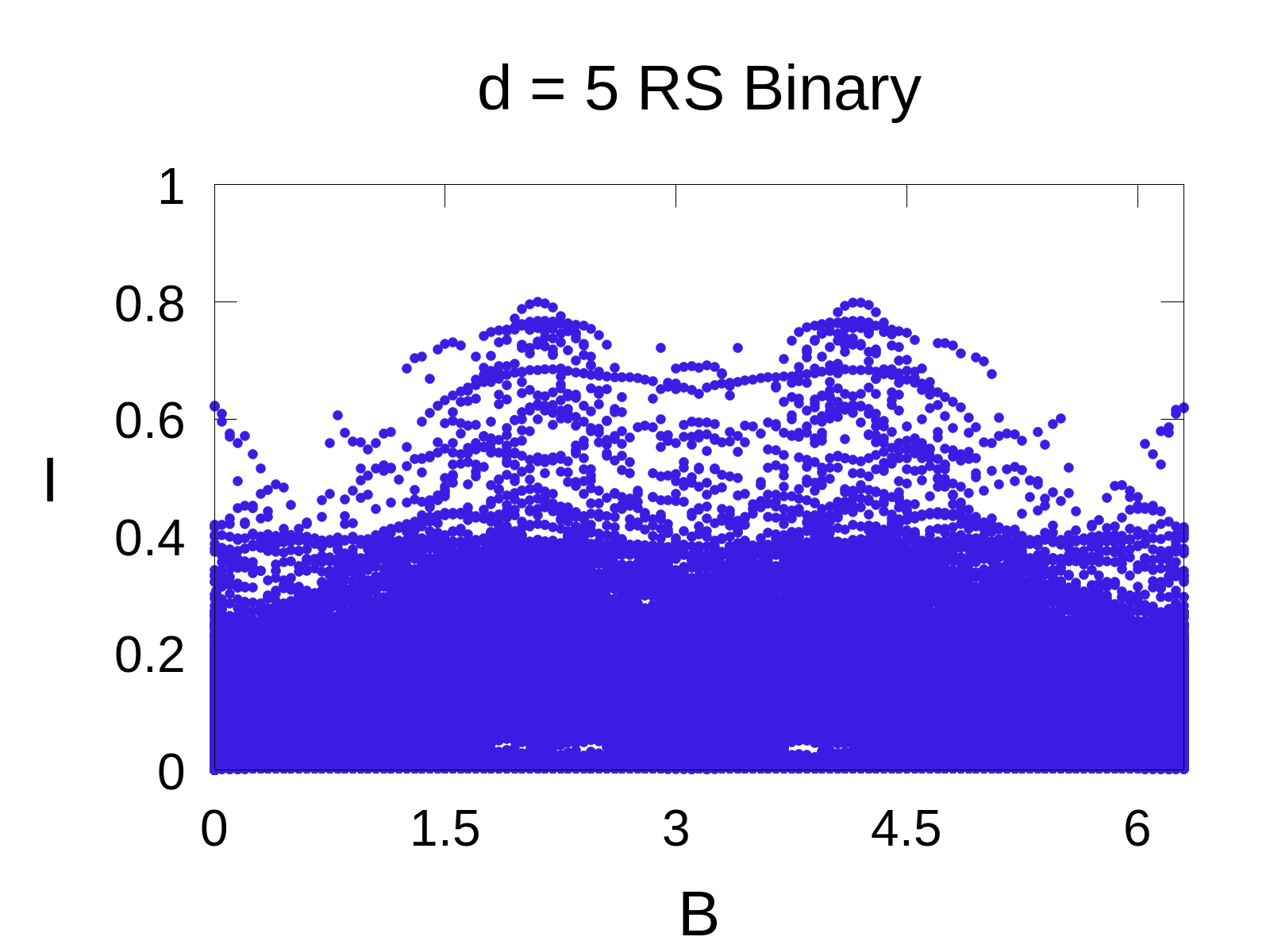}
		\subcaption{}
		\label{sq3}
	\end{subfigure}%
	\begin{subfigure}{.25\textwidth}
		\centering
		\includegraphics[width=\linewidth]{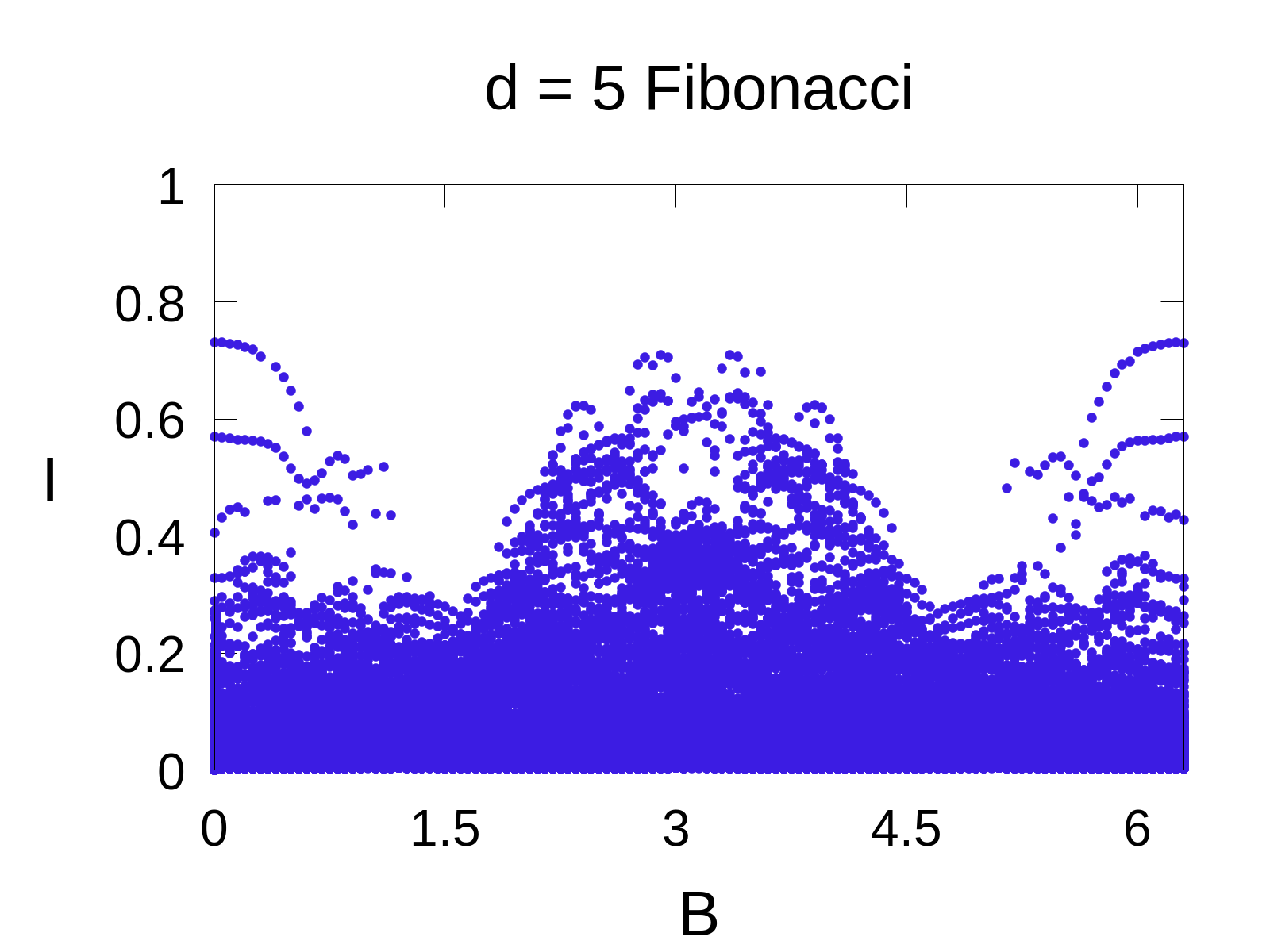}
		\subcaption{}
		\label{sq4}
	\end{subfigure}%
	
	\caption{IPR as a function of the magnetic field $B$ for different quasiperiodic disorders. The upper panel shows the results for the weak-disorder case ($d = 2$), while the lower panel corresponds to the stronger disorder ($d = 5$). As expected, the IPR increases with disorder strength, with the highest IPR values indicating the most strongly localized states.
	}
	\label{IPR4}
\end{figure*}

\begin{figure*}[]
	\centering
	\begin{subfigure}{.25\textwidth}
		\centering
		\includegraphics[width=\linewidth]{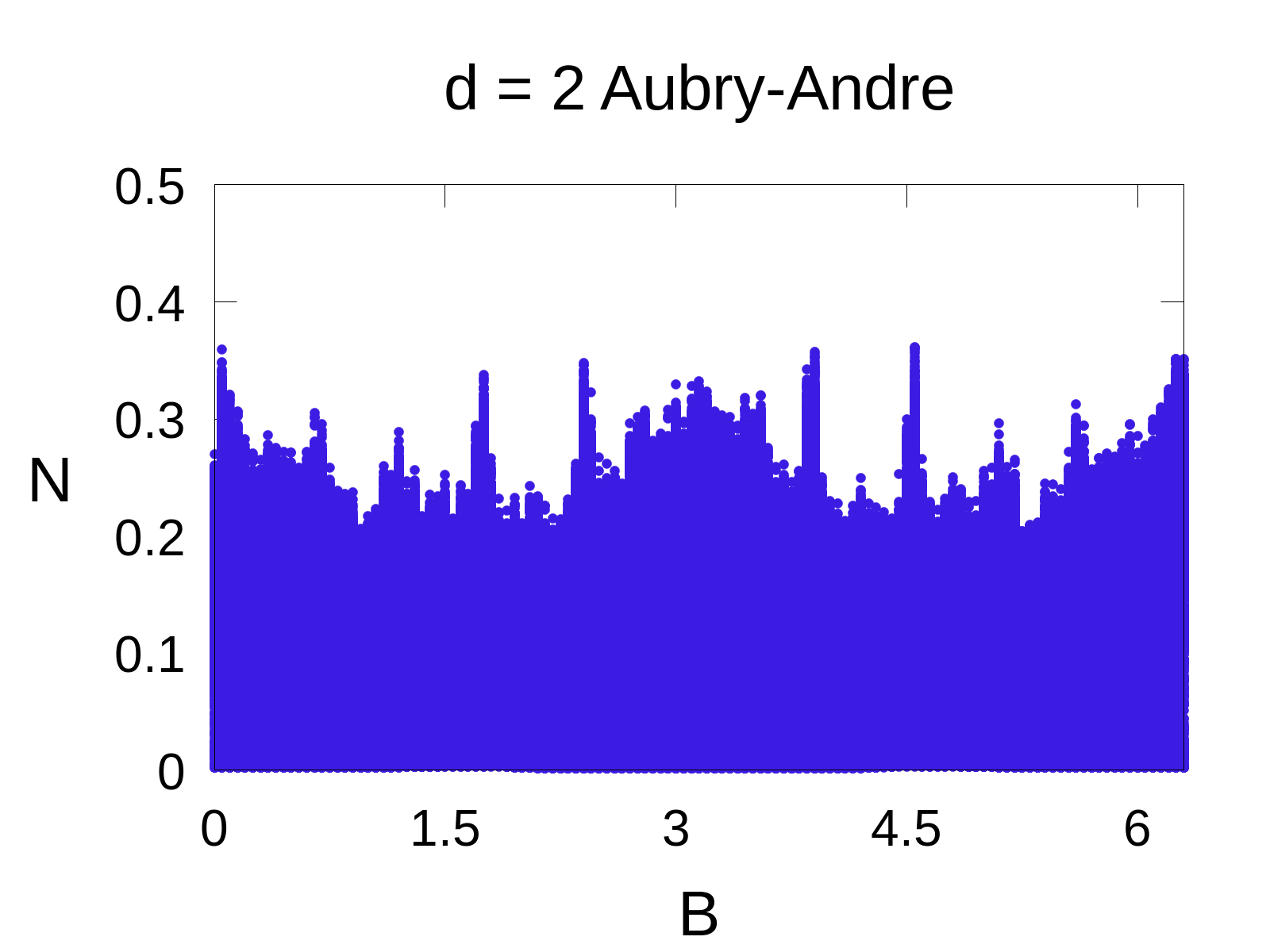}
		\subcaption{}
		\label{sq1}
	\end{subfigure}%
	\begin{subfigure}{.25\textwidth}
		\centering 
		\includegraphics[width=\linewidth]{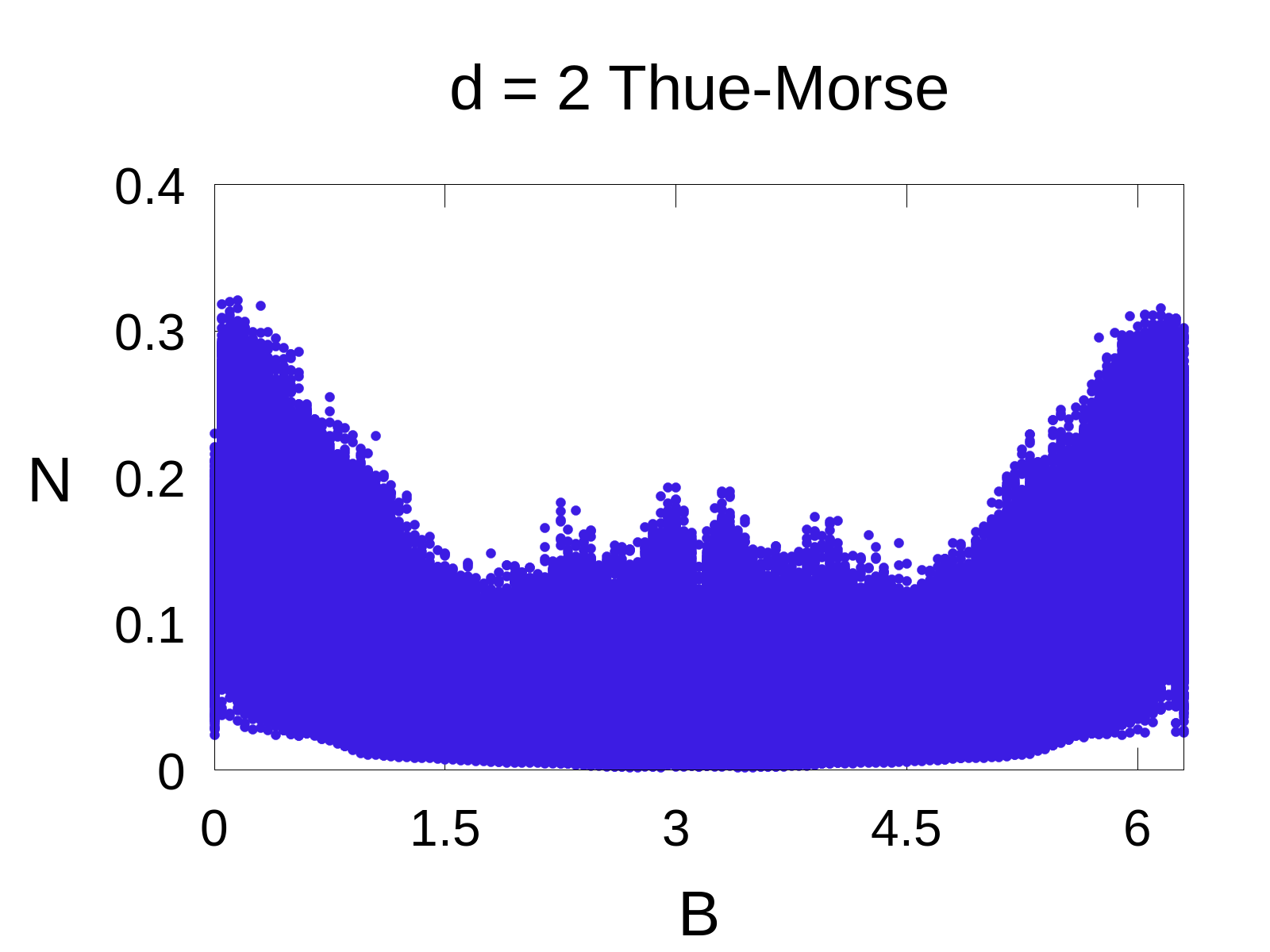}
		\subcaption{}
		\label{sq2}
	\end{subfigure}%
	\begin{subfigure}{.25\textwidth}
		\centering
		\includegraphics[width=\linewidth]{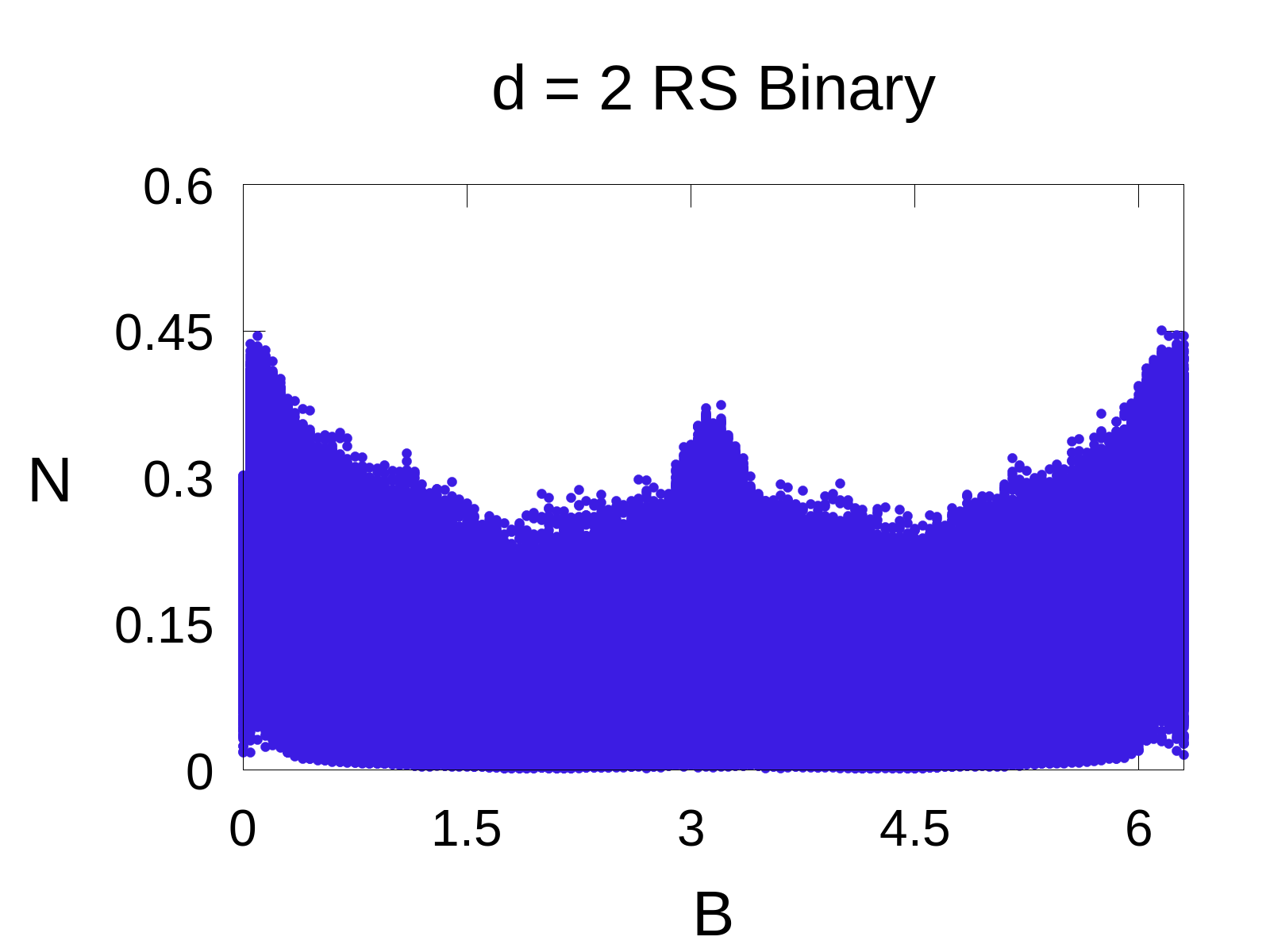}
		\subcaption{}
		\label{sq3}
	\end{subfigure}%
	\begin{subfigure}{.25\textwidth}
		\centering
		\includegraphics[width=\linewidth]{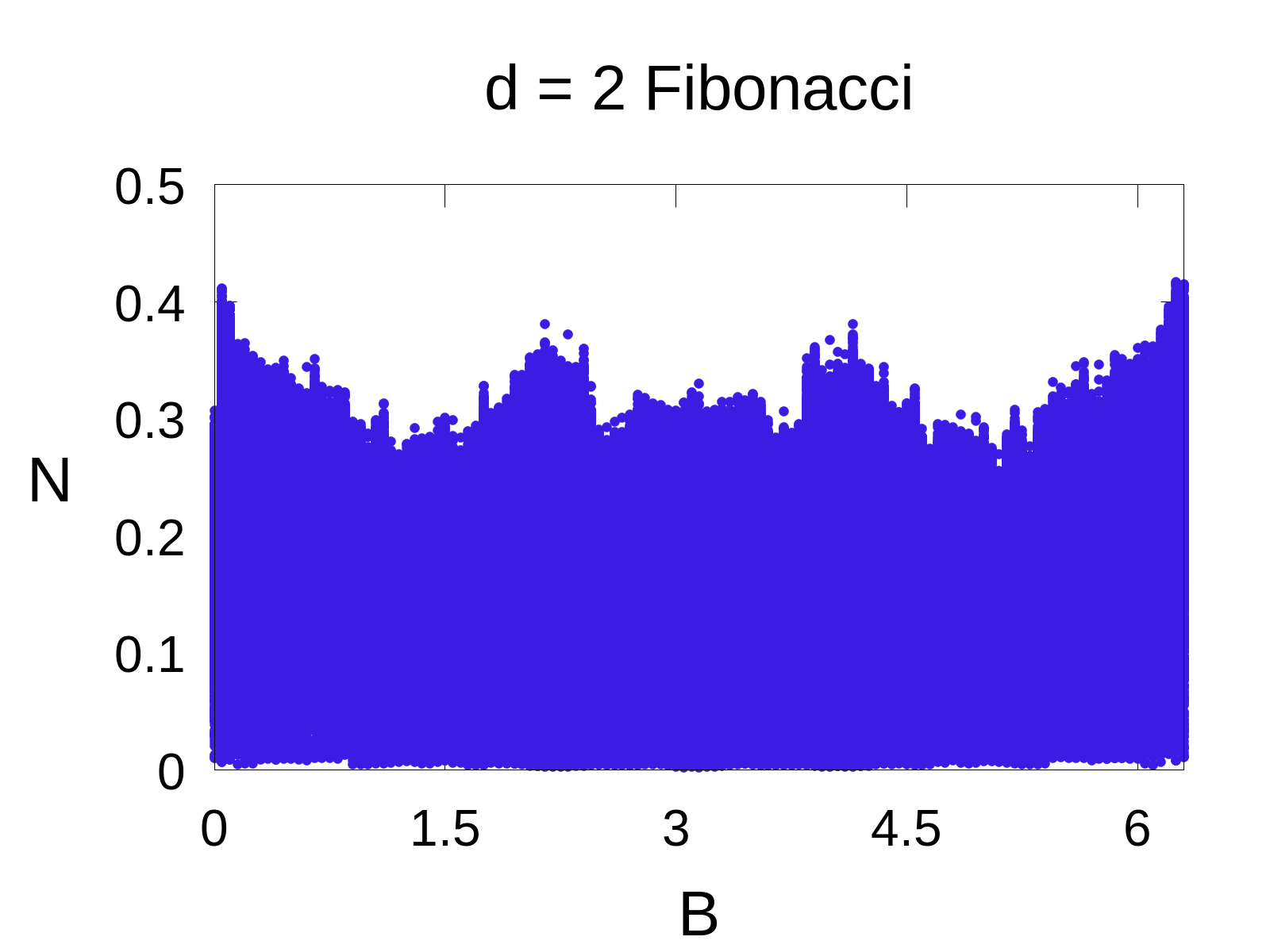}
		\subcaption{}
		\label{sq4}
	\end{subfigure}%

	\begin{subfigure}{.25\textwidth}
		\centering
		\includegraphics[width=\linewidth]{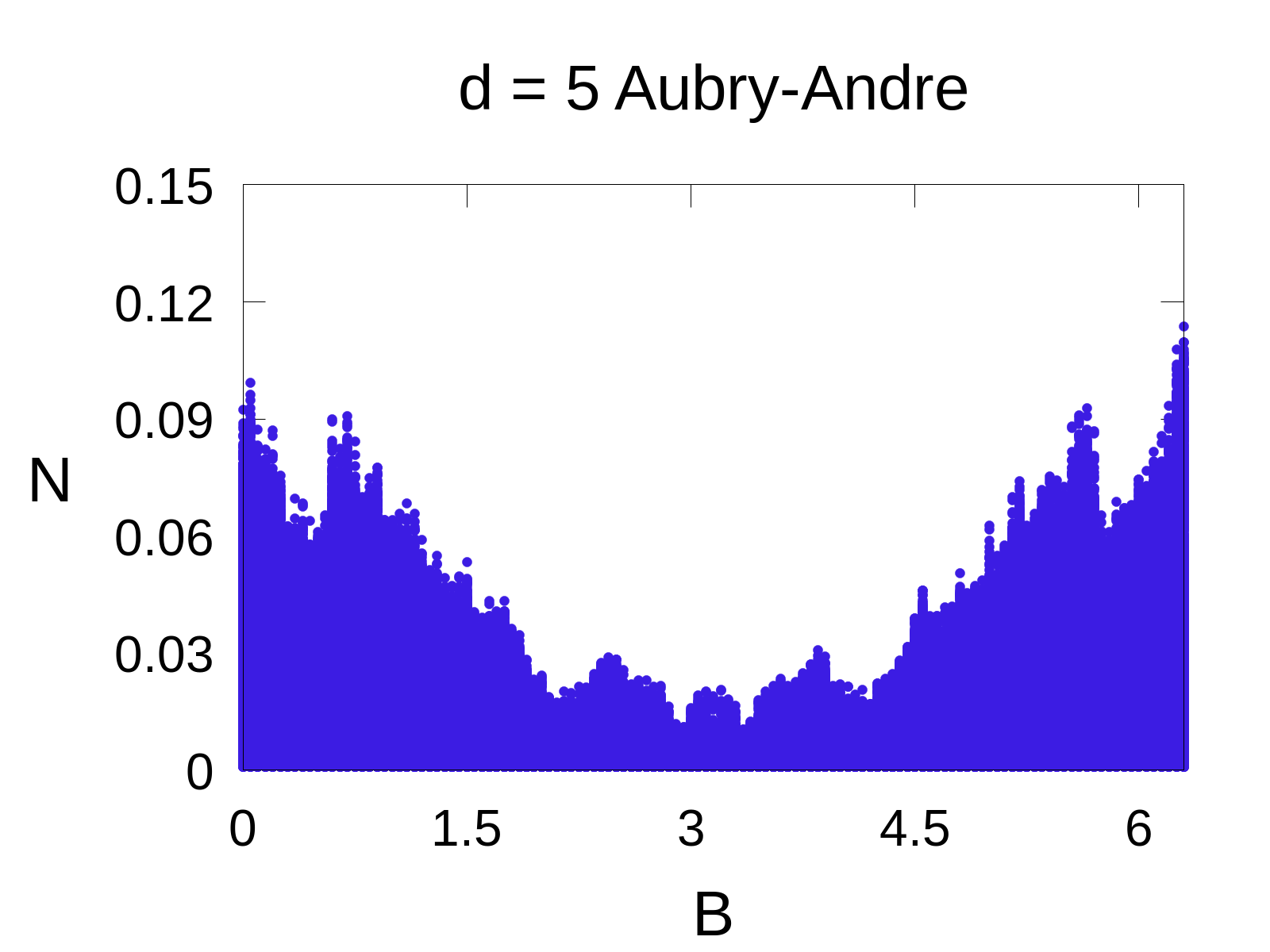}
		\subcaption{}
		\label{sq1}
	\end{subfigure}%
	\begin{subfigure}{.25\textwidth}
		\centering 
		\includegraphics[width=\linewidth]{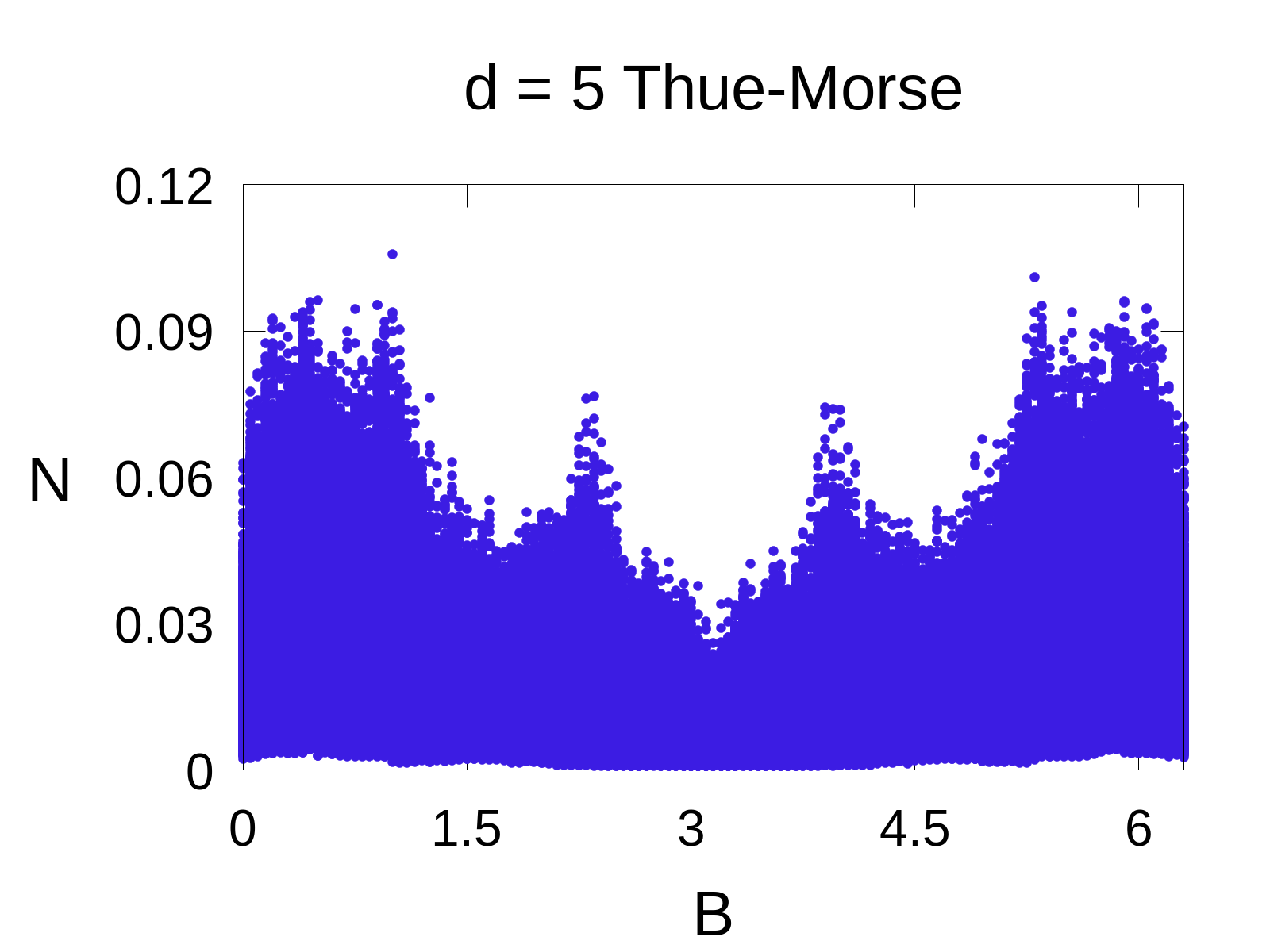}
		\subcaption{}
		\label{sq2}
	\end{subfigure}%
	\begin{subfigure}{.25\textwidth}
		\centering
		\includegraphics[width=\linewidth]{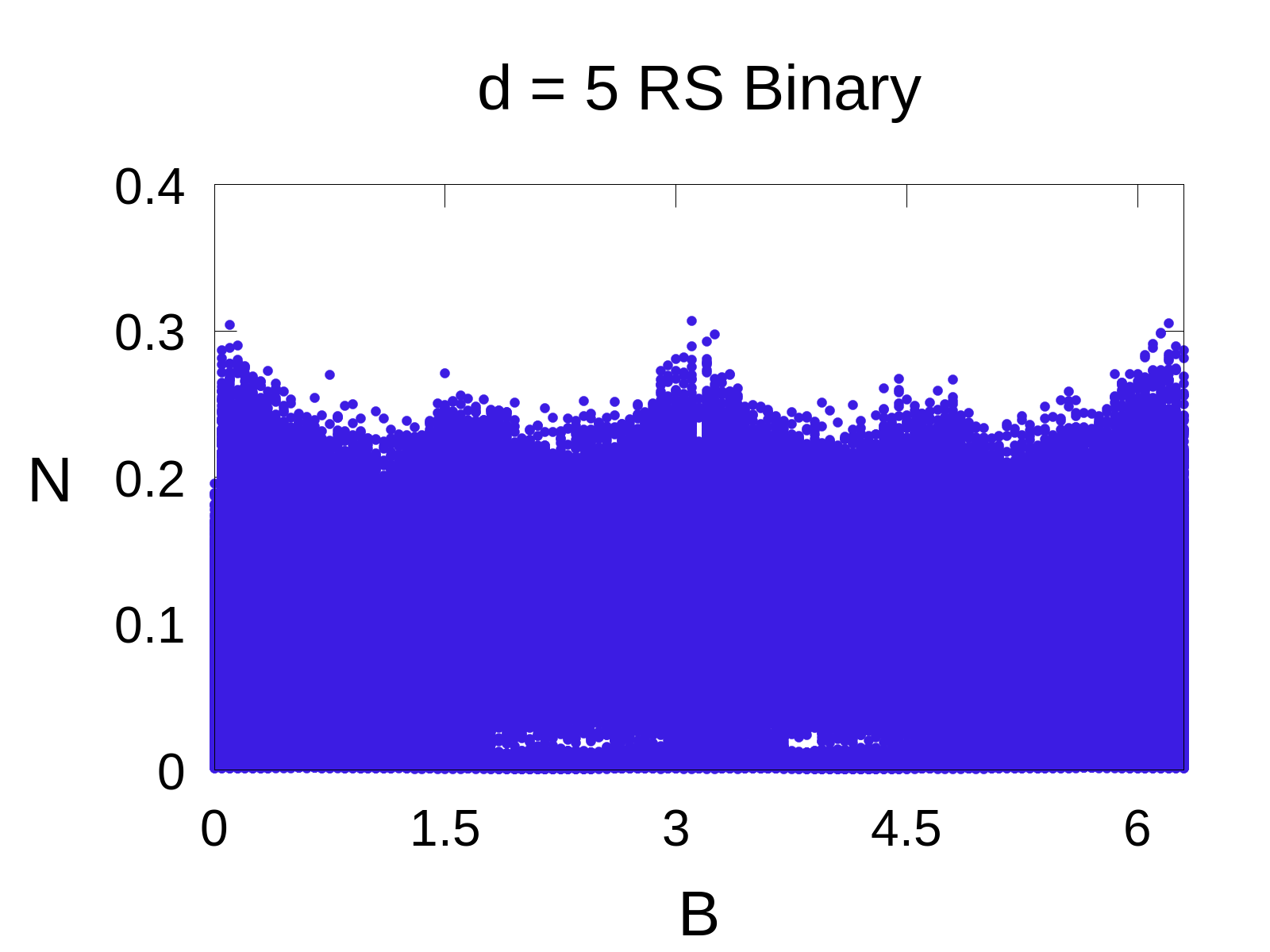}
		\subcaption{}
		\label{sq3}
	\end{subfigure}%
	\begin{subfigure}{.25\textwidth}
		\centering
		\includegraphics[width=\linewidth]{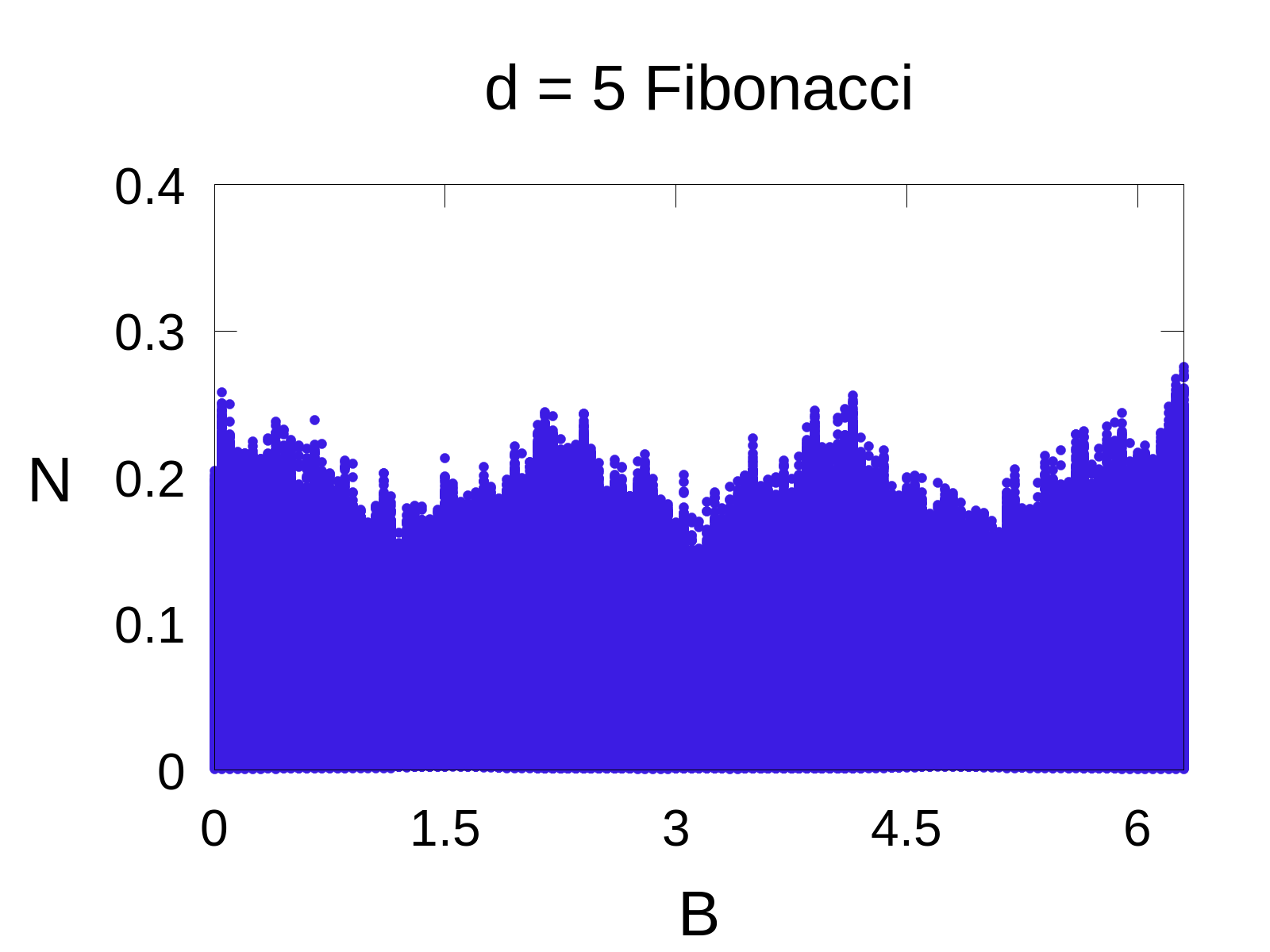}
		\subcaption{}
		\label{sq4}
	\end{subfigure}%
	
	\caption{NPR plotted as a function of the magnetic field $B$ for various quasiperiodic disorders. The upper panel corresponds to the weak-disorder case ($d = 2$), and the lower panel presents the results for the stronger disorder ($d = 5$). A stronger disorder strength generally suppresses the NPR, reflecting enhanced 
   localization. Among all disorder types, the AA potential consistently produces 
		the lowest NPR values, indicating the strongest degree of localization.
	}
	\label{NPR40}
\end{figure*}

\begin{figure*}[!htb]
	\centering
	\begin{subfigure}{.24\textwidth}
		\centering
		\includegraphics[width=\linewidth]{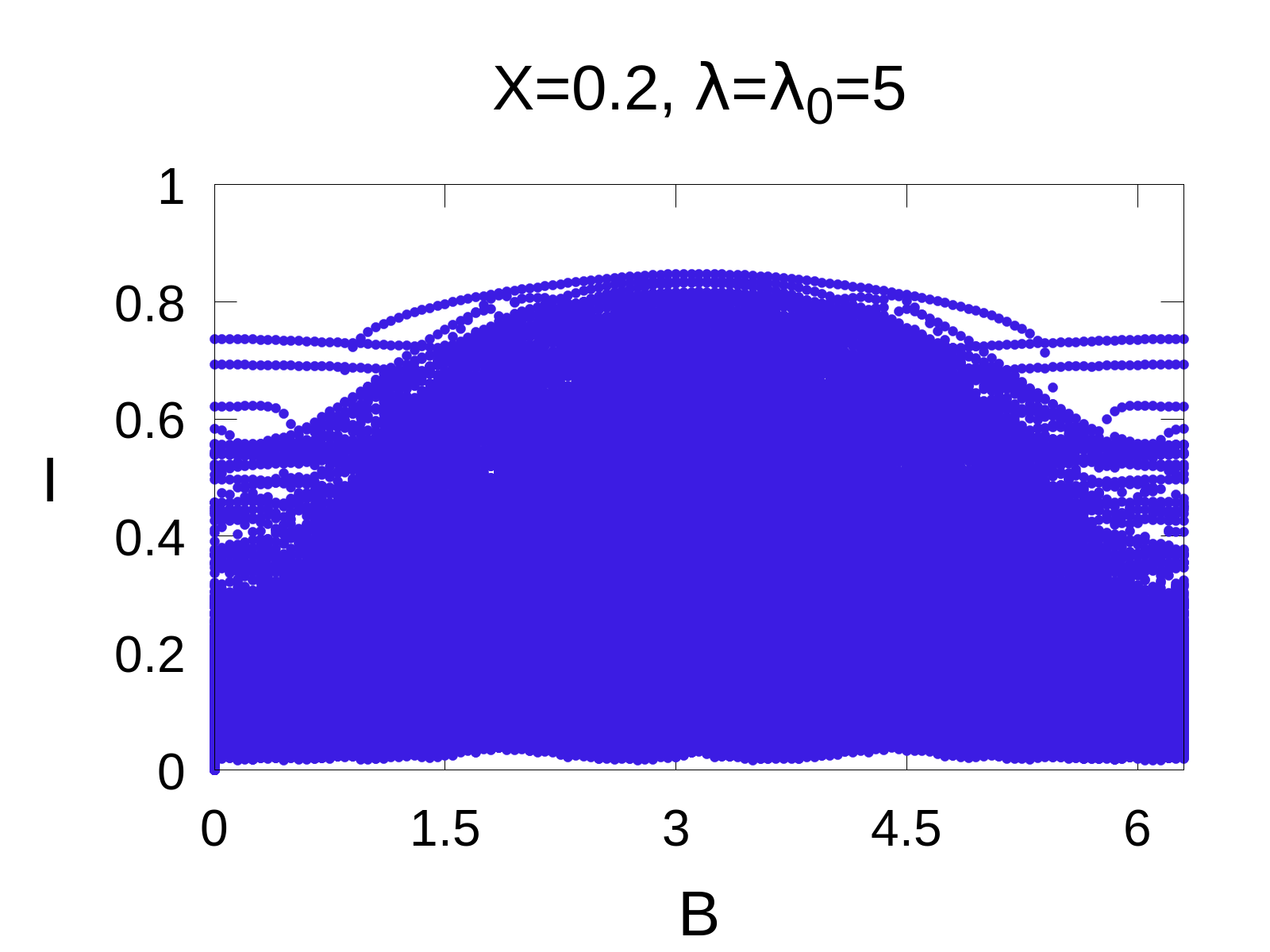}
		\subcaption{}
		\label{sq1}
	\end{subfigure}%
	\begin{subfigure}{.24\textwidth}
		\centering 
		\includegraphics[width=\linewidth]{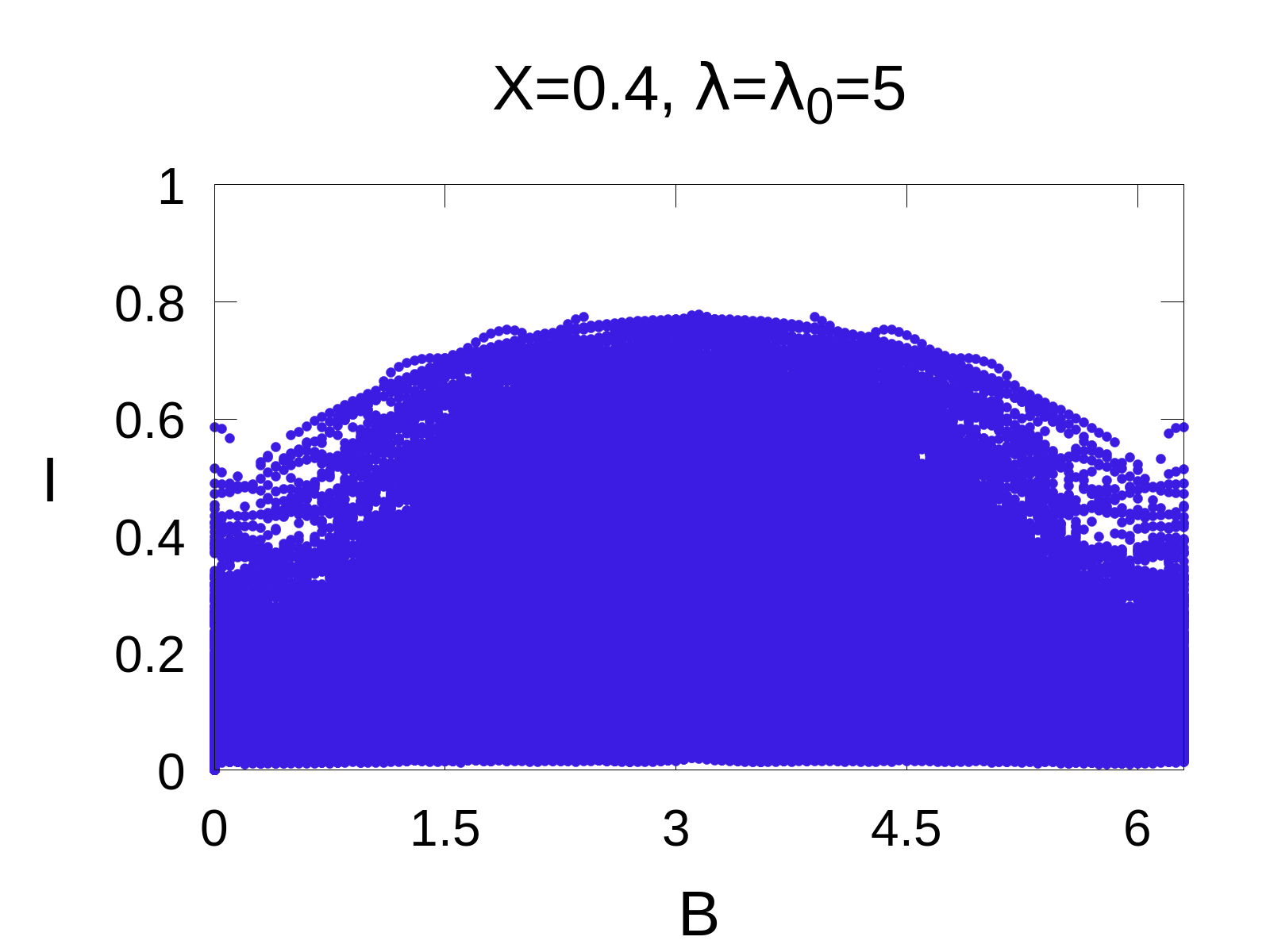}
		\subcaption{}
		\label{sq2}
	\end{subfigure}%
	\begin{subfigure}{.24\textwidth}
		\centering
		\includegraphics[width=\linewidth]{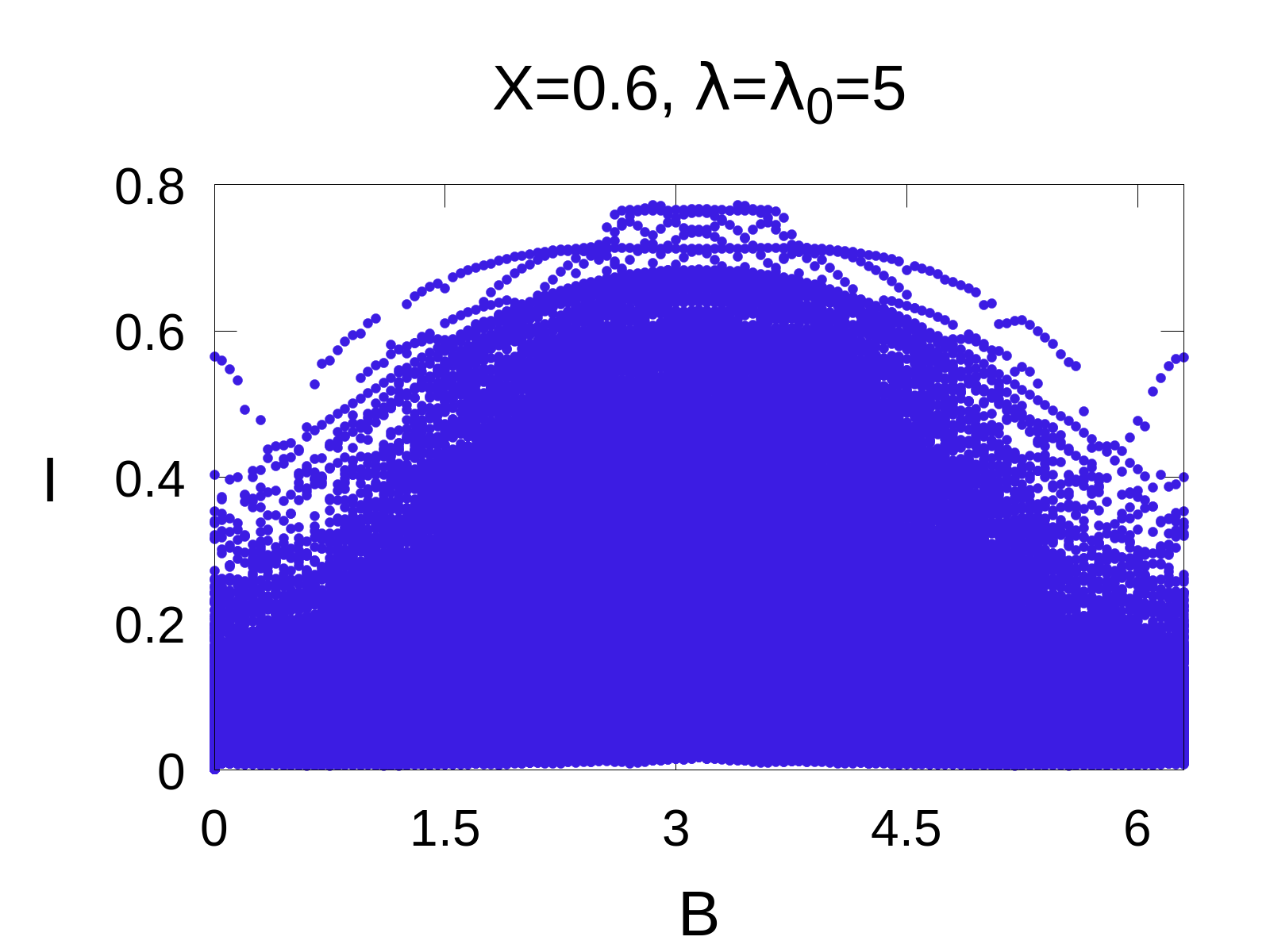}
		\subcaption{}
		\label{sq3}
	\end{subfigure}%
	\begin{subfigure}{.24\textwidth}
		\centering
		\includegraphics[width=\linewidth]{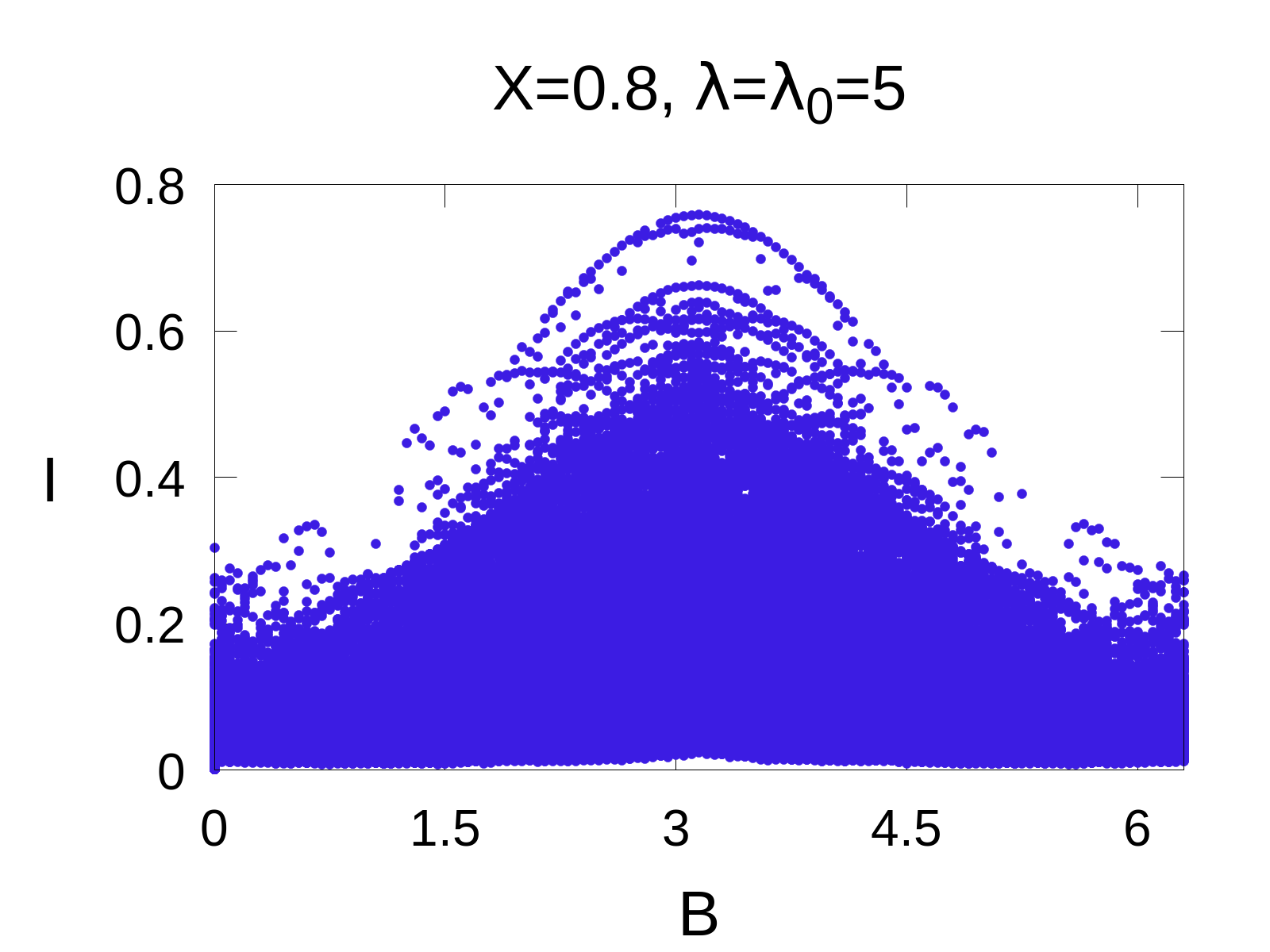}
		\subcaption{}
		\label{sq4}
	\end{subfigure}%

	
	\begin{subfigure}{.24\textwidth}
		\centering
		\includegraphics[width=\linewidth]{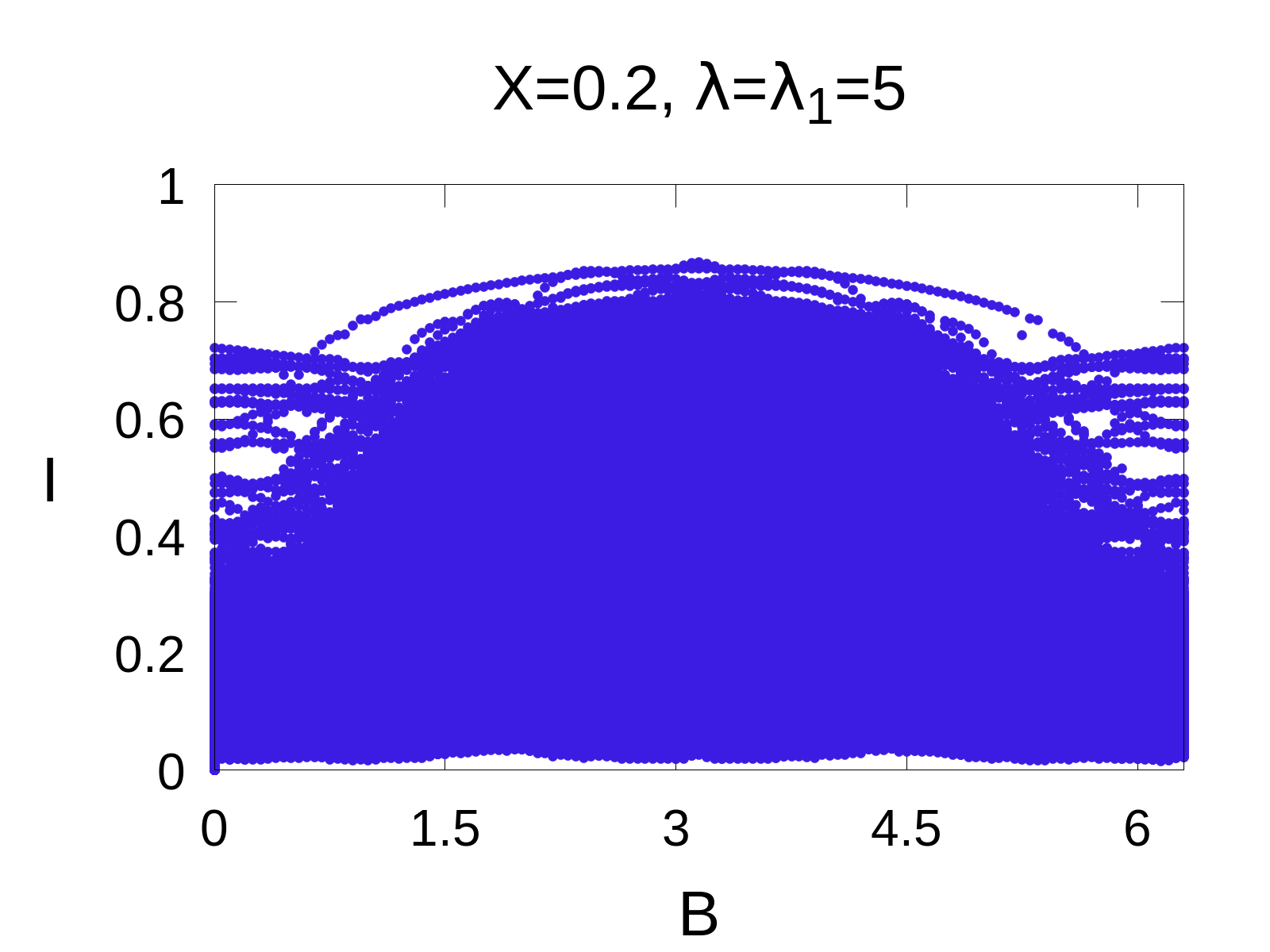}
		\subcaption{}
		\label{sq1}
	\end{subfigure}%
	\begin{subfigure}{.24\textwidth}
		\centering 
		\includegraphics[width=\linewidth]{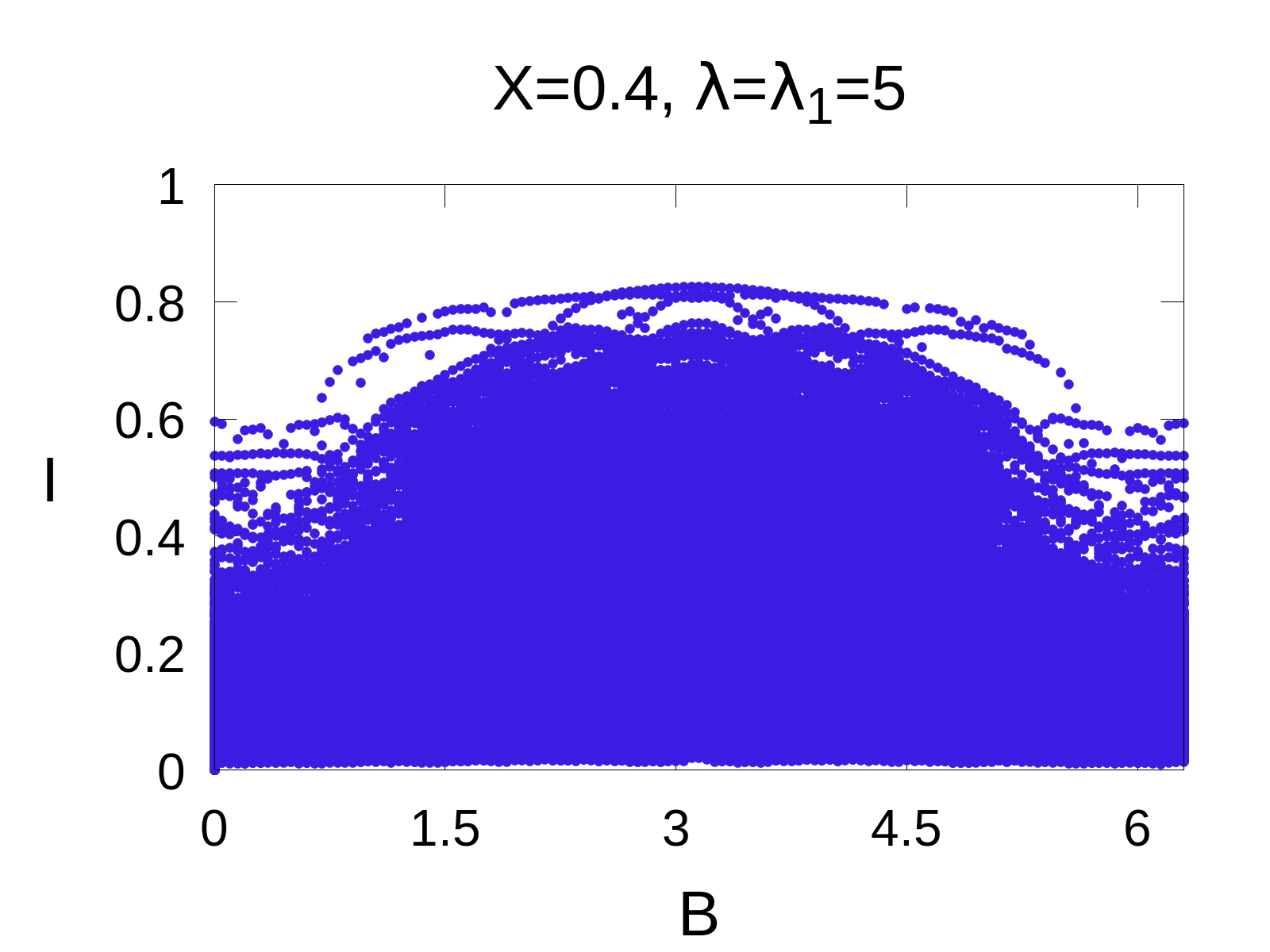}
		\subcaption{}
		\label{sq2}
	\end{subfigure}%
	\begin{subfigure}{.24\textwidth}
		\centering
		\includegraphics[width=\linewidth]{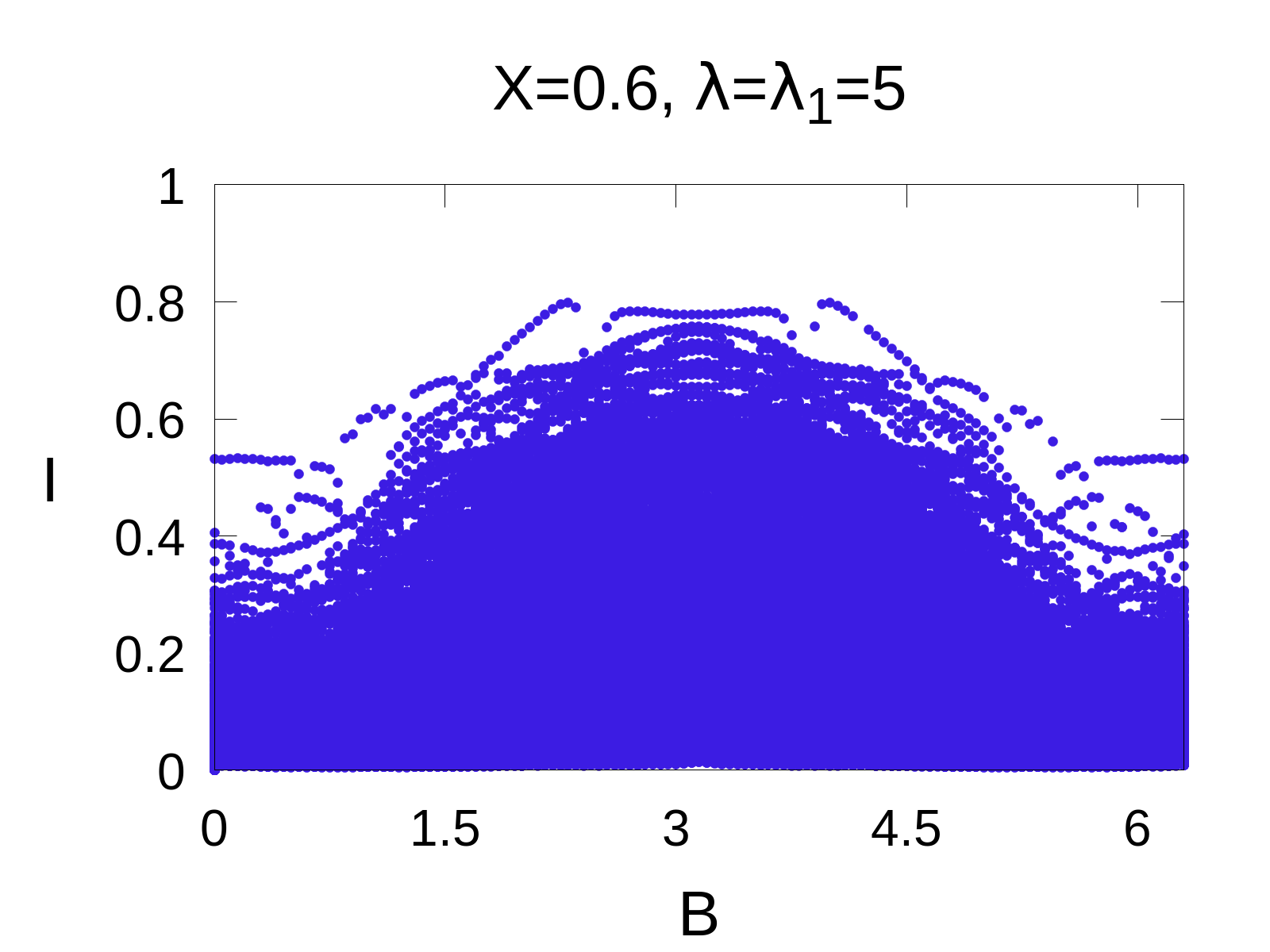}
		\subcaption{}
		\label{sq3}
	\end{subfigure}%
	\begin{subfigure}{.24\textwidth}
		\centering
		\includegraphics[width=\linewidth]{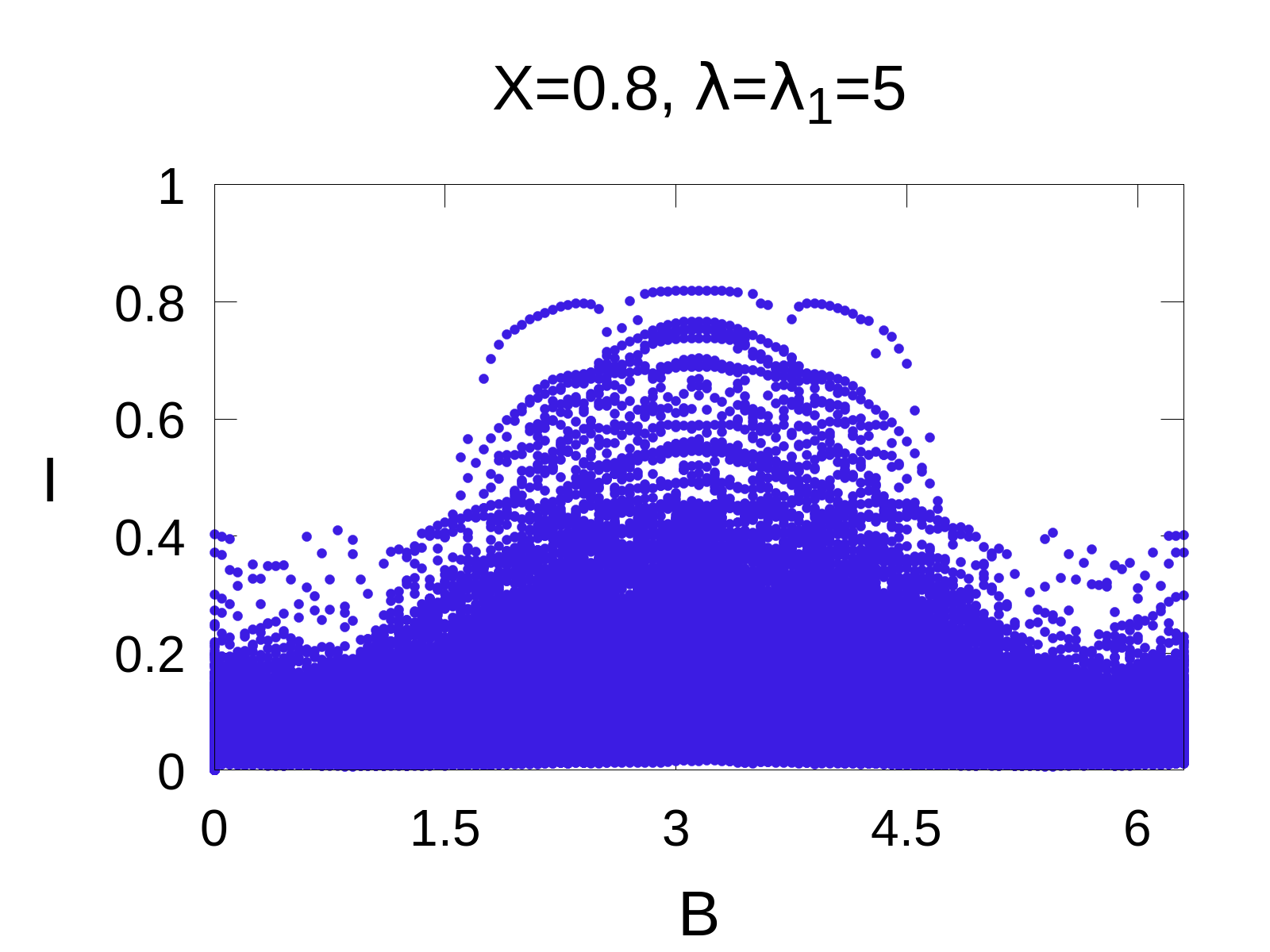}
		\subcaption{}
		\label{sq4}
	\end{subfigure}%


	\begin{subfigure}{.24\textwidth}
		\centering
		\includegraphics[width=\linewidth]{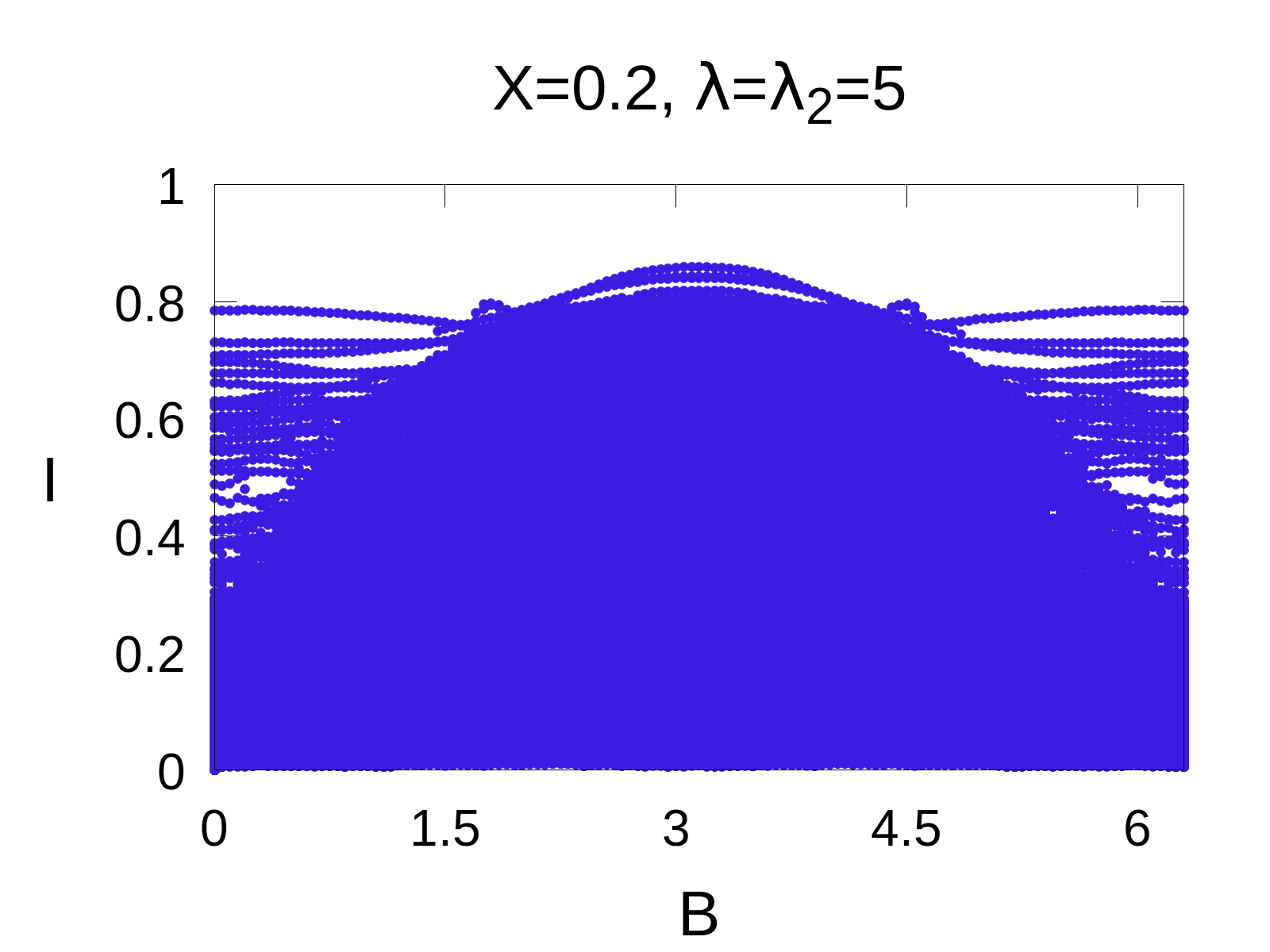}
		\subcaption{}
		\label{sq1}
	\end{subfigure}%
	\begin{subfigure}{.24\textwidth}
		\centering 
		\includegraphics[width=\linewidth]{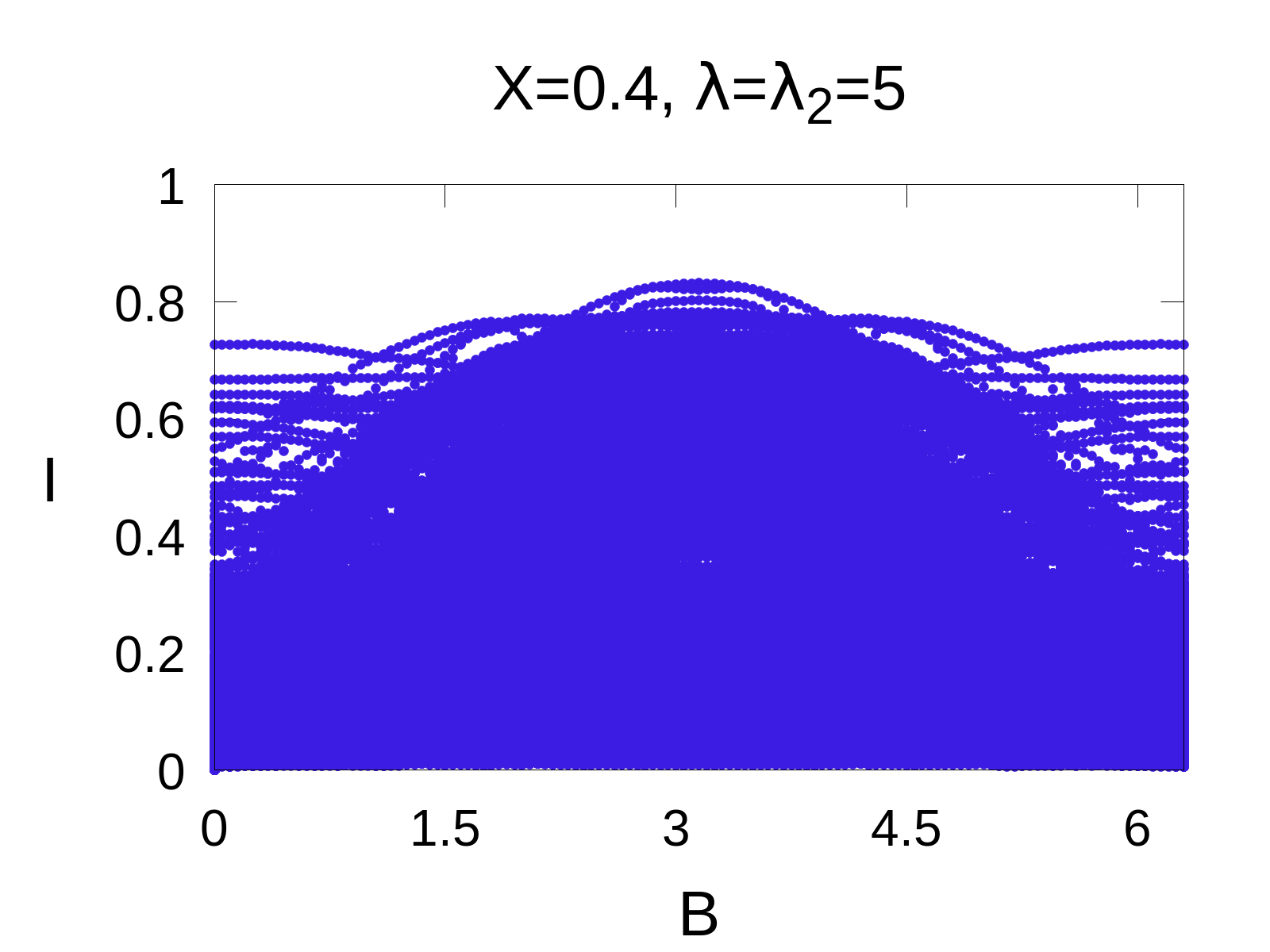}
		\subcaption{}
		\label{sq2}
	\end{subfigure}%
	\begin{subfigure}{.24\textwidth}
		\centering
		\includegraphics[width=\linewidth]{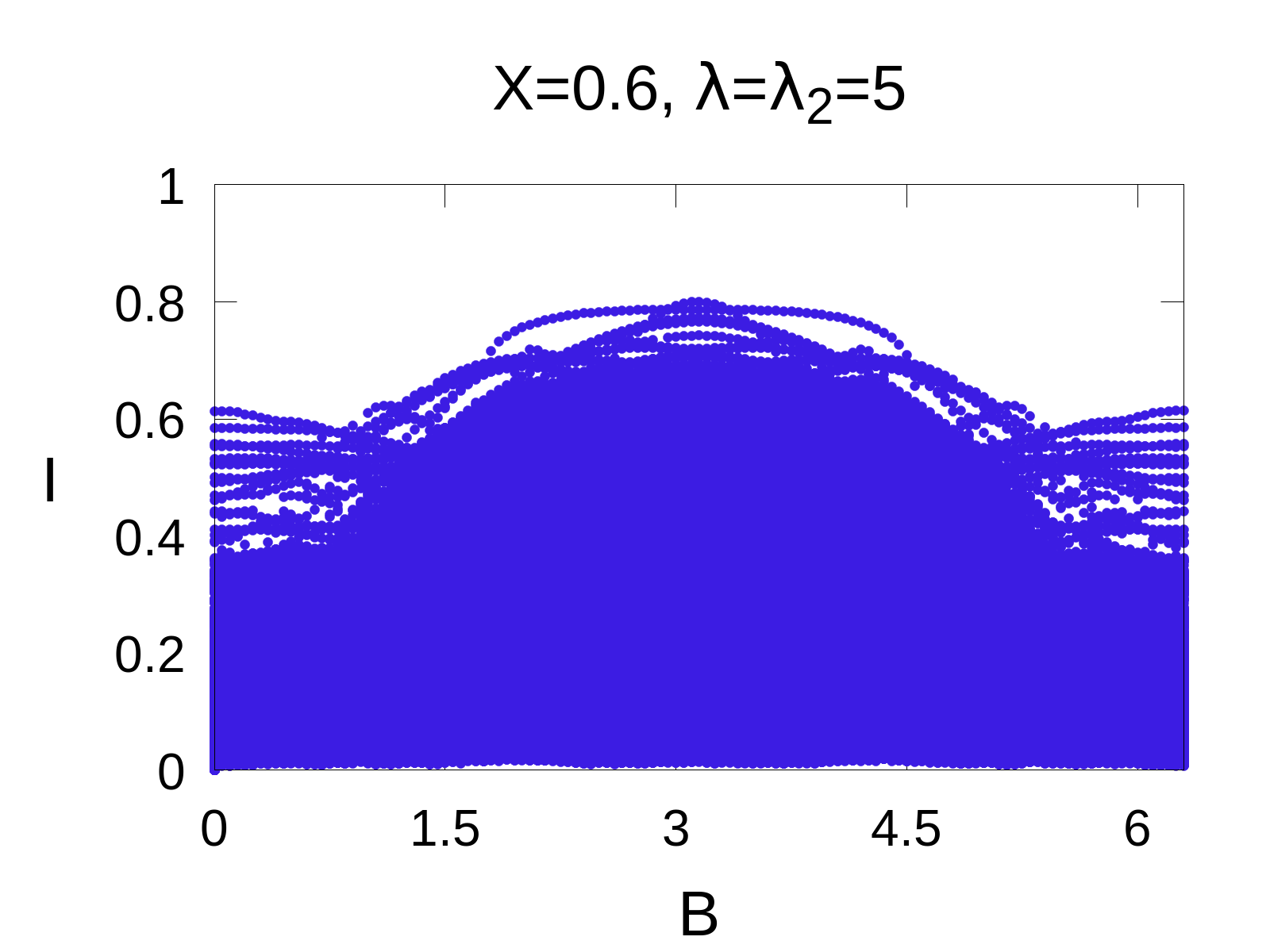}
		\subcaption{}
		\label{sq3}
	\end{subfigure}%
	\begin{subfigure}{.24\textwidth}
		\centering
		\includegraphics[width=\linewidth]{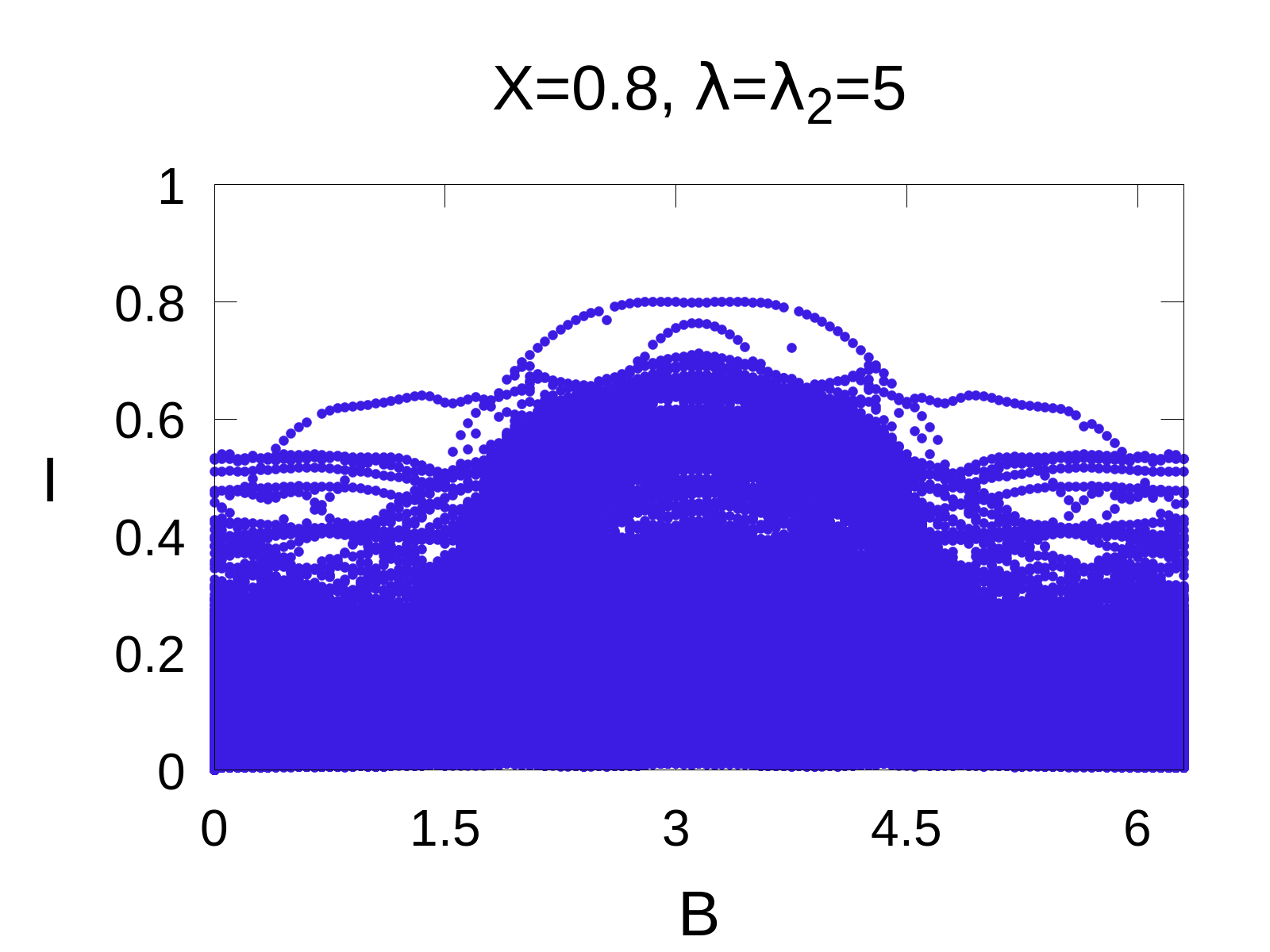}
		\subcaption{}
		\label{sq4}
	\end{subfigure}%

	\caption{IPR $I$ as a function of the magnetic field $B$, illustrating the combined effects of the Aubry--André potential with different quasiperiodic disorders. The first row shows the interpolation between AA and Thue--Morse, the middle row corresponds to AA and Rudin--Shapiro, and the last row displays the AA–Fibonacci interpolation. The interpolation parameter $X$ varies from $X = 0.2$, where the AA component is dominant, to $X = 0.8$, where the corresponding quasiperiodic disorder becomes dominant.}
	\label{ATIP}
\end{figure*}

\begin{figure*}[]
	\centering
	\begin{subfigure}{.24\textwidth}
		\centering
		\includegraphics[width=\linewidth]{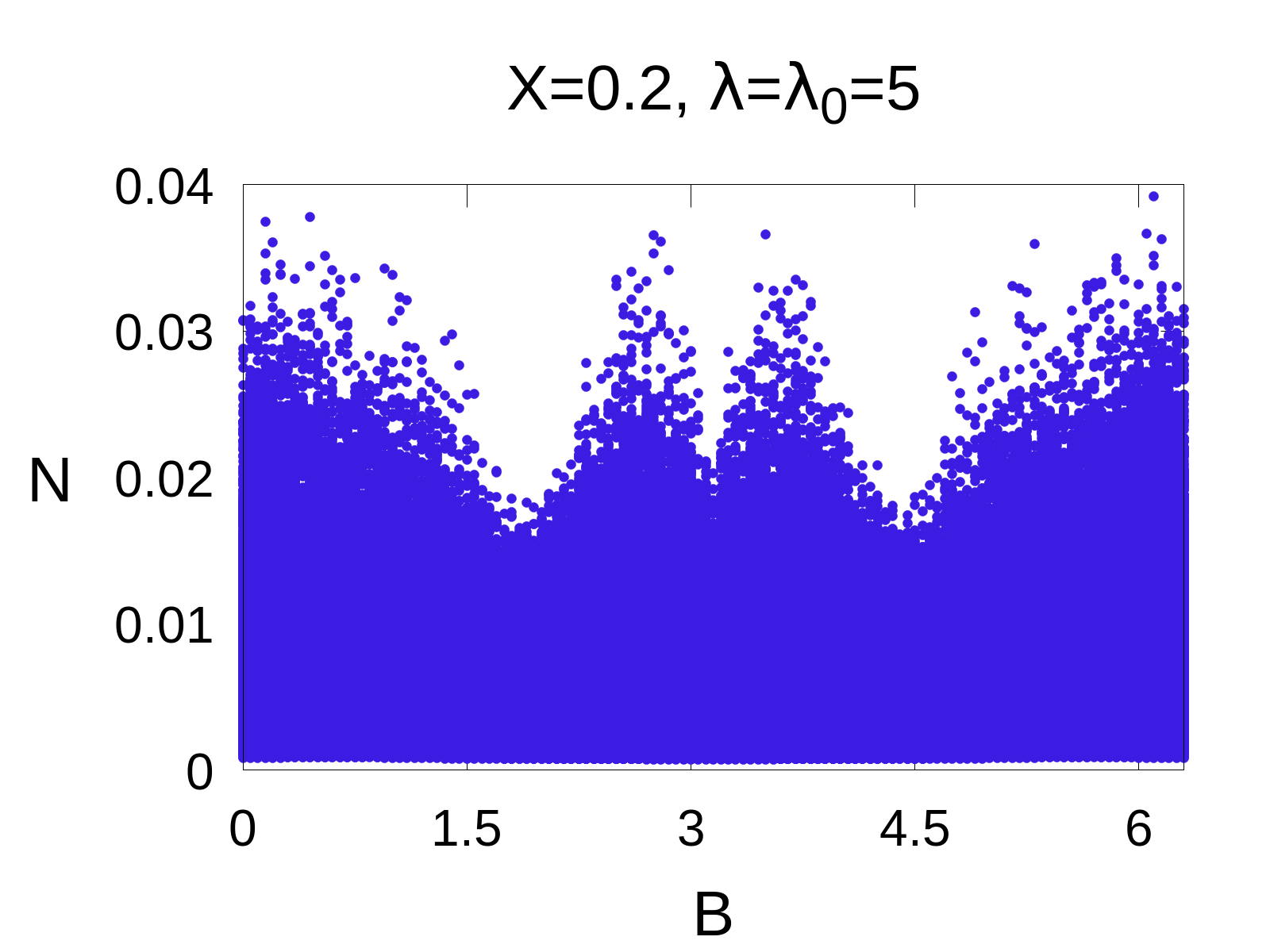}
		\subcaption{}
		\label{sq1}
	\end{subfigure}%
	\begin{subfigure}{.24\textwidth}
		\centering 
		\includegraphics[width=\linewidth]{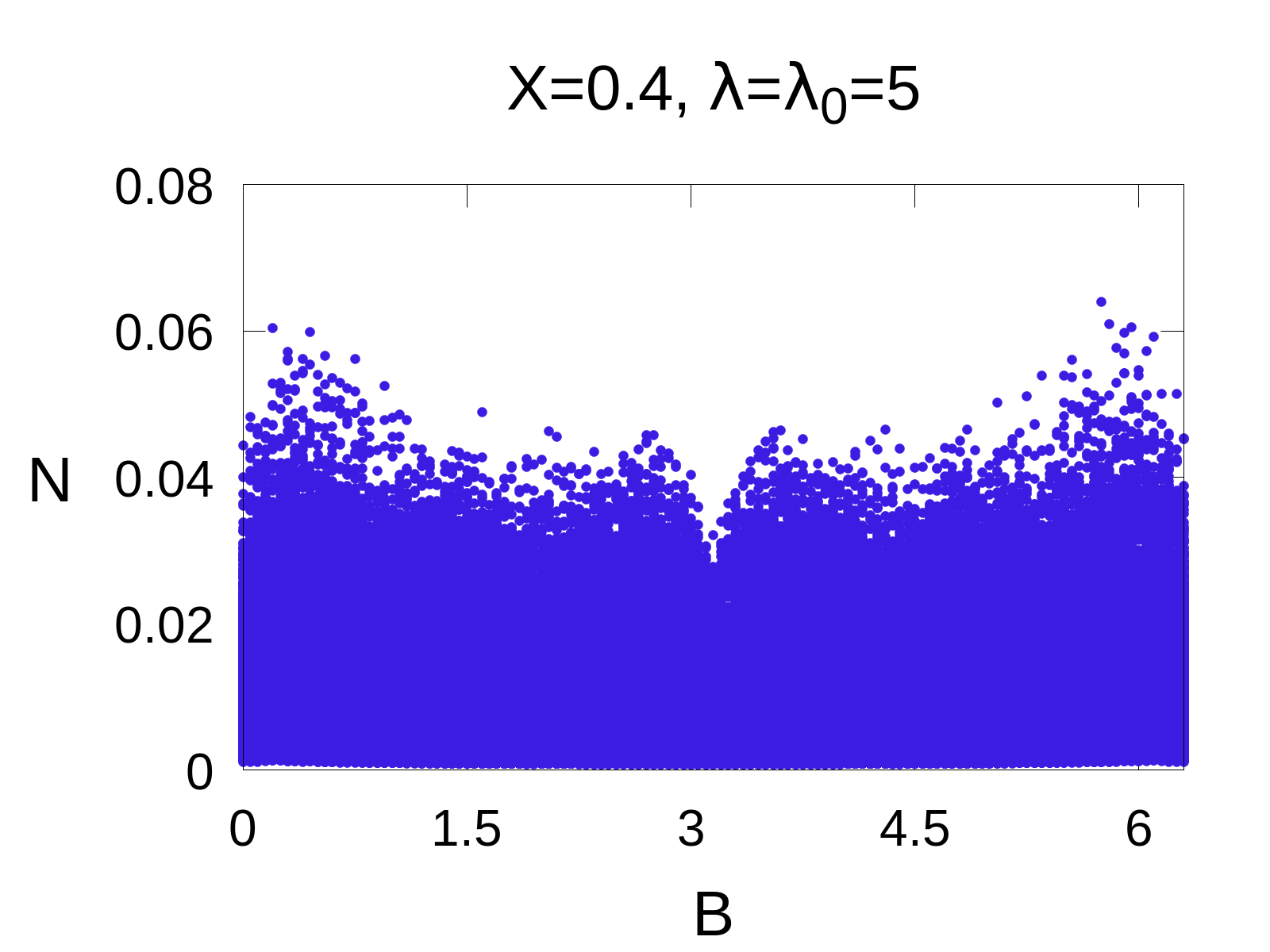}
		\subcaption{}
		\label{sq2}
	\end{subfigure}%
	\begin{subfigure}{.24\textwidth}
		\centering
		\includegraphics[width=\linewidth]{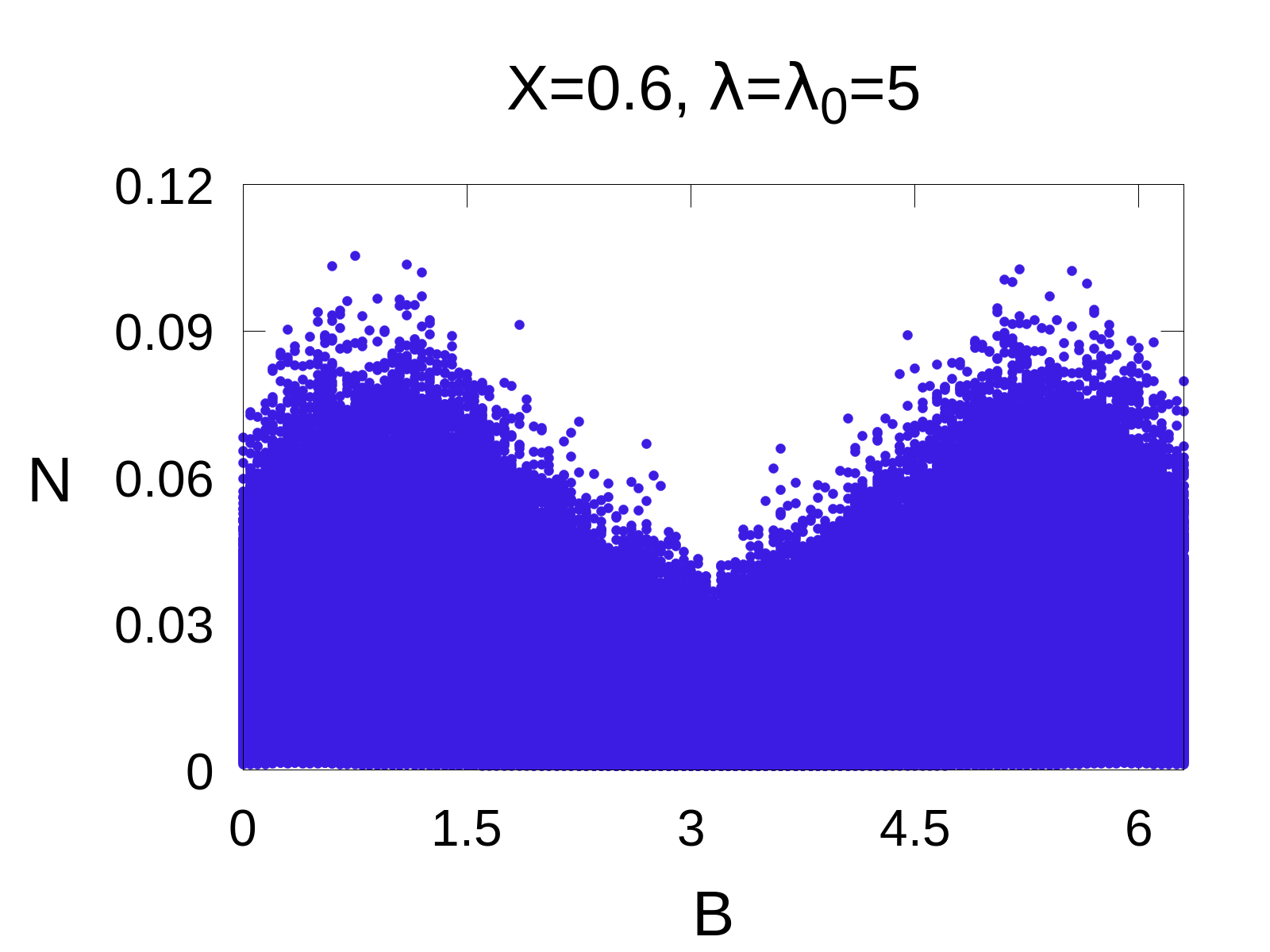}
		\subcaption{}
		\label{sq3}
	\end{subfigure}%
	\begin{subfigure}{.24\textwidth}
		\centering
		\includegraphics[width=\linewidth]{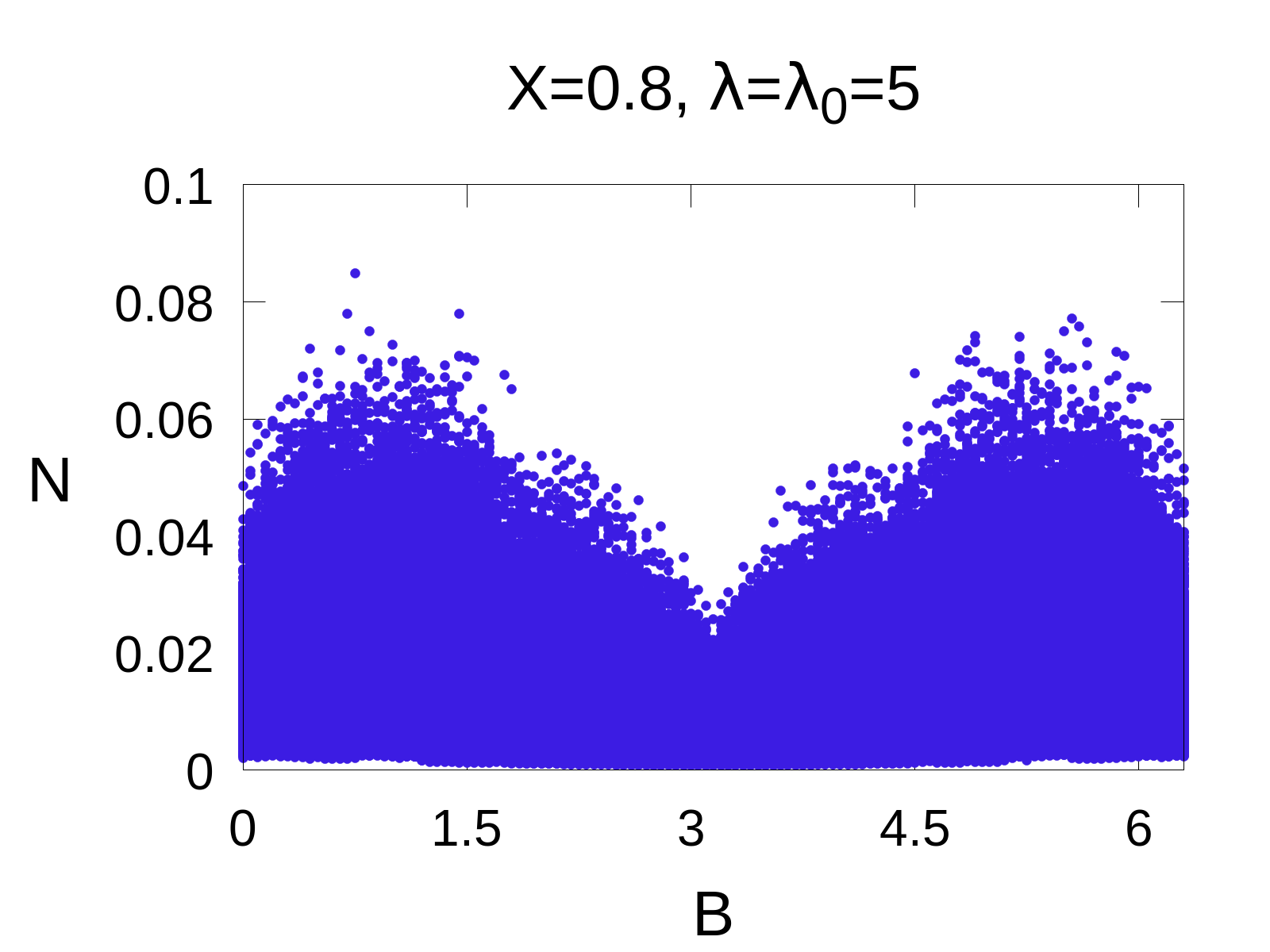}
		\subcaption{}
		\label{sq4}
	\end{subfigure}%

	\begin{subfigure}{.24\textwidth}
		\centering
		\includegraphics[width=\linewidth]{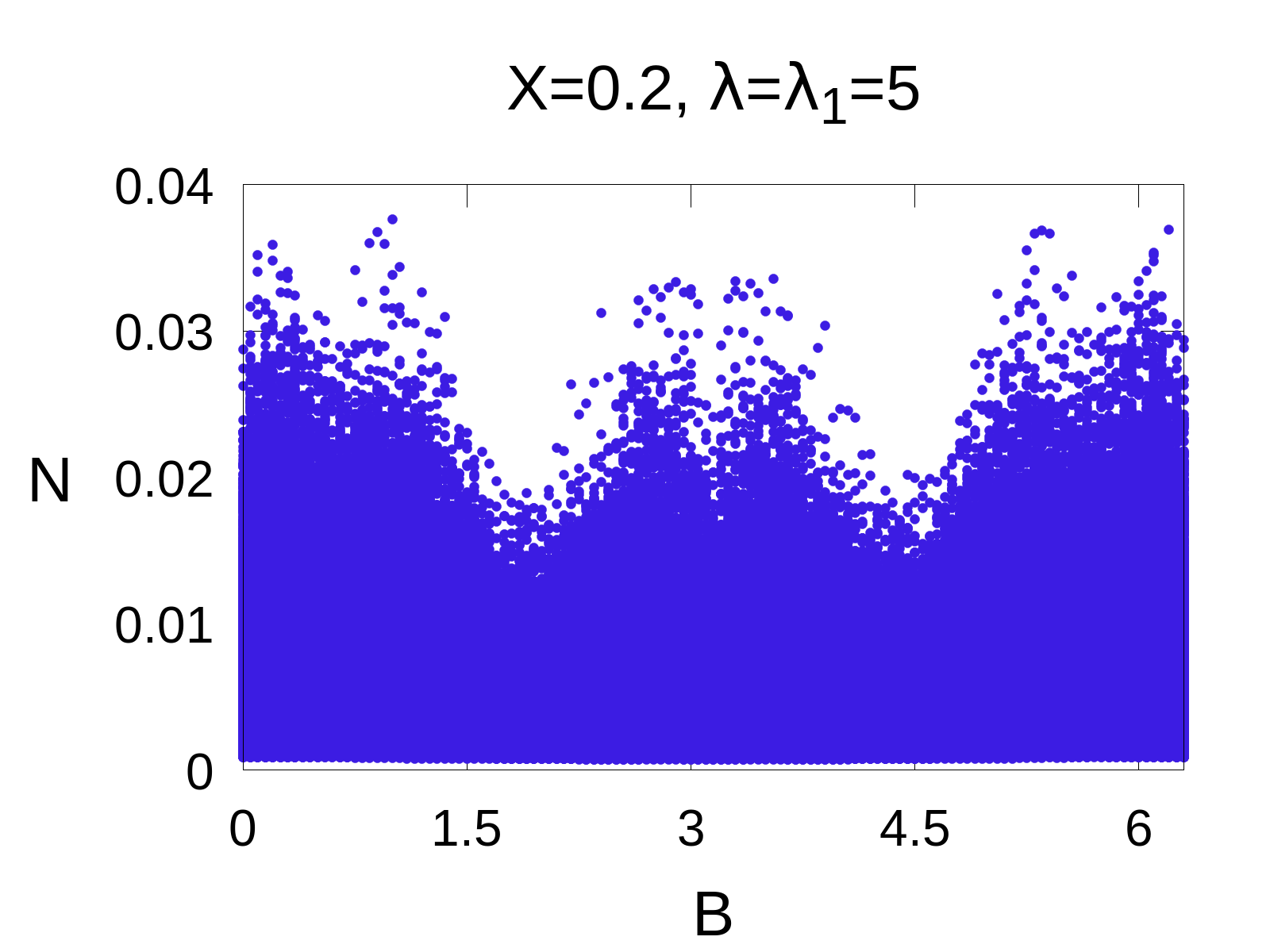}
		\subcaption{}
		\label{sq1}
	\end{subfigure}%
	\begin{subfigure}{.24\textwidth}
		\centering 
		\includegraphics[width=\linewidth]{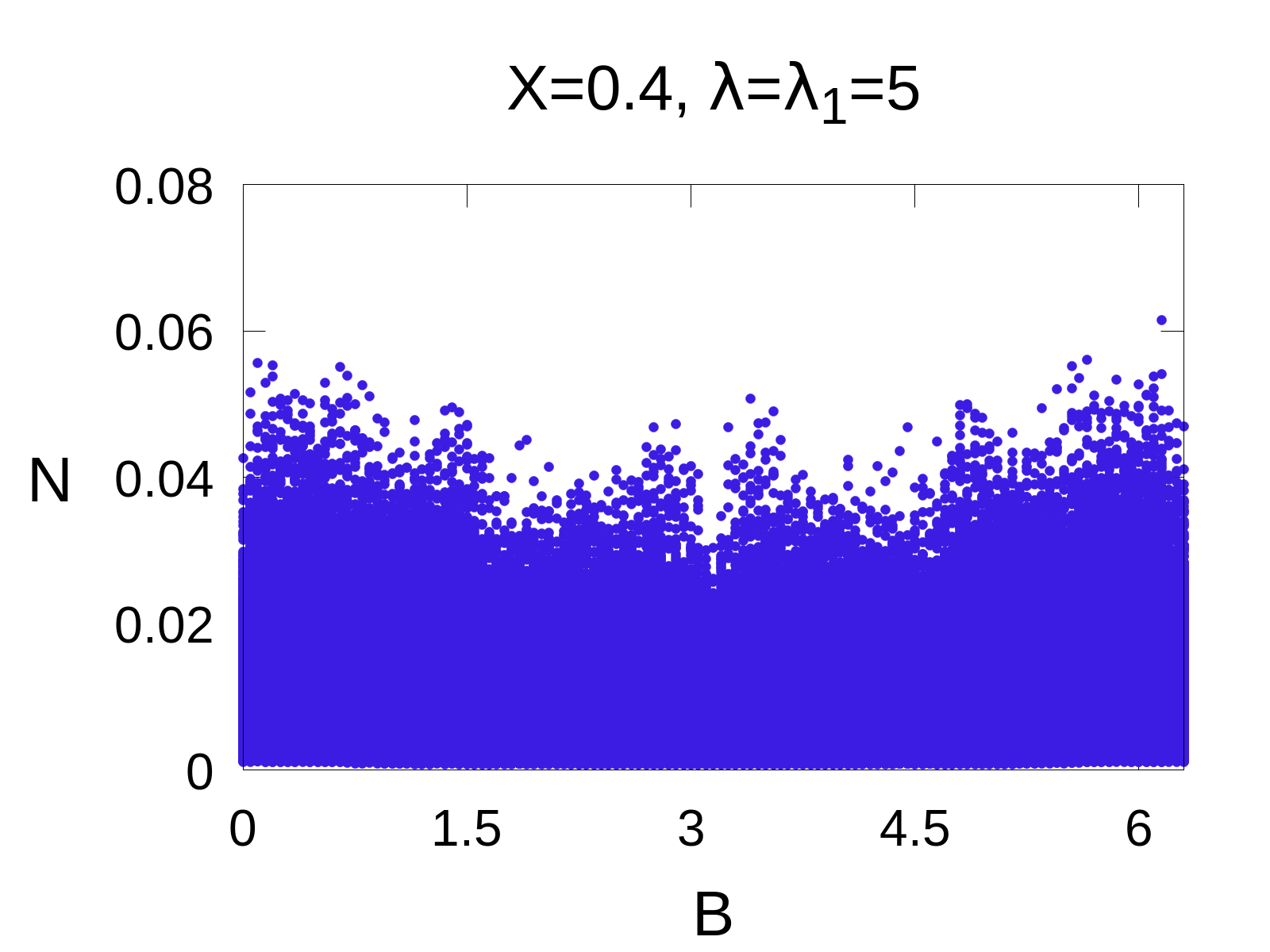}
		\subcaption{}
		\label{sq2}
	\end{subfigure}%
	\begin{subfigure}{.24\textwidth}
		\centering
		\includegraphics[width=\linewidth]{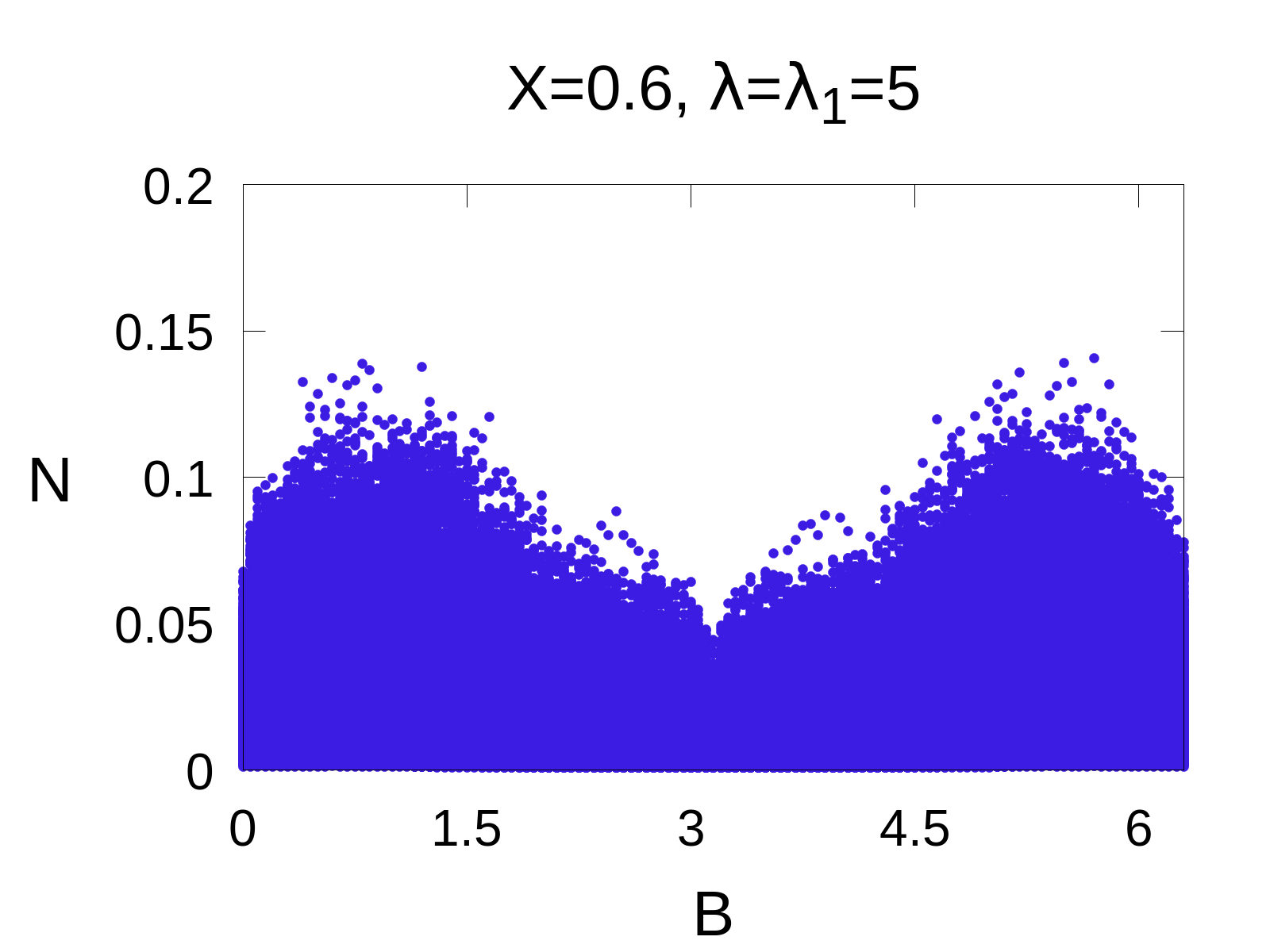}
		\subcaption{}
		\label{sq3}
	\end{subfigure}%
	\begin{subfigure}{.24\textwidth}
		\centering
		\includegraphics[width=\linewidth]{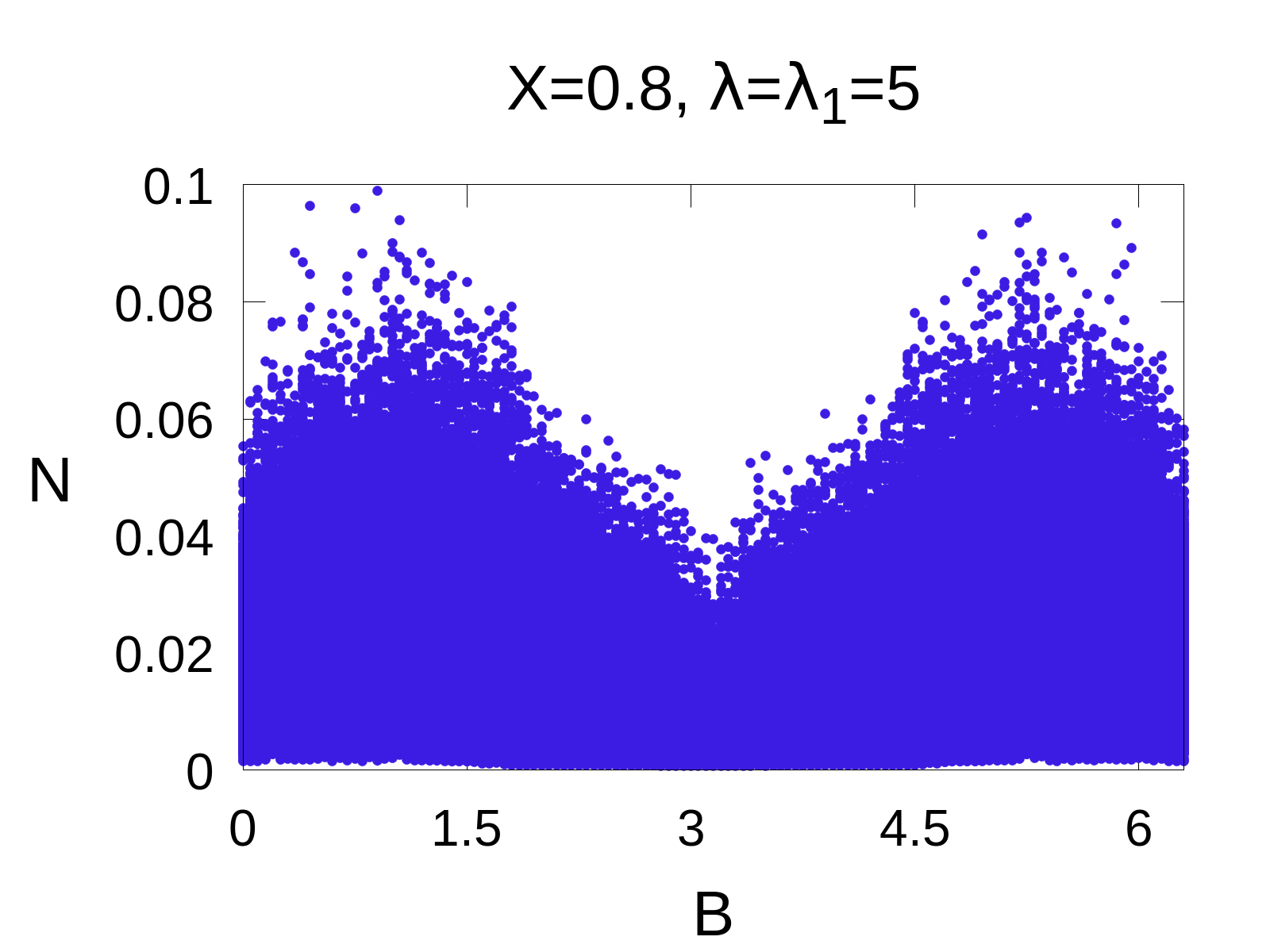}
		\subcaption{}
		\label{sq4}
	\end{subfigure}%
	
	
	\begin{subfigure}{.24\textwidth}
		\centering
		\includegraphics[width=\linewidth]{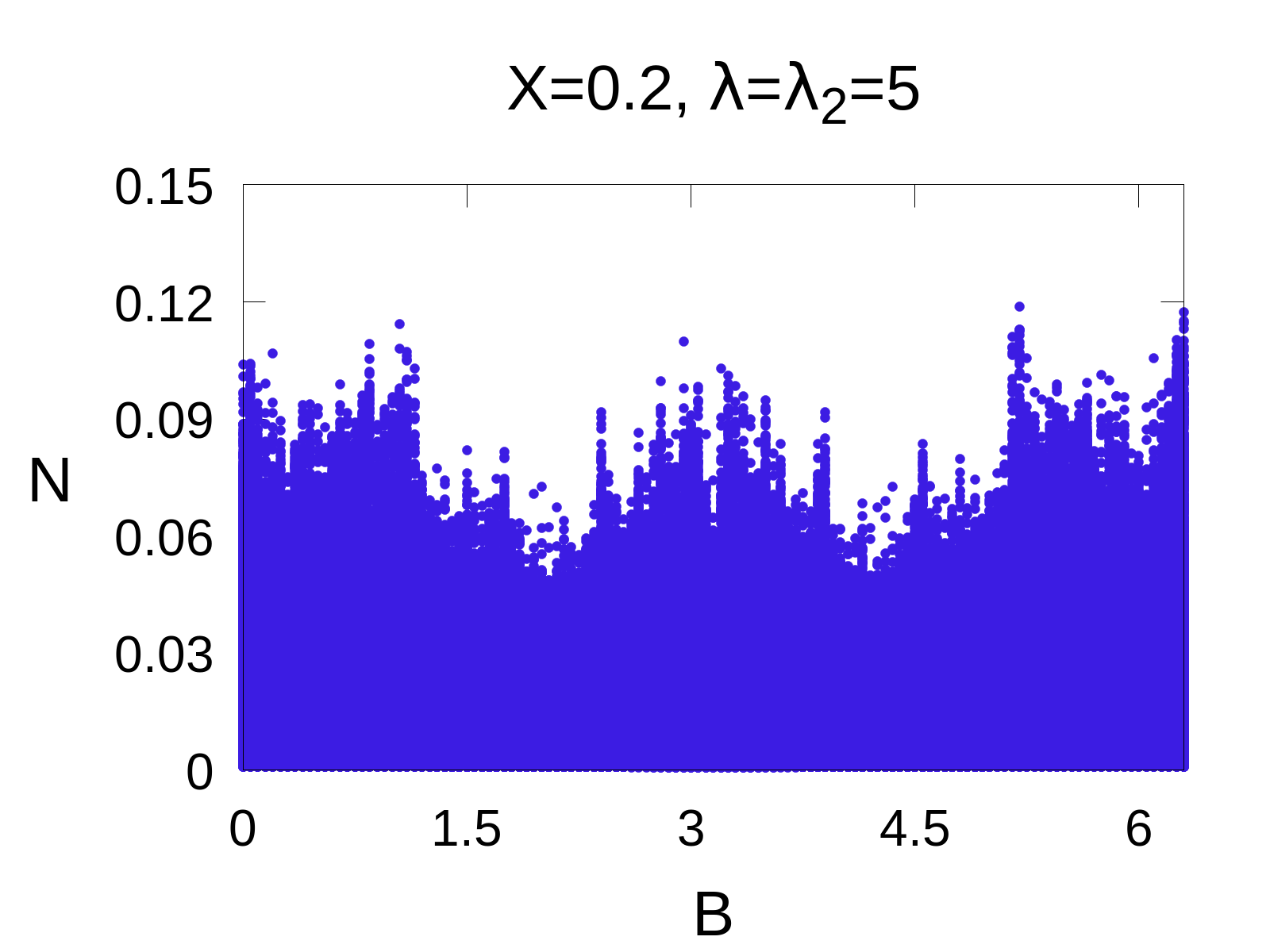}
		\subcaption{}
		\label{sq1}
	\end{subfigure}%
	\begin{subfigure}{.24\textwidth}
		\centering 
		\includegraphics[width=\linewidth]{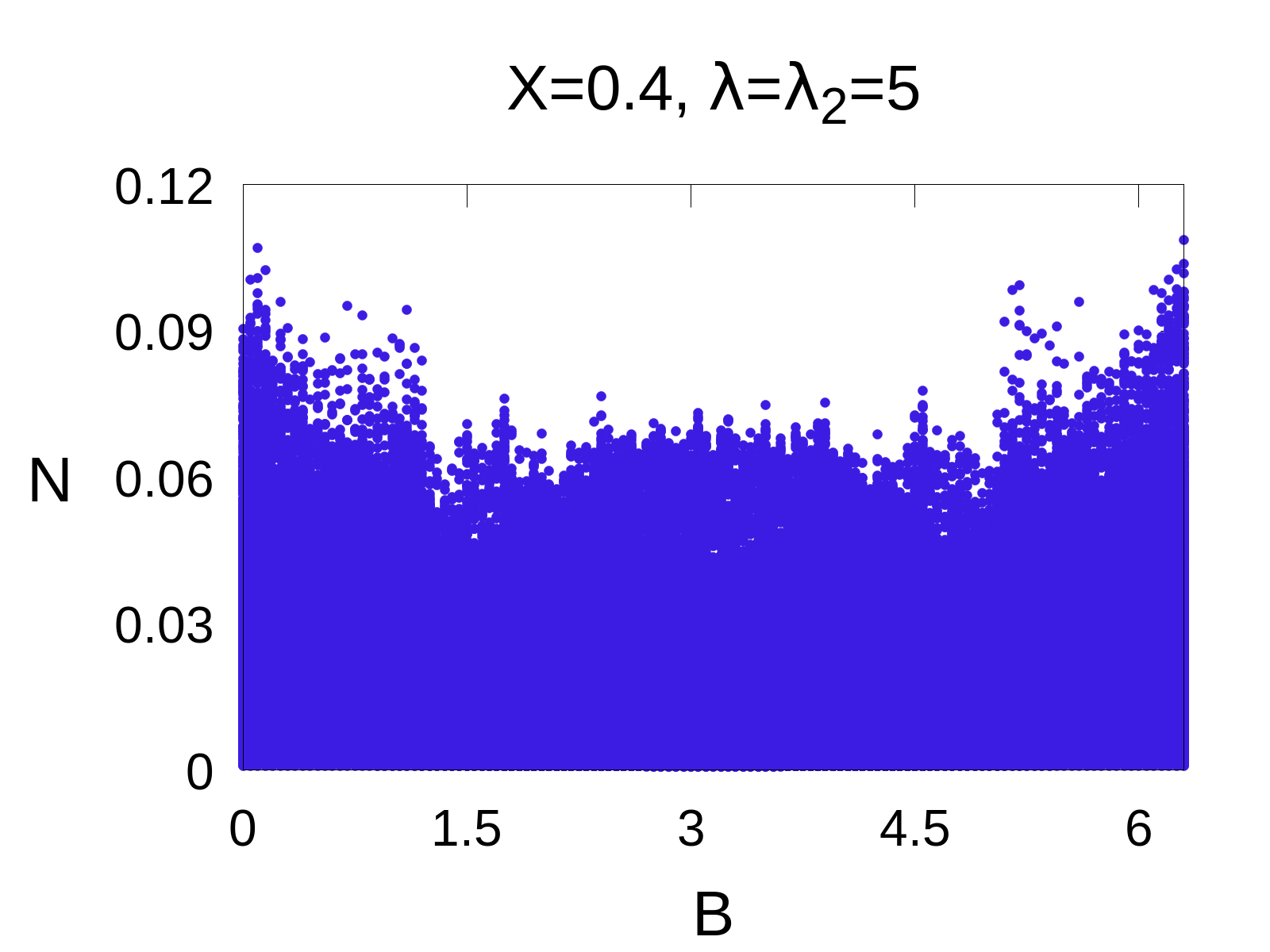}
		\subcaption{}
		\label{sq2}
	\end{subfigure}%
	\begin{subfigure}{.24\textwidth}
		\centering
		\includegraphics[width=\linewidth]{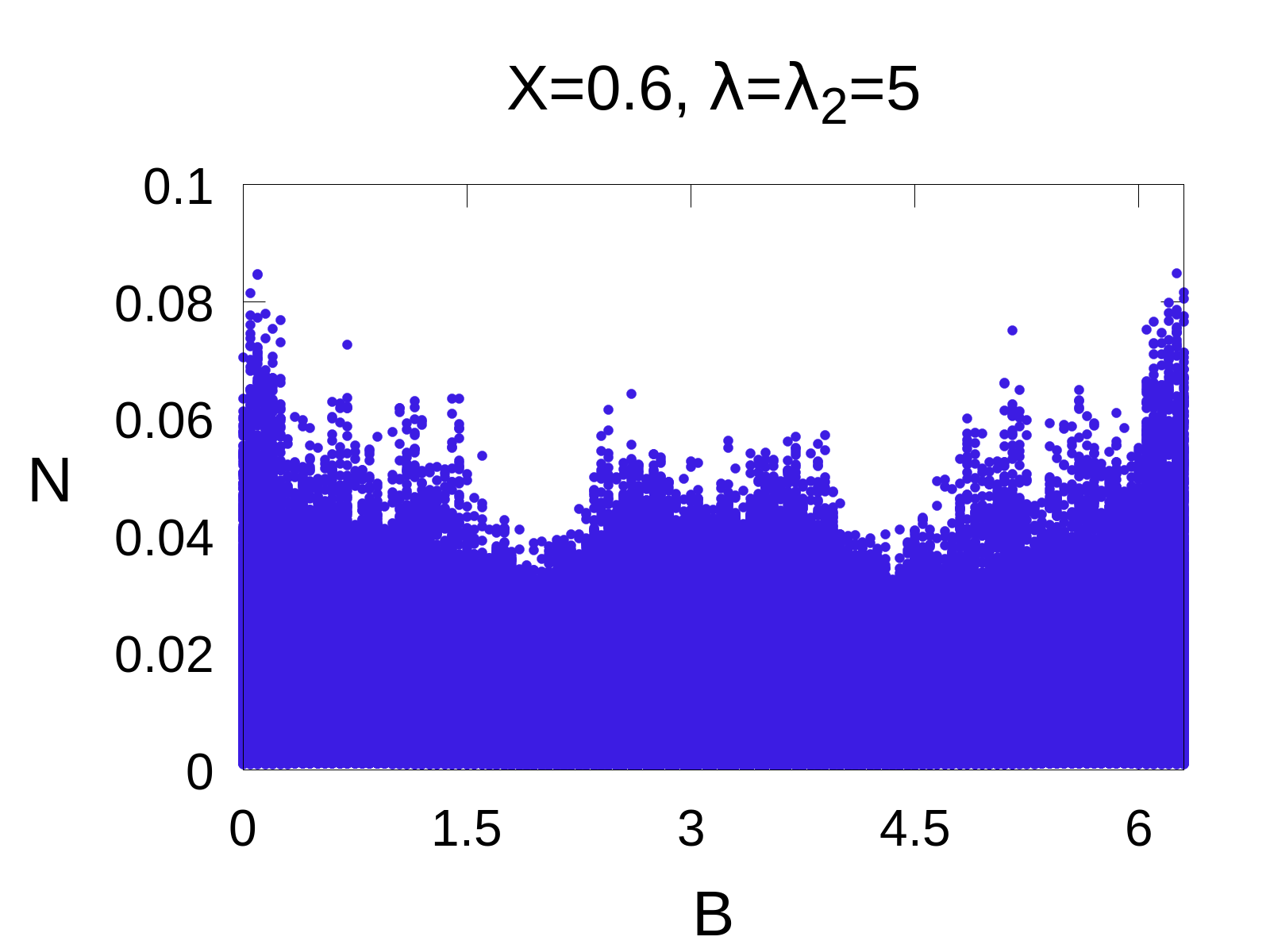}
		\subcaption{}
		\label{sq3}
	\end{subfigure}%
	\begin{subfigure}{.24\textwidth}
		\centering
		\includegraphics[width=\linewidth]{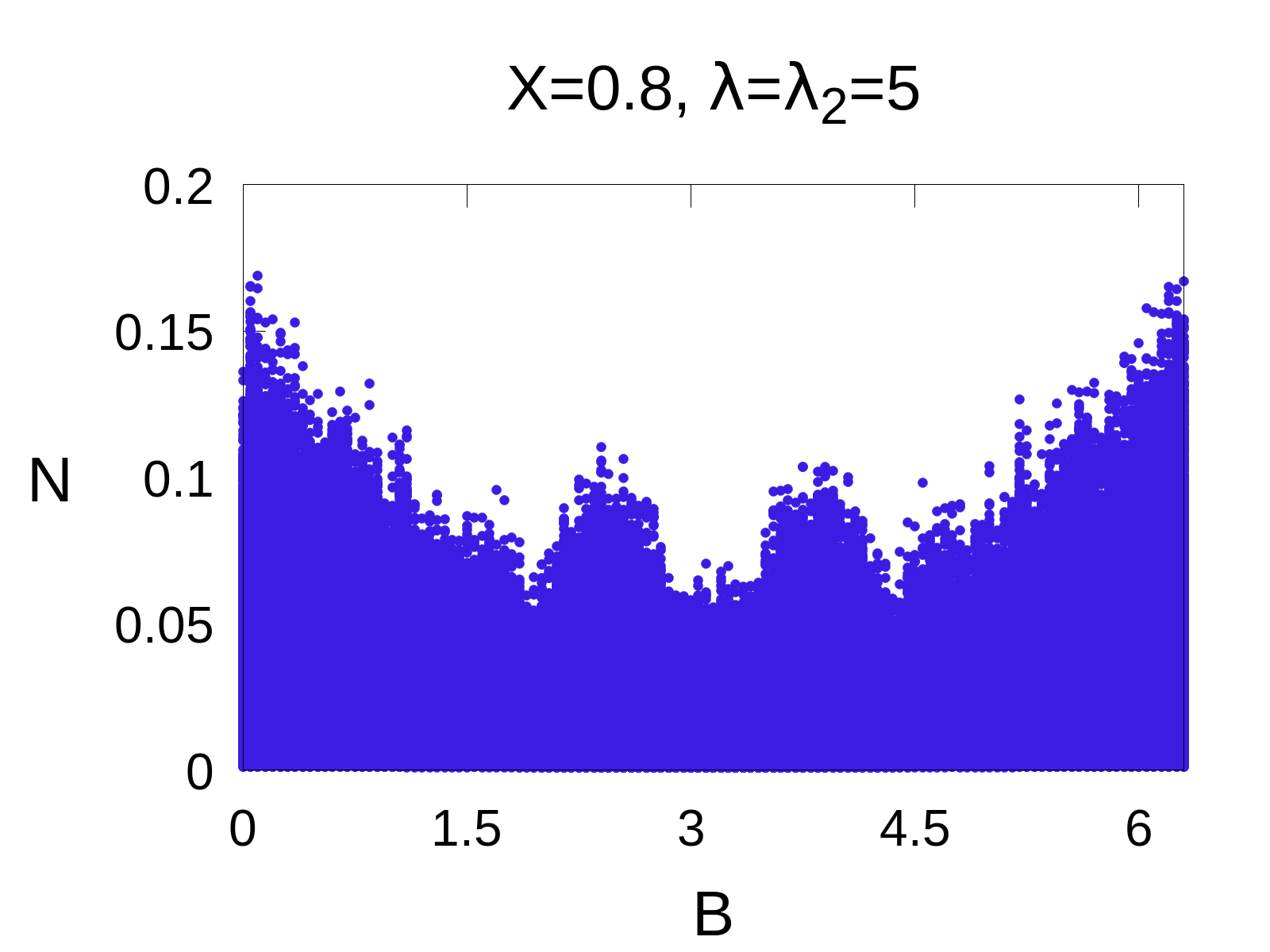}
		\subcaption{}
		\label{sq4}
	\end{subfigure}%

	\caption{NPR $N$ as a function of the magnetic field $B$, illustrating the combined effects of the Aubry--André potential with different quasiperiodic disorders. The first row shows the interpolation between AA and Thue--Morse, the middle row corresponds to AA and Rudin--Shapiro, and the last row displays the AA–Fibonacci interpolation.}
	\label{ATNP}
\end{figure*}

Figs.~\ref{IPR4} and \ref{NPR40} display the IPR and NPR as functions of the 
magnetic field for all the quasiperiodic disorders discussed above, shown for 
two disorder strengths, $d = 2$ and $d = 5$, respectively. Consistent with the behavior observed in the average IPR, the AA disorder produces the largest IPR values, indicating the strongest localization. This effect is particularly pronounced near $B = \pi$. In contrast, the corresponding NPR results shown in Fig.~\ref{NPR40} exhibit the opposite trend, where lower NPR values signal enhanced localization. Among all the 
disorders, the RS binary potential yields the highest NPR values, suggesting that 
its spectrum contains a mixture of localized and extended states. Figs.~\ref{ATIP} and \ref{ATNP} show the IPR and NPR as functions of the magnetic field $B$ for the interpolation between the AA potential and other quasiperiodic disorders. When the AA component dominates, the plots exhibit a pronounced blue region, indicating larger IPR values and therefore stronger localization, particularly around $B = \pi$.


\clearpage
\bibliographystyle{unsrt}

\end{document}